\documentclass[english,12pt,a4paper]{book}
\usepackage[utf8]{inputenc}
\usepackage[T1]{fontenc}
\usepackage[english]{babel}
\usepackage[default,oldstyle, scale=.95]{opensans} 
\usepackage{amsmath}
\usepackage{amsfonts}
\usepackage{amssymb}
\usepackage{caption}
\usepackage{xcolor} 
\definecolor{Prune}{RGB}{99,0,60} 
\definecolor{B1}{RGB}{49,62,72} 
\definecolor{C1}{RGB}{124,135,143}
\definecolor{D1}{RGB}{213,218,223}
\definecolor{A2}{RGB}{198,11,70}
\definecolor{B2}{RGB}{237,20,91}
\definecolor{C2}{RGB}{238,52,35}
\definecolor{D2}{RGB}{243,115,32}
\definecolor{A3}{RGB}{124,42,144}
\definecolor{B3}{RGB}{125,106,175}
\definecolor{C3}{RGB}{198,103,29}
\definecolor{D3}{RGB}{254,188,24}
\definecolor{A4}{RGB}{0,78,125}
\definecolor{B4}{RGB}{14,135,201}
\definecolor{C4}{RGB}{0,148,181}
\definecolor{D4}{RGB}{70,195,210}
\definecolor{A5}{RGB}{0,128,122}
\definecolor{B5}{RGB}{64,183,105}
\definecolor{C5}{RGB}{140,198,62}
\definecolor{D5}{RGB}{213,223,61}

\definecolor{blue}{rgb}{0,0,0} 

\usepackage{mdframed}
\usepackage{multirow} 
\usepackage{multicol} 
\usepackage{tikz}
\usepackage{graphicx}
\usepackage[absolute]{textpos} 
\usepackage{colortbl}
\usepackage{array}
\usepackage{geometry}
\usepackage{titlesec}
\usepackage{hyperref}

\usepackage{mathtools}
\usepackage{physics}
\usepackage{xspace}
\usepackage{mwe}
\usepackage{dsfont}

\usetikzlibrary{positioning}
\usetikzlibrary{decorations.pathmorphing, decorations.shapes, fadings}
\usetikzlibrary{arrows.meta, shapes.symbols}

\hypersetup{ 
    colorlinks=true,
    linkcolor=black,
    urlcolor=A5}

\newcommand{\RDBM}{Resetting Dyson Brownian motion\xspace}
\newcommand{\DBM}{Dyson Brownian motion\xspace}
\newcommand{\OU}{Ornstein-Uhlenbeck\xspace}
\newcommand{\VBM}{Vicious Brownian motion\xspace}
\newcommand{\NESS}{non-equilibrium steady state\xspace}
\newcommand{\RMT}{random matrix theory\xspace}
\newcommand{\EVS}{extreme value statistics\xspace}
\newcommand{\MFPT}{mean first passage time\xspace}

\begin{document}

\begin{titlepage}

\newgeometry{left=6cm,bottom=2cm, top=1cm, right=1cm}

\tikz[remember picture,overlay] \node[opacity=1,inner sep=0pt] at (-13mm,-135mm){\includegraphics{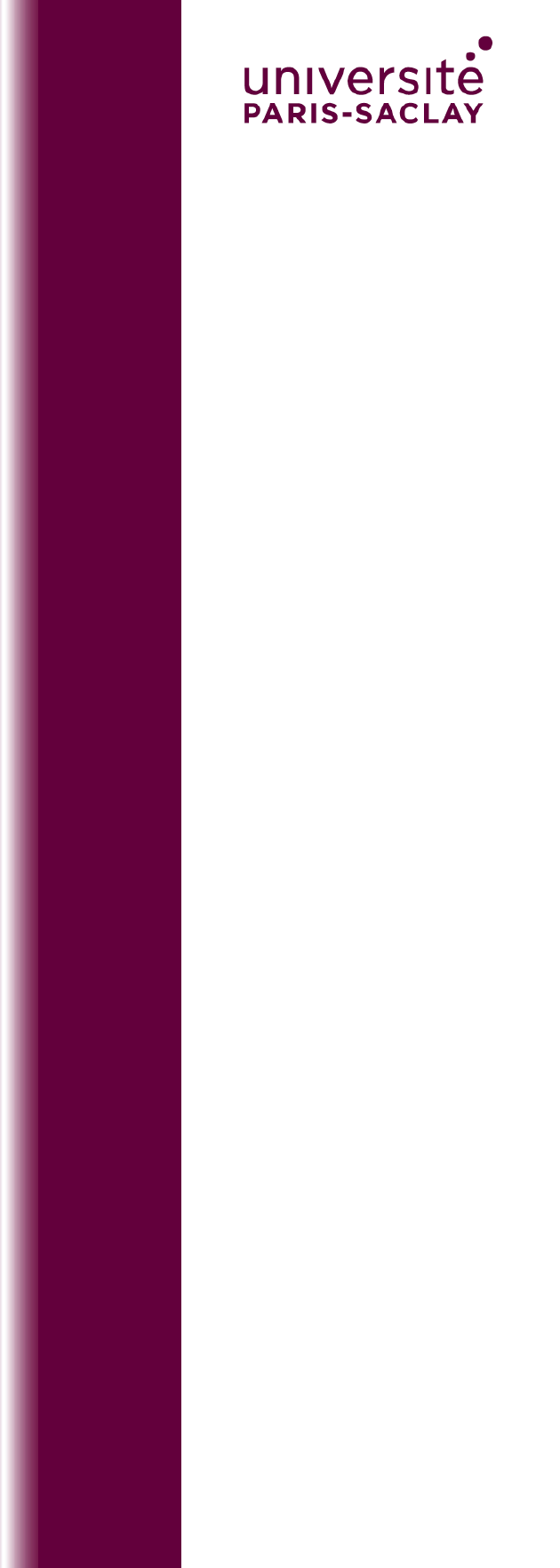}};


\color{white}

\begin{picture}(0,0)
\put(-152,-743){\rotatebox{90}{\Large \textsc{THESE DE DOCTORAT}}} \\
\put(-120,-743){\rotatebox{90}{NNT : 2020UPASA001}}
\end{picture}
 


\flushright
\vspace{10mm} 
\color{Prune}

\fontsize{22}{26}\selectfont
  \Huge Strongly correlated stochastic systems \\

\normalsize
\color{black}
\Large{\textit{Systèmes stochastiques avec correlations fortes}} \\

\fontsize{8}{12}\selectfont

\vspace{1.5cm}

\normalsize
\textbf{Thèse de doctorat de l'université Paris-Saclay} \\

\vspace{6mm}

\small École doctorale n$^{\circ}$564, Physique en Ile-de-France (PIF)\\
\small Spécialité de doctorat: Physique\\
\small Graduate School : Physique. Référent : Faculté des sciences d’Orsay \\
\vspace{6mm}

\footnotesize Thèse préparée dans l'unité de recherche \textbf{LPTMS (Université Paris-Saclay, CNRS)}, sous la direction de \textbf{Satya N. MAJUMDAR}, Directeur de Recherche. \\
\vspace{15mm}

\textbf{Thèse soutenue à Paris-Saclay, le 11 Juin 2025, par}\\
\bigskip
\Large {\color{Prune} \textbf{Marco BIROLI}} 

\vspace{\fill} 

\bigskip

\flushleft
\small {\color{Prune} \textbf{Composition du jury}}\\
{\color{Prune} \scriptsize {Membres du jury avec voix délibérative}} \\
\vspace{2mm}
\scriptsize
\begin{tabular}{|p{7cm}l}
\arrayrulecolor{Prune}
\textbf{Olivier BENICHOU} &   Président\\ 
Directeur de Recherche, Sorbonne Université & \\
\textbf{Malte HENKEL} &  Rapporteur \& Examinateur \\ 
Professeur de Recherche, Université de Lorraine   &   \\ 
\textbf{Pierpaolo VIVO} &  Rapporteur \& Examinateur \\ 
Reader, King's College  &   \\ 
\textbf{Jean Philippe BOUCHAUD} & Examinateur \\
Directeur de Recherche, CFM & \\
\textbf{Léticia CUGLIANDOLO} &  Examinatrice \\ 
Professeure de Recherche, Sorbonne Université   &   \\ 
\textbf{Shlomi REUVENI} &  Examinateur \\ 
Associate Professor, Tel Aviv University   &   \\ 
\textbf{Satya N. MAJUMDAR} & Directeur de thèse \\
Directeur de Recherche CNRS, LPTMS (Université
Paris-Saclay) &

\end{tabular} 

\end{titlepage}

\thispagestyle{empty}
\newgeometry{top=1.5cm, bottom=1.25cm, left=2cm, right=2cm}

\noindent 
\includegraphics[height=2.45cm]{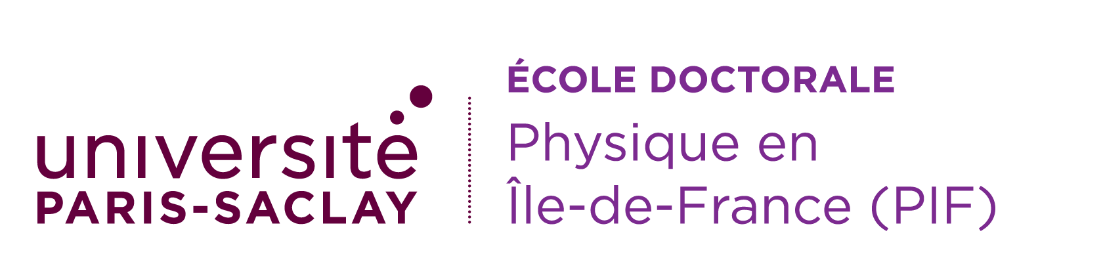}
\vspace{1cm}

\small

\begin{mdframed}[linecolor=Prune,linewidth=1]

\textbf{Titre:} Système stochastiques avec correlations fortes

\noindent \textbf{Mots clés:} Processus stochastiques, états hors-equilibre, statiques des valeurs extrêmes, correlations fortes, theory des matrices aléatoires, processus de recherche

\vspace{-.5cm}
\begin{multicols}{2}
\noindent \textbf{Résumé:}
Cette thèse développe des outils analytiques exacts pour l’étude de systèmes stochastiques avec fortes corrélations, en s’intéressant particulièrement aux statistiques extrêmes, aux statistiques d’écarts et aux statistiques de comptage dans des systèmes à plusieurs corps. Une contribution centrale est l’introduction des variables conditionnellement indépendantes et identiquement distribuées, des variables aléatoires qui deviennent indépendantes une fois conditionnées sur des paramètres latents. Cette structure émerge naturellement dans les systèmes soumis à des mécanismes de réinitialisation stochastique, qui induisent des corrélations à longue portée.

Nous obtenons des expressions universelles exactes pour plusieurs observables dans divers modèles, tels que le mouvement brownien, les vols de Lévy, les particules balistiques, ou encore le gaz de Dyson, soumis à différentes formes de réinitialisation. Nous montrons que la réinitialisation donne naissance à des états stationnaires hors d’équilibre, solvables analytiquement.

Ces prédictions théoriques sont confrontées à des résultats numériques et expérimentaux, notamment pour des particules diffusives dans un piège harmonique à raideur fluctuante. Enfin, des applications à l’optimisation de la recherche sont étudiées, identifiant des régimes où la réinitialisation améliore ou dégrade l’efficacité du premier passage, et proposant des protocoles alternatifs basés sur la repliement plus performants que la réinitialisation classique.
\end{multicols}

\end{mdframed}

\vspace{8mm}

\begin{mdframed}[linecolor=Prune,linewidth=1]

\textbf{Title:} Strongly correlated stochastic systems


\begin{multicols}{2}
\noindent \textbf{Abstract:} 

This thesis develops exact analytical tools to study strongly correlated stochastic systems, with a focus on extreme value statistics, gap statistics, and full counting statistics in multi-particle processes. A central contribution is the universal characterization of conditionally independent identically distributed variables—random variables that become independent upon conditioning on latent parameters. This structure arises naturally in systems with stochastic resetting, a mechanism that generates strong long-range correlations while retaining analytical tractability.

Using this framework, we derive universal closed-form expressions for several observables across diverse models, including Brownian motion, Lévy flights, ballistic particles, and Dyson Brownian motion, under various resetting protocols. In particular, we demonstrate that resetting induces analytically tractable non-equilibrium steady states.

Theoretical predictions are supported by numerical comparisons and experimental comparisons in systems such as diffusive particles in switching harmonic traps. Applications to search optimization are also explored, identifying regimes where resetting enhances or impairs first-passage efficiency, and proposing rescaling-based protocols that outperform traditional resetting.

\end{multicols}
\end{mdframed}


\thispagestyle{empty}
\newgeometry{top=1.5cm, bottom=4cm, left=2cm, right=2cm}

\,\vspace{6cm}

\noindent {\it To Megi, the love of my life, \\
who had to reset three times \\
to ask me out.}

\titleformat{\chapter}[hang]{\bfseries\Large\color{Prune}}{\thechapter\ -}{.1ex}
{\vspace{0.1ex}
}
[\vspace{1ex}]
\titlespacing{\chapter}{0pc}{0ex}{0.5pc}

\titleformat{\section}[hang]{\bfseries\normalsize}{\thesection\ .}{0.5pt}
{\vspace{0.1ex}
}
[\vspace{0.1ex}]
\titlespacing{\section}{1.5pc}{4ex plus .1ex minus .2ex}{.8pc}

\titleformat{\subsection}[hang]{\bfseries\small}{\thesubsection\ .}{1pt}
{\vspace{0.1ex}
}
[\vspace{0.1ex}]
\titlespacing{\subsection}{3pc}{2ex plus .1ex minus .2ex}{.1pc}

\newgeometry{top=4cm, bottom=4cm, left=2cm, right=2cm}

\tableofcontents

\newgeometry{top=4cm, bottom=4cm, left=4cm, right=4cm}

\chapter{Acknowledgements} \label{ch:aknowledgments}
First and foremost, I am deeply grateful to my advisor, Satya N. Majumdar, for his constant encouragement over the years, his patience with my manuscripts, and for offering the perfect balance between guidance and independence throughout this thesis. I am equally indebted to Grégory Schehr, who acted as a second mentor and always took the time to discuss and guide me as if I were one of his own students.

\vspace{0.2cm}

Special thanks go to Hernán Larralde, Manas Kulkarni, Yannick Feld, Alexander Hartmann, and Francesco Mori for their invaluable collaborations and stimulating discussions. A special mention to Francesco, who kindly supervised my internship prior to the doctorate, and thanks to whom I inherited an amazing apartment.

\vspace{0.2cm}

To the members of the jury — Olivier Bénichou, Malte Henkel, Pierpaolo Vivo, Jean-Philippe Bouchaud, Léticia Cugliandolo, and Shlomi Reuveni — thank you for taking the time to evaluate my thesis. I am especially grateful to Malte Henkel and Pierpaolo Vivo for reviewing the manuscript in detail.

\vspace{0.2cm}

To Sergio Ciliberto, thank you for all the experimental discussions, which helped ground parts of this thesis in reality.

\vspace{0.2cm}

A heartfelt thank you to the LPTMS family, and in particular to Alberto Rosso, Claudine, and Delphine, whose incredible efficiency and support — far beyond what was required — made dealing with bureaucracy almost enjoyable. Thank you to all the students of LPTMS — Francesco, Sap, Vicio, Ana, Saverio, Maximilien, Alberto, Andrea, Benoit, Charbel, Flavio, Federico, Guido, Louis, Luca, Oscar, Pawat, Pietro, Romain, Vincent — for making the lab such a vibrant and enjoyable place to work. Thanks also to Léo, Mathis, Giuseppe, and Ivan, who wrestled with similar problems and were always around to chat.

\vspace{0.2cm}

I'm grateful to Alessandro, Matteo, Giulia, Sara, Jeanne, Jules, Adrien, Ghjulia, Jean, Carlos, and Ayoub, who accompanied me on this academic journey and have been incredible friends. A special mention to Matteo and Alessandro, who have come to feel like brothers — and to Alessandro, for all the laughs and memories in our office.

\vspace{0.2cm}

Thank you to my non-academic friends — Alexandre, Alexis, Axel, Florian, Gabriel, Grégoire, Ivan, Julien, Thomas, and Vincent — for the amazing vacations and shared moments. Special thanks to Alexis for organizing many of them.

\vspace{0.2cm}

I’m also grateful to Ricardo Marino, for his friendship and guidance — especially when it came to navigating the chaos of postdoc applications alongside real life, and his repeated invitations for good food and beers which were just what I needed.

\vspace{0.2cm}

To my family, I thank you from the bottom of my heart. To my parents, who instilled in me the drive to pursue my passions and gave me the freedom to explore, and to my brothers, for their unconditional love and support. I am also grateful to my grandparents, who taught me the value of hard work — especially nonno Marco, whom I have always looked up to. This thesis is, in part, for you.

\vspace{0.2cm}

And finally, to Megi, my muse — whose love has shaped me into the man I am today. You've always pushed me to overcome my limits and strive for the best. These achievements would have been impossible without you.

\chapter{List of publications} \label{ch:publications}
\begin{enumerate}
    \item {\it Extreme statistics and spacing distribution in a Brownian gas correlated by resetting.} M. Biroli, H. Larralde, S.N. Majumdar and G. Schehr, Physical Review Letters {\bf 130} (20), 207101 (2023). {\color{blue} arXiv:2211.00563}
    \item {\it Critical number of walkers for diffusive search processes with resetting.}
    M. Biroli, S. N. Majumdar and G. Schehr,
    Physical Review E {\bf 107} (6), 064141 (2023). {\color{blue} arXiv:2303.18012}
    \item {\it Exact extreme, order, and sum statistics in a class of strongly correlated systems.}
    M. Biroli, H. Larralde, S.N. Majumdar and G. Schehr,
    Physical Review E {\bf 109} (1), 014101 (2024). {\color{blue} arXiv:2307.15351}
    \item {\it Dynamically emergent correlations between particles in a switching harmonic trap.}
    M. Biroli, M. Kulkarni, S.N. Majumdar and G. Schehr,
    Physical Review E {\bf 109} (3), L032106 (2024). {\color{blue} arXiv:2312.02570}
    \item {\it Resetting by rescaling: Exact results for a diffusing particle in one dimension.}
    M. Biroli, Y. Feld, A.K. Hartmann, S.N. Majumdar and G. Schehr,
    Physical Review E {\bf 110} (4), 044142 (2024). {\color{blue} arXiv:2406.08387}
    \item {\it Resetting Dyson Brownian motion},
    M. Biroli, S.N. Majumdar, G. Schehr, Physical Review E {\bf 112} (1), 014101 (2025).
    arXiv:2503.14733
    \item {\it Experimental evidence for strong emergent correlations between particles in a switching trap}, M. Biroli, S. Ciliberto, M. Kulkarni, S. N. Majumdar, A. Petrosyan, G. Schehr, arXiv preprint arXiv:2508.07199 (2025).
\end{enumerate}

\vspace{0.2cm}

There is also one other relevant work, anterior to the thesis.

\begin{enumerate}
    \item {\it Number of distinct sites visited by a resetting random walker.}
    M. Biroli, F. Mori and S.N. Majumdar,
    Journal of Physics A: Mathematical and Theoretical {\bf 55} (24), 244001 (2022). {\color{blue} arXiv:2202.04906}
\end{enumerate}

\chapter{Résumé en français} \label{ch:resume}
Du mouvement chaotique des molécules dans un gaz aux fluctuations des marchés financiers, en passant par les changements apparemment imprévisibles de la météo, l'aléa est une caractéristique inévitable du monde naturel. Malgré le désordre apparent qu'il engendre, le hasard peut être exploité comme un outil puissant pour comprendre des systèmes complexes comportant un grand nombre de degrés de liberté. Même lorsque la dynamique microscopique sous-jacente est déterministe, elle peut devenir si compliquée qu'il n'est ni faisable ni instructif d'en suivre chaque détail. Dans ce cas, les modèles stochastiques constituent une abstraction efficace, remplaçant la complexité microscopique par un bruit effectif. Cette approche est à la base d'un large éventail de théories physiques, allant de la théorie cinétique des gaz~\cite{K07} à l'approximation de Wigner pour les espacements de niveaux d’énergie atomiques~\cite{M91,F10,LNV18}. Le pari de la physique statistique est de considérer le hasard non pas seulement comme un reflet de notre ignorance ou incertitude, mais comme un outil permettant de capturer les propriétés macroscopiques émergentes issues d'interactions microscopiques complexes.

\vspace{0.2cm}

La première question naturelle à poser lorsqu'on étudie de tels systèmes stochastiques consiste à décrire leur comportement typique. Puisque le bruit est le moteur sous-jacent du modèle, il faut étudier les réponses typiques, c'est-à-dire celles qui seront le plus souvent observées expérimentalement. Le résultat le plus célèbre qui fonde de nombreuses descriptions statistiques modernes est le théorème central limite, qui précède de loin la théorie moderne des probabilités et a donné lieu à d'innombrables applications~\cite{Fi10}. Cependant, une dépendance excessive à ce théorème et à sa description des comportements typiques est également ce qui, selon Mandelbrot, a conduit au krach boursier de 1998 : `` Août 1998 n'aurait tout simplement jamais dû se produire ; selon les modèles standards de l'industrie financière, une telle séquence d'événements était si improbable qu'elle était impossible. Les théories standards [...] estimaient les chances de cet [...] effondrement à une sur 20 millions — un événement que, si vous investissiez quotidiennement pendant près de 100 000 ans, vous ne verriez même pas une fois. ''~\cite{B10}.

\vspace{0.2cm}

Bien que rares, les événements extrêmes peuvent avoir des conséquences dévastatrices, et leur description précise est donc d'une valeur inestimable. C'est la motivation fondamentale des statistiques de valeurs extrêmes. En effet, dans de nombreux contextes comme l'ingénierie~\cite{G58}, les sciences environnementales~\cite{KPN02}, l'informatique~\cite{KM00, MK00, MK02, MK03, MDK05}, la finance~\cite{MB08, EKM97} ou la physique~\cite{FC15,D81,BM97,BBP07}, pour n'en citer que quelques-uns, comprendre les événements extrêmes est une question cruciale. L'une des contributions majeures à la théorie des valeurs extrêmes est le théorème de Fisher-Tippett-Gnedenko, qui caractérise universellement la statistique des valeurs extrêmes d'un ensemble de $N$ variables aléatoires indépendantes et identiquement distribuées. La loi du maximum d'un tel ensemble converge, lorsque $N$ est grand, vers l'une des trois lois limites possibles : Gumbel, Fréchet ou Weibull, selon le comportement de la queue de distribution. Ces résultats ont été généralisés aux variables indépendantes mais non identiquement distribuées~\cite{W88, A78, A84, DR84, SW87} et à un grand ensemble de variables faiblement corrélées, où, dans la limite $N \to \infty$, on retrouve les résultats du cas indépendant~\cite{MPS20}.

\vspace{0.2cm}

En revanche, l'étude de variables aléatoires corrélées, c'est-à-dire non indépendantes, est nettement plus complexe, et aucune description universelle analogue à celle des variables indépendantes \cite{G58} ne peut être donnée. Les systèmes corrélés doivent être étudiés au cas par cas, et seuls quelques-uns peuvent être traités analytiquement de manière approfondie. Les exemples les plus connus sont peut-être les statistiques extrêmes des interfaces fluctuantes dans les classes d'universalité Edwards-Wilkinson ou Kardar-Parisi-Zhang~\cite{RCPS01, GHPR03,MC04,MC05}, la description des valeurs propres extrêmes en théorie des matrices aléatoires~\cite{TW94,TW96,DM06,DM08,MV09,MS14} ou encore le modèle d'énergie aléatoire généralisé pour les systèmes désordonnés~\cite{D81, DG86}. Pourtant, de nombreux systèmes physiques présentent de fortes corrélations, et l'absence de caractérisation universelle pour ces systèmes limite considérablement les avancées dans ce domaine.

\vspace{0.2cm}

Les événements extrêmes sont intimement liés aux problèmes de premier passage, qui quantifient la probabilité et le temps mis par un processus stochastique pour atteindre un certain état pour la première fois. Ces problèmes apparaissent naturellement dans un large éventail de contextes, allant du déclenchement de l'activité neuronale~\cite{GM64,SG13,T88}, à l'évasion d'une particule d'un puits de potentiel, en passant par le temps nécessaire à un chercheur pour trouver une cible dans un milieu désordonné~\cite{MRO14, VK92}. Au cœur, les problèmes de premier passage relèvent des statistiques extrêmes, puisque les trajectoires pertinentes correspondent souvent à des événements rares mais décisifs. L'application la plus intuitive des propriétés de premier passage est la description des processus de recherche. Ceux-ci sont omniprésents, aussi bien dans la nature que dans les activités humaines~\cite{B12,AD68,MRO14}, qu'il s'agisse de la quête de nourriture chez les animaux~\cite{BC09, VDLRS11} ou de la liaison de protéines~\cite{BWH81, CBM04,GMKC18,C19}. Comprendre les propriétés statistiques des temps de premier passage est crucial pour évaluer l'efficacité et les limites des stratégies de recherche.

\vspace{0.2cm}

Dans cette thèse, nous étudions diverses classes de processus stochastiques tels que le mouvement brownien, les vols de Lévy ou le gaz logarithmique de Dyson. L'objectif principal est d'analyser ces processus en présence de fortes corrélations de longue portée et d'obtenir, si possible, des descriptions générales des systèmes fortement corrélés. Comme nous le verrons au Chapitre~\ref{ch:ciid}, nous avons identifié et caractérisé une famille entière de variables aléatoires fortement corrélées : les variables conditionnellement indépendantes et identiquement distribuées. Au Chapitre~\ref{ch:sim-reset}, nous montrerons que ces variables apparaissent naturellement dans de nombreux systèmes physiques, et permettent une description exacte de problèmes qui seraient autrement d'une grande complexité. Nous mettrons l'accent sur les résultats analytiques exacts, dans l'espoir que ces systèmes idéalisés fournissent des lois universelles. Une partie de notre travail a également consisté à proposer des systèmes réalisables expérimentalement. En un second temps, nous avons étudié les propriétés de premier passage de certains de ces modèles stochastiques, dans le but de comprendre comment ils peuvent améliorer les processus de recherche, et d'utiliser cette compréhension pour concevoir des processus de recherche plus rapides. Nous introduisons maintenant brièvement le principal modèle stochastique sous-jacent à la majorité des résultats de cette thèse.

\vspace{0.2cm}

Le mouvement brownien est l'un des premiers modèles stochastiques formalisés, et probablement le premier processus stochastique en temps continu à avoir été défini rigoureusement, profondément lié à la fois à la physique et à la théorie des probabilités. Il tire son nom de Robert Brown, qui observa en 1827 le mouvement erratique de grains de pollen en suspension dans l'eau. Ce phénomène resta inexpliqué jusqu'au début du XX\textsuperscript{e} siècle, lorsque Smoluchowski proposa un modèle cinématique de la diffusion : ce mouvement aléatoire résulte d'une infinité de chocs microscopiques, chacun infinitésimal, dus aux collisions avec les molécules d'eau. Ce concept d'une somme infinie de déplacements infinitésimaux s'avère extrêmement général : Louis Bachelier fut le premier~\cite{B1900} à utiliser le mouvement brownien pour modéliser les fluctuations de prix en finance. Quelques années plus tard, Einstein~\cite{E56} et Smoluchowski~\cite{S06} introduisirent indépendamment le mouvement brownien dans la communauté physique, le premier avec une approche thermodynamique, le second avec une perspective mécanique microscopique.

\vspace{0.2cm}

Au cours du siècle dernier, le mouvement brownien est devenu un outil fondamental en physique, en finance et en informatique, grâce à sa simplicité et à sa solvabilité analytique. Il fournit un exemple rare de variables aléatoires fortement corrélées dont les statistiques extrêmes sont connues exactement. Par exemple, le maximum (ou minimum) d'un mouvement brownien unidimensionnel est connu analytiquement~\cite{MPS20}. Pour une revue comprehensive nous vous referons à Ref.~\cite{MPS20}. Dans cette thèse, nous nous concentrons principalement sur des systèmes à plusieurs particules, et étudions leur comportement près des maxima (statistiques extrêmes) ainsi qu'au sein du régime typique (bulk). Les observables considérées incluent les positions des particules, les écarts entre particules successives, et le nombre de particules dans une région donnée (full counting statistics). Ces observables permettent de décrire complètement un gaz de particules, et s'inspirent d'applications concrètes : par exemple, en théorie des matrices aléatoires, les écarts correspondent à ceux des niveaux d'énergie dans des noyaux lourds~\cite{W51,M91}, et les statistiques de comptage sont liées à l'intrication quantique dans une chaîne de spins unidimensionnelle~\cite{TBFDS21}.

\vspace{0.2cm}

Une variante récente du mouvement brownien, devenue populaire au cours de la dernière décennie, est le mouvement brownien avec réinitialisation stochastique (voir~\cite{EMS20} pour une revue). Ce mécanisme consiste à interrompre la dynamique naturelle du processus à des instants aléatoires et à le redémarrer depuis une position ou une distribution prédéfinie. Le modèle le plus simple est celui d'une réinitialisation avec taux de Poisson $r$~\cite{EM11PRL,EM11JPhysA}. Ce processus apparemment simple présente déjà une richesse phénoménologique : existence d'un état stationnaire hors d'équilibre~\cite{EM11PRL}, réduction du temps moyen de premier passage~\cite{EM11PRL}, et existence d'un taux de réinitialisation optimal minimisant ce temps~\cite{EM11PRL}. L'état stationnaire hors d'équilibre est particulièrement intéressant~\cite{EM11PRL, MC16, EM16}, car encore mal compris~\cite{HP10,HHL08}, et la possibilité d'accélérer les processus de recherche conduit à des transitions de phase dynamiques~\cite{KMSS14,KG15,CM15}. Ces ingrédients, combinés à la solvabilité analytique du modèle, expliquent l'intérêt croissant pour la réinitialisation, avec des applications allant de l'informatique à la chimie ou l'écologie~\cite{EMS20}.

\vspace{0.2cm}

Dans cette thèse, nous utilisons la réinitialisation à la fois comme outil pour étudier et concevoir des algorithmes de recherche plus rapides, et comme mécanisme pour générer des corrélations fortes à longue portée dans des systèmes multiparticules. Bien que ces modèles de réinitialisation aient initialement conduit à la description plus générale en termes de variables conditionnellement indépendantes, nous avons choisi de ne pas présenter les résultats dans l'ordre chronologique. Nous commençons par le formalisme général, que nous appliquons ensuite à une grande variété de systèmes stochastiques.

\chapter{Introduction} \label{ch:intro}
From the chaotic motion of molecules in a gas to fluctuations in financial markets and seemingly unpredictable shifts in weather, randomness is an inescapable feature of the natural world. Despite its apparent disorder, randomness can be harnessed as a powerful tool to understand complex systems with many degrees of freedom. Even when the underlying microscopic dynamics are deterministic, they can become so intricate that it is neither feasible nor informative to track every detail. In such cases, stochastic models serve as a powerful abstraction, replacing microscopic intricacies with effective noise. This approach underlies a wide range of physical theories, from the kinetic theory of gases \cite{K07} to the Wigner surmise of the gaps in atomic energy levels \cite{M91,F10,LNV18}. The gambit of statistical physics is to view randomness not merely as a reflection of ignorance or uncertainty, but rather as a tool that allows us to capture the emergent macroscopic features arising from complex microscopic interactions.

\vspace{0.2cm}

The first and most logical question to ask when studying such stochastic systems is to try to describe the typical behavior of the system. Since noise is the underlying motor beneath the model, we have to study the typical responses. These are the ones that will be measured most often experimentally. The most famous result that underlies several modern statistical descriptions is the Central Limit Theorem, which long predates modern probability theory and has led to innumerable applications and results \cite{Fi10}. However, overreliance on the Central Limit Theorem and its description of typical behaviors is also what led to the 1998 stock market crash according to Mandelbrot: ``August 1998 simply should never have happened; it was, according to the standard models of the financial industry, so improbable a sequence of events as to have been impossible. The standard theories, [...] would estimate the odds of that [...] collapse at one in 20 million—an event that, if you traded daily for nearly 100,000 years, you would not expect to see even once.'' \cite{B10}

\vspace{0.2cm}

Although incredibly rare, extreme events can {\color{blue}lead to} devastating consequences, and hence their precise description is invaluable. This is the core motivation behind the importance of {\color{blue}extreme-value} statistics. Indeed, in many contexts, such as engineering \cite{G58}, environmental sciences \cite{KPN02}, computer science \cite{KM00, MK00, MK02, MK03, MDK05}, finance \cite{MB08, EKM97} or physics \cite{FC15,D81,BM97,BBP07}, to cite but a few, understanding extreme events is a matter of crucial importance. One of the most important contributions to the field of {\color{blue}extreme-value} theory was the Fisher-Tippett-Gnedenko theorem, which universally characterized the {\color{blue}extr\-eme-value} statistics of a set of $N$ identically distributed independent random variables. The distribution of the maximum of $N$ independently identically distributed random variables converges in the large $N$ limit to one of three possible limiting forms: Gumbel, Fréchet, or Weibull according to the behavior of the tail of the distribution of the variables. These results have been generalized for non-identically distributed independent variables \cite{W88, A78, A84, DR84, SW87} and for a large set of weakly correlated identically distributed random variables, where the results, in the large $N$ limit, reduce to the independent case \cite{MPS20}.

\vspace{0.2cm}

On the other hand, the study of correlated random variables, i.e. non-independent random variables, is significantly more challenging and no universal description as the one of independent variables \cite{G58} can be made. Correlated systems have to be studied on a case-by-case basis, and only a handful can be studied to great analytical detail. The most well-known examples are perhaps the {\color{blue}extreme-value} statistics of fluctuating interfaces in the Edward-Wilkinson or Kardar-Parisi-Zhang universality classes \cite{RCPS01, GHPR03,MC04,MC05}, the description of the maximal (or minimal) eigenvalues in random matrix theory \cite{TW94,TW96,DM06,DM08,MV09,MS14} and the generalized random energy model for disordered systems \cite{D81, DG86}. Yet, many physical systems present strong correlations, and a lack of universal characterization for such systems
greatly hinders progress in this area.

\vspace{0.2cm}

Extreme events are closely tied to first-passage problems, which quantify the probability and time with which a stochastic process reaches a specified state for the first time. These problems naturally arise in a wide range of contexts, from the triggering of neuronal firing \cite{GM64,SG13,T88}, the escape of a particle from a potential well, to the time it takes for a searcher to find a target in a disordered medium \cite{MRO14, VK92}. At their core, first-passage problems are intimately related to extremal statistics, since the relevant paths are often rare but decisive events. The most intuitive application of first-passage properties is the description of search processes. Search processes are ubiquitous in both natural and human behavior \cite{B12,AD68,MRO14}: from foraging animals \cite{BC09, VDLRS11} to protein binding \cite{BWH81, CBM04,GMKC18,C19}. Understanding the statistical properties of first-passage times is crucial to understand the efficiency and limitations of search strategies.

\vspace{0.2cm}

In this thesis, we {\color{blue} shall} study a variety of stochastic processes such as Brownian motion, Lévy flights, or Dyson's log gas. If any of these processes are unfamiliar, do not fret, we will introduce each one properly in due time. The principal aim of this thesis {\color{blue} is} to study these stochastic processes in the presence of strong long-range correlations and to try and obtain some general descriptions of strongly correlated systems. As we will see in Chapter \ref{ch:ciid}, we successfully identified and characterized an entire family of strongly correlated random variables: namely conditionally independent identically distributed random variables. We will see in Chapter \ref{ch:sim-reset} that these conditionally independent random variables appear naturally in a variety of physical systems and allow for the exact description of what would otherwise be extremely intricate problems. However, we mainly focused on exact analytical results in the hope that these idealized systems could provide universal results. A portion of our work was also dedicated to creating experimentally realizable systems. In a second, alternative direction, we investigated the first-passage properties of some of these stochastic models. With the purpose of understanding better how they could improve search processes and finally using our understanding of these systems to engineer a faster search process. Let us now briefly introduce the main stochastic model underlying most of the results in the thesis. 

\vspace{0.2cm}

Brownian motion is one of the first stochastic models, and arguably the first continuous-time stochastic process to be rigorously defined and deeply linked with both physics and probability theory. It is named after Robert Brown, who in 1827 observed the erratic motion of pollen in a water solution. However, Robert Brown could not understand the origin of this seemingly random motion. To obtain an explanation, we have to wait for the next century, when Smoluchowski presented a kinematic model of diffusion. The erratic motion was the result of infinitely many and infinitely small kicks from collisions with water molecules. This abstract concept of infinitely many, infinitely small shifts is not confined to the description of pollen in water but is extremely general. Louis Bachelier was the first \cite{B1900} to use Brownian motion to model price fluctuations in finance. A couple of years later, Einstein \cite{E56} and Smoluchowski \cite{S06} independently introduced Brownian motion to the physics community.  Einstein approached Brownian motion from a thermodynamic angle, whereas Smoluchowski provided the microscopic mechanical perspective described above.

\vspace{0.2cm}

In the past century, Brownian motion has become a fundamental model and tool in physics, finance, and computer science due to its simplicity and analytical tractability. Brownian motion trajectories are a rare example of strongly correlated random variables whose {\color{blue}extreme-value} statistics can be studied analytically. For example, the maximum (or minimum) of a one-dimen\-sional Brownian motion is known exactly \cite{MPS20}. For a comprehensive review, see Ref. \cite{MPS20} and for a beginner-friendly book see Ref. \cite{MSbook}. In this thesis, we will mainly focus on multiparticle systems. We will study the behavior of our systems both near the maxima, i.e. their {\color{blue}extreme-value} statistics, and in the bulk. The typical observables we will compute are the positions of the particles, gaps between successive particles, and counting the number of particles in a given region, i.e. the full counting statistics. These observables allow us to fully describe our gas of particles and are inspired by known practical applications. For example, in the context of random matrix theory, gaps can be linked to gaps in the energy spectra of heavy nuclei \cite{W51,M91}, and full counting statistics can be related to the entanglement of a one-dimensional spin chain \cite{TBFDS21}.

\vspace{0.2cm}

A recent variant of Brownian motion which gained popularity in the last decade is stochastically resetting Brownian motion, see Ref. \cite{EMS20} for a review. Stochastically resetting a process consists of interrupting its natural dynamics at random times and restarting it from a pre-defined position or distribution. The simplest model of stochastic resetting is a Brownian motion resetting with a Poissonian rate $r$ \cite{EM11PRL,EM11JPhysA}. This `simple' process already has a very rich physical behavior. Such as the presence of a non-equilibrium steady state at long times \cite{EM11PRL}, a lower mean first-passage time \cite{EM11PRL} than Brownian motion and the existence of an optimal reset rate minimizing the mean first-passage time \cite{EM11PRL}. The existence of a non-equilibrium steady state is of particular interest \cite{EM11PRL, MC16, EM16} since they are still poorly understood \cite{HP10,HHL08}, and the possible speedup of the search process also leads to interesting dynamical phase transitions \cite{KMSS14,KG15,CM15}. These key ingredients and resetting's analytical tractability are a core reason for its wide-spread interest, with applications ranging from computer science to chemistry or ecology \cite{EMS20}. 

\vspace{0.2cm}

In this thesis, we will make use of resetting both to study and engineer faster search algorithms and as a way to generate strong long-range correlations in multiparticle systems. Although these resetting models were the ones that originally led us to the more general description in terms of conditionally independent variables, we chose not to present these results chronologically. Rather, we will first present the general formalism, which we later apply to a large variety of stochastic systems. The thesis is structured as follows.

\section{Overview of the thesis}

Chapter \ref{ch:main} provides an overview of the main results derived in this thesis. The aim is not to provide a comprehensive review of every result in the thesis but to present the core results, which I, the author, believe to be the most salient outcomes of the thesis.

\vspace{0.2cm}

Chapter \ref{ch:extreme} introduces all the mathematical baggage necessary to understand the thesis. We define all the observables that will be often studied throughout the thesis and recall known results from the literature for a sequence of independent identically distributed random variables. Namely, we will recall the central limit theorem, the behavior of {\color{blue}extreme-value} statistics (i.e. the maximum or minimum of the sequence), namely the Fisher-Tippett-Gnedenko theorem, the order statistics, the statistics of the gaps, and the full counting statistics.

\vspace{0.2cm}

Chapter \ref{ch:ciid} presents the formalism that we derived during the thesis which underlies several of the stochastic models which we have studied. We derive the equivalent universal results for the above-mentioned observables which are known for independent variables, for these conditionally independent random variables. 

\vspace{0.2cm}

In Chapter \ref{ch:sim-reset} we study a series of multiparticle stochastic systems that use resetting, more precisely {\it simultaneous} resetting, to generate strong long-range correlations between the particles. We will look at a simultaneously resetting gas of Brownian motions, Lévy flights, or ballistic particles. These will allow us to showcase the usefulness of the conditionally independent formalism presented in Chapter \ref{ch:ciid}. At the end of Chapter \ref{ch:sim-reset} we will study another example of simultaneous resetting where the conditionally independent structure is not as aparent. Specifically, we will look at a gas of one-dimensional particles which simultaneously reset whenever any of them reach a certain target. 

\vspace{0.2cm}

In Chapter \ref{ch:ou-switch}, we introduce an experimental protocol that has been used to model resetting. This protocol is the inspiration for the more physically accurate model of resetting we consider. Instead of considering instantaneous resetting to a fixed position, we will study a gas of particles confined in a harmonic trap whose stiffness alternates between a wide (i.e. very free) and tight (i.e. very confined) value. Although not obvious at all a priori, conditional independence will emerge in this process, allowing us to re-use the formalism introduced in Chapter \ref{ch:ciid}. We will also show some preliminary experimental results that confirm our theoretical predictions.

\vspace{0.2cm}

In Chapter \ref{ch:dyson} we move away from this conditionally independent formalism and enter the realm of random matrix theory. We study a gas of diffusing particles with pairwise logarithmic repulsion, on top of which, we will apply simultaneous resetting to induce strong long-range correlations as was done in Chapter \ref{ch:sim-reset}. However, unlike the models studied in Chapter \ref{ch:sim-reset}, the underlying reset-free dynamics are already correlating the particles. Hence, we will not be able to resort to the conditionally independent formalism from Chapter \ref{ch:ciid}. Instead we need to use known results from Random Matrix Theory, which we also recall at the start of Chapter \ref{ch:dyson}. The stochastic process presented in Chapter \ref{ch:dyson} is a rare example of a stochastic process with competing (attractive and repulsive) strong long-range correlations, which can nevertheless be studied in great analytical detail.

\vspace{0.2cm}

In Chapter \ref{ch:search} we recall some known results in the literature which describe why, how, and when the introduction of stochastic resetting can be used to speed up search processes. Subsequently, we will study how the correlations induced from simultaneous resetting can affect the search process. We will see that resetting is not always optimal and that the resetting-induced correlations influence this transition from optimal to suboptimal.

\vspace{0.2cm}

Finally, in Chapter \ref{ch:rescaling} with the intention of engineering a faster (than resetting) search process, we introduce a variant of stochastic resetting. Instead of resetting to the origin, the position of a diffusing particle is rescaled by a constant factor $-1 < a < 1$. We will see that for $-1 < a < 0$ we can indeed obtain a lower mean first-passage time than the usual stochastically resetting Brownian motion model.

\chapter{Main Results} \label{ch:main}
\begin{figure}
    \centering
    \includegraphics[width=\textwidth]{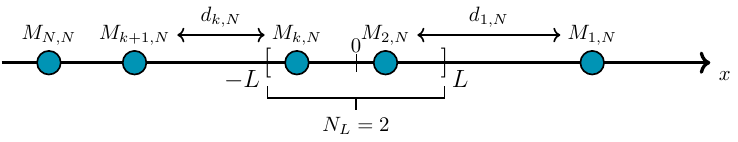}
    \caption{A visual representation of the main observables studied in the thesis. {\color{blue} Namely, the extreme-value statistics $M_{1, N}$ (or $M_{N, N}$), the order statistics $M_{k, N}$, the gap statistics $d_{k, N}$ and the full-counting statistics $N_L$.}} \label{fig:main-res-observables}
\end{figure}

Before we provide the main results obtained during the thesis, we have to define just a couple of core concepts. In this thesis, we investigate mainly multiparticle systems $X_1(t), \cdots, X_N(t)$ in their steady state. Hence, we drop the time dependency $X_i(t) \equiv X_i$. We studied the following observables that are sketched in Fig. \ref{fig:main-res-observables}:
\begin{itemize}
    \item the center of mass, or sample mean
    \begin{equation}
        C_N = \frac{1}{N} \sum_{i = 1}^N X_i
    \end{equation}
    \item the average density of particles
    \begin{equation}
        \rho(x) = \left\langle \frac{1}{N} \sum_{i = 1}^N \delta(x - X_i) \right\rangle
    \end{equation}
    \item the {\color{blue}extreme-value} statistics, such as the maximum
    \begin{equation}
        M_{1, N} = \max_{i} X_i
    \end{equation}
    \item more generally the order statistics $\min_i X_i = M_{N, N} \leq M_{N-1, N} \leq \cdots \leq M_{1, N} = \max_i X_i$, i.e. the study of the $M_{k, N}$ or in other words the $k$-th particle counting from the right.
    \item the gap statistics $d_k = M_k - M_{k+1}$ both for $k$ of order one, i.e. close to the global maximum, and for $k = \alpha N$ in the bulk.
    \item the full counting statistics
    \begin{equation}
        N_L = \# \{ X_i : X_i \in [-L, L] \}\;,
    \end{equation}
    i.e. the number of particles in a box $[-L, L]$ centered around the origin.
\end{itemize}
The main stochastic models considered in the thesis are
\begin{itemize}
    \item \emph{Brownian motion and the \OU process.} We consider a sto\-chastic particle $X(t)$ with the following dynamics
    \begin{equation}
        \dv{X(t)}{t} = -\mu X(t) + \sqrt{2 D} \, \eta(t) \;,
    \end{equation}
    where $\eta(t)$ is a Gaussian white noise with zero-mean $\langle \eta(t) \rangle = 0$ and delta-correlations $\langle \eta(t) \eta(t') \rangle = \delta(t - t')$. This is an \OU process, which when setting $\mu = 0$ corresponds to a Brownian motion.
    \item \emph{Poissonian resetting.} Poissonian resetting consists in modifying the dynamics of a generic process $X(t)$ as follows
    \begin{equation}
        X(t + \dd t) = \begin{dcases}
            {\rm usual~dynamics~from~~} X(t) &{\rm ~~with~probability~~} 1 - r \dd t \\
            X(0) &{\rm~~with~probability~~} r \dd t \;.
        \end{dcases}
    \end{equation}
    For example, for the above-mentioned Brownian motion
    \begin{equation}
        X(t + \dd t) = \begin{dcases}
            X(t) + \sqrt{2 D \dd t} \; \eta(t) &\mbox{~~with~probability~~} 1 - r \dd t \\
            X(0) &\mbox{~~with~probability~~} r \dd t \;.
        \end{dcases}
    \end{equation}
\end{itemize}

\section{Conditionally independent identically distributed random variables \cite{BLMS24}}

In Chapter \ref{ch:ciid} we consider $N$ random variables $X_1, \cdots, X_N$ which are independent when conditioned on $M$ latent random variables $Y_1, \cdots, Y_M = \vec{Y}$, i.e.
\begin{align}
    {\rm Prob.}[X_1, \cdots, X_N] &= \int \dd^M \vec{y} \; {\rm Prob.}[\vec{Y} = \vec{y}] \prod_{i = 1}^N {\rm Prob.}[X_i = x_i] \\
    &\equiv \int \dd^M \vec{y} \; h(\vec{y}) \prod_{i = 1}^N p(x_i | \vec{y}) \;.
\end{align}
For such conditionally independent random variables we derive universal {\color{blue} clos\-ed} {\color{blue}form} expressions for the above-mentioned observables. Assuming that $p(x | \vec{y})$ is zero-mean (which is often the case, for other considerations check the detailed results in Chapter \ref{ch:ciid}) the center of mass is distributed as 
\begin{equation}
    {\rm Prob.}[C_N = c] \underset{N \to \infty}{\longrightarrow}  \sqrt{N} \, {\cal P} \left( c \sqrt{N}\right)
\end{equation}
where the scaling function ${\cal{P}}(z)$ is given by
\begin{equation} 
   {\cal{P}}(z) = \frac{1}{\sqrt{2 \pi}}\int \frac{\dd^M  \vec{y}}{\sigma(\vec{y})} \; h(\vec{y}) \exp \left(-\frac{z^2}{2\sigma^2(\vec{y})}\right) \;,
\end{equation}
and 
\begin{equation}
    \sigma^2(\vec{y}) = \int \dd x \; x^2 p(x | \vec{y}) \;.
\end{equation}
The order statistics $M_{k, N}$ for $k = \alpha N$ are given by 
\begin{equation} \label{eq:main-res-order}
    {\rm Prob.}[M_{k, N} = w] \underset{N \to \infty}{\longrightarrow} \int \dd^M\vec{y} \; h(\vec{y}) \delta[w - q(\alpha, \vec{y})] \;,
\end{equation}
where $q(\alpha, \vec{y})$ is the $\alpha$-quantile of the conditional distribution defined as
\begin{equation}
    \alpha = \int_{q(\alpha, \vec{y})}^{+\infty} \dd x \; p(x | \vec{y}) \;.
\end{equation}
The {\color{blue}extreme-value} statistics $M_{k, N}$ for $k \sim \mathcal{O}(1)$ are also given by Eq. (\ref{eq:main-res-order}) unless $p(x | \vec{y})$ has fat power-law tails, i.e.
\begin{equation}
    p(x | \vec{y}) \underset{|x| \to +\infty}{\longrightarrow} \frac{A(\vec{y})}{|x|^{1 + \mu(\vec{y})}} \;.
\end{equation}
In which case the {\color{blue}extreme-value} statistics are given by
\begin{equation} 
    {\rm Prob.}[M_{k, N} \leq w] = 1 - \frac{1}{\Gamma(k)} \int_0^\infty \dd t \, e^{-t}\,t^{k-1} \int \dd \vec{y} \, h(\vec{y}) \,  \Theta\left(\lambda_N(w,\vec{y})-t\right) \;,
\end{equation}
where 
\begin{equation} 
    \lambda_N(w, \vec{y}) =  \frac{A(\vec{y}) N}{ \mu(\vec{y}) w^{\mu(\vec{y})} }  \;.
\end{equation}
The gap statistics $d_{k, N} = M_{k, N} - M_{k+1, N} \geq 0$ in the bulk, i.e. $k = \alpha N$ are given by
\begin{equation} \label{eq:main-res-gaps-bulk}
    {\rm Prob.}[d_{k, N} = g] {\color{blue} \underset{N \to +\infty}{\longrightarrow}} \int \dd^M \vec{y} \; h(\vec{y}) \, N p(q(\alpha, \vec{y}) | \vec{y}) e^{- N p(q(\alpha, \vec{y}) | \vec{y}) g} \;.
\end{equation}
If $p(x | \vec{y})$ has exponential tails, then the gap statistics at the edge, i.e., when $k \sim \mathcal{O}(1)$, are also given by Eq.~(\ref{eq:main-res-gaps-bulk}). Finally, the full counting statistics are distributed as 
\begin{equation}
    {\rm Prob.}[N_L = n] {\color{blue} \underset{N \to +\infty}{\longrightarrow}} \int \dd^M \vec{y} \; h(\vec{y}) \delta\left[n - N \int_{-L}^{L} \dd x \; p(x | \vec{y})\right] \;.
\end{equation}

\section{Simultaneously resetting Brownian motion \cite{BLMS23,BLMS24}}

\begin{figure}
    \centering
    \includegraphics[width=0.7\textwidth]{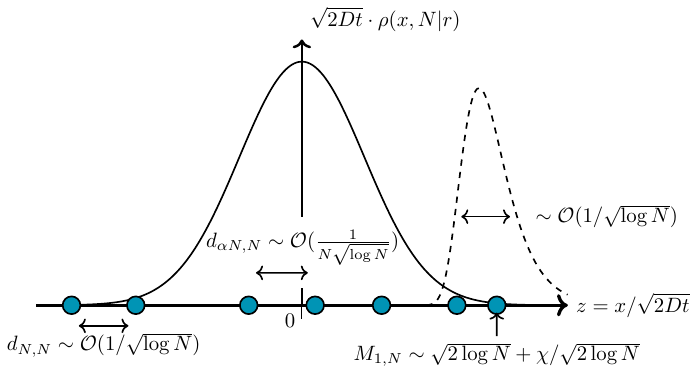}
    \caption{{\color{blue}A summary of the behavior of some key observables of a gas of Brownian motions. Unlike all the resetting processes, free Brownian motion does not admit a stationary state at long times. To obtain static distributions as shown in the figure we have to rescale particle positions by $\sqrt{2 D t}$. The scaled average density profile is plotted with a black line. The distribution of the position $M_{1, N}$ of the rightmost particle is shown schematically by a dashed curve. The centered and rescaled maximum $\chi$ has a Gumbel distribution $p(\chi) = e^{-\chi - e^{-\chi}}$.}} \label{fig:key-freebm}
\end{figure}

\begin{figure}
    \centering
    \includegraphics[width=0.7\textwidth]{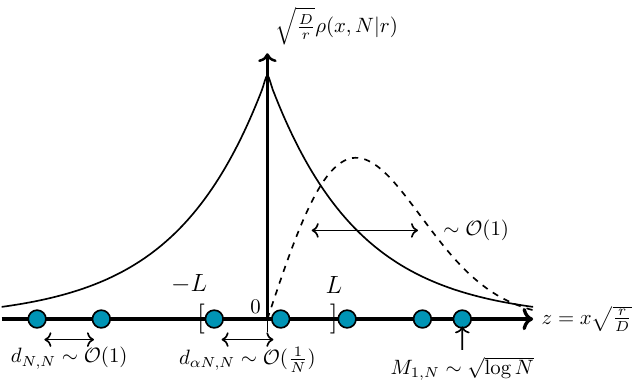}
    \caption{A summary of the behavior of some key observables in the non-equilibrium steady state of a simultaneously resetting gas of Brownian motions. The scaled average density profile is plotted with a black line. The distribution of the position $M_{1, N}$ of the rightmost particle is shown schematically by a dashed curve.} \label{fig:key-simbm}
\end{figure}

In Chapter \ref{ch:sim-reset} in Section \ref{subsec:BM-simreset} we study $N$ independently diffusing Brownian motions $X_1(t), \cdots, X_N(t)$ which are simultaneously reset with rate $r$. {\color{blue} Unlike a free gas of $N$ Brownian motions (depicted in Fig.~\ref{fig:key-freebm}) which does not reach a steady state at long times.} We show that at long times the process reaches a non-equilibrium steady state
\begin{equation}
    {\rm Prob.}[X_1 = x_1, \cdots, X_N = x_N] = r \int_0^{+\infty} \dd \tau\; e^{-r \tau} \prod_{i = 1}^{N} \frac{1}{\sqrt{4 \pi D \tau}} e^{-\frac{x_i^2}{4 D \tau}} \;.
\end{equation}
We compute exactly the behavior of all the above mentioned observables {\color{blue} in the large $N$ limit}; some of the key results are sketched in Fig.~\ref{fig:key-simbm}. Notably, the maximum $M_{1, N}$ is distributed as 
\begin{equation}
    {\rm Prob.}[M_{1, N} = w] \underset{N \to \infty}{\longrightarrow} \sqrt{\frac{r}{4 D \log N}} \; f\left( w \cdot \sqrt{\frac{r}{4 D \log N}} \right) \;,
\end{equation}
where 
\begin{equation}
    f(z) = 2 z e^{-z^2} \;.
\end{equation}
In fact, all the order statistics $M_{k, N}$ have the same scaling function $f(z)$ but with different scale factors. {\color{blue} Notice that the maximum is distributed over the entire positive half-line, unlike the reset-free gas (see Fig.~\ref{fig:key-freebm}) whose maximum is concentrated around $\sqrt{4 D t \log N}$. The distribution is completely modified, not belonging to any of the known Gumbel/Fréchet/Weibull classes. The gas is also more `spread-out' with gaps typically of order $1$ at the edge and $1/N$ in the bulk.}

\begin{figure}
    \centering
    \includegraphics[width=0.7\textwidth]{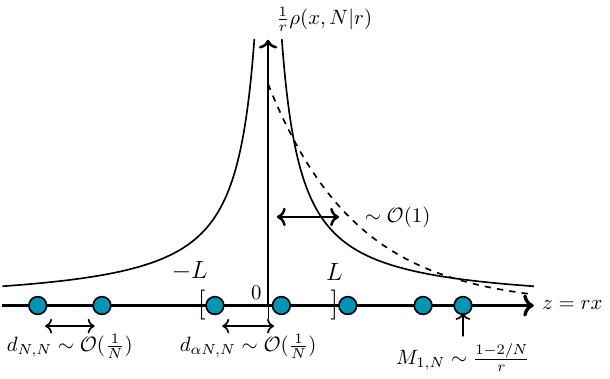}
    \caption{A summary of the behavior of some key observables in the non-equilibrium steady state of a simultaneously resetting gas of ballistic motions. The scaled average density profile is plotted with a black line. The distribution of the position $M_{1, N}$ of the rightmost particle is shown schematically by a dashed curve.} \label{fig:key-sim-ballistic}
\end{figure}

\section{Simultaneously resetting Ballistic particles \cite{BLMS24}}

In Chapter \ref{ch:sim-reset} in Section \ref{subsec:ballistic-simreset} we consider $N$ ballistic particles $X_1(t)$, $\cdots$, $X_N(t)$ each with a random velocity distributed uniformly in $[-1, 1]$ that are reset simultaneously with rate $r$. At each reset event, new velocities are re-sampled. This process also reaches a non-equilibrium steady state at long times
\begin{equation}
    {\rm Prob.}[X_1 = x_1, \cdots, X_N = x_N] = r \int_0^{+\infty} \dd \tau \; e^{-r\tau} \prod_{i = 1}^N \mathds{1}_{-1 \leq \frac{x_i}{\tau} \leq 1} \;.
\end{equation}
We compute exactly the behavior of all the above-mentioned observables, some of the key results are sketched in Fig.~\ref{fig:key-sim-ballistic}. The average density {\color{blue}, for any $N$,} is surprisingly non-trivial
\begin{equation}
    \rho(x) = - \frac{r}{2} {\rm Ei}(-r|x|) \;,
\end{equation}
where ${\rm Ei}(z)$ is the exponential integral function. On the other hand, the maximum $M_{1, N}$ - and more generally all the order statistics $M_{k, N}$ - are distributed exponentially {\color{blue}, in the large $N$ limit}. 
{\color{blue} Notice that the maximum is distributed over the entire positive half-line and is typically much closer $\sim \mathcal{O}(1)$ to the origin compared to the diffusing particles $\sim \mathcal{O}(\sqrt{\log N})$, with particles evenly spaced $\sim 1/N$ throughout the gas.}

\section{Simultaneously resetting Lévy flights \cite{BLMS24}}

\begin{figure}
    \centering
    \includegraphics[width=0.7\textwidth]{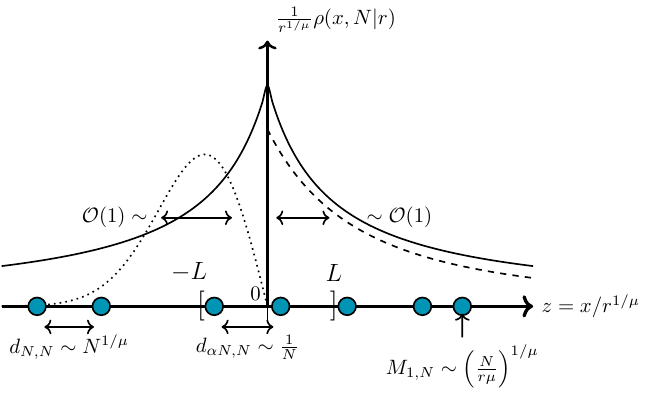}
    \caption{A summary of the behavior of some key observables in the non-equilibrium steady state of a simultaneously resetting gas of Lévy flights {\color{blue} with $\mu = 1$}. The scaled average density profile is plotted with a black line. The scaling function of the position $M_{1, N}$ of the rightmost particle is shown schematically by a dashed curve and the scaling function of the order statistics $M_{k, N}$ in the bulk are plotted with a dotted line.} \label{fig:key-sim-levy}
\end{figure}

In Chapter \ref{ch:sim-reset} in Section \ref{subsec:levy-simreset} we consider $N$ Lévy flights $X_1(t), \cdots, X_N(t)$ (for a detailed description of what is a Lévy flight refer to Section \ref{subsec:levy-simreset} itself) simultaneously resetting with rate $r$. At long times this system also admits a \NESS
\begin{equation}
    {\rm Prob.}[X_1 = x_1, \cdots, X_N = x_N] = r \int_0^{+\infty}\dd \tau\; e^{-r \tau} \prod_{i = 1}^N \frac{1}{\tau^{1/\mu}} \mathcal{L}_\mu \left(\frac{x_i}{\tau^{1/\mu}}\right) \;,
\end{equation}
where $\mathcal{L}_\mu(z)$ is a Lévy stable law of index $\mu$. We compute exactly the behavior of all the observables mentioned above, some of the key results are sketched in Fig.~\ref{fig:key-sim-levy}. The average density {\color{blue} for any $N$} is fat-tailed
\begin{equation}
    \rho(x) \underset{|x| \to \infty}{\longrightarrow} \frac{r^{1/\mu}}{2 \pi} \frac{1}{| r^{1/\mu} x |^{1 + \mu}} \;,
\end{equation}
and the {\color{blue} large $N$ limit of the} order statistics $M_{k, N}$ in the bulk, i.e. for $k = \alpha N$, are given by 
\begin{equation}
    {\rm Prob.}[M_{k, N} = w] \underset{N \to \infty}{\longrightarrow} \frac{r^{1/\mu}}{\beta_\mu} \; f_\mu\left(w \cdot \frac{r^{1/\mu}}{\beta_\mu}\right) \;,
\end{equation}
where $\beta_\mu$ is a scaling constant which we detail in Section \ref{subsec:levy-simreset} and the scaling function $f_\mu(z)$ defined for $z \geq 0$ is given by
\begin{equation}
    f_\mu(z) = \mu z^{\mu - 1} e^{-z^\mu} \;.
\end{equation}
Notice that for $\mu = 2$ we recover the Brownian result. However, the maximum, unlike in the previous two cases, does not have the same scaling form as the order statistics. Instead
\begin{equation}
    {\rm Prob.}[M_{1, N} = w] \underset{N \to \infty}{\longrightarrow} \left( \frac{r \mu}{G N} \right)^{1/\mu} \mathcal{S}_{\mu}\left[w \left(\frac{r \mu}{G N}\right)^{1/\mu}\right] \;,
\end{equation}
where $G$ is once again some scaling constant and the scaling function $\mathcal{S}_\mu(z)$ defined for $z \geq 0$ is given by
\begin{equation}
    \mathcal{S}_\mu(z) = \frac{\mu z^{\mu - 1}}{1 + z^\mu} \;.
\end{equation}
{\color{blue} Notice in Fig.~\ref{fig:key-sim-levy} that the first gap is as big as the maximum $\sim \mathcal{O}(N^{1/\mu})$. This is symptomatic of heavy-tailed distributions, extreme events are so probable that the maximum completely dominates any other typical value.}

\section{First-Passage simultaneous resetting [unpublished]}

\begin{figure}
    \centering
    \includegraphics[width=0.7\textwidth]{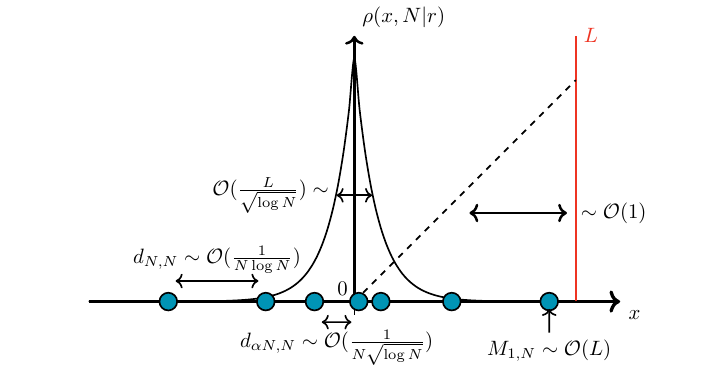}
    \caption{A summary of the behavior of some key observables in the non-equilibrium steady state of a gas of Brownian motions which reset simultaneously whenever any of them reach a target located at $L > 0$ (sketched in red). The scaled average density profile is plotted with a black line. The distribution of the position $M_{1, N}$ of the rightmost particle is shown schematically by a dashed curve.} \label{fig:key-sim-fpt}
\end{figure}

In Chapter \ref{ch:sim-reset} in Section \ref{subsec:fpt-simreset} we consider a gas of $N$ Brownian motions $X_1(t), \cdots, X_N(t)$ and whenever {\it any} of them reach a fixed target located at $L > 0$ they {\it all} reset to the origin. In the large $N$ limit, this system also reaches a \NESS at long times
\begin{equation}
    {\rm Prob.}[\vec{X} = \vec{x}] \underset{N \to \infty}{\longrightarrow} \int_1^{+\infty} \dd u \; \frac{2}{u^3} \prod_{i = 1}^N \frac{u}{L} \sqrt{\frac{\log N}{\pi}} \exp[ - \frac{u^2 \log N x_i^2}{L^2} ] \;.
\end{equation}
We compute exactly the behavior of all the above mentioned observables, some of the key results are sketched in Fig.~\ref{fig:key-sim-fpt}. Strikingly nearly all particles seem to condense on a length scale $\sim L / \sqrt{\log N}$ around the origin except for the {\color{blue}extreme-value}s which are distributed as
\begin{equation}
    {\rm Prob.}[M_{1, N} = w] \underset{N \to \infty}{\longrightarrow} \frac{2 w}{L^2} \mathds{1}_{0 \leq w \leq L} \;.
\end{equation}
{\color{blue} Notice that the maximum is typically `stuck' to the hard edge $x = L$ of the gas, meaning that the gas typically operates on the verge of resetting keeping most particles concentrated in a region $\sim 1/\sqrt{\log N}$ around the origin. Physically, this means that the gas is reset frequently by a spurious trajectory shooting straight for the edge while the other particles diffuse normally around the origin.}

\section{Non-interactive diffusive particles in a switching harmonic trap \cite{BKMS24}}

\begin{figure}
    \centering
    \includegraphics[width=0.49\textwidth]{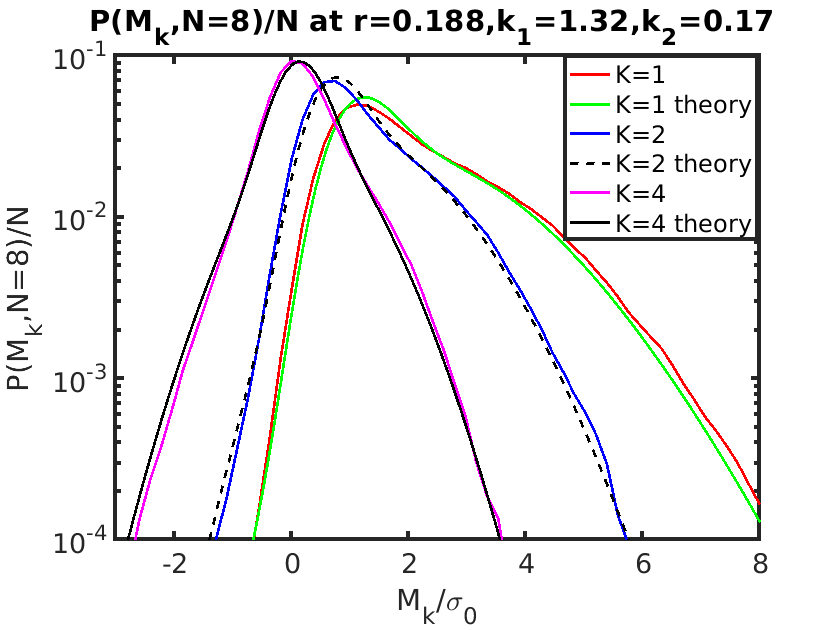}
    \includegraphics[width=0.49\textwidth]{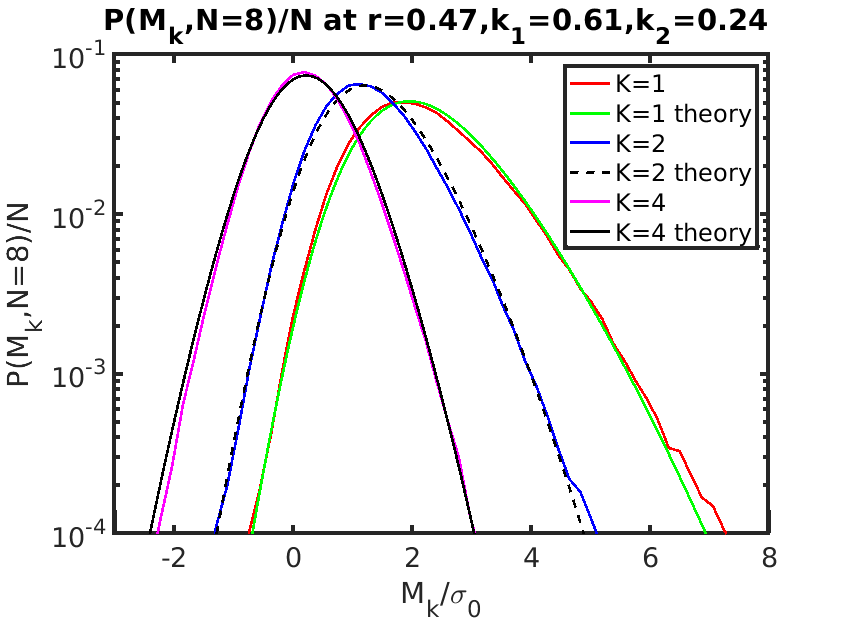}
    \includegraphics[width=0.49\textwidth]{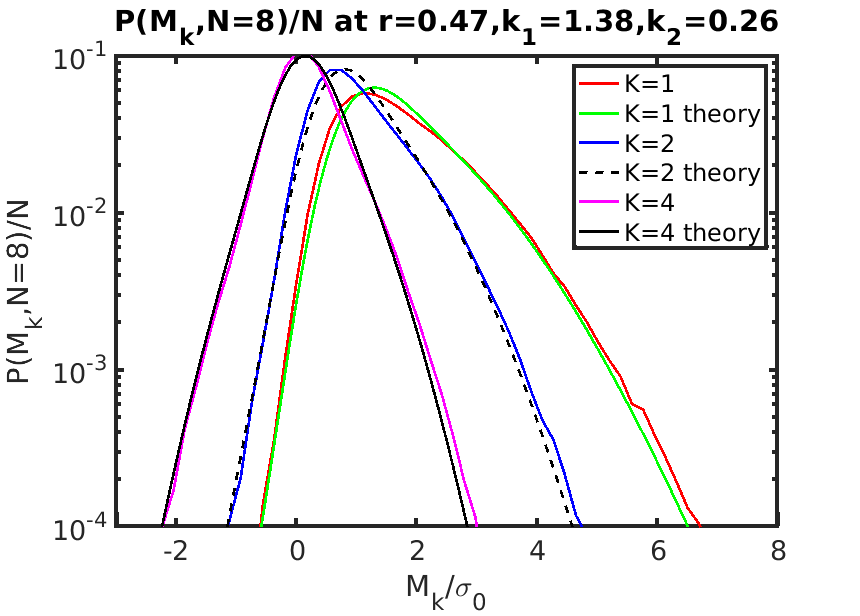}
    \includegraphics[width=0.49\textwidth]{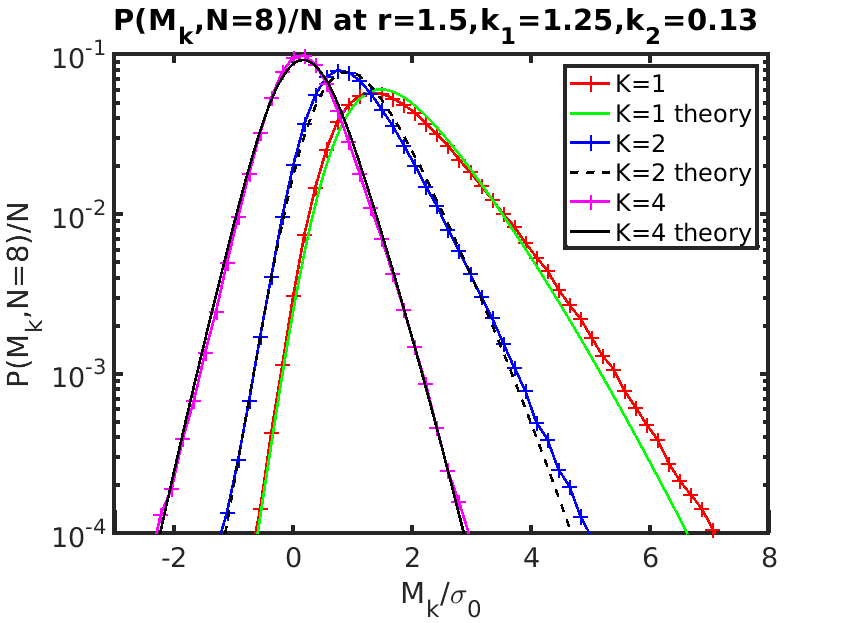}
    \caption{The experimental measurements for the distribution of the maximum $M_{1, N}$ for 2,4 and 8 particles are compared
    with the theoretical predictions obtained which follow from the analytical treatment in Chapter \ref{ch:ou-switch}. These are experimental results from Ref.~\cite{BCKMPS25}. Each panel corresponds to different resetting rates $r$, and stiffnesses $k_1, k_2$ and each line corresponds to different number of {\color{blue} particles} $N$. Theoretical predictions and experimental results match perfectly.} \label{fig:main-res-exp-ou}
\end{figure}

In Chapter \ref{ch:ou-switch} we study a gas of $N$ independently diffusing particles $X_1(t)$, $\cdots$, $X_N(t)$ in a harmonic trap of stiffness $\mu(t)$ which switches between two stiffness $\mu_1 > \mu_2$ with rate $r_1$ to go from $\mu_1$ to $\mu_2$ and rate $r_2$ when going from $\mu_2$ to $\mu_1$. This model was inspired by experimental considerations and completely unexpectedly the long time \NESS presented a conditionally independent structure
\begin{equation}
{\rm Prob.}[X_1 = x_1, \cdots, X_N = x_N] = \int_0^1 {\rm d}u\, h(u) \prod_{i=1}^N p_0(x_i, u) \;,
\end{equation}
where 
\begin{equation}
h(u) = \frac{c\,r_H}{4} u^{R_1 - 1} (1 - u)^{R_2 - 1} \left[ \frac{1 - u}{\mu_1} + \frac{u}{\mu_2} \right] 
\end{equation}
with $c$ being some constant, $r_H = 2 \, r_1 r_2/(r_1+r_2)$ and 
\begin{equation} 
    p_0(x, u) = \frac{1}{\sqrt{2 \pi V(u)}} \; e^{- \frac{x^2}{2 V(u)}} \;,
\end{equation}
where 
\begin{equation}
    V(u) = D\,\left( \frac{u}{\mu_2} + \frac{1 - u}{\mu_1} \right) \;.
\end{equation}
Thanks to this conditionally independent form we were able to compute the same observables as described above which have very non-trivial expressions. Most importantly, as can be seen in Fig. \ref{fig:main-res-exp-ou}, our theoretical predictions perfectly match the experimental data obtained from the setup that inspired this process. {\color{blue} We will briefly discuss the experimental setup used to obtain these figures in Section \ref{sec:experimental-setup}.}

\section{{\color{blue} Logarithmically} repelling diffusive particles with simultaneous resetting \cite{BMS25}}

\begin{figure}
    \includegraphics[width=0.5\textwidth]{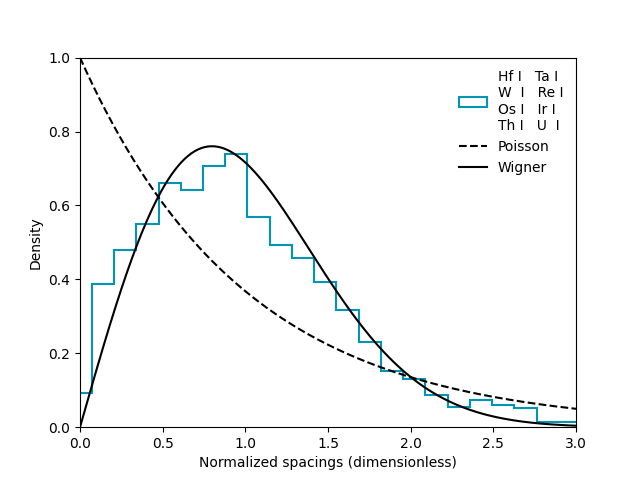}%
    \includegraphics[width=0.5\textwidth]{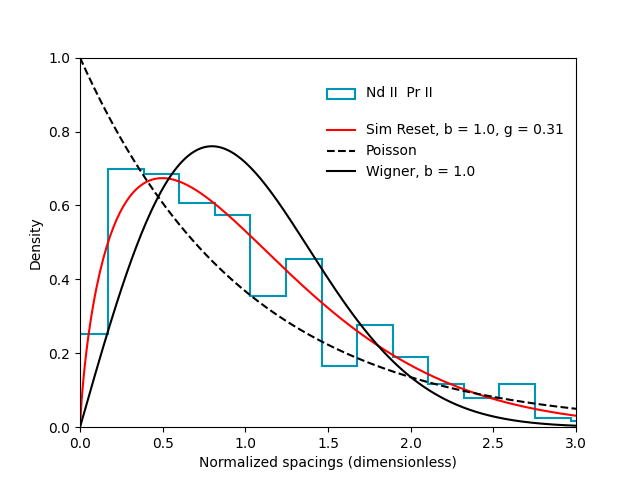}
    \caption{Experimental data (obtained from the Atomic Spectral Database \cite{ASD}) for the level spacings of Hafnium (Hf I), Tantalum (Ta I), Tungsten (W I), Rhenium (Re I), Osmium (Os I), Iridium (Ir I), Thorium (Th I), Uranium (U I), the ion of Neodymium (Nd II) and the ion of Praseodymium (Pr II) compared to the independent hypothesis (Poisson) or the Wigner surmise for the gaussian orthogonal ensemble ($\beta = 1$) and the gap statistics for a simultaneously resetting log-gas ({\color{blue}with} $\beta = 1$ and $\gamma = \mu/r = 0.31$). 
    } \label{fig:main-atomic}
\end{figure}

In Chapter \ref{ch:dyson} we consider $N$ diffusive particles inside of a harmonic trap of stiffness $\mu$ which repel each other with a pairwise logarithmic repulsion, on top of which we apply simultaneous resetting with Poissonian rate $r$. In the long-time limit it admits a \NESS with strong long-range attractive and repulsive correlations competing with each other. As we will see, this process is intricately linked to random matrix theory and exploiting this link we are able to study several observables analytically. We find that the {\color{blue}extreme-value} statistics are given by
\begin{equation}
    {\rm Prob.}[M_{1, N} = w] \underset{N \to \infty}{\longrightarrow} \sqrt{\frac{\mu}{N D}} \, f\left(w \cdot \sqrt{\frac{\mu}{N D}}, \frac{\mu}{r}\right) \;,
\end{equation}
where the normalized scaling function $f(z, \gamma)$ is supported over $z \in [0, \sqrt{2}]$ (strictly for $\gamma > 0$) and is given by
\begin{equation}
    f(z, \gamma) = \frac{z}{2 \gamma} \left(1 - \frac{z^2}{2}\right)^{\frac{1}{2\gamma} - 1} \;.
\end{equation}
For those familiar with random matrix theory, {\color{blue} the extreme-value statistics described above} are extremely different from the usual behavior typically described by the Tracy-Widom distribution {\color{blue} \cite{M91,F10,G58,TW94,TW96}}. The spacing distribution will also be significantly modified from the usual Wigner surmise. Furthermore, being able to tune the competition between the attractive and repulsive correlations we are able to obtain spacing distributions which are able to fit atomic spacings that could not be described by the Wigner surmise, see Fig. \ref{fig:main-atomic}.

\section{Critical number of walkers for diffusive search processes with resetting \cite{BMS23}}

In Chapter \ref{ch:search} we study the {\color{blue}first-passage} properties of $N$ diffusing particles with resetting. Specifically, we look at whether Poissonian resetting can lower the mean {\color{blue}first-passage} time to a fixed target, i.e. can resetting improve the search process. We see that if resetting is performed independently for each particle, then it will be useful as long as $N \leq 7$, if $N > 7$ resetting hinders the search. On the other hand, if the resetting is performed simultaneously for all particles, then the resetting will be useful as long as $N \leq 6$, if $N > 6$ simultaneous resetting will hinder the search process.

\section{Resetting by rescaling \cite{BFHMS24}}

Finally, the last model considered in this thesis is a generalization of resetting motivated by practical applications. Instead of resetting to the origin, a particle $X(t)$ is rescaled $X(t) \to a X(t)$ with a constant factor $a \in ]-1, 1[$ at random Poissonian times. The usual resetting process can be recovered by setting $a = 0$. We study the mean {\color{blue}first-passage} time of this system to a fixed target at $L > 0$ and show that for negative $a$, that is, $-1 < a < 0$, the mean {\color{blue}first-passage} time is lower than that of resetting, that is, $a = 0$. The lowest mean {\color{blue}first-passage} time is achieved for $a \to -1^{+}$, the caveat being that when $a \to -1^{+}$ we must also take the reset rate $r \to +\infty$.

\chapter{Extreme value statistics, Order statistics, Gap statistics and Full Counting Statistics} \label{ch:extreme}
In this chapter, we introduce the main class of observables studied throu\-ghout the thesis. We will start by reviewing the known results for independent identically distributed random variables in Section \ref{sec:iid}. We will then move to correlated random variables in Section \ref{sec:correlated}, extending independent identically distributed results from Section \ref{sec:iid} to weakly correlated variables in Section \ref{subsec:weak}. Finally, in Section \ref{subsec:strong} we introduce random walks which are a special solvable case of strongly correlated variables and the backbone of all the models presented in this thesis.

\vspace{0.2cm}

The aim of this chapter is not to provide a comprehensive review of all existing results in the literature, more so to recall the main results laying the foundations for the results in the upcoming chapters. For a more complete description of the topic we recommend Ref. \cite{MS24}. \vspace{0.2cm}

\vspace{0.2cm}

\begin{figure}
    \centering
    \includegraphics[width=\textwidth]{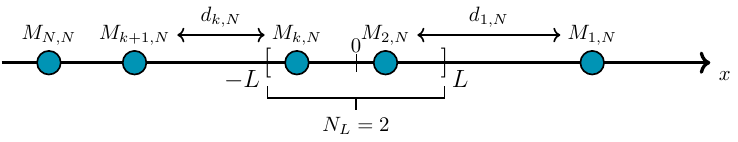}
    \caption{{\color{blue} We recall here Fig.~\ref{fig:main-res-observables} for easy access.} A sketch of the observables studied. We represent the set of random variables $\{X_1, \cdots, X_N\}$ as the positions of $N$ one-dimensional particles on the line. The observables studied are the $k$-th maximum $M_{k, N}$, the gaps $d_{k, N} = M_{k, N} - M_{k + 1, N}$ and the full counting statistics $N_L$ which are the number of particles in a box $[-L, L]$ around the origin.} \label{fig:observables}
\end{figure}

We start by introducing the observables which we will study in this chapter. Let $X_1, \cdots, X_N$ be $N$ random variables. For our purposes, we will suppose that $X_i$ are real or integer valued random variables, and in physical models they will always represent the positions of one-dimensional particles on a line or lattice. Hence we commonly refer to the set of random variables $\{X_1, \cdots, X_N\}$ as `the gas' or `the gas of particles'. A sketch of this representation is provided in Fig. \ref{fig:observables}. We will denote by 
\begin{equation} \label{eq:def-max}
M_{1, N} = \max_{i = 1, \cdots, N} X_i
\end{equation}
the maximum and 
\begin{equation} \label{eq:def-min}
M_{N, N} = \min_{i = 1, \cdots, N} X_i
\end{equation}
the minimum. 
{\color{blue} These are known as the extreme-value statistics and they are of critical importance in a variety of applications. Although their behavior represents rare/extreme configurations of the gas, extreme events can also have extreme consequences, such as overpopulation of a species leading to an extinction or a stress reaching critical levels and triggering an earthquake.}
More generally, we denote by $M_{k, N}$ the $k$-th maximum, i.e. the maximum of $\{X_1, \cdots, X_N\}$ excluding $\{M_{1,N}, \cdots, M_{k - 1, N}\}$. Observables $M_{1, N} \geq M_{2, N} \geq \cdots \geq M_{N, N}$ are the order statistics of $\{X_1, \cdots, X_N\}$. Looking at $M_{k, N}$ for values of $k$ close to 1 (or $N$), we are probing the gas close to its maximum (resp. minimum), i.e. we are studying the {\color{blue}extreme-value} statistics of the gas. The other observables that we will study are the gaps 
\begin{equation} \label{eq:def-gaps}
d_{k, N} = M_{k, N} - M_{k + 1, N} \geq 0
\end{equation} 
between two successive particles, and the full counting statistics $N_L$ corresponding to the number of particles in a box $[-L, L]$ around the origin, i.e. 
\begin{equation} \label{eq:def-fcs}
N_L = \# \{ X_i : X_i \in [-L, L] \} \;.
\end{equation}
All these observables are sketched in Fig. \ref{fig:observables}.

\section{Independent identically distributed random variables} \label{sec:iid}

We first consider the case of $N$ independent identically distributed random variables and recall some known results from probability theory \cite{MS24, G58,ABN92,ND03,MPS20,SM14,G43}. Let $\{X_1, \cdots, X_N\}$ be $N$ independent random variables with law ${\rm Prob.}[X_i = x] = p(x)$, i.e.
\begin{equation} \label{eq:def-independence}
    {\rm Prob.}[X_1 = x_1, \cdots, X_N = x_n] = \prod_{i = 1}^N {\rm Prob.}[X_i = x_i] = \prod_{i = 1}^N p(x_i)\;.
\end{equation}

\subsection{The scaled sum}

Perhaps the most well-known result for independent identically distribut\-ed random variables is the central limit theorem which characterizes the distribution of the scaled sum of $N$ independent identically distributed random variables in the large $N$ limit. If the probability distribution function $p(x)$ has a finite first and second moment
\begin{equation} \label{eq:def-mean-X}
m = \langle X_i \rangle = \int \dd x \; x\, p(x) \;,
\end{equation}
and 
\begin{equation} \label{eq:def-sigma-X}
\sigma^2 = \langle X_i^2 \rangle - \langle X_i \rangle^2 = \int \dd x\; x^2\, p(x) - \left(\int \dd x\; x\, p(x) \right)^2 \;,
\end{equation}
where the integrals are over the support of $p(x)$. The central limit theorem states that, for large $N$, the rescaled sum (which physically would correspond to the ``center of mass'' and is sometimes referred to as the sample mean) defined as 
\begin{equation}\label{eq:CN-def}
    C_N = \frac{1}{N} \sum_{i = 1}^N X_i\;,
\end{equation}
converges to a Gaussian distribution  
\begin{equation} \label{eq:clt}
{\rm Prob.}[C_N = c] \underset{N \to +\infty}{\longrightarrow} \sqrt{\frac{N}{2 \pi \sigma^2}} \exp\left( - N \frac{(c - m)^2}{2 \sigma^2} \right) \;,
\end{equation}
centered at $m$ with a variance $\sigma^2/N$. When the probability density function $p(x)$ has a diverging first or second moment, e.g., when $p(x)$ is heavy-tailed, such as $p(x) \sim x^{-1-\mu}$ for large $x$ with $0 < \mu <2$, the centered and scaled sum converges to a L\'evy stable law with index $\mu$~\cite{F71, F91}. 

\subsection{The maximum}
We start by looking at the maximum $M_{1, N} = \max_{i = 1,\cdots, N} X_i$. The cumulative distribution of the maximum is particularly simple, for the maximum to be lower than a threshold, all the particles must be below the threshold
\begin{equation} \label{eq:max-to-X}
    {\rm Prob.}[M_{1, N} < w] = {\rm Prob.}[X_1 < w, X_2 < w, \cdots, X_N < w].
\end{equation}
Since the $X_i$'s are independent we can use Eq. (\ref{eq:def-independence}) to simplify Eq. (\ref{eq:max-to-X}) to
\begin{equation} \label{eq:max-iid-integral-form}
    {\rm Prob.}[M_{1, N} < w] = \left( \int_{-\infty}^{w} \dd x \; p(x)  \right)^N \;.
\end{equation}
A careful analysis \cite{G58} of the asymptotic behavior of Eq. (\ref{eq:max-iid-integral-form}) when $N \to +\infty$ shows that the maximum can be expressed as
\begin{equation} \label{eq:evs-universal-form}
    M_{1, N} \sim a_N + b_N \chi \;, \mbox{~~when~~} N \to +\infty \;,
\end{equation}
where $a_N$ and $b_N$ are scaling coefficients and $\chi \sim \mathcal{O}(1)$ is a random variable. The coefficient $a_N$ encodes the typical value of the maximum and $b_N$ is the scale of the fluctuations of the maximum. While $a_N$ and $b_N$ depend on the details of the distribution $p(x)$ the random variable $\chi$ has only three possible universal distributions: Gumbel, Fréchet and Weibull. The shape of the tails of $p(x)$ will dictate which of the three $\chi$ belongs to. This is the famous Fisher-Tippet-Gnedenko theorem.
\begin{enumerate}
    \item[I] {\bf The Gumbel universality class.\\} 
    If the probability density $p(x)$ decays faster than any power-law for large $x$, i.e. if
    \begin{equation} \label{eq:gumbel-trail-definition}
        \mbox{when~~} x \gg 1 \mbox{~~then~~} p(x) \ll x^{-\mu} \mbox{~~for any~~} \mu > 0 \;,
    \end{equation}
    then $p(x)$ is said to belong to the Gumbel universality class. In this case $a_N$ and $b_N$ are given by
    \begin{equation} \label{eq:gumbel-coefs}
        \frac{1}{N} = \int_{a_N}^{+\infty} \dd x \; p(x) \mbox{~~and~~} b_N = N \int_{a_N}^{+\infty} \dd x \; (x - a_N) p(x) \;,
    \end{equation}
    and $\chi$ follows a Gumbel distribution, namely
    \begin{equation} \label{eq:gumbel-cumulative}
        {\rm Prob.}[\chi < z] = G_{\rm I}(z) = e^{- e^{-z}} .
    \end{equation}
    \item[II] {\bf The Fréchet universality class.\\}
    If the probability density $p(x)$ has an unbounded support and decays as a power law for large $x$, i.e. if
    \begin{equation} \label{eq:frechet-tail-definition}
        \mbox{when~~} x \gg 1 \mbox{~~then~~} p(x) \simeq x^{-1-\mu} \mbox{~~for a given~~} \mu > 0 \;,
    \end{equation}
    then $p(x)$ is said to belong to the Fréchet universality class. In this case
    \begin{equation} \label{eq:frechet-coefs}
        a_N = 0 \mbox{~~and~~} \int_{b_N}^{+\infty} \dd x \; p(x) = \frac{1}{N} \;,
    \end{equation}
    and $\chi$'s cumulative distribution is given by
    \begin{equation} \label{eq:frechet-cumulative}
        {\rm Prob.}[\chi < z] = G_{\rm II}(z) = \Theta(z) e^{-z^{-\mu}} \;,
    \end{equation}
    where $\Theta(z)$ is the Heaviside distribution.
    \item[III] {\bf The Weibull universality class.\\}
    If the probability density $p(x)$ has a support bounded from above by $x^\star$ and decays as a power-law when approaching $x^\star$, i.e. if
    \begin{equation} \label{eq:weibull-tail-definition}
        \mbox{when~~} x^\star - x \ll 1 \mbox{~~then~~} p(x) \simeq (x^\star - x)^{\mu - 1} \mbox{~~for a given~~} \mu > 0\;, 
    \end{equation}
    then $p(x)$ is said to belong to the Weibull universality class. In this case
    \begin{equation} \label{eq:weibull-coefs}
        a_N = x^\star \mbox{~~and~~} \int_{x^\star - b_N}^{x^\star} \dd x \; p(x) = \frac{1}{N} \;,
    \end{equation}
    and $\chi$'s cumulative distribution is given by
    \begin{equation} \label{eq:weibull-cumulative}
        {\rm Prob.}[\chi < z] = G_{\rm III}(z) = \begin{dcases}
        1 & \mbox{~~if~~} z > 0 \\
        e^{-|z|^\mu} &\mbox{~~otherwise.}
        \end{dcases}
    \end{equation}
\end{enumerate}

\begin{figure}
    \centering
    \includegraphics[width = 0.5\textwidth]{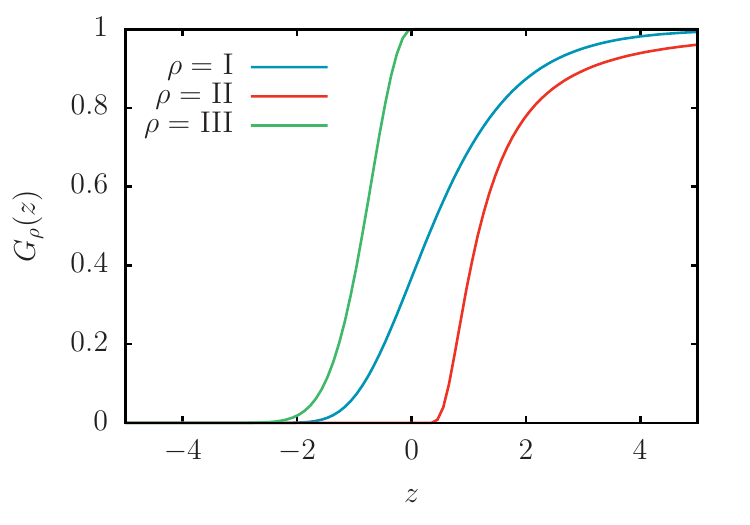}\hfill
    \includegraphics[width = 0.5\textwidth]{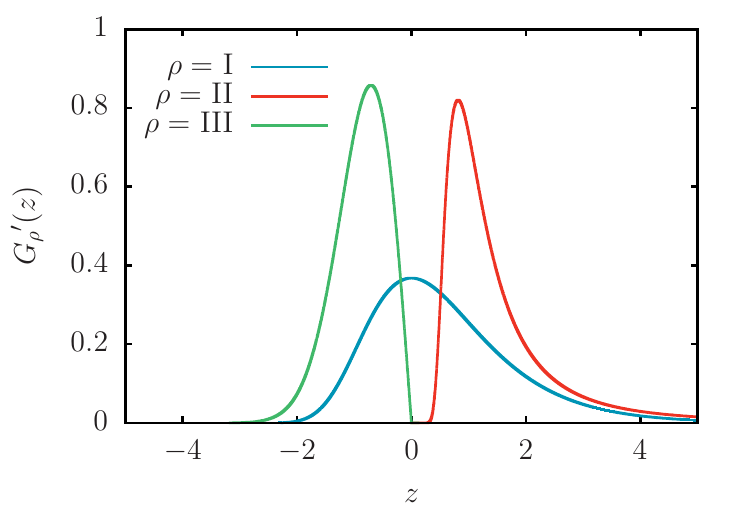}
    \caption{A plot of the Gumbel, Fréchet and Weibull universal {\color{blue}extreme-value} statistics cumulative distributions $G_{\rm I}(z)$, $G_{\rm II}(z)$ and $G_{\rm III}(z)$ on the left, and their derivatives, i.e. the probability density functions, on the right. The equations of the cumulative distributions $G_\rho(z)$ are given in Eqs. (\ref{eq:gumbel-cumulative}), (\ref{eq:frechet-cumulative}) and (\ref{eq:weibull-cumulative}) respectively. Both the Fréchet and Weibull distributions are plotted with $\mu = 2$.} \label{fig:evs-universal-functions}
\end{figure} 

Note that if $p(x)$ has a support bounded above by $x^\star$ but decays faster than any power law when approaching $x^\star$ then the distribution of the maximum is once again given by the Gumbel distribution but the integrals defining the {\color{blue} scale-coefficient in Eq. (\ref{eq:gumbel-coefs}) need} to be adapted to account for the finite support. Furthermore, these descriptions only account for the typical fluctuations of the maximum. To describe atypical fluctuations of the maximum one has to resort to the theory of large deviations (see. Ref. \cite{V15} and the lecture notes \cite{T24,V24,E24} for a pedagogical review). 

\subsection{The order statistics}
A natural generalization is to consider the more general order statistics instead of only the maximum \cite{ABN92,ND03}, i.e. studying the behavior of the $k$-th maximum $M_{k, N}$. The probability density function of $M_{k, N}$ can be expressed simply in terms of $p(x)$ as
\begin{align} \label{eq:order-iid-integral-form-1}
    {\rm Prob.}[M_{k, N} = w] = &\binom{N}{k-1} (N - k + 1) p(w) \\
    &\quad \times \left(\int_{w}^{+\infty} \dd x \; p(x)\right)^{k - 1} \left( \int_{-\infty}^{w} \dd x \; p(x) \right)^{N - k} \;.
    \label{eq:order-iid-integral-form-2}
\end{align}
This equation can be understood as follows. For the $k$-th maximum to be at $w$, we must choose $k - 1$ particles to be above $w$, which gives rise to the binomial term in Eq. (\ref{eq:order-iid-integral-form-1}) and the first integral term in Eq. (\ref{eq:order-iid-integral-form-2}). Then we must choose one of the remaining $(N - k + 1)$ particles to place at $w$, which corresponds to the two remaining terms in Eq. (\ref{eq:order-iid-integral-form-1}). Finally, we have to place the remaining $N - k$ particle below $w$ which gives rise to the second and last integral term in Eq. (\ref{eq:order-iid-integral-form-2}).

\vspace{0.2cm}

To extract the asymptotic behavior of Eqs.~(\ref{eq:order-iid-integral-form-1}-\ref{eq:order-iid-integral-form-2}) we need to distinguish whether we are probing the edge, i.e. close to the {\color{blue}extreme-value}s $k \sim \mathcal{O}(1)$ or in the bulk, i.e. $k = \alpha N$ with $\alpha \in ]0, 1[$. If we place ourselves in the bulk, i.e we take $k = \alpha N$ then we re-write Eqs.~(\ref{eq:order-iid-integral-form-1}-\ref{eq:order-iid-integral-form-2}) as
\begin{equation} \label{eq:iid-order-alpha}
{\rm Prob.}[M_{\alpha N, N} = w] = \frac{\Gamma(N+1)}{\Gamma(\alpha N) \Gamma(N(1 - \alpha) + 1)} \frac{p(w)}{\int_{w}^{+\infty} p(x) \dd x} e^{- N \Phi_\alpha(w)} \;,
\end{equation}
where we re-wrote the combinatorial factor using the Gamma function $\Gamma(z)$ and we introduced the function 
\begin{equation} \label{eq:def_phi}
\Phi_\alpha(w) = -\alpha \ln\left( \int_{w}^{+\infty} p(x) \dd x \right) - (1 - \alpha) \ln \left( \int_{-\infty}^{w} p(x) \dd x \right) \;.
\end{equation}
So far, we have not taken the large $N$ limit. When $N \to \infty$, the probability density function in Eq. (\ref{eq:iid-order-alpha}) gets sharply concentrated around the value $w = q(\alpha)$ that minimizes $\Phi_\alpha(w)$. Setting $\Phi_\alpha'(w = q(\alpha)) = 0$ yields 
\begin{equation} \label{eq:def_wstar}
\int_{q(\alpha)}^{+\infty} p(x) \; \dd x = \alpha \;.
\end{equation} 
Notice that this corresponds to the definition of the $\alpha$-quantile of the distribution $p(x)$. Expanding $\Phi_\alpha(w)$ around $q(\alpha)$ up to quadratic order, one finds that for large $N$ and close to $q(\alpha)$, the probability density function defined in Eq.~(\ref{eq:iid-order-alpha}) simplifies to
\begin{equation} \label{eq:iid-order-bulk-gaussian}
\text{Prob.}[M_{\alpha N, N} = w] \underset{N \to \infty}{\longrightarrow} \sqrt{\frac{N \left[p(q(\alpha))\right]^2 }{2 \pi \alpha (1 - \alpha)}} \exp\left( - \frac{N \left[p(q(\alpha))\right]^2}{2 \alpha(1 - \alpha)} [w - q(\alpha)]^2 \right) \;.
\end{equation}
Eq. (\ref{eq:iid-order-bulk-gaussian}) is simply a Gaussian distribution centered around the $\alpha$-quantile $q(\alpha)$ with variance 
\begin{equation} \label{eq:iid-order-variance-bulk}
{\rm Var}[M_{\alpha N, N}] = \frac{\alpha(1-\alpha)}{N \left[p(q(\alpha))\right]^2} \;. 
\end{equation}
In other words, the appropriately centered and scaled order statistics will have a Gaussian distribution
\begin{equation}
    M_{\alpha N, N} \sim a_{\alpha N, N} + b_{\alpha N, N} \; \eta \;, 
\end{equation}
where 
\begin{equation} \label{eq:bulk-coefs}
    \alpha = \int_{a_{\alpha N, N}}^{+\infty} \dd x\; p(x) \mbox{~~and~~} b_{\alpha N, N} = \sqrt{\frac{\alpha(1 - \alpha)}{N p(a_{\alpha N, N})^2}}
\end{equation}
and $\eta \sim \mathcal{N}(0, 1)$, i.e.
\begin{equation}
    {\rm Prob.}[\eta] = \frac{1}{\sqrt{2 \pi}} e^{-\frac{\eta^2}{2}}\;.
\end{equation}
On the other hand, if we place ourselves at the edge, i.e. $k \sim \mathcal{O}(1)$, a careful analysis \cite{ABN92,ND03} of the asymptotic behavior of Eqs. (\ref{eq:order-iid-integral-form-1}-\ref{eq:order-iid-integral-form-2}) in the large $N$ limit reveals that the $k$-th maximum can be expressed as
\begin{equation} \label{eq:order-universal-form}
    M_{k, N} \sim a_{N} + b_{N} \chi_k, \mbox{~~when~~} N \to +\infty \;.
\end{equation}
The scale coefficients $a_{N}$ and $b_{N}$ are the same as the ones defined in Eqs.~(\ref{eq:gumbel-coefs}, \ref{eq:frechet-coefs}, \ref{eq:weibull-coefs}), and surprisingly $\chi_k$ has once again a universal distribution. Given that $p(x)$ is of class $\rho$ ($\rho = $ I, II, III for the Gumbel, Fréchet and Weibull respectively) then the cumulative distribution of $\chi_k$ is given by
\begin{equation} \label{eq:order-cumulative}
    {\rm Prob.}[\chi_k < z] = G^k_{\rho}(z) = G_\rho(z) \sum_{j = 0}^{k-1} \frac{\left[ - \ln G_\rho(z) \right]^j}{j!} \;,
\end{equation}
where $G_\rho(z) = G^1_\rho(z)$ are given in Eq. (\ref{eq:gumbel-cumulative}), Eq. (\ref{eq:frechet-cumulative}) and Eq. (\ref{eq:weibull-cumulative}) for $\rho = $ I, II, and III respectively. For independent identically distributed random variables the order statistics present a significantly different behavior at the edge of the gas, i.e. when $k \sim 1$ compared to the bulk of the gas, i.e. when $k \sim N/2$. The expression in Eq.~(\ref{eq:order-cumulative}) nicely interpolates between the edge behavior where it is a slight modification of the universal {\color{blue}extreme-value} statistics distribution, and the Gaussian behavior in the bulk. Taking the appropriate {\color{blue}asymptotes} of Eq. (\ref{eq:order-cumulative}) yields
\begin{equation} \label{eq:order-cumulative-edge}
    G^k_\rho(z) \simeq \frac{1}{\Gamma(k)} \int_{-\log G_\rho(z)}^{+\infty} \dd t\; e^{-t} t^{k - 1}, \mbox{~~when~~} k \sim 1
\end{equation} 
and 
\begin{equation} \label{eq:order-cumulative-bulk}
    G^k_\rho(z) \simeq \frac{1}{2} \left( 1 + {\rm erf}\left(\frac{z}{\sqrt{2}}\right) \right), \mbox{~~when~~} k \sim N/2 \;. 
\end{equation}

\subsection{The gap statistics}
The next observables we will look at are the gap statistics $d_{k, N} = M_{k, N} - M_{k+1, N} > 0$, i.e. the distance between the $k$-th and $(k+1)$-th maxima. The simplest way to write the probability density function of $d_{k, N}$ is to first write the joint distribution of $M_{k, N}$ and $M_{k+1, N}$ which is given by
\begin{align}
    {\rm Prob.}[M_{k, N} = u, \, &M_{k+1, N} = v] = \binom{N}{k-1} (N-k+1) (N-k) p(u) p(v) \label{eq:joint-order-iid-integral-form-1}\\
    & \times \left(\int_{u}^{+\infty} \dd x \; p(x) \right)^{k-1} \left(\int_{-\infty}^{v} \dd x \; p(x) \right)^{N - k - 1} \;. \label{eq:joint-order-iid-integral-form-2}
\end{align}
This formula has a similar interpretation to Eqs. (\ref{eq:order-iid-integral-form-1}-\ref{eq:order-iid-integral-form-2}), with the exception that we must now choose one other particle among the top $N - k$ particles to place at $v$ and the other particles have to be either above $u$ or below $v$ instead of being above/below $w$ as in Eq. (\ref{eq:order-iid-integral-form-2}). The distribution of the gap statistics can readily be obtained from Eqs. (\ref{eq:joint-order-iid-integral-form-1}-\ref{eq:joint-order-iid-integral-form-2}) by performing a change of variable and integrating out the extra degree of freedom, yielding
\begin{equation} \label{eq:gap-from-joint}
    {\rm Prob.}[d_{k, N} = g] = \int_{-\infty}^{+\infty} \dd v \; {\rm Prob.}[M_{k, N} = v + g, \, M_{k+1, N} = v] \;.
\end{equation}
A difference has to be made on whether we are looking at the edge of the gas, i.e. $k \sim \mathcal{O}(1)$, or if we place ourselves in the bulk, i.e. $k = \alpha N$ with $\alpha \in [0, 1]$. In the bulk, i.e. when $k = \alpha N$ a similar Laplace expansion of Eq.~(\ref{eq:gap-from-joint}) as the one we did previously for the order statistics yields
\begin{equation} \label{eq:iid-gap-bulk}
    {\rm Prob.}[d_{\alpha N, N} = g] = N p(a_{\alpha N, N}) e^{- N p(a_{\alpha N, N}) g} \;,
\end{equation}
where $a_{\alpha N, N}$ is the same one as the one defined in Eq.~(\ref{eq:bulk-coefs}). On the other hand, at the edge of the gas, i.e. when $k \sim \mathcal{O}(1)$, a careful asymptotic study \cite{SM14} of Eqs. (\ref{eq:joint-order-iid-integral-form-1}-\ref{eq:joint-order-iid-integral-form-2}) in Eq. (\ref{eq:gap-from-joint}) when taking the $N \to +\infty$ limit shows that 
\begin{equation} \label{eq:gaps-universal-form}
    d_{k, N} \sim b_{N} \Delta_{k, N}, \mbox{~~when~~} N \to +\infty \;,
\end{equation}
where $b_{N}$ is the same scale factor as the one in Eq. (\ref{eq:order-universal-form}) and $\Delta_{k, N}$ is a random variable of order 1 with universal distribution
\begin{equation} \label{eq:gaps-iid-cumulative}
    {\rm Prob.}[\Delta_{k, N} < z] = \frac{\Theta(z)}{(k - 1)!} \int_{-\infty}^{+\infty} \dd x \; G_\rho'(x) [-\ln G_\rho(x)]^{k-1} \left[ 1 - \frac{G_\rho(x - z)}{G_\rho(x)} \right] \;.
\end{equation}
although Eq. (\ref{eq:gaps-iid-cumulative}) may be daunting, when $\rho = I$ (i.e. the Gumbel class) Eq. (\ref{eq:gaps-iid-cumulative}) simplifies to an exponential distribution
\begin{equation} \label{eq:gaps-iid-cumulative-gumbel}
    {\rm Prob.}[\Delta_{k, N} = z] = \Theta(z) k e^{- k z} \mbox{~~when~~} \rho = {\rm I}.
\end{equation}
Hence in the Gumbel case the gaps are distributed identically in the bulk and at the edge. On the other hand, for the Weibull case ($\rho = {\rm III}$) we get
\begin{equation}\label{eq:gaps-iid-cumulative-weibull}
    {\rm Prob.}[\Delta_{k, N} = z] = \frac{\mu^2}{(k - 1)!} \int_0^{+\infty} \dd x \; (x + z)^{\mu - 1} e^{-(x + z)^{\mu}} x^{\mu k - 1} \;,
\end{equation}
where $\mu$ is the exponent of the power-law tail minus one, as defined in Eq.~(\ref{eq:weibull-tail-definition}). In the special case when $\mu = 1$, Eq.~(\ref{eq:gaps-iid-cumulative}) also simplifies to an exponential distribution
\begin{equation} \label{eq:gaps-iid-cumulative-weibull-mu1}
    {\rm Prob.}[\Delta_{k, N} = z] = \Theta(z) e^{-z} \mbox{~~when~~} \mu = 1 \mbox{~~and~~} \rho = {\rm II} \;. 
\end{equation}
Finally, for the Fréchet case ($\rho = {\rm II}$) there is no simplification, but the explicit form of the gap distribution is given by
\begin{equation} \label{eq:gaps-iid-cumulative-frechet}
    {\rm Prob.}[\Delta_{k, N} = z] = \frac{\mu^2}{(k - 1)!} \int_{0}^{+\infty} \dd x \; e^{-x^{-\mu}} x^{-\mu - 1} (x + z)^{-\mu k - 1} \;,
\end{equation}
where $\mu$ is the exponent of the power law tail minus one, as defined in Eq.~(\ref{eq:frechet-tail-definition}).

\subsection{The Full Counting Statistics}

The last observable that we are going to study are the full counting statistics $N_L = \#\{X_i : X_i \in [-L, L]\}$, i.e. the number of particles in a box $[-L, L]$ around the origin. We can write the distribution of $N_L$ in terms of $p(x)$ as
\begin{equation} \label{eq:fcs-iid-form}
    {\rm Prob.}[N_L = n] = \binom{N}{n} \left( \int_{-L}^{L} \dd x \; p(x) \right)^n \left( 1 - \int_{-L}^{L} \dd x \; p(x) \right)^{N - n} \;,
\end{equation}
which is a simple binomial distribution where a success corresponds to a particle landing in $[-L, L]$. The large $N$ asymptotic of a binomial distribution is known to be Gaussian, hence we obtain easily 
\begin{equation} \label{eq:fcs-iid-scale}
    N_L \sim a_N + b_N \eta\;, \mbox{~~when~~} N \to +\infty \;,
\end{equation}
where 
\begin{equation} \label{eq:fcs-iid-scale-factor}
    a_N = N \int_{-L}^{L} \dd x \; p(x) \mbox{~~and~~} b_N = \sqrt{N} \int_{-L}^{L} \dd x \; p(x) \left(1 - \int_{-L}^{L} \dd x \; p(x) \right) \;,
\end{equation}
and $\eta \sim \mathcal{N}(0, 1)$ simply has a Gaussian distribution
\begin{equation} \label{eq:fcs-iid-gaussian}
    {\rm Prob.}[\eta]  = \frac{1}{\sqrt{2 \pi}} e^{-\frac{\eta^2}{2}} \;.
\end{equation}
Note, however, that this Gaussian asymptotic is only valid for typical fluctuations of the full counting statistics. The description of atypically large fluctuations requires the machinery of large deviation theory, which is out of the scope of this thesis.

\section{Correlated random variables} \label{sec:correlated}

In the previous Section we looked at independent identically distributed random variables and the universal description of several of their observables. The story becomes considerably more complicated when we introduce correlations between the random variables, i.e. they are not independent anymore. However, there are some special cases where we can still make analytical progress. If the correlations are weak, we might still be able to use the independent results \cite{MPS20}. In Section \ref{subsec:weak} we will detail what is meant by `weak' and show how those cases reduce to the previously studied independent identically distributed random variables. \vspace{0.2cm}

When the variables are strongly correlated universal analytical descriptions like the ones for independent identically distributed random variables are impossible, only case-by-case studies can be done. One famous family of strongly correlated random variables that can be studied analytically is random walks, which are also the underlying structure for all the models studied in this thesis. We will present random walks and their general analytical description in Section \ref{subsec:strong}.

\subsection{Weak correlations} \label{subsec:weak}

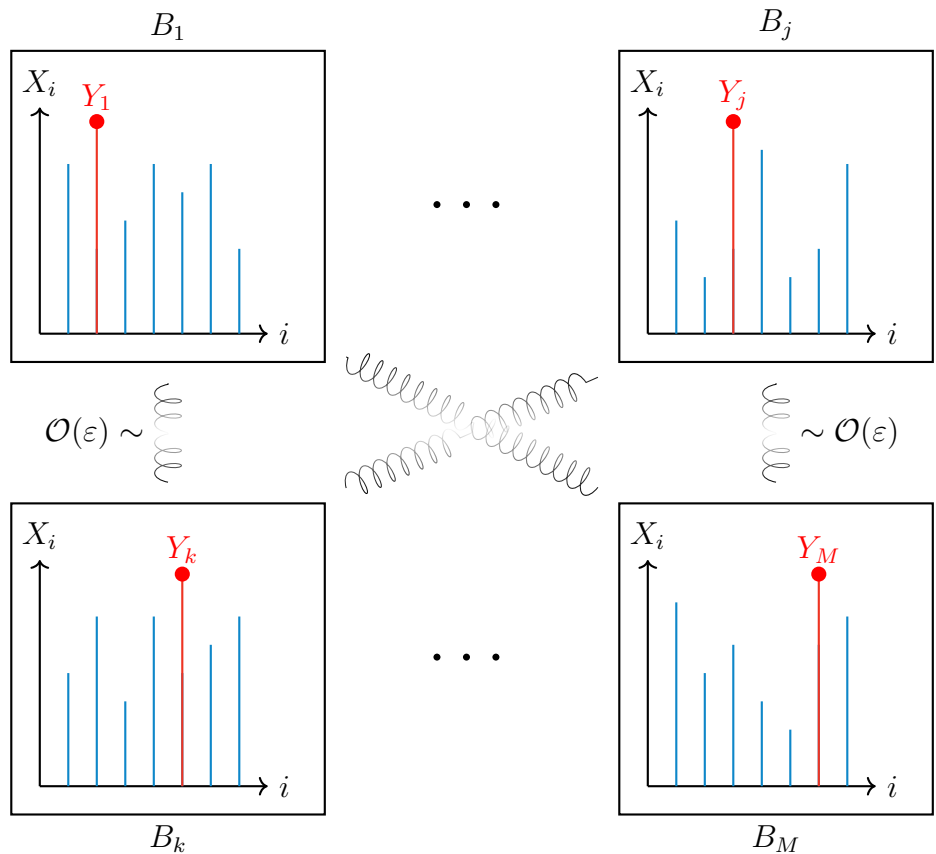
\begin{figure}
    \centering
    \begin{tikzpicture}
    \tikzset{bluebar/.style={color=B4, thick}, redbar/.style={color=C2, thick}, boxstyle/.style={draw=black, thick, minimum width=5cm, minimum height=5cm, inner sep=5pt}}

    \begin{scope}[shift={(0, 0)}, scale=0.75]
        \draw[->, thick] (0,0) -- (4,0) node[right] {$i$};
        \draw[->, thick] (0,0) -- (0,4) node[above] {$X_i$};

        \foreach \x/\h in {0.5/2, 1.0/3, 1.5/1.5, 2.0/3, 2.5/2.0, 3.0/2.5, 3.5/3}
            \draw[bluebar] (\x,0) -- (\x,\h);
        
        \foreach \x/\h/\label in {2.5/3.75/Y_k}
            \draw[redbar] (\x,0) -- (\x,\h) node[above, red] {$\label$} node[circle,fill=red,inner sep=2pt] {};

        \draw[thick] (-0.5,-0.5) rectangle (5,5);

        \node[below] at (2.25, -0.5) {$B_k$};
        \node[above] (rectSWtop) at (2.25, 5)  {};
        \node[right] (rectSWright) at (5, 2.25) {};
        \node[above right] (rectSWcornerNE) at (5, 5) {};
    \end{scope}

    \begin{scope}[shift={(0, 6)}, scale = 0.75]
        \draw[->, thick] (0,0) -- (4,0) node[right] {$i$};
        \draw[->, thick] (0,0) -- (0,4) node[above] {$X_i$};

        \foreach \x/\h in {0.5/3, 1.0/1.5, 1.5/2.0, 2.0/3, 2.5/2.5, 3.0/3.0, 3.5/1.5}
            \draw[bluebar] (\x,0) -- (\x,\h);
        
        \foreach \x/\h/\label in {1.0/3.75/Y_1}
            \draw[redbar] (\x,0) -- (\x,\h) node[above, red] {$\label$} node[circle,fill=red,inner sep=2pt] {};

        \draw[thick] (-0.5,-0.5) rectangle (5,5);
        \node[above] at (2.25, 5) {$B_1$};
        \node[below] (rectNWbottom) at (2.25, -0.5)  {};
        \node[right] (rectNWright) at (5, 2.25) {};
        \node[above right] (rectNWcornerSE) at (5, -0.5) {};
    \end{scope}

    \begin{scope}[shift={(8, 0)}, scale = 0.75]
        \draw[->, thick] (0,0) -- (4,0) node[right] {$i$};
        \draw[->, thick] (0,0) -- (0,4) node[above] {$X_i$};

        \foreach \x/\h in {0.5/3.25, 1.0/2.0, 1.5/2.5, 2.0/1.5, 2.5/1.0, 3.0/2.5, 3.5/3}
            \draw[bluebar] (\x,0) -- (\x,\h);
        
        \foreach \x/\h/\label in {3.0/3.75/Y_M}
            \draw[redbar] (\x,0) -- (\x,\h) node[above, red] {$\label$} node[circle,fill=red,inner sep=2pt] {};

        \draw[thick] (-0.5,-0.5) rectangle (5,5);
        
        \node[below] at (2.25, -0.5) {$B_M$};
        \node[above] (rectSEtop) at (2.25, 5)  {};
        \node[left] (rectSEleft) at (-0.5, 2.25) {};
        \node[above left] (rectSEcornerNW) at (-0.5, 5) {};
    \end{scope}

    \begin{scope}[shift={(8, 6)}, scale = 0.75]
        \draw[->, thick] (0,0) -- (4,0) node[right] {$i$};
        \draw[->, thick] (0,0) -- (0,4) node[above] {$X_i$};

        \foreach \x/\h in {0.5/2, 1.0/1.0, 1.5/1.5, 2.0/3.25, 2.5/1.0, 3.0/1.5, 3.5/3}
            \draw[bluebar] (\x,0) -- (\x,\h);
        
        \foreach \x/\h/\label in {1.5/3.75/Y_j}
            \draw[redbar] (\x,0) -- (\x,\h) node[above, red] {$\label$} node[circle,fill=red,inner sep=2pt] {};

        \draw[thick] (-0.5,-0.5) rectangle (5,5);
        
        \node[above] at (2.25, 5) {$B_j$};
        \node[below] (rectNEbottom) at (2.25, -0.5)  {};
        \node[left] (rectNEleft) at (-0.5, 2.25) {};
        \node[below left] (rectNEcornerSW) at (-0.5, -0.5) {};
    \end{scope}

    \definecolor{springcolor}{rgb}{0,0,0} 
    \path (rectSWtop) -- (rectNWbottom) coordinate[midway] (M);
    \draw[decorate,decoration={coil,aspect=0.4,amplitude=5pt,segment length=2mm}, opacity=1, color=springcolor, path fading=north] 
        (rectSWtop) -- (M);
    \draw[decorate,decoration={coil,aspect=0.4,amplitude=5pt,segment length=2mm}, opacity=1, color=springcolor, path fading=south] 
        (M) -- (rectNWbottom);
    \node[left, xshift=-5pt] at (M) {$\mathcal{O}(\varepsilon) \sim$};

    \path (rectSEtop) -- (rectNEbottom) coordinate[midway] (M);
    \draw[decorate,decoration={coil,aspect=0.4,amplitude=5pt,segment length=2mm}, opacity=1, color=springcolor, path fading=north] 
        (rectSEtop) -- (M);
    \draw[decorate,decoration={coil,aspect=0.4,amplitude=5pt,segment length=2mm}, opacity=1, color=springcolor, path fading=south] 
        (M) -- (rectNEbottom);
    \node[right, xshift=5pt] at (M) {$\sim \mathcal{O}(\varepsilon)$};

    \path (rectSEcornerNW) -- (rectNWcornerSE) coordinate[midway] (M);
    \draw[decorate,decoration={coil,aspect=0.4,amplitude=5pt,segment length=2mm}, opacity=1, color=springcolor, path fading=north] 
        (rectSEcornerNW) -- (M);
    \draw[decorate,decoration={coil,aspect=0.4,amplitude=5pt,segment length=2mm}, opacity=1, color=springcolor, path fading=south] 
        (M) -- (rectNWcornerSE);

    \path (rectSWcornerNE) -- (rectNEcornerSW) coordinate[midway] (M);
    \draw[decorate,decoration={coil,aspect=0.4,amplitude=5pt,segment length=2mm}, opacity=1, color=springcolor, path fading=north] 
        (rectSWcornerNE) -- (M);
    \draw[decorate,decoration={coil,aspect=0.4,amplitude=5pt,segment length=2mm}, opacity=1, color=springcolor, path fading=south] 
        (M) -- (rectNEcornerSW);

    \path (rectSWright) -- (rectSEleft) coordinate[midway] (M);
    \node at (M) {\Huge $\cdots$};

    \path (rectNWright) -- (rectNEleft) coordinate[midway] (M);
    \node at (M) {\Huge $\cdots$};

\end{tikzpicture}
    \caption{A sketch of weakly correlated variables $\{X_1, \cdots, X_N\}$ split in $M$ nearly independent blocks $B_1, \cdots, B_M$. Variables within each block have correlations of order $\mathcal{O}(1)$ but correlations across blocks (represented by faded springs) are vanishingly small. The maximum of each block $B_j$ is denoted by $y_j$ and shown in red.} \label{fig:weak-correlations}
\end{figure}

The first step to study `weak' correlations is to define what is meant by weak. Consider $N$ identically distributed random variables $\{X_1, X_2, \cdots, X_N\}$ which can be separated in $M$ non-overlapping blocks $B_1, \cdots, B_M$ such that their connected correlator is given by 
\begin{equation} \label{eq:weak-corerlations}
    \left\langle X_i X_j \right\rangle  - \left\langle X_i \right\rangle \left\langle X_j \right\rangle = \begin{dcases}
        \mathcal{O}(1) &\mbox{~~if~~} i, j \mbox{~~are in the same block~~} B_k \\
        \mathcal{O}(\varepsilon) &\mbox{~~if~~} i \in B_k, j \in B_{k'} \mbox{~~and~~} k \neq k' \;,
    \end{dcases}
\end{equation}
with $\varepsilon \ll 1$. A sketch of this definition is given in Fig. \ref{fig:weak-correlations}. We denote by $Y_j = \max_{i \in B_j} X_i$ the maximum of the $j$-th block. Since different blocks are essentially uncorrelated, {\color{blue} we can approximate} the variables $Y_1, \cdots, Y_M$ {\color{blue} as being} uncorrelated and identically distributed. Furthermore, the global maximum $y = \max_{i = 1, \cdots, N} X_i$ can be re-expressed as
\begin{equation} \label{eq:max-reparametrization-trick}
    Y = \max_{j = 1, \cdots, M} Y_j \;.
\end{equation}
Since {\color{blue} we approximate} $\{Y_1, \cdots, Y_M\}$ {\color{blue}as} uncorrelated identically distributed random variables if $M \gg 1$ we can use the known {\color{blue}extreme-value} statistics results presented in Section \ref{sec:iid}. Therefore, the global maximum $Y$ must still be distributed as one of the three universal {\color{blue}extreme-value} statistics functions: Gumbel, Fréchet, or Weibull. The actual distribution depends on the tails of the distribution of $Y_j$, which is generically difficult to compute since $Y_j$ is the maximum of $N/M$ strongly correlated random variables. However, even if the full distribution might not be computable extracting the behavior of its tail is considerably simpler and can often be guessed. In subsection \ref{subsec:strong} we will study the \OU process which is a concrete example of these weakly correlated variables. \vspace{0.2cm}

Note also that if the $X_i$'s are Gaussian then Berman showed \cite{B64} that if the connected correlator $C_{i,j} = \left\langle X_i X_j \right\rangle - \left\langle X_i \right\rangle \left\langle X_j \right\rangle$ can be expressed as $C_{i, j} = c(|i - j|)$ with $c(x)$ a function that decays faster than $1/\ln x$ for large $x$ then the distribution of $Y = \max_{i = 1, \cdots, N} X_i$ is still given by a Gumbel distribution.

\section{Strong correlations} \label{subsec:strong}

The analytical description of strongly correlated random variables is difficult, and no general description exists. In the absence of a general theory, we have to resort to studying a variety of exactly solvable special cases to elucidate some of the behaviors of strongly correlated random variables. Random walks are one of the rare examples of strongly correlated variables for which analytical computations can be performed. \vspace{0.2cm}

In this Section, we will start by introducing random walks in Subsection \ref{subsec:random-walks}. After that, we will study the {\color{blue}extreme-value} statistics of a one-dimensional Brownian motion in Subsection \ref{subsec:Brownian-motion}. {\color{blue}Subsequently,} we will then look at the \OU process in Subsection \ref{subsec:OU-process}, which describes the motion of a diffusing particle in a harmonic trap. The {\color{blue}extreme-value} statistics of the \OU process can be computed explicitly and are also a practical example of the heuristic weak-correlations argument presented in Subsection \ref{subsec:weak}. 

\subsection{Random walks} \label{subsec:random-walks}

From now on the random variables $X_i$ will represent the position of a particle at discrete time step $i$, starting from $X_0 = 0$. At each time step $n$, a random shift $\eta_n$ is sampled from a distribution $f(\eta)$ and added to the previous position. Thus the positions of the walker undergoes the Markovian dynamics
\begin{equation} \label{eq:def-rw}
    X_{n + 1} = X_{n} + \eta_n, \mbox{~~where~~} \eta_n \sim f(\eta) \mbox{~~and~~} X_0 = 0 \;.
\end{equation}
The general random walk model presented in Eq. (\ref{eq:def-rw}) can describe discrete jumps, with 
\begin{equation} \label{eq:discrete-jumps}
    f(\eta) = \frac{1}{2} \delta(\eta - 1) + \frac{1}{2} \delta(\eta + 1) \;.
\end{equation}
As well as continuous jumps with 
\begin{equation} \label{eq:continuous-jumps}
    f(\eta) = \frac{1}{\sqrt{2 \pi}} e^{-\eta^2/2} \mbox{~~or~~} f(\eta) = \frac{1}{\pi (1 + \eta^2)} \;,
\end{equation}
for example. A central quantity in the study of random walks is the propagator $p_n(x) = {\rm Prob.}[X_n = x]$ denoting the probability of finding the particle at position $x$ after $n$ steps. From Eq. (\ref{eq:def-rw}) we can obtain a recursive equation for the propagator
\begin{equation} \label{eq:rec-propagator}
    \begin{dcases}
        p_{n+1}(x) = \int_{-\infty}^{+\infty} \dd y \; p_n(y) f(x - y)\\
        p_0(x) = \delta(x) \;.
    \end{dcases}    
\end{equation}
This formula can be understood as follows, in order for the particle to be at $x$ at time $n+1$ it must have been at $y$ at time $n$ and made a jump $\Delta = x - y$ from $y$ to $x$ to reach $x$ at time $n+1$. We then need to integrate over all possible previous positions. The convolution structure of Eq. (\ref{eq:rec-propagator}) suggests the use of Fourier transforms which leads to 
\begin{equation} \label{eq:rw-propagator}
    p_n(x) = \int_{-\infty}^{+\infty} \frac{\dd k}{2 \pi} \left[\hat{f}(k)\right]^n e^{-i k x} \;,
\end{equation}
where $\hat{f}(k)$ is the Fourier transform of the jump distribution, defined as 
\begin{equation} \label{eq:jump-fourier}
    \hat{f}(k) = \int_{-\infty}^{+\infty} \dd \eta \; f(\eta) e^{i k \eta} \; .
\end{equation}
Note that since $f(\eta)$ is a distribution the normalization condition imposes $\hat{f}(0) = 1$. For non-pathological jump distributions the Fourier transform will generally behave as
\begin{equation} \label{eq:jump-small-fourier}
    \hat{f}(k) \sim 1 - |\alpha k|^{\mu} \mbox{~~when~~} k \to 0 \;,
\end{equation}
where $\alpha > 0$ is some coefficient characterizing the length scale of a typical jump. The coefficient $0 < \mu \leq 2$ is called the Lévy index and characterizes the tail of the jump distribution. Distributions $f(\eta)$ which decay exponentially or faster than exponentially as $|\eta| \to +\infty$ have a Levy-index $\mu = 2$. For example, for the jump distribution given in Eq. (\ref{eq:discrete-jumps}) as well as the Gaussian distribution given in Eq. (\ref{eq:continuous-jumps}) the small-$k$ asymptotic is given by
\begin{equation} \label{eq:small-fourier-mu2}
    \hat{f}(k) \sim 1 - \frac{k^2}{2} \mbox{~~when~~} k \to 0 \;.
\end{equation}
If the distribution decays as a power-law $f(\eta) \propto |\eta|^{-1 - p}$ for large $|\eta| \to +\infty$ then if $p \geq 2$ the distribution will still have a Lévy index $\mu = 2$, but if $0 < p < 2$ the distribution will have a Lévy index $\mu = p$. For example, the Cauchy distribution in Eq. (\ref{eq:continuous-jumps}) decays as $f(\eta) \propto |\eta|^{-2}$ when $|\eta| \to +\infty$ and indeed the small-$k$ asymptotic of the Fourier transform of the distribution is given by 
\begin{equation} \label{eq:small-fourier-mu1}
    \hat{f}(k) \sim 1 - k \mbox{~~when~~} k \to 0 \;.
\end{equation}
Generally, taking the large-$n$ limit of Eq. (\ref{eq:rw-propagator}) we see that the integral is dominated for values of $\hat{f}(k)$ close to one, i.e. for $k \to 0$. Then using the expansion in Eq. (\ref{eq:jump-small-fourier}) in Eq. (\ref{eq:rw-propagator}) we obtain
\begin{equation}\label{eq:large-n-rw-propagator}
    p_n(x) \sim \int_{-\infty}^{+\infty} \frac{\dd k}{2 \pi} e^{-n |\alpha k|^\mu - i k x} \;,
\end{equation}
in the large-$n$ limit. Rescaling the integration variable by $\alpha n^{1/\mu}$ we can re-write Eq. (\ref{eq:large-n-rw-propagator}) in a scaling form
\begin{equation} \label{eq:propagator-levy}
    p_n(x) \sim \frac{1}{\alpha \, n^{1/\mu}} \mathcal{L}_\mu \left( \frac{x}{\alpha \, n^{1/\mu}} \right) \mbox{~~when~~} n \to +\infty \;,
\end{equation}
where $\mu$ is the Lévy index of the jump distribution $f(\eta)$ and $\mathcal{L}_\mu(z)$ is the Lévy stable distribution defined as
\begin{equation} \label{eq:levy-stable}
    \mathcal{L}_\mu(z) = \int_{-\infty}^{+\infty} \frac{\dd k}{2 \pi} e^{- i k z - |k|^\mu} \;.
\end{equation}
Although the integral in Eq. (\ref{eq:levy-stable}) cannot generically be computed explicitly, for $\mu = 2$ and $\mu = 1$ it simplifies to the Gaussian and Cauchy distributions respectively, which are given in Eq. (\ref{eq:continuous-jumps}). \vspace{0.2cm}

Random walks are generically divided into three categories according to their Lévy index $\mu$. 
\begin{enumerate}
    \item If $\mu = 2$, then $\sigma^2 = \langle \eta^2 \rangle - \langle \eta\rangle^2$ is finite and after a large number of steps the random walk converges to a Brownian motion, i.e. the propagator is Gaussian and we can take a continuous time limit.
    \item If $1 \leq \mu < 2$, then $\sigma^2$ is infinite and these walks converge to the Lévy distribution given in Eq. (\ref{eq:propagator-levy}) which does not further {\color{blue}simplify}. These random walks are usually called recurrent Lévy flights. They are said to be recurrent because they will revisit sites infinitely many times with probability one.
    \item If $0 < \mu < 1$, then $\sigma^2$ is infinite and $\langle \eta \rangle$ is also infinite. Once again, these walks converge to the Lévy distribution given in Eq. (\ref{eq:propagator-levy}) which does not simplify. These random walks are usually called transient Lévy flights. They are said to be transient because they will revisit sites infinitely many times with probability zero.
\end{enumerate}
The {\color{blue}extreme-value} statistics of Brownian motion or Lévy flights are significantly different and considerably harder for Lévy flights. For simplicity, we will present the main results for the Brownian motion. 

\subsection{One-dimensional Brownian motion} \label{subsec:Brownian-motion}

Consider a random walk as defined in Eq. (\ref{eq:def-rw}) with a jump distribution of Lévy index $\mu = 2$. Using Eq. (\ref{eq:propagator-levy}) we know that after many steps the propagator will be given by
\begin{equation} \label{eq:discrete-BM}
    p_n(x) \sim \sqrt{\frac{1}{2 \pi n \alpha^2}} \; e^{ - \frac{x^2}{2 n \alpha^2}} \;.
\end{equation}
We are now going to take the continuous space and time limit by introducing
\begin{equation} \label{eq:diffusion-constant}
    n = t / \delta t \mbox{~~and~~} D = \frac{\alpha^2}{2 \delta t}
\end{equation}
and taking the $\delta t \to 0, \alpha \to 0$ limit while keeping $D$ fixed which yields
\begin{equation} \label{eq:propagator-BM}
    p(x, t) = \lim_{\delta t, \alpha \to 0} p_{t/\delta t}(x) = \frac{1}{\sqrt{4 \pi D t}} e^{- \frac{x^2}{4 D t}} \;.
\end{equation}
Notice that we replaced the $\sim$ symbol with an equality because in the limit described above the convergence becomes exact, i.e. we are doing infinitely many infinitely small jumps. The propagator given in Eq. (\ref{eq:propagator-BM}) is the propagator of the Brownian motion. \vspace{0.2cm}

\begin{figure}
    \centering
    \includegraphics[width=0.7\textwidth]{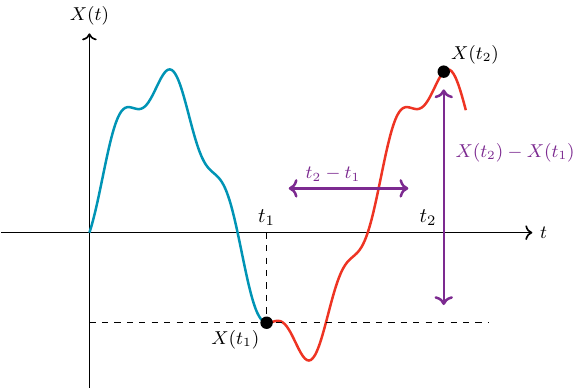}
    \caption{A sketch of a one-dimensional Brownian motion and the geometric arguments involved in computing the two point connected correlator. All paths going from $x = 0$ at $t = 0$ to $X(t_1)$ at $t = t_1$ to $X(t_2)$ at $t = t_2$ can be split in two segments: $[0, t_1]$ and $[t_1, t_2]$ plotted in blue and red respectively. Since Brownian motion is a Markov process, i.e. it has no memory, the red segment is independent from the blue segment. The probability of the blue segment is given by the standard propagator $p(X(t_1), t_1)$ characterizing the probability of reaching $X(t_1)$ at time $t_1$ starting from $x = 0$ at time $t = 0$. Using the time and space translation invariance of Brownian motion the probability of the red segment can be recast as the probability of reach $X(t_2) - X(t_1)$ at time $t_2 - t_1$ having started at $x = 0$ at time $t = 0$, as sketched by the two purple arrows. } \label{fig:BM-correlator}
\end{figure}

Brownian motion is usually defined directly in continuous time and space through the stochastic ordinary differential equation
\begin{equation} \label{eq:BM-SODE}
    \begin{dcases}
        \dv{X(t)}{t} = \sqrt{2 D} \; \eta(t) \\
        X(0) = 0
    \end{dcases} \;,
\end{equation}
where $X(t)$ is the random variable corresponding to the position of the particle at time $t$ and $\eta(t)$ represents a Gaussian white noise with zero-mean $\langle \eta(t) \rangle = 0$ and delta-correlations $\langle \eta(t) \eta(t') \rangle = \delta(t - t')$. Let $p(x, t)$ be the propagator of the Brownian motion described in Eq. (\ref{eq:BM-SODE}), the Fokker-Planck partial differential equation for $p(x, t)$ is
\begin{equation} \label{eq:BM-FP}
    \begin{dcases}
        \pdv{p(x, t)}{t} = D \pdv[2]{p(x, t)}{x} \\
        p(x, 0) = \delta(x) \;.
    \end{dcases}
\end{equation}
The partial differential equation in Eq. (\ref{eq:BM-FP}) can be solved exactly and leads to the Gaussian expression in Eq. (\ref{eq:propagator-BM}). The strong correlations in Brownian motion can be explicited by looking at the correlator
\begin{equation} \label{eq:def-BM-correlator}
    \langle X(t_1) X(t_2) \rangle = \int_{-\infty}^{+\infty} \dd x_1 \int_{-\infty}^{+\infty} \dd x_2 \; x_1 x_2 {\rm Prob.}[X(t_1) = x_1, X(t_2) = x_2] \;.
\end{equation}
Using the Markovian property of Brownian motion, i.e. the steps taken at every time step have no memory of the past history of the walk, we can split the probability in the integral as
\begin{equation} \label{eq:BM-correlator-int}
    \langle X(t_1) X(t_2) \rangle = \int_{-\infty}^{+\infty} \dd x_1\; \int_{-\infty}^{+\infty} \dd x_2\;  x_1 p(x_1, t) x_2 p(x_2 - x_1, t_2 - t_1) \;,
\end{equation}
where we assumed without loss of generality that $t_1 < t_2$. This factorization can be understood as follows and is sketched in Fig. \ref{fig:BM-correlator}. The trajectories that reach $x_1$ at time $t_1$ and then go to $x_2$ at time $t_2$ can be divided into two segments $[0, t_1]$ and $[t_1, t_2]$. During the first segment, we have to reach $x_1$ at time $t_1$ starting from the origin at time 0, which is given by the propagator $p(x_1, t_1)$. Using the time-translation and space translation invariance of the system, we can rephrase the second segment as having to reach $x_2 - x_1$ at time $t_2 - t_1$ starting from the origin at time 0, which is given by the propagator $p(x_2 - x_1, t_2 - t_1)$. Replacing $p(x, t)$ in Eq. (\ref{eq:BM-correlator-int}) with its expression in Eq. (\ref{eq:propagator-BM}) we can perform the integral explicitly yielding
\begin{equation} \label{eq:BM-correlator}
    \langle X(t_1) X(t_2) \rangle = 2 D \min (t_1, t_2) \;. 
\end{equation}
This process is clearly strongly correlated since there exist non-negligible correlations between any pair of positions $X(t_1)$ and $X(t_2)$ as long as $\min (t_1, t_2) \gtrsim 1/(2 D)$.

\vspace{0.2cm}

\begin{figure}
    \centering
    \includegraphics[width = 0.5\textwidth]{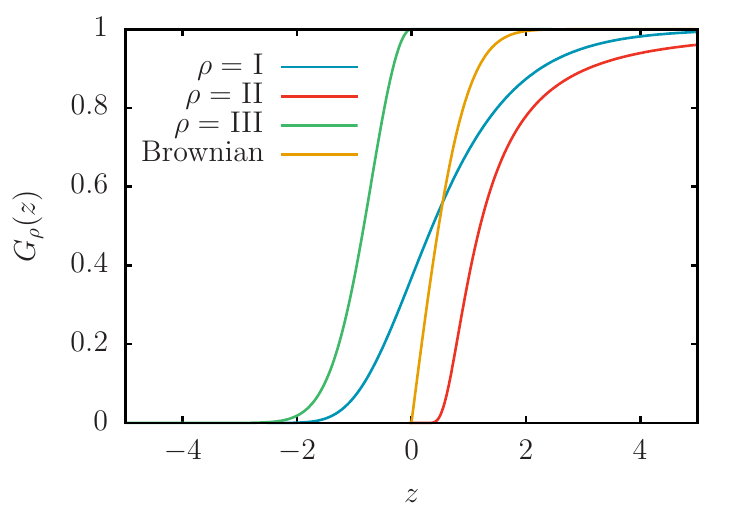}\hfill
    \includegraphics[width = 0.5\textwidth]{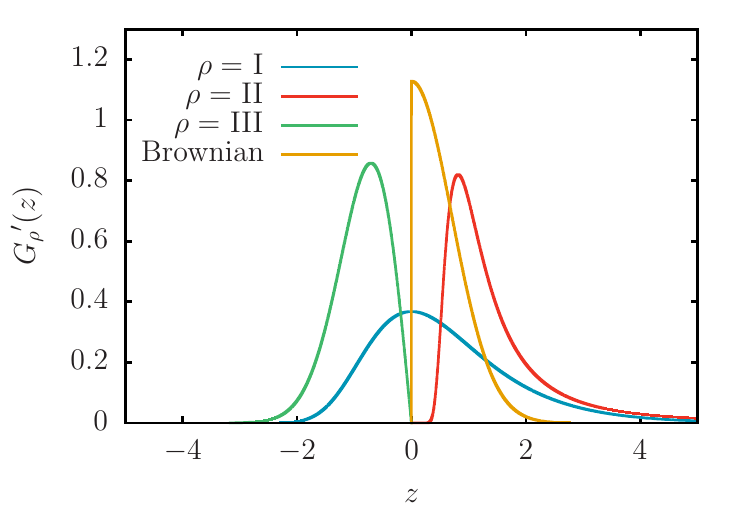}
    \caption{{\color{blue}A plot of the Gumbel, Fréchet, Weibull universal {\color{blue}extreme-value} statistics $G_{\rm I}(z)$, $G_{\rm II}(z)$, $G_{\rm III}(z)$ compared to the extreme-value statistics $Q(z, t)$ of Brownian motion. The cumulative distributions $G_\rho(z)$ and $Q(z, t)$ are shown on the left, and their derivatives, i.e. the probability density functions, on the right. The equations of the cumulative distributions $G_\rho(z)$ and $Q(z, t)$ are given in Eqs. (\ref{eq:gumbel-cumulative}), (\ref{eq:frechet-cumulative}), (\ref{eq:weibull-cumulative}) and (\ref{eq:cumulative-BM}) respectively. Both the Fréchet and Weibull distributions are plotted with $\mu = 2$.}} \label{fig:evs-universal-vs-BM}
\end{figure}

We are now going to study the {\color{blue}extreme-value} statistics of Brownian motion. Specifically, we are going to look at the maximum $M(t)$ up to time $t$, i.e.
\begin{equation} \label{eq:def-BM-max}
    M(t) = \max_{0 \leq \tau \leq t} X(\tau) \;.
\end{equation}
As in the independent identically distributed case, it is convenient to introduce the cumulative distribution of the maximum
\begin{equation}\label{eq:def-cum-BM-max}
    Q(z, t) = {\rm Prob.}[M(t) \leq z] = {\rm Prob.}[X(\tau) \leq z, 0 \leq \tau \leq t] \;.
\end{equation}
As we have seen from Eq. (\ref{eq:BM-correlator}) the different positions of a random walk are strongly correlated, hence we cannot simplify Eq. (\ref{eq:def-cum-BM-max}) as
\begin{equation} \label{eq:non-independence-BM-max}
    Q(z, t) = {\rm Prob.}[X(\tau) \leq z, 0 \leq \tau \leq t] \neq \prod_{\tau = 0}^{t} {\rm Prob.}[X(\tau) \leq z] \;,
\end{equation}
where we clearly abused the product notation to convey the point. Instead, we re-phrase the probability in Eq. (\ref{eq:def-cum-BM-max}) by introducing the constrained propagator $p_{\leq z}(x, t)$ characterizing the probability of reaching $x$ at time $t$, having started at $x = 0$ at $t = 0$ while staying below $z$. Using $p_{\leq z}(x, t)$ we can re-write Eq. (\ref{eq:def-cum-BM-max}) as
\begin{equation} \label{eq:cumulative-integral-form}
    Q(z, t) = \int_{-\infty}^{z} \dd x\; p_{\leq z}(x, t) \;,
\end{equation}
i.e. we are summing over all possible endpoints of all the paths which stay below $z$. The constrained propagator $p_{\leq z}(x, t)$ follows the same partial differential equation as the unconstrained propagator $p(x, t)$, i.e.
\begin{equation} \label{eq:PDE-constrained-propagator}
    \begin{dcases}
        \pdv{p_{\leq z}(x ,t)}{t} = D \pdv[2]{p_{\leq z}(x ,t)}{x} \\
        p_{\leq z}(x, 0) = \delta(x) \;,
    \end{dcases}
\end{equation}
but with different boundary conditions
\begin{equation} \label{eq:BC-constrained-propagator}
    p_{\leq z}(x \to -\infty, t) = 0 \mbox{~~and~~} p_{\leq z}(z, t) = 0 \;,
\end{equation}
which ensure that although the particle {\color{blue} is not bounded by below}, it is {\color{blue} constrained to remain below} $z$. This partial differential equation can be solved through the method of images \cite{MPS20,R01,M10,BMS13} yielding
\begin{equation} \label{eq:constrained-propagator}
    p_{\leq z}(x, t) = \frac{1}{\sqrt{4 \pi D t}} \left[ e^{-\frac{x^2}{4 D t}} - e^{- \frac{(x - 2 z)^2}{4 D t}} \right] \;.
\end{equation}
Using this form in Eq. (\ref{eq:cumulative-integral-form}) we can compute the integral explicitly 
\begin{equation} \label{eq:cumulative-BM}
    Q(z, t) = \int_{-\infty}^z \dd x \; \frac{1}{\sqrt{4 \pi D t}} \left[ e^{-\frac{x^2}{4 D t}} - e^{- \frac{(x - 2 z)^2}{4 D t}} \right] = {\rm erf}\left(\frac{z}{\sqrt{4 D t}}\right) \;,
\end{equation}
where ${\rm erf}(z)$ is the error complementary function defined as
\begin{equation} \label{eq:def-erf}
    {\rm erf}(z) = \frac{2}{\sqrt{\pi}} \int_0^z \dd u \; e^{-u^2} \;.
\end{equation}
Notice that the maximum is not distributed according to any of the three universal distributions (Gumbel, Fréchet, Weibull) of independent identically distributed random variables presented in Section \ref{sec:iid}. {\color{blue}See Fig.~\ref{fig:evs-universal-vs-BM} for a graphical comparison.} This may be the simplest example of an observable of a strongly correlated system which we can characterize exactly analytically. The actual probability density function of the maximum can be obtained by taking a derivative of Eq. (\ref{eq:cumulative-BM}) which yields
\begin{equation} \label{eq:maximum-BM}
    {\rm Prob.}[M(t) = z] = \pdv{Q(z, t)}{z} = \frac{\Theta(z)}{\sqrt{\pi D t}} e^{- \frac{z^2}{4 D t}} \;,
\end{equation}
where $\Theta(z)$ is the Heaviside distribution and subsequently the average value of the maximum can be obtained
\begin{equation}
    \langle M(t) \rangle = \int_{-\infty}^{+\infty} \dd z \; {\rm Prob.}[M(t) = z] = 2 \sqrt{\frac{D t}{\pi}} \;.
\end{equation}
{\color{blue} For comparison, suppose instead that the maximum $M(t)$ was the maximum of $N \propto t$ independent identically distributed Gaussian random variables of variance $\sigma$, then we would have
\begin{equation} \label{eq:maximum-BM-2}
    \langle M(t) \rangle \sim \sqrt{2 \sigma^2 \log t} \;.
\end{equation}
Clearly the behavior in Eq.~(\ref{eq:maximum-BM}) is significantly different to its independent counterpart in Eq.~(\ref{eq:maximum-BM-2}).
}

\subsection{One-dimensional \OU process} \label{subsec:OU-process}

\begin{figure}
    \centering
    \includegraphics[width=0.7\textwidth]{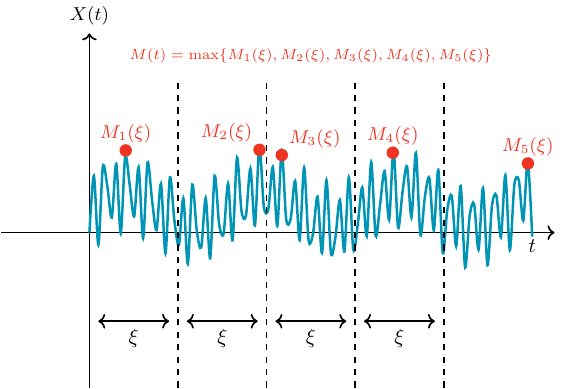}
    \caption{Sketch of the blocking argument used to compute the maximum of an \OU process using the weakly correlated random variables assumption. The \OU process (sketched in blue) is split in $N = t / \xi$ blocks of size $\xi \sim 1/\mu$ where $\mu > 0$ is the stiffness of the confining potential. Each block is approximately independent from the others. Then the overall maximum $M(t)$ can be re-phrased as the maximum of each block $M_i(\xi)$ (sketched in red) which are approximately independent.} \label{fig:OU-weak}
\end{figure}

We now add an ingredient to the Brownian motion defined in Eq. (\ref{eq:BM-SODE}) presented in the previous section: a harmonic confining potential. The stochastic ordinary differential equation becomes
\begin{equation} \label{eq:OU-SODE}
    \begin{dcases}
        \dv{X(t)}{t} = -\mu X(t) + \sqrt{2 D} \; \eta(t) \\
        X(0) = 0
    \end{dcases} \;,
\end{equation}
where $X(t)$ is the random variable corresponding to the position of the particle at time $t$ and $\eta(t)$ is a Gaussian white noise with zero-mean $\langle \eta(t) \rangle = 0$ and delta-correlations $\langle \eta(t) \eta(t') \rangle = \delta(t - t')$. Let $p_\mu(x, t)$ be the propagator associated with Eq. (\ref{eq:OU-SODE}). The Fokker-Planck equation of the \OU process is given by 
\begin{equation} \label{eq:OU-FP}
    \begin{dcases}
        \pdv{p_\mu(x, t)}{t} = D \pdv[2]{p_\mu(x, t)}{x} + \mu \pdv{}{x} \left[x\, p_\mu(x, t) \right] \\
        p_\mu(x, t = 0) = \delta(x) 
    \end{dcases} \;.
\end{equation}
This partial differential equation can be solved, noticing from Eq.~(\ref{eq:OU-SODE}) that $X(t)$ is a linear combination of Gaussian random variables and therefore must be Gaussian as well. Using the Gaussian form as an Ansatz in Eq.~(\ref{eq:OU-FP}) we obtain
\begin{equation} \label{eq:OU-propagator}
    p_\mu(x, t) = \sqrt{\frac{\mu}{2 \pi D (1 - e^{-2 \mu t})}} \exp[ -\frac{\mu \, x^2 }{2 D (1 - e^{-2\mu t})} ] \;.
\end{equation}
To compute the two point connected correlator, a similar argument to the one used for Brownian motion in Fig. \ref{fig:BM-correlator} can be made for the Ornstein-Uhlenbeck process with the slight difference that we do not have space translational invariance anymore. Doing so results in 
\begin{equation} \label{eq:connected-correlator-OU}
    \langle X(t_1) X(t_2) \rangle = \frac{D}{\mu} \left[ e^{-\mu |t_1 - t_2|} - e^{-\mu (t_1 + t_2)} \right] \;.
\end{equation}
In the $\mu \to 0^+$ limit we recover the Brownian motion result given in Eq. (\ref{eq:BM-correlator}). However, for any finite $\mu > 0$ the correlations behave very differently from the Brownian case, the two point correlation function decays exponentially fast in time. Any two points in a given path which are more than $t_2 - t_1 \gg 1 / \mu$ time apart are effectively nearly uncorrelated. Therefore, we can apply the weak correlation argument presented in subsection \ref{subsec:weak}. We can split any given path up to time $t$ in temporal blocks of size $\xi \sim 1/\mu$ as depicted in Fig. \ref{fig:OU-weak}. Positions within the same block are strongly correlated however correlations across blocks are nearly vanishing, and the overall maximum $M(t)$ can be rephrased as 
\begin{equation} \label{eq:max-OU-reframing}
    M(t) = \max \{ M_1(\xi), M_2(\xi), \cdots, M_N(\xi) \} \;,
\end{equation}
where $M_i(\xi)$ is the maximum of the $i$-th block of duration $\xi$ and $N = t / \xi$. Note that each of the $M_i(\xi)$ are independent and identically distributed following the weak correlation argument presented above. Hence, in the limit $N \gg 1$, i.e. $t \gg \xi$, the maximum $M(t)$ will belong to one of the three known universal distributions: Gumbel, Fréchet or Weibull. Given that the fastest motion in our system comes from the diffusion, we can reasonably guess that the tails of the maxima $M_i(\xi)$ cannot decay slower than exponentially and therefore $M(t)$ must have a Gumbel distribution when $t \gg 1/\mu$. In fact, the full distribution of $M(t)$ can be computed exactly analytically \cite{MPS20}, and indeed $M(t)$ has a Gumbel distribution for $t \gg 1 / \mu$. Specifically,
\begin{equation} \label{eq:max-OU}
    \langle M(t) \rangle \sim \begin{cases}
    \sqrt{t} &\mbox{~~for~~} t \ll 1/\mu\\
    \sqrt{\ln t} &\mbox{~~for~~} t \gg 1 /\mu 
    \end{cases}
\end{equation}
showing that the maximum typically behaves as the strongly correlated normal Brownian motion when $t \ll 1 /\mu$, since the system hasn't yet had the time to feel the effect of the confining potential. It then transitions to a weakly correlated regime where the confining potential effectively decorrelates sections of the trajectory by erasing memory exponentially fast and therefore leading to a Gumbel distribution.

\chapter{Conditionally independent identically distributed random variables \cite{BLMS24}} \label{ch:ciid}
One of the central difficulties in studying the statistical properties of many-particle systems is the presence of correlations. For independent identically distributed random variables, the behavior of key observables such as the maximum, the order statistics, or the number of particles in a given region (i.e. full counting statistics) is well understood and governed by universal results, which we presented in Section \ref{sec:iid}. However, as we saw in Chapter \ref{ch:extreme}, introducing correlations dramatically complicates the picture. Even weak correlations can render analytical progress quite delicate, and in the presence of strong correlations, exact results are rare and there exist only a few exactly solvable cases~\cite{MPS20, D85, DG86, TW94, TW96, DM01, BC06, MC04, MC05, SM06}.

\vspace{0.2cm}

In this chapter, we explore a particularly interesting class of strongly correlated random variables, namely conditionally independent identically distributed random variables. These are variables that are not independent in general but become so once we condition on some latent, hidden variables. Although this structure may seem artificial at first glance, we will see that it naturally arises in a variety of physical systems. More importantly, it allows for exact analytical results and often reveals universal {\color{blue}behavior} despite strong correlations.

\vspace{0.2cm}

Conditionally independent random variables have already appeared sporadically in several areas. In random matrix theory, superstatistical ensembles \cite{BC03, BCS05,AM05,BCP08,AAV09} model complex fluctuations by introducing a fluctuating inverse temperature, effectively making the matrix entries conditionally independent. {\color{blue} Since then several generalizations were developed \cite{B09}. For example, classifying the extreme-value statistics in three universality classes according to the distribution of the inverse temperature $\beta$: the $\chi^2$ class, the inverse-$\chi^2$ class and the log-normal class \cite{B09, RB14,HW04,DGKJS16}. On another note,} in neuroscience, correlated firing patterns across neurons are often explained as responses to a shared latent stimulus \cite{T88}. Even in statistics and machine learning, this idea underlies factor models, principal component analysis, and topic models, where observed correlations are explained by a small number of hidden factors.

\vspace{0.2cm}

From a physical perspective, these latent variables often correspond to global environmental parameters, hidden couplings, or shared external fields. From a mathematical viewpoint, they allow us to rewrite the joint distribution of a set of random variables as a mixture of independent ensembles, allowing us to exploit the known universal results of independent identically distributed random variables.

\vspace{0.2cm}

The aim of this chapter is to present a universal characterization of the behavior of some key observables for conditionally independent identically distributed random variables. We will look at the {\color{blue}extreme-value} statistics, the order statistics, the gap statistics, and the full counting statistics. Remarkably, despite the presence of strong correlations, we will see that for many of the classical results known for independent identically distributed variables, we are able to derive equivalent relations in the conditionally independent case. We will derive these results systematically and discuss how this framework not only captures previously studied models, but also provides a natural language for new, experimentally relevant systems — some of which we will explore in later chapters.

\vspace{0.2cm}

Consider a set of $N$ independent and identically distributed random variables $X_1, \cdots, X_N = \vec{X}$ with a common distribution that contains a set of parameters $Y_1, Y_2, \cdots, Y_M = \vec{Y}$ which themselves are random variables with their own distribution. An example of the cases $N=1$ and $M=1$, is when $X$ refers to the energy of a gas and $Y$ refers to the temperature or the magnetic field. In this case, the statistics of $X$, averaged over the distribution of $Y$, these are the "superstatistics" to which we referred previously, which have been studied in various contexts~\cite{BC03, BCS05,AM05,BCP08,AAV09,B09, RB14,HW04,DGKJS16}. For fixed values of these parameters, the $X_i$ variables are statistically independent with a joint distribution  
\begin{equation}
{\rm Prob.}[\vec X = \vec x | \vec{Y} = \vec{y}] = \prod_{i = 1}^N {\rm Prob.}[X_i = x_i | \vec{Y} = \vec{y}] \;. \label{eq:P_joint_IID}
\end{equation}
Hence we call these $X_i$ variables conditionally independent and identically distributed random variables. When one integrates over the $Y_i$ variables, the joint distribution of $X_1, X_2, \cdots, X_N$ is no longer factorizable 
\begin{equation}\label{eq:def-ciid}
{\rm Prob.}[\vec{X} = \vec{x}] = \int \dd^M \vec{Y} \; \left\{\prod_{i = 1}^N {\rm Prob.}[X_i = x_i |\vec{Y} = \vec{y}]\right\} {\rm Prob.}[\vec{Y} = \vec{y}] \;.
\end{equation}
Thus, the $X_i$'s become correlated since they share the same set of parameters $\vec{Y}$. 

\vspace{0.2cm}

A physical example of such a system is provided by a simplified version of models of ``diffusing diffusivity'' where $N$ Brownian motions $X_1(t), \cdots, X_N(t)$ share a common diffusion constant $Y_1 = D$ \cite{CS14,CSMS16} which does not evolve in time but is initialized randomly. In this specific case, Eq. (\ref{eq:def-ciid}) would read
\begin{equation}\label{eq:gauss_example}
{\rm Prob.}[\vec{X}(t) = \vec{x}] = \int \dd D \; \left\{\prod_{i = 1}^N \frac{1}{\sqrt{4 \pi D t}} \exp[ - \frac{x_i^2}{4 D t} ]\right\} {\rm Prob.}[D] \;.
\end{equation}
This example clearly demonstrates that the joint distributions of the $X_i$'s do not factorize, and hence the $X_i$ variables are correlated. Many other physical examples will be discussed in this thesis in Chapters \ref{ch:sim-reset} and \ref{ch:ou-switch}, but this family of systems describes a wide variety of physical problems. We provide here a short non-exhaustive list of such problems so that the reader may grasp the type of problems which can be described by this formalism. Conditionally independent random variables can be used to describe:
\begin{itemize}
\item the statistics of $N$ independent stochastic processes measured after a random time $Y = T$. 
\item the statistics of $N$ experimental observations $X_1$, $\cdots$, $X_N$ that depend on some experimental parameters $Y_1$, $\cdots$, $Y_M$ that may have some non-negligible uncertainties $\sigma_1$, $\cdots$, $\sigma_M$. Suppose we know ${\rm Prob.}[X_i | \vec{Y}]$, we can model ${\rm Prob.}[\vec{Y}]$ by $M$ Gaussians centered around their expected values and with variances $\vec{\sigma}$.
\item the statistics of $N$ independent random particles $X_1, \cdots, X_N$ evolving in an energy landscape $E(Y_1, \cdots, Y_M)$ that depends on some random parameters $Y_1, \cdots, Y_M$ (magnetization, temperature, etc.) which are randomly initialized according to some distribution at $t = 0$.
\end{itemize}

\vspace{0.2cm}

A natural question is whether it is possible to generalize the well-known results which we have seen in Section \ref{sec:iid} for independent identically distributed random variables, such as the sum or the {\color{blue}extreme-value} statistics, to the conditionally independent identically distributed random variables. For example, for $N$ independent identically distributed random variables, we have seen in Section \ref{sec:iid} that the rescaled sum, defined in Eq.~(\ref{eq:CN-def}) and reminded here for convenience, $C_N = \frac{1}{N} \sum_{i=1}^N X_i$ converges to a Gaussian random variable in the large $N$ limit. How does the central limit theorem get modified for conditionally independent identically distributed random variables? Similarly, as mentioned in Section \ref{sec:iid}, the {\color{blue}extreme-value} statistics of $N$ independent identically distributed random variables, appropriately centered and scaled, converge for large $N$, to one of the three limiting distributions Gumbel, Fréchet and Weibull. Are there similar universal limiting distributions for conditionally independent identically distributed random variables? 

\vspace{0.2cm}

Once these general results are elucidated, it is natural to look for examples in physical systems where the conditionally independent identically distributed random variables arise naturally. This will be the topic of Chapter \ref{ch:sim-reset}. For readers who may prefer seeing first the physical motivation and then the mathematical generalization, I recommend starting by reading the first example of Chapter \ref{ch:sim-reset} and then coming back to the current section. 

\vspace{0.2cm}

The starting point for all derivations in this section is the definition of conditionally independent identically distributed random variables given in Eq. (\ref{eq:def-ciid}). To keep similar notations as the ones used in Section \ref{sec:iid} we introduce 
\begin{equation}\label{eq:def-p-h}
    p(x | \vec{y}) = {\rm Prob.}\left[X = x \,|\, \vec{Y} = \vec{y} \right] \mbox{~~and~~} h(\vec{y}) = {\rm Prob.}\left[ \vec{Y} = \vec{y} \right] \;,
\end{equation}
notice that we dropped the $i$ index in ${\rm Prob.}[X_i = x \,|\, \vec{Y} = \vec{y}]$ since all $X_i$ are identically distributed. Using Eq. (\ref{eq:def-p-h}) we can rewrite Eq. (\ref{eq:def-ciid}) compactly as
\begin{equation} \label{eq:def-ciid-compact}
    {\rm Prob.}\left[\vec{X} = \vec{x}\right] = \int \dd^M \vec{y} \; \left\{ \prod_{i = 1}^N p\left(x_i | \vec{y}\right) \right\} h\left(\vec{y}\right) \;.
\end{equation}
It is clear from Eq. (\ref{eq:def-ciid-compact}) that the joint probability density function of the $X_i$s is not factorizable, in general. Hence, the $X_i$'s are correlated. Choosing $h\left(\vec{y}\right)$, one can generate a wide class of such correlated variables. To analyze the nature of the correlations, it is convenient to compute the connected correlator $\langle X_i^n X_j^n \rangle - \langle X_i^n \rangle \langle X_j^n \rangle$, for a generic $n$. To compute this, we first need the $n$-th moment $\langle X_i^n \rangle$, which is simply
\begin{equation} \label{eq:Xin}
\langle X_i^n \rangle = \int \dd^N \vec{x} \; x_i^n \; {\rm Prob.} \left[\vec{X} = \vec{x}\right] = \int \dd^N \vec{x} \; x_i^n \int \dd^M \vec{y} \; h\left(\vec{y}\right) \prod_{k = 1}^N p\left(x_k | \vec{y}\right) \;.
\end{equation}
Permuting the integrals (which in all non-pathological examples can always be done) and using the normalization of the distributions, i.e.
\begin{equation} \label{eq:normalization-p}
    1 = \int \dd x \; p(x | \vec{y}) \;,
\end{equation}
we can simplify Eq. (\ref{eq:Xin}) significantly
\begin{equation} \label{eq:n-moment-ciid}
    \left\langle X_i^n \right\rangle = \int \dd^M\vec{y} \; h(\vec{y}) \int \dd x \; x^n p(x | \vec{y}) = \int \dd^M \vec{y} \; h(\vec{y}) \left\langle X^n | \vec{y} \right\rangle \;.
\end{equation}
Consequently, the connected correlator is given by
\begin{equation} \label{eq:connected-ciid}
\langle X_i^n X_j^n \rangle - \langle X_i^n \rangle \langle X_j^n \rangle = \int \dd^M \vec{y}\; h\left(\vec{y}\right)  \int \dd^N \vec{x}\;  x_i^n x_j^n \prod_{k = 1}^N p(x_k | \vec{y})   -  \langle X_i^n \rangle \langle X_j^n \rangle   \;.
\end{equation}
Once again, commuting the integrals and using Eq. (\ref{eq:normalization-p}) we can simplify the connected correlator to
\begin{equation} \label{eq:connected-ciid-2}
    \langle X_i^n X_j^n \rangle - \langle X_i^n \rangle \langle X_j^n \rangle = \int \dd^M \vec{y}\; h(\vec{y}) \langle X^n | \vec{y} \rangle^2 - \left( \int \dd^M \vec{y} \; h(\vec{y}) \langle X^n | \vec{y} \rangle \right)^2 \;,
\end{equation}
which can be re-written as
\begin{align} 
    \langle X_i^n X_j^n \rangle &- \langle X_i^n \rangle \langle X_j^n \rangle = \nonumber\\
    &\int \dd^M \vec{y_1} \dd^M \vec{y_2} \; h(\vec{y_1}) h(\vec{y_2}) \langle X^n | \vec{y_1} \rangle \langle X^n | \vec{y_2}\rangle \left(\delta(\vec{y_1} - \vec{y_2}) - 1\right)\label{eq:connected-ciid-3}
\end{align}
For a generic $h(\vec{y})$, this connected correlator is nonzero, indicating
the presence of all-to-all correlations between the $X_i$ variables.

\section{The scaled sum}\label{subsec:sum-ciid}

We start by studying the sum of such conditionally independent identically distributed random variables to explore if there is an equivalent to the central limit theorem or L\'evy stable theorem for the conditionally independent case. We will see shortly that this is not the case for conditionally independent random variables. We start with the joint distribution in Eq. (\ref{eq:def-ciid-compact}). The probability density function of the sample mean, or equivalently the center of mass, 
\begin{equation} \label{eq:def-CN}
    C_N =\frac{1}{N} \sum_{i=1}^N X_i    
\end{equation}
can be written as
\begin{equation} \label{eq:PC}
{\rm Prob.}[C_N = c] = \int \dd^N {\vec{x}} \; \delta\left(c - \frac{1}{N} \sum_{i=1}^N x_i\right) \; \int \dd^M \vec{y}\; \left\{\prod_{i = 1}^N p(x_i | \vec{y})\right\} h(\vec{y}) \;.
\end{equation}
Taking the Fourier transform with respect to $c$, one gets
\begin{eqnarray}\label{eq:PC_TF}
\int \dd c \; {\rm Prob.}[C_N = c] \, e^{i k c} = \int \dd^M \vec{y} \; \left[\hat p\left( \frac{k}{N}\Big \vert \vec{y}\right)\right]^N h(\vec{y}) \;,
\end{eqnarray}
where $\hat{p}(k|\vec{y})$ is the Fourier transform of the conditional distribution function $p(x|\vec{y})$ with respect to $x$
\begin{equation} \label{eq:def-fourier-p}
    \hat{p}(k | \vec{y}) = \int \dd x \; p(x | \vec{y}) e^{i k x} \;.
\end{equation}
To {\color{blue}analyze} the large $N$ behavior of the Fourier transform in Eq. \ref{eq:PC_TF}, we need to {\color{blue}analyze} the small $k$ behavior of $\hat p(k\vert \vec{y})$. Assume that $p(x|\vec{y})$ admits a finite first moment 
\begin{equation} \label{eq:ciid-first-moment}
m(\vec{y}) = \int \dd^N x \; x \, p(x | \vec{y})
\end{equation}
and second moment, 
\begin{equation} \label{eq:ciid-second-moment}
    \sigma^2(\vec{y}) = \int \dd x \; x^2 \, p(x | \vec{y}) - \left(\int \dd x \; x \, p(x | \vec{y})\right)^2  \;.
\end{equation}
Then the logarithm of its Fourier transform admits the small $k$ expansion  
\begin{eqnarray}\label{eq:log_p}
\log \hat p(k|\vec{y}) = i \, m(\vec{y})\,k - \frac{\sigma^2(\vec{y})}{2} k^2 + O(k^3) \;. 
\end{eqnarray}
Consequently, exponentiating Eq. (\ref{eq:log_p}) and keeping terms up to order $O(k^2)$ inside the exponential, one gets
\begin{eqnarray} \label{exp_p}
\hat{p}(k|\vec{Y})\sim e^{i \,m(\vec{Y})\,k -\frac{1}{2}\sigma^2(\vec{Y})k^2} \;.
\end{eqnarray}
Substituting this in Eq. (\ref{eq:PC_TF}), and inverting the Fourier transform, we get, for large $N$,
\begin{equation} \label{eq:dnC_theta}
   {\rm Prob.}[C_N = c] \approx \frac{1}{2\pi} \int_{-\infty}^{\infty} \dd k \int \dd^M \vec{y}\; h(\vec{y}) \exp[i m(\vec{y}) k -\frac{1}{2 N}\sigma^2(\vec{y})k^2-i k c] \;.
\end{equation}
If $m(\vec{y})$ is not a constant function of $\vec{y}$ then we can drop the quadratic term $O(k^2/N)$ in Eq. (\ref{eq:dnC_theta}). Then, to leading order for large $N$, the distribution ${\rm Prob.}[C_N = c]$ becomes independent of $N$ with a limiting form  
\begin{equation}
{\rm Prob.}[C_N = c] \underset{N \to \infty}{\longrightarrow} \int \dd^M \vec{y}\; \delta(m(\vec{y}) - c) \; h(\vec{y})\;. \label{eq:CLT_u}
\end{equation}
In particular, the moments of this limiting distribution can be calculated easily from Eq. (\ref{eq:CLT_u}) leading to
\begin{equation} \label{eq:moment-center-ciid}
    \langle C^n\rangle \underset{N \to \infty}{\longrightarrow} \int \dd^M \vec{y} \; m^n(\vec{y}) \, h(\vec y) \;.
\end{equation} 
On the other hand, if $m(\vec{y}) = m$ is a constant, then one can shift $C_N$ by $m$ and rescale it by $\sqrt{N}${\color{blue}, in other words, apply the central limit theorem on the conditional variable $C_N | \vec{y}$.} It is then easy to see that ${\rm Prob.}[C_N = c]$ converges to a scaling form
\begin{eqnarray} \label{eq:scaling_C}
{\rm Prob.}[C_N = c] \underset{N \to \infty}{\longrightarrow}  \sqrt{N} \, {\cal P} \left( \left(c-m\right)\sqrt{N}\right)
\end{eqnarray}
where the scaling function ${\cal{P}}(z)$ is given by
\begin{equation} \label{eq:P_of_Z1}
   {\cal{P}}(z)  = \frac{1}{2\pi} \int_{-\infty}^{\infty} \dd \tilde k \int \dd^M \vec{y} \; h(\vec{y}) e^{-\frac{1}{2}\sigma^2(\vec{y})\tilde k^2 - i\tilde k z}  \;.
\end{equation}
Performing the integral over $\tilde k$, {\color{blue}we find} 
\begin{equation} \label{eq:P_of_Z2}
   {\cal{P}}(z) = \frac{1}{\sqrt{2 \pi}}\int \frac{\dd^M  \vec{y}}{\sigma(\vec{y})} \; h(\vec{y}) \exp \left(-\frac{z^2}{2\sigma^2(\vec{y})}\right) \;.
\end{equation}
This is the main exact result for the limiting distribution of the scaled sum in the case where $m(\vec{y})$ is independent of $\vec{y}$. {\color{blue} Notice the Gaussian form in the integrand which is the consequence of having applied the central limit theorem to the conditional center of mass $C_N | \vec{y}$. However, the non-conditional} limiting distribution ${\cal{P}}(z)$ {\color{blue} is clearly non-Gaussian for generic $h(\vec{y})$ and $\sigma(\vec{y})$, its exact form will depend} on the details of $h(\vec{y})$ and $\sigma(\vec{y})$. The moments of the scaling variable $z$ can be computed from Eq. (\ref{eq:P_of_Z2}) leading to
\begin{eqnarray}\label{eq:moments_Z}
\langle z^{2n}\rangle = \frac{\Gamma(2n)}{2^n \Gamma(n)} \int \dd^M {\vec{y}} \sigma^{2n}(\vec{y})\, h(\vec{y})  \quad {\rm and} \quad  \langle z^{2n+1}\rangle=0 \;,
\end{eqnarray}
for any $n = 0, 1, 2, \cdots$. Using this result together with the scaling form in Eq. (\ref{eq:scaling_C}) we get
\begin{align}
&\langle (C_N-m)^{2n}\rangle = \frac{\Gamma(2n)}{(2N)^n \Gamma(n)} \int^M \dd {\vec{y}} \sigma^{2n}(\vec{y})\, h(\vec{y})  \nonumber \\
\quad {\rm and} \quad  &\langle (C_N-m)^{2n+1}\rangle=0 \;. \label{moments_C}
\end{align}
So far, we have assumed that the two first moments of $p(x\vert \vec{y})$ are finite. In case they are divergent, one can perform a similar analysis 
for conditionally independent random variables as in the case of independent identically distributed L\'evy variables. For simplicity, we assume that the variable $X | \vec{Y}$ is symmetric with zero mean and the conditional probability density function $p(x|\vec{y})$ has a power law tail $p(x|\vec{y}) \sim 1/x^{1+\mu}$ for large $x$, with $0 < \mu < 2$. We also assume for simplicity that $\mu$ is independent of $\vec{y}$. In this case, one can approximate the small $k$ behavior of the Fourier transform $\hat p(k\vert \vec{y}) \approx e^{- \,|b(\vec{y})\, k|^{\mu}}$ where $b(\vec{y})$ is a scale factor. Substituting this in Eq.~(\ref{eq:PC_TF}) and inverting the Fourier transform, we get
\begin{eqnarray}\label{eq:PC_TF_mu}
{\rm Prob.}[C_N = c] \approx \int_{-\infty}^\infty \frac{\dd k}{2 \pi}  \int \dd^M \vec{y} \; h(\vec{y}) e^{- i k c} e^{-  N^{1-1/\mu} |b(\vec{y})\, k|^\mu} \;.
\end{eqnarray}
Performing the change of variable $k = \tilde k N^{1-1/\mu}$, one finds that ${\rm Prob.}[C_N = c]$ takes the scaling form
\begin{eqnarray}\label{scaling_mu} 
{\rm Prob.}[C_N = c] \approx N^{1-1/\mu} \; \tilde{\cal P}_\mu\left(\frac{c}{N^{1/\mu - 1}}\right)
\end{eqnarray}
where the scaling function $\tilde{\cal P}_\mu(z)$ reads
\begin{eqnarray}\label{scaling_mu2}
\tilde{\cal P}_\mu(z) = \int_{-\infty}^\infty \frac{\dd \tilde k}{2 \pi} \int  \dd^N \vec{y}\; h(\vec{y}) e^{- i \tilde k z} e^{- |b(\vec{y})\,\tilde k|^\mu} \;.
\end{eqnarray}
Finally, performing the integral over $\tilde k$, it can be expressed in the compact form
\begin{eqnarray}\label{eq:scaling_mu3}
\tilde{\cal P}_\mu(z) = \int \frac{\dd^M \vec{y}}{b(\vec{y})} \, h(\vec{Y}) \,{\cal L}_{\mu} \left( \frac{z}{b(\vec{y})}\right)  \;,
\end{eqnarray}
where ${\cal L}_{\mu}(z)$ is the L\'evy stable distribution as stated in Eq.~(\ref{eq:levy-stable}). In the case $\mu = 2$, ${\cal L}_2(z)$ $=$ $e^{-z^2/4}/(2 \sqrt{\pi})$ and Eq. (\ref{eq:scaling_mu3}) gives back (\ref{eq:P_of_Z2}) with $\sigma(\vec{y})$ $=$ $\sqrt{2}\,b(\vec{y})$. 

\section{The order statistics}\label{subsec:order-ciid}

In this section, we provide a complete characterization of the order statistics and {\color{blue}extreme-value} statistics for conditionally independent identically distributed random variables. From the joint probability distribution in Eq.~(\ref{eq:def-ciid-compact}) the probability density function of the $k$-th maximum $M_{k, N}$ is given by
\begin{equation} \label{eq:total_prob}
    {\rm Prob.}[M_{k, N} = w] = \int \dd^M \vec{y} \; h(\vec{y}) {\rm Prob.}[M_{k, N} = w \, | \, \vec{y}]
\end{equation}
The conditional distribution ${\rm Prob.}[M_{k, N} = w \, | \, \vec{y}]$ of the order statistics is given by the independent identically distributed result presented in Eq.~(\ref{eq:iid-order-bulk-gaussian})
\begin{align} \label{eq:Mk_gauss}
\text{Prob.}[M_{k, N} = w | \vec{y}] \underset{N\to\infty}{\longrightarrow} \sqrt{\frac{N \left[p (q \,|\,  \vec{y})\right]^2 }{2 \pi \alpha (1 - \alpha)}} \exp\left( - \frac{N \left[p(q \,|\, \vec{y})\right]^2}{2 \alpha(1 - \alpha)} [w - q]^2 \right) \;,
\end{align}
where $\alpha = k/N$ and $q \equiv q(\alpha, \vec{y})$ is the $\alpha$-quantile of the conditional distribution $p(x | \vec{y})$ defined implicitly via
\begin{equation} \label{eq:def-q-ciid}
    \alpha = \int_{q(\alpha, \vec{y})}^{+\infty} \dd x \; p(x | \vec{y}) \;.
\end{equation}
For brevity, we suppressed the explicit dependence on $\alpha$ and $\vec{y}$ of $q \equiv q(\alpha, \vec{y})$ in Eq.~(\ref{eq:Mk_gauss}). Eq. (\ref{eq:Mk_gauss}) is simply a Gaussian distribution centered around $q$ with variance 
\begin{equation} \label{eq:variance_main}
{\rm Var}[M_{k, N} | \vec{y}] = \frac{\alpha(1-\alpha)}{N \left[p(q \, | \, \vec{y})\right]^2} \;. 
\end{equation}
For the order statistics in the bulk where $\alpha \sim \mathcal{O}(1)$, it follows from Eq. (\ref{eq:def-q-ciid}) that $q(\alpha, \vec{Y})$ is also of order $\mathcal{O}(1)$. Consequently, from Eq. (\ref{eq:variance_main}), we find that ${\rm Var}[M_{k, N} | \vec{y}] \sim \mathcal{O}(1/N)$. Under these conditions, in the large $N$ limit, the probability density function in Eq. (\ref{eq:Mk_gauss}) becomes sharply peaked and can be approximated by a Dirac delta function centered at $q(\alpha, \vec{y})$. Note that this is true regardless of the underlying distribution $p(x | \vec{y})$. Substituting the Dirac delta function back into Eq. (\ref{eq:total_prob}), we find that the probability density function of the $k$-th maximum in the bulk converges, in the large $N$ limit, to an $N$-independent limiting form given by
\begin{equation} \label{eq:res_bulk_integral_form}
\text{Prob.}[M_{k, N} = w] \underset{N \to \infty}{\longrightarrow} \int \dd \vec{y} \; \; h(\vec{y}) \, \delta[w - q(\alpha, \vec{y})] \;.
\end{equation}
This result characterizes the order statistics in the bulk for conditionally independent identically distributed random variables. 

\section{The {\color{blue}extreme-value} statistics} \label{subsec:max-ciid}

However, for the order statistics near the edge, where $\alpha = \mathcal{O}(1/N)$, we need to carefully study the dependence on $N$ of the quantile $q(\alpha, \vec{y})$ and the variance~${\rm Var}[M_{k, N} | \vec{y}]$. From Eq. (\ref{eq:def_wstar}) we see that when $\alpha \sim \mathcal{O}(1/N) \ll 1$, the quantile $q(\alpha, \vec{y})$ depends on the tail of $p(x | \vec{y})$ for large $x$. We therefore analyze separately the three classes of tails that lead to the three universality classes of the {\color{blue}extreme-value} statistics in the independent identically distributed case. We start by studying the Gumbel class, where the distribution $p(x \, | \vec{y})$ decays faster than a power law for large $X$. Then we will look at the Weibull class, where the support of the distribution $p(x | \vec{y})$ is bounded above and approaches its upper bound as a power law. Finally, we study the Fréchet class, where the distribution $p(x  | \vec{y})$ has an unbounded support and decays as a power-law for large $x$. We will see that while the bulk result in Eq. (\ref{eq:res_bulk_integral_form}) can be extrapolated to the edge (where $k = \mathcal{O}(1)$) for the Gumbel and Weibull classes; the same cannot be done for the Fréchet case. In the Fréchet case, the order statistics at the edge needs to be analyzed separately. 

\subsubsection{The Gumbel case}

We start by studying random variables belonging to the Gumbel class, e.g., when the tail of the distribution $p(x | \vec{y})$ decays as 
\begin{equation} \label{eq:gumbel_tail}
p(x | \vec{y}) \; \underset{{x} \gg 1}{\longrightarrow} \;  A(\vec{y}) e^{- B(\vec{y}) {x}^{\mu(\vec{y})} } \;,
\end{equation}
where $A(\vec{y}), B(\vec{y}), \mu(\vec{y}) > 0$ are positive real functions of $\vec{y}$. For simplicity, we will drop the explicit dependence on $\vec{y}$ of $A(\vec{y}), B(\vec{y})$ and $\mu(\vec{y})$ and simply denote them by $A, B$ and $\mu$. In order to probe the behavior of the extremes near the edge, i.e., of $M_{k,N}$ for $k \sim \mathcal{O}(1)$ we start by studying the dependence on $N$ of the quantile $q(\alpha, \vec{y})$ when $\alpha = k/N = \mathcal{O}(1/N)$. From Eq. (\ref{eq:def_wstar}) we see that when $\alpha \ll 1$ then necessarily $q(\alpha, \vec{y}) \gg 1$. Hence, in Eq. (\ref{eq:def_wstar}), we can replace the integrand by its tail behavior given by Eq. (\ref{eq:gumbel_tail}) and obtain
\begin{align} \label{eq:compute_quantile_gumbel_1}
\alpha \approx \int_{q(\alpha, \vec{Y})}^{+\infty}  \; A \, e^{-B x^{\mu}} \; \dd x = \frac{A}{\mu B^{1/\mu}} \; \int_{B \, q^\mu(\alpha, \vec{Y}) }^{+\infty} \; t^{1/\mu - 1} \, e^{-t} \; \dd t \\
\underset{q(\alpha, \vec{Y}) \gg 1}{\approx} \;\;  
 \frac{A}{B \, \mu} \, q^{1 - \mu}(\alpha, \vec{Y})\, e^{-B \, q^{\mu}(\alpha, \vec{Y})}  \;{.} \label{eq:compute_quantile_gumbel_2}
\end{align}
where we approximated the integral by its value at its lower bound which is the leading order contribution due to the exponential decay of the integrand. Since we are at the edge, where $\alpha = k/N \ll 1$ and $q(\alpha, \vec{y}) \gg 1$, the leading order solution for $q(\alpha, \vec{y})$ from Eq. (\ref{eq:compute_quantile_gumbel_1}-\ref{eq:compute_quantile_gumbel_2}) is given by 
\begin{equation} \label{eq:qaY_gumbel_tail}
q(\alpha, \vec{Y}) \underset{\alpha \to 0}{ \approx} \frac{1}{B^{1/\mu}} {\log^{1/\mu}}\left( \frac{A}{B \alpha \mu} \right) \;.
\end{equation}
Replacing $\alpha = k/N$ with $k \sim \mathcal{O}(1)$ fixed and taking $N \to +\infty$ we obtain
\begin{equation} \label{eq:q_alpha_gumbel}
q(\alpha, \vec{Y}) \approx \frac{1}{B^{1/\mu}} {\log^{1/\mu}}\left( N \frac{A}{B  k \mu} \right) \sim \mathcal{O}\left[ {\log^{1/\mu}}(N) \right] \;,
\end{equation}
thus justifying a posteriori that $q(\alpha, \vec{Y}) \gg 1$ for large $N$. We now investigate how the variance ${\rm Var}[M_{k, N} | \vec{y}]$ in Eq. (\ref{eq:variance_main}) depends on $N$ for large $N$. To do so, we substitute Eq. (\ref{eq:q_alpha_gumbel}) back into Eq. (\ref{eq:gumbel_tail}) yielding
\begin{equation} \label{eq:pqY_gumbel_tail}
p(  q \, | \, \vec{y} ) \underset{\alpha \ll 1}{ \approx} \frac{B k \mu}{N} \;.
\end{equation}
Using Eq. (\ref{eq:pqY_gumbel_tail}) in Eq. (\ref{eq:variance_main}), we get that the variance is given by
\begin{equation}
{\rm Var}[M_{k, N} | \vec{y}] = \frac{\alpha (1 - \alpha)}{N \left[  p(q\,  | \, \vec{y}) \right]^2 }  \approx \frac{k/N}{N \left( \frac{B k \mu}{N} \right)^2}  \approx \frac{1}{{B^2} \mu^2 k} \sim \mathcal{O}(1)\;.
\end{equation}
Thus, the variance ${\rm Var}[M_{k, N} | \vec{y}] = \mathcal{O}(1)$ while from Eq. (\ref{eq:q_alpha_gumbel}) the mean $q(\alpha, \vec{y})$ $=$ $\mathcal{O}( [\log N]^{1/\mu} )$. Hence, we can write
\begin{equation}\label{eq:def_etak}
M_{k, N} \approx q(\alpha, \vec{y})+ \eta_k\;,
\end{equation}
where $\eta_k$ is a random variable of order $\mathcal{O}(1)$. Therefore, in the large $N$ limit, the fluctuations are negligible compared to the mean for any positive $A(\vec{y})$, $B(\vec{y})$ and $\mu(\vec{y})$. For large $N$, the $k$-th maximum $M_{k, N}$ effectively concentrates on its mean value $M_{k, N} \approx q(\alpha, \vec{y}) \sim {\log^{1/\mu}}(N)$.
Thus, the approximation used to obtain Eq. (\ref{eq:res_bulk_integral_form}) is still valid even at the edge, i.e. for $\alpha \sim \mathcal{O}(1/N)$. Hence, for the Gumbel class, Eq. (\ref{eq:res_bulk_integral_form}) characterizes both the order statistics in the bulk as well as at the edge for conditionally independent identically distributed random variables. 

\subsubsection{The Weibull case}

We now turn our attention to conditionally independent variables belonging to the Weibull class. In this case, the support of the distribution $p(x | \vec{y})$ is bounded above by $x^\star(\vec{y})$, and approaches this bound with a power-law tail given by
\begin{equation} \label{eq:tail_weibull}
p(x | \vec{y}) {\underset{x \to x^\star(\vec{y})}{\approx}}  A(\vec{y}) (x^\star(\vec{y})- x)^{\mu(\vec{y}) - 1} \;,
\end{equation}
for $x$ close to $x^\star(\vec{Y})$, where $A(\vec{y}), x^\star(\vec{y}), \mu(\vec{y}) > 0$ are positive real functions of $\vec{y}$, where for simplicity we will drop the explicit dependence on $\vec{y}$. We start by studying the dependence on $N$ of the quantile $q(\alpha, \vec{y})$. Once again, to probe the {\color{blue}extreme-value} statistics we have to fix $k \sim \mathcal{O}(1)$ while taking $N \to +\infty$ to study the behavior close to the global maximum $M_{1,N}$. Hence $\alpha\sim\mathcal{O}(1/N)\ll 1$ and $q(\alpha, \vec{y})$ must be approaching $x^\star$ from below. This justifies using the tail expression given in Eq. (\ref{eq:tail_weibull}) inside the integral in Eq. (\ref{eq:def_wstar}). Hence,
\begin{equation}
\alpha = \int_{q(\alpha, \vec{y})}^{+\infty}\dd x\; p(x | \vec{y})  = A \int_{q(\alpha, \vec{y})}^{x^\star}\dd x\; (x^\star - x)^{\mu - 1}  = \frac{A}{\mu} \, (x^\star - q(\alpha, \vec{y}))^{\mu} \;,
\end{equation}
which gives
\begin{equation} \label{eq:q_alpha_weibull}
q(\alpha, \vec{y}) = x^\star - \left(\frac{\alpha \mu}{A}\right)^{1/\mu} \;.
\end{equation}
In the limit of $N\to+\infty$, keeping $k \sim \mathcal{O}(1)$ fixed, $\alpha = k/N \to 0$ and therefore $q(\alpha, \vec{y}) \sim \mathcal{O}(1)$. We now study the dependence on $N$ of the variance ${\rm Var}[M_{k, N} | \vec{y}]$. Plugging Eq. (\ref{eq:q_alpha_weibull}) back into Eq. (\ref{eq:tail_weibull}) we get
\begin{equation}
p(q \, | \, \vec{y}) \approx A^{1/\mu} (\alpha \mu)^{1 - 1/\mu} \;.
\end{equation}
Thus, the variance in Eq. (\ref{eq:variance_main}), is given by
\begin{align} \label{eq:var_alpha_weibull_1}
{\rm Var}[M_{k, N} | \vec{y}] = \frac{\alpha (1 - \alpha)}{ N \left[p(q \,|\, \vec{Y})\right]^2 } \approx \frac{k/N (1 - k/N)}{ N A^{2\mu} (k \mu/N)^{2 - 2/\mu} } \\
\approx \frac{k}{A^{2\mu} (k \mu)^{2 - 2/\mu} } N^{-2\mu} \sim \mathcal{O}(N^{-2\mu})  \;, \label{eq:var_alpha_weibull_2}
\end{align}
where we used $\alpha = k/N$ and took the {\color{blue}extreme-value} statistics limit of $N \to +\infty$ keeping $k = \mathcal{O}(1)$ fixed. From Eqs. (\ref{eq:q_alpha_weibull}) and (\ref{eq:var_alpha_weibull_1}-\ref{eq:var_alpha_weibull_2}) we see that the mean is of order $\mathcal{O}(1)$ while the variance is of order $\mathcal{O}(N^{-2\mu})$. Hence, the approximation that the Gaussian is sharply peaked holds, and the result derived in Eq. (\ref{eq:res_bulk_integral_form}) is valid for any $k$. 
Hence, for the Weibull class, Eq. (\ref{eq:res_bulk_integral_form}) also fully characterizes both the order statistics in the bulk and at the edge for conditionally independent identically distributed random variables. 

\subsubsection{The Fréchet case}

Finally, we turn our attention to conditionally independent variables belonging to the Fréchet class. The conditional distribution $p(x  | \vec{y})$ has an unbounded support with a power-law tail given by
\begin{equation} \label{eq:tail_frechet}
p(x | \vec{y}) \underset{x \gg 1}{\approx}  \frac{A(\vec{y})}{{x}^{1 + \mu(\vec{y})}} \;,
\end{equation}
where $A(\vec{y}) > 0$ and $2 > \mu(\vec{y}) > 0$ are positive real functions of $\vec{y}$. Once more, we drop the explicit dependence on $\vec{y}$ of $A(\vec{y})$ and $\mu(\vec{y})$ for brevity and we will restore it later. 
To probe the {\color{blue}extreme-value} statistics, we have to take the limit of $N \to +\infty$ keeping $k = \mathcal{O}(1)$ fixed to study the behavior close to the global maximum $M_{1,N}$. In this limit, $\alpha = k/N = \mathcal{O}(1/N) \ll 1$, which from Eq. (\ref{eq:def_wstar}) implies $q(\alpha, \vec{y}) \gg 1$. Therefore, the integrand in Eq. (\ref{eq:def_wstar}) can be replaced by its tail behavior given in Eq. (\ref{eq:tail_frechet}). Then,
\begin{equation}
\alpha \approx \int_{q(\alpha, \vec{y})}^{+\infty} A \frac{\dd x}{x^{1 + \mu}} \approx \frac{A}{\mu} {q^{-\mu}}(\alpha, \vec{y}) \;,
\end{equation}
hence
\begin{equation}
q(\alpha, \vec{y}) \approx \left(\frac{A}{\mu \alpha}\right)^{1/\mu} \;.
\end{equation}
Taking, $\alpha = k/N$ and the limit $N \to +\infty$ keeping $k = \mathcal{O}(1)$ fixed results in
\begin{equation} \label{eq:q_alpha_frechet}
q(\alpha, \vec{y}) \approx \left(\frac{A N}{\mu k}\right)^{1/\mu}  \sim \mathcal{O}(N^{1/\mu})\;.
\end{equation}
This characterizes the dependence on $N$ of the quantile $q(\alpha, \vec{y})$. We now turn our attention to the dependence on $N$ of the variance ${\rm Var}[M_{k, N} | \vec{y}]$. 
Plugging Eq. (\ref{eq:q_alpha_frechet}) back in Eq. (\ref{eq:tail_frechet}) we get
\begin{equation}
p( q \,|\, \vec{y}) \underset{\alpha \ll 1}{\approx}  A^{1/\mu} \left(\alpha\right)^{1/\mu + 1} \approx \frac{A^{1/\mu} k^{1/\mu + 1}}{N^{1/\mu + 1}} ,
\end{equation}
where we used $\alpha = k/N$. Then, the variance in Eq. (\ref{eq:variance_main}) is given by
\begin{equation}
{\rm Var}[M_{k, N} | \vec{y}] = \frac{\alpha(1- \alpha)}{N \left[p(q \, | \, \vec{y})\right]^2} \sim  \mathcal{O}(N^{2/\mu}) \;.
\end{equation}
Thus the width of the fluctuations is of the same order as the mean value for large $N$, since ${\rm Var}[M_{k, N} | \vec{y}]/q(\alpha, \vec{y})^2 \sim \mathcal{O}(1)$ and hence the distribution does not concentrate on its mean for large $N$ -- contrary to the Gumbel and the Weibull class. Thus in the Fréchet case, we need to analyze the extremes at the edge directly from the limiting behavior given in Eq. (\ref{eq:order-universal-form}), instead of extrapolating the bulk result to the edge. We recall the result in Eq. (\ref{eq:order-universal-form}) for $\rho = {\rm II}$ corresponding to the Fréchet class. It reads, for $k \sim O(1)$, 
\begin{equation} \label{eq:EVS-frechet-ciid}
{\rm Prob.}[M_{k, N} \leq w \,|\, \vec{y}] \approx G^{k}_{{\rm II}}\left( \frac{w - a_N}{b_N} \right) \;,
\end{equation}
where $G^{k}_{{\rm II}}(z)$ is given in Eq. (\ref{eq:order-cumulative-edge}). From Eq. (\ref{eq:frechet-coefs}), we know that the scale factors are given by
\begin{equation} \label{eq:aN_bN_frechet}
a_N = 0 \mbox{~~and~~} \int_{b_N}^{+\infty} p(x | \vec{y}) \dd x = \frac{1}{N} \;.
\end{equation}
Comparing Eq. (\ref{eq:aN_bN_frechet}) with Eq. (\ref{eq:def_wstar}) we see immediately that 
\begin{equation}
b_N = q(1/N, \vec{y}) \equiv q_N(\vec{y}) \;.
\end{equation}
Putting the scale factors obtained in Eq. (\ref{eq:aN_bN_frechet}) in Eq. (\ref{eq:EVS-frechet-ciid}) we get
\begin{equation} \label{eq:Mk_2_Gk}
{\rm Prob.}[M_{k, N} \leq w \,|\, \vec{y} ] \underset{N \to +\infty}{\longrightarrow} \frac{1}{\Gamma(k)} \int_{ w^{-\mu} q_N(\vec{Y})^\mu }^{+\infty} \dd t \; e^{-t} t^{k-1} \;.
\end{equation} 
where we used $G_{\rm II}(z) = \Theta(z)\,e^{-z^{-\mu}}$ from Eq. (\ref{eq:frechet-cumulative}). 

\vspace{0.2cm}

Once again, in the $N \gg 1$ limit, we have $q_N(\vec{y}) \gg 1$, so we can plug the tail behavior given in Eq. (\ref{eq:tail_frechet}) into Eq. (\ref{eq:aN_bN_frechet}) and solve for $q_N(\vec{y})$, as was done before for $q(\alpha, \vec{y})$. This yields
\begin{equation} \label{eq:bny}
q_N(\vec{y}) \simeq \left( \frac{A N}{\mu} \right)^{1/\mu} \;.
\end{equation}
Now, using Eq. (\ref{eq:Mk_2_Gk}) in Eq. (\ref{eq:total_prob}), we get
\begin{align} \label{eq:intermed_1-1}
{\rm Prob.}[M_{k, N} \leq w]  = \int \dd \vec{y} \; h(\vec{y}) {\rm Prob.}[M_{k, N} \leq w \, | \, \vec{y}] \\
= \frac{1}{\Gamma(k)} \int \dd \vec{y} \; h(\vec{y}) \int_{w^{-\mu} q_N(\vec{y})^\mu}^{+\infty} \dd t \; e^{-t}t^{k-1} \;. \label{eq:intermed_1-2}
\end{align}
We denote the lower bound of the integral over $t$ by 
\begin{equation} \label{eq:def_lambda}
\lambda_N(w, \vec{y}) = \left(\frac{q_N(\vec{y})}{w}\right)^{\mu(\vec{y})} =  \frac{A(\vec{y}) N}{ \mu(\vec{y}) w^{\mu(\vec{y})} }  \;,
\end{equation}
where we restored the explicit dependence of $A$ and $\mu$ on $\vec{y}$. Next we express the integral over $t$ as
\begin{align} \label{decomp-1}
\int_{\lambda_N(w, \vec{y})}^{+\infty} \dd t \; e^{-t}\, t^{k-1} &= \Gamma(k) - \int_0^{\lambda_N(w, \vec{y})} \dd t \; e^{-t}t^{k-1}  \\
&= \Gamma(k) - \int_0^{\infty} \, \dd t \, \Theta\left(\lambda_N(w,\vec{y})-t\right) \,  e^{-t}\,t^{k-1} \;.\label{decomp-2}
\end{align}
Substituting this integral in Eq. (\ref{eq:intermed_1-1}-\ref{eq:intermed_1-2}) and using the fact that $h(\vec{y})$ is normalized to unity, we get, 
\begin{equation} \label{eq:intermed_2}
{\rm Prob.}[M_{k, N} \leq w] = 1 - \frac{1}{\Gamma(k)} \int_0^\infty \dd t \, e^{-t}\,t^{k-1} \int \dd \vec{y} \, h(\vec{y}) \,  \Theta\left(\lambda_N(w,\vec{y})-t\right) \;.\qquad 
\end{equation}
Thus interpreting $\lambda_N(w, \vec{y})$ as a random variable, we get
\begin{equation}\label{eq:res:frechet}
{\rm Prob.}[M_{k, N} \leq w] = 1 - \frac{1}{\Gamma(k)} \int_0^{\infty} \dd t \; {\rm Prob.}\left[\lambda_N(w, \vec{y}) \geq t\right]\; e^{-t} \, t^{k-1} \;, \quad
\end{equation}
where 
\begin{equation} \label{dist_lambda}
{\rm Prob.}\left[\lambda_N(w, \vec{y}) \geq t\right] = \int \dd \vec{y} \, h(\vec{y}) \,  \Theta\left(\lambda_N(w,\vec{y})-t\right) \;.
\end{equation}
Given the distribution $h(\vec{y})$ of $\vec{y}$, it follows that $A(\vec{y})$ and $\mu(\vec{y})$ are random variables in Eq. (\ref{eq:def_lambda}). We first need to calculate the distribution of $\lambda_N(w,\vec{y})$ defined in (\ref{eq:def_lambda}) using Eq. (\ref{dist_lambda}). Finally, we need to substitute this cumulative distribution function of $\lambda_N(w,\vec{y})$ in Eq. (\ref{eq:res:frechet}) to compute the cumulative distribution of $M_{k,N}$. Hence, Eq. (\ref{eq:res:frechet}) characterizes the order statistics at the edge for the conditionally independent identically distributed random variables belonging to the Fréchet class.  

\section{The gap statistics}\label{subsec:gap-ciid}

We now turn our attention to the gap statistics, that is, the description of $d_{k,N} = M_{k,N} - M_{k+1, N} \geq 0$. Similarly to what we did for the order statistics we exploit the conditionally independent structure
\begin{equation}\label{eq:ciid-gap-renewal}
    {\rm Prob.}[d_{k, N} = g] = \int \dd^M \vec{y} \; h(\vec{y}) {\rm Prob.}[d_{k, N} = g \, | \, \vec{y}] \;,
\end{equation}
to relate the gap statistics to their independent equivalent. From Section \ref{sec:iid}, we know that if the underlying conditional distribution $p(x | \vec{y})$ belongs to the Gumbel class, we do not need to separate the behavior in the bulk from the one at the edge, since the gaps are distributed everywhere exponentially. Using Eq.~(\ref{eq:iid-gap-bulk}) we can re-write Eq.~(\ref{eq:ciid-gap-renewal}) as
\begin{equation} \label{eq:ciid-gaps-bulk}
    {\rm Prob.}[d_{k, N} = g] \underset{N \to \infty}{\longrightarrow} \int \dd^M \vec{y} \; h(\vec{y}) \; N p(q | \vec{y}) e^{-N p(q | \vec{y}) g} \;,
\end{equation}
where $q \equiv q(\alpha, \vec{y})$ is the $\alpha$-quantile of the conditional distribution defined in Eq. (\ref{eq:def-q-ciid}). On the other hand, for the Fréchet and Weibull class, the behavior in the bulk is not the same as that at the edge. In the bulk, the gaps are distributed exponentially and are therefore also described by Eq.~(\ref{eq:ciid-gaps-bulk}). However, at the edge, we need to use Eq.~(\ref{eq:gaps-iid-cumulative-weibull}) and Eq.~(\ref{eq:gaps-iid-cumulative-frechet}). When the conditional distribution $p(x | \vec{y})$ belongs to the Weibull class, i.e. Eq.~(\ref{eq:tail_weibull}) holds, we get
\begin{align} \label{eq:ciid-gaps-weibull}
    {\rm Prob.}[d_{k, N} = g] = \int &\dd^M \vec{y} \; \Bigg[\frac{h(\vec{y})}{b_N(\vec{y})} \frac{\mu(\vec{y})^2}{(k-1)!} \\
    \times & \int_0^{+\infty} \dd x\; \left(x + \frac{g}{b_N(\vec{y})}\right)^{\mu(\vec{y}) - 1} e^{-\left(x + \frac{g}{b_N(\vec{y})}\right)^{\mu(\vec{y})}} x^{\mu(\vec{y}) k - 1}\Bigg] \;, \nonumber
\end{align}
where $b_N(\vec{y})$ is the scale factor defined in Eq.~(\ref{eq:weibull-coefs}) which we remind for convenience
\begin{equation}
    \int_{x^\star(\vec{y}) - b_N(\vec{y})}^{x^\star(\vec{y})} \dd x \; p(x | \vec{y}) = \frac{1}{N} \;.
\end{equation}
For the special case, $\mu(\vec{y}) = 1$ Eq.~(\ref{eq:ciid-gaps-weibull}) simplifies to
\begin{equation} \label{eq:ciid-gaps-weibull-mu1}
    {\rm Prob.}[d_{k, N} = g] = \int \dd^M \vec{y} \; \frac{h(\vec{y})}{b_N(\vec{y})} e^{-g / b_N(\vec{y})} \;.
\end{equation}
For conditional distributions belonging to the Fréchet universality class, i.e. when Eq.~(\ref{eq:tail_frechet}) holds, we must use Eq.~(\ref{eq:gaps-iid-cumulative-frechet}) in Eq.~(\ref{eq:ciid-gap-renewal}) yielding
\begin{align} \label{eq:ciid-gaps-frechet}
    {\rm Prob.}[d_{k, N} = g] = \int &\dd^M \vec{y} \Bigg[\frac{h(\vec{y})}{b_N(\vec{y})} \frac{\mu(\vec{y})^2}{(k - 1)!}  \\
    \times &\int_0^{+\infty} \dd x \; e^{-x^{-\mu(\vec{y})}} x^{-\mu(\vec{y})-1} (x + \frac{g}{b_N(\vec{y})})^{-\mu(\vec{y}) k - 1} \Bigg] \;,\nonumber
\end{align}
which also simplifies when $\mu(\vec{y}) = 1$ in terms of the confluent hypergeometric $U$ function
\begin{equation}
    {\rm Prob.}[d_{k, N} = g] = \int \dd^M \vec{y} \; \frac{h(\vec{y})}{b_N(\vec{y})} k (k+1) \left(\frac{b_N(\vec{y})}{g}\right)^{1 + k} U\left(k+1, 0, \frac{b_N(\vec{y})}{g}\right) \;. 
\end{equation}
We remind again the expression of the Fréchet scale factor $b_N(\vec{y})$ {\color{blue}which was given in} Eq.~(\ref{eq:frechet-coefs}) for completeness
\begin{equation}
    \int_{b_N(\vec{y})}^{+\infty} \dd x \; p(x | \vec{y}) = \frac{1}{N} \;.
\end{equation}

\section{The full counting statistics}\label{subsec:fcs-ciid}

Finally, the last observable we will be looking at are the full counting statistics. Identically to the previous subsection \ref{subsec:gap-ciid}, we start from the conditional distribution given in Eq. (\ref{eq:def-ciid-compact}) and exploit the conditionally independent structure
\begin{equation}
    {\rm Prob.}[N_L = n] = \int \dd^M \vec{y} \; h(\vec{y}) {\rm Prob.}[N_L = n \, | \, \vec{y}] \;,
\end{equation}
allowing us to use the exact independent form derived in Eq. (\ref{eq:fcs-iid-scale}). From Eq.~(\ref{eq:fcs-iid-scale}) we know
\begin{equation}
    N_L | \vec{y} \sim a_N(\vec{y}) + b_N(\vec{y}) \eta \;, \mbox{~~when~~} N \to +\infty \;,
\end{equation}
where 
\begin{equation}
    a_N(\vec{y}) = N \int_{-L}^{+L} \dd x \; p(x | \vec{y}) 
\end{equation}
and
\begin{equation}
    b_N(\vec{y})^2 = N \left(\int_{-L}^{L} \dd x \; p(x | \vec{y})\right) \left(1 - \int_{-L}^{L} \dd x \; p(x | \vec{y}) \right) \;.
\end{equation}
We immediately see that 
\begin{equation}
    \frac{b_N(\vec{y})}{a_N(\vec{y})} = \frac{1}{\sqrt{N}} \sqrt{\frac{1 - \int_{-L}^{L} \dd x \; p(x | \vec{y})}{\int_{-L}^{L} \dd x \; p(x | \vec{y})}} \stackrel{N \to \infty}{\longrightarrow} 0 \;,
\end{equation}
where we assumed that we are not in a pathological case where $a_N(\vec{y}) = 0$. Therefore, in the large $N$ limit, the random fluctuations of the full counting statistics become negligible compared to its mean. Hence, similarly to what was done for the order statistics, the full counting statistics of conditionally independent identically distributed random variables are given by
\begin{equation} \label{eq:ciid-fcs}
    {\rm Prob.}[N_L = n] \underset{N\to\infty}{\longrightarrow} \int \dd^M \vec{y} \; h(\vec{y}) \; \delta\left[ n - N \int_{-L}^{L} \dd x \; p(x | \vec{y}) \right] \;.
\end{equation}


\chapter{Resetting-induced strong correlations} \label{ch:sim-reset}
In Chapter \ref{ch:extreme} we lay the foundations and recall the main results which are known for independent and weakly correlated variables. In Chapter \ref{ch:ciid}, we introduced the mathematical formalism which underlies several of the models studied in this thesis, namely, conditionally independent identically distributed random variables. Stochastically resetting systems are a family of processes for which this structure arises naturally, and were the original inspiration from which this generalization came. Therefore, we will start by defining and introducing some of the general results for resetting stochastic systems in Section \ref{sec:intro-resetting}. After which we will look at 4 systems with stochastic resetting in Section \ref{sec:ciid-reset} which can be solved analytically exactly using conditionally independent identically distributed random variables. Namely, we will study simultaneously resetting Brownian motions, ballistic motions, and Lévy flights in Subsections \ref{subsec:BM-simreset}, \ref{subsec:ballistic-simreset} and \ref{subsec:levy-simreset}, respectively. Each of these examples was chosen specifically to showcase a different class of conditionally independent variables. Namely, Gumbel, Weibull and Fréchet for Brownian motion, ballistic motion, and Lévy flights, respectively. Finally, in Subsection \ref{subsec:fpt-simreset}, we will look at an alternative way to use simultaneous resetting to generate correlations. Instead of resetting at random Poissonianly distributed times, we consider $N$ Brownian motion which reset simultaneously whenever {\it any} of them cross a fixed barrier $L > 0$.

\section{A brief overview of resetting} \label{sec:intro-resetting}

\begin{figure}
    \centering
    \includegraphics[width=0.7\textwidth]{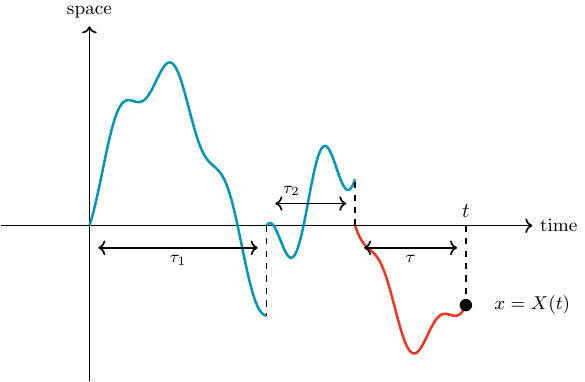}
    \caption{A schematic representation of a one-dimensional resetting Brownian motion ending at $x$ at time $t$ with two resetting events at time $\tau_1$ and $\tau_1 + \tau_2$ respectively. The intervals $\tau_i$ are distributed exponentially with rate $r$, i.e. from the distribution $h(\tau) = r e^{-r \tau}$. The time interval between the last resetting event and the final time $t$ is denoted by $\tau$. Outside of the resetting events the particle undergoes a typical Brownian motion with diffusion constant $D$. Since Brownian motion is a Markov process the last segment (in red) is independent from the previous two (in blue). Hence the probability to reach $x$ at time $t$ can be obtained from the reset-free probability of the last (red) segment to reach $x$ in time $\tau$ as explained after Eq. (\ref{eq:resetting-renewal}).} \label{fig:resetting-sketch}
\end{figure}

The aim of this Section is not to provide a comprehensive review of all existing results in the resetting literature, more so to introduce the main definitions and results which will be useful in the subsequent Sections. For a review of the resetting literature, we recommend Ref. \cite{EMS20,PKR22,GJ22}.

\vspace{0.2cm}

Stochastic resetting has become a significant research topic in statistical physics, with broad applications across multiple disciplines. It plays a crucial role in areas such as search algorithms in computer science, foraging behavior in ecology, reaction-diffusion systems in chemistry, and gene transcription in biology \cite{EMS20,PKR22,GJ22}. Stochastic resetting involves randomly interrupting a system's natural evolution and restarting it instantly from either its initial state or a predetermined configuration. The time intervals between resets typically follow a Poisson distribution, although other mechanisms, such as periodic resetting, have also been explored. A key consequence of stochastic resetting is that resetting events disrupt detailed balance, leading the system to a non-equilibrium stationary state \cite{EMS20,PKR22,GJ22,EM11PRL,EM11JPhysA}. Understanding the characteristics of such a non-equilibrium steady state and its spatial properties has attracted considerable interest, both theoretically \cite{EMS20,PKR22,GJ22} and experimentally \cite{TPSRR20,BBPMC20,FBPCM21}. One of the simplest theoretical frameworks \cite{EM11PRL,EM11JPhysA} examines a single particle undergoing diffusion in a $d$-dimensional space while experiencing stochastic resetting at a constant rate $r > 0$, i.e. Poissonian resetting. 
Following this foundational model, numerous other theoretical studies have explored different variations of single-particle stochastic dynamics under several resetting conditions \cite{RUK14,BSS14,RRU15,MSS15,PKE16,R16,MV16,NG16,PR17,BEM17,EM18,CS18,MVB18,PKR19,MCM19,PKR20,BS20,BS20b,BRR20,B20,P20,BMS22,MMS22,SBS22,BM23,MOK23}.

\vspace{0.2cm}

\begin{figure}
    \centering
    \includegraphics[width=0.5\textwidth]{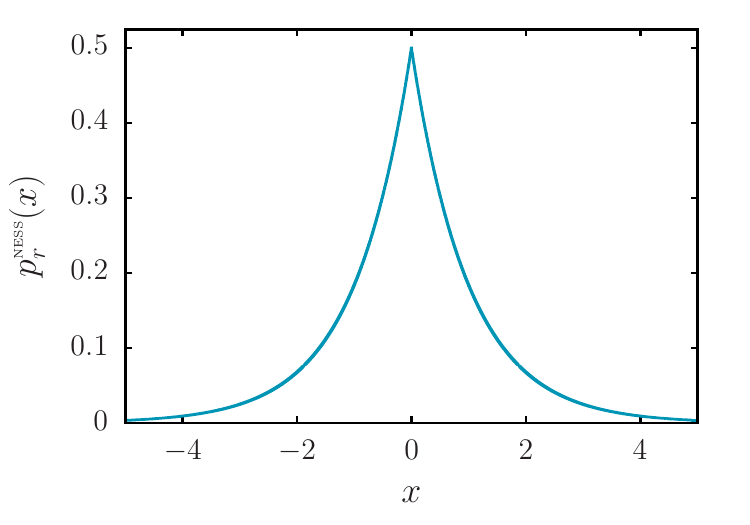}\hfill%
    \includegraphics[width=0.5\textwidth]{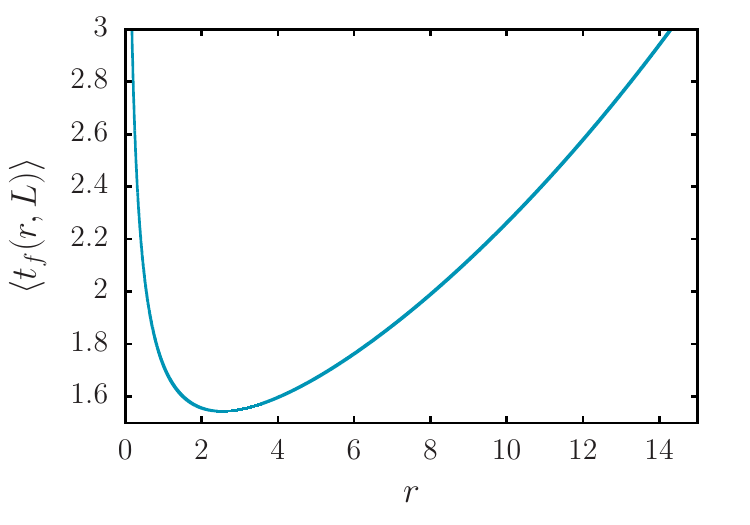}
    \caption{{\bf Left.} A plot of the probability density function $p_r^{\rm NESS}(x)$ given in Eq. (\ref{eq:resetting-ness}) of finding a resetting Brownian motion at $x$ in the long-time limit once the non-equilibrium steady state has been reached. The plot was obtained with $\sqrt{r/D} = 1$. {\bf Right.} A plot of the mean first-passage time $\langle t_f(r, L) \rangle$, given in Eq.~(\ref{eq:mean-first-passage-resetting}), as a function of the resetting rate $r$. The plot was obtained by setting $L = D = 1$.} \label{fig:resetting-NESS}
\end{figure}

To familiarize the reader with the main concepts and analytical techniques pertaining to theoretical resetting we will now re-derive some of the most well-known results for the foundational model presented in Refs. \cite{EM11PRL,EM11JPhysA}. Consider a single diffusing one-dimensional particle whose position at time $t$ we denote by $X(t)$, i.e.
\begin{equation} \label{eq:diffusion-reset}
    \dv{X(t)}{t} = \sqrt{2 D} \, \eta(t) \;,
\end{equation}
where $D > 0$ is the diffusion constant and $\eta(t)$ is a $\delta$-correlated Gaussian white noise with $\langle \eta(t) \rangle = 0$ and $\langle \eta(t) \eta(t') \rangle = \delta(t - t')$. This is the simple Brownian motion model studied in the previous chapter in Section \ref{sec:iid}. As stated previously, resetting a system consists in interrupting its natural dynamics and restarting it from a predetermined distribution. In the simplest case, we interrupt the dynamics given in Eq. (\ref{eq:diffusion-reset}) at any infinitesimal step $\dd t$ with a given probability $r \dd t > 0$ and restart from the origin. Hence, modifying the dynamics to 
\begin{equation} \label{eq:reset-dynamics}
    X(t + \dd t) = \begin{dcases}
        X(t) + \sqrt{2 D \dd t} \, \eta(t) &\mbox{~~with probability~~} 1 - r \dd t\\
        0 &\mbox{~~with probability~~} r \dd t \;.
    \end{dcases} 
\end{equation}
A sketch of a typical resetting Brownian motion trajectory is given in Fig. \ref{fig:resetting-sketch}. Consistently with the notations in the previous chapter, we denote by $p_r(x, t$ $|$ $x_0, t_0)$ the probability density function corresponding to the particle being located at $x$ at time $t$ given that it undergoes a Brownian motion with resetting rate $r$ starting from $x_0$ at time $t_0$. The Fokker-Planck equation correspond to the stochastic differential equation in Eq. (\ref{eq:reset-dynamics}) can be obtained from the forward master equation of $p(x, t | x_0, t_0)$ and is given by
\begin{equation} \label{eq:resetting-FP}
    \pdv{p_r(x, t | x_0, t_0)}{t} = D \pdv[2]{p_r(x, t | x_0, t_0)}{x} - r p_r(x, t | x_0, t_0) + r \delta(x) \;,
\end{equation}
and can be interpreted as follows. The first two terms correspond to the usual diffusive behavior (refer to Eq. (\ref{eq:BM-FP})). Then at every time-step, the process may `die' with rate $r$ and re-appear at the origin. Hence, every point in space acts as a probability sink, leading to the $-r p_r(x, t | x_0, t_0)$ term, with the corresponding source term at the origin $r \delta(x)$. The Fokker-Planck equation in Eq. (\ref{eq:resetting-FP}) can be solved directly through a series of analytical manipulations, but a simpler renewal argument can lead us directly to the solution
\begin{equation}\label{eq:resetting-renewal}
    p_r(x, t | x_0, t_0) = e^{-r t} p_0(x, t | x_0, t_0) + r \int_0^{t-t_0} \dd \tau \; e^{-r\tau} p_0(x, \tau | 0, 0) \;,
\end{equation}
where $p_0(x, t | x_0, t_0) \equiv p_0(x - x_0, t - t_0)$ is the reset-free propagator, i.e. the propagator of Brownian motion which we gave in Eq. (\ref{eq:propagator-BM}) and remind here
\begin{equation}\label{eq:reminder-propagator-BM}
    p_0(x, t) = \frac{1}{\sqrt{4 \pi D t}} e^{-\frac{x^2}{4 D t}} \;.
\end{equation}
The expression in Eq. (\ref{eq:resetting-renewal}) can be understood as follows. To reach $x$ at time $t$ starting from $x_0$ at time $t_0$ the resetting Brownian motion can either: 
\begin{itemize}
    \item {\bf Never reset.} The probability of never resetting is $e^{-r t}$ and the probability of reaching $x$ at time $t$ without resetting is given by $p_0(x, t | x_0, t_0)$ which leads to the first term in Eq. (\ref{eq:resetting-renewal}).
    \item {\bf Reset at least once.} If the process resets at least once, as sketched in Fig.~\ref{fig:resetting-sketch} we denote by $t - \tau$ the time at which the last resetting event occurred. Since Brownian motion is a Markov process, the probability of reaching $x$ at time $t$ does not depend on the history of the process before the last resetting event. Resetting once occurs with probability $r \dd \tau$ and reaching $x$ in a reset-free duration $\tau$ happens with probability $p_0(x, \tau | 0, 0)$. Integrating all possible values of $\tau$ leads to the second term in Eq.~(\ref{eq:resetting-renewal}).
\end{itemize}
At long times, the resetting Brownian motion reaches a non-equilibrium stea\-dy state, which can be obtained by taking the $t \to +\infty$ limit of Eq. (\ref{eq:resetting-renewal}). As $t \to +\infty$, the first term drops out and the bounds of the integral in the second term can be approximated by $[0, +\infty[$ up to an exponentially small error leading to
\begin{equation} \label{eq:resetting-ness}
    p_r(x) \equiv p_r(x, t \to +\infty) = r \int_0^{+\infty} \dd \tau \; e^{-r \tau} p_0(x, \tau) = \frac{1}{2}\sqrt{\frac{r}{D}} e^{- |x| \sqrt{\frac{r}{D}}} \;. 
\end{equation}
A plot of the probability distribution is given in the left panel of Fig. \ref{fig:resetting-NESS}. Notice the cusp at $x = 0$, which is a result of the resetting events circulating probability from everywhere in space to the origin.

\vspace{0.2cm}


We will now consider the resetting diffusion as a search process by introducing a target site at $L > 0$ and assuming that the process stops whenever it reaches the target. Let $t_f(r, L)$ denote the time at which the process reaches $L$ for the first time and terminates, i.e. the first-passage time at $L$. The main observable of interest is the mean {\color{blue}first-passage} time $\langle t_f(r, L) \rangle$ characterizing the expected time it would take for the search to be successful. In one-dimension the problem can be simplified since the first-passage at $L$ can be directly related to the behavior of the maximum $M(t)$ of the walk up to time $t$. The same observable which was studied in Section \ref{subsec:strong}. The process will not reach the target so long as $M(t) < L$ and it will reach the target for the first time whenever $M(t)$ crosses $L$. Secondly, notice that the renewal argument presented in Fig.~(\ref{fig:resetting-sketch}) and after Eq.~(\ref{eq:resetting-renewal}) is a perfect use case of the weakly correlated random variables introduced in Section \ref{subsec:weak}. The maximum $M(t)$ can be re-casted as the maximum of the maxima $m_i(\tau_i)$ of each non-resetting segment of duration $\tau_i$ as presented in Fig.~(\ref{fig:resetting-sketch}) and Fig~(\ref{fig:OU-weak}). Since the process is renewed at every reset all the $m_i(\tau_i)$ are perfectly independent. The distribution of the $m_i$ is known since they are the maximum of a normal diffusing walk, which we studied in Section \ref{subsec:strong}. We recall the result given in Eq.~(\ref{eq:maximum-BM}) for simplicity
\begin{equation} \label{eq:maximum-resetting-segment-BM}
    {\rm Prob.}[m_i(\tau_i) = z] = \frac{\Theta(z)}{\sqrt{\pi D \tau_i}} e^{-\frac{z^2}{4 D \tau_i}} \;.
\end{equation}
The distribution in Eq.~(\ref{eq:maximum-resetting-segment-BM}) clearly has exponential tails and a path of duration $t$ is typically split in $N = r t$ segments. Hence, from the general arguments presented in Section~\ref{subsec:weak}, we know that $M(t)$ will have a Gumbel distribution in the $N \gg 1$, i.e. $t \gg 1/r$ limit
\begin{equation} \label{eq:maximum-asymptotic-resetting}
    {\rm Prob.}[M(t) \leq z] \underset{t \gg 1/r}{\approx} \exp[ - r t \exp( - z \sqrt{\frac{r}{D}} )] \;.
\end{equation}
However, to obtain the mean first-passage time $\langle t_f(r, L) \rangle$ we need the behavior of $M(t)$ at all times, not only in the $t \gg 1/r$ limit. To do so, we introduce the survival probability
\begin{equation}
    Q_r(x, t) = {\rm Prob.}[X(0) = x \mbox{~~and~~} M(t) < L] \;,
\end{equation}
characterizing the probability that the resetting process does not find the target before time $t$, having started at $x$ at $t = 0$. The distribution of the {\color{blue}first-passage} time $t_f(r, L)$ can be easily obtained from the survival probability
\begin{equation} \label{eq:first-passage-surival}
    {\rm Prob.}[t_f(r, L) = t] = - \pdv{Q_r(0, t)}{t} \;,
\end{equation}
since the paths which reach $L$ for the first time at time $t$ correspond exactly to the paths which `die' in $[t, t + \dd t]$. Using Eq.~(\ref{eq:first-passage-surival}) we obtain a simple expression for the mean first-passage time
\begin{equation} \label{eq:mean-first-passage-int-survival}
    \langle t_f(r, L) \rangle = \int_0^{+\infty} \dd t \; t \left( - \pdv{Q_r(0, t)}{t} \right) = \int_0^{+\infty} \dd t\; Q_r(0, t) \;. 
\end{equation}
The survival probability $Q_r(x, t)$ has a similar renewal form as the one presented in Eq~(\ref{eq:resetting-renewal})
\begin{equation} \label{eq:resetting-survival-renewal}
    Q_r(x, t) = e^{-r t} Q_0(x, t) + r \int_0^{t} \dd \tau \; e^{-r \tau} Q_0(0, \tau) Q_r(x, t - \tau) \;,
\end{equation}
with a similar interpretation. The resetting Brownian motion can either never reset, leading to the first term, or reset at least once. If it resets at least once, which happens with probability $r \dd \tau$, let $\tau$ denote the duration of the last non-resetting interval. Then for the process to survive up to time $t$, it must have survived up to time $t-\tau$ in the presence of resetting, i.e. $Q_r(x, t - \tau)$, and it must {\color{blue}survive} for the remaining duration $\tau$ having re-started from the origin in the absence of resetting $Q_0(0, \tau)$. {\color{blue} For those familiar with these type of integral equations, notice that Eq.~(\ref{eq:resetting-survival-renewal}) is a Volterra integral equation of the second kind.} Furthermore, leveraging the space-translation invariance of Brownian motion we have
\begin{equation} \label{eq:survival-space-invariance}
    Q_0(x, t) = {\rm Prob.}[X(0) = x \mbox{~~and~~} M(t) < L | r = 0] = {\rm Prob.}[M(t) < L - x | r = 0]
\end{equation}
allowing us to use our previously obtained result in Eq~(\ref{eq:cumulative-BM}) for the survival probability of {\color{blue}one-dimensional} Brownian motion. Namely,
\begin{equation} \label{eq:survival-resetfree}
    Q_0(x, t) = {\rm erf}\left( \frac{L - x}{\sqrt{4 D t}} \right) \;.
\end{equation}
To solve the integral equation in Eq.~(\ref{eq:resetting-survival-renewal}) we need to perform a Laplace transform to simplify the {\color{blue}convolutional} structure. The Laplace transform is defined as
\begin{equation}
    \tilde{Q}_r(x, s) = \int_0^{+\infty} \dd t \; e^{-s t} Q_r(x, t) 
\end{equation}
and allows us to simplify Eq.~(\ref{eq:resetting-survival-renewal}) to 
\begin{equation} \label{eq:laplace-survival-reset-to-resetfree}
    \tilde{Q}_r(x, s) = \frac{\tilde{Q}_0(x, s + r)}{1 - r \tilde{Q}_0(0, s + r)} \;.
\end{equation}
Furthermore, we can readily compute the Laplace transform of Eq.~(\ref{eq:survival-resetfree}) which yields
\begin{equation} \label{eq:laplace-survival-resetfree}
    \tilde{Q}_0(x, s) = \frac{1 - e^{-\sqrt{\frac{s}{D}} (L - x) }}{s} \;.
\end{equation}
Using Eq.~(\ref{eq:laplace-survival-resetfree}) in Eq.~(\ref{eq:laplace-survival-reset-to-resetfree}) we obtain
\begin{equation} \label{eq:laplace-reset-survival}
    \tilde{Q}_r(x, s) = \frac{1 - e^{-\sqrt{\frac{r + s}{D}} \, (L - x) }}{s + r \, e^{-\sqrt{\frac{r + s}{D}} \, L}} \;.
\end{equation}
Using Eq.~(\ref{eq:mean-first-passage-int-survival}) we now can immediately obtain the mean first-passage time
\begin{equation} \label{eq:mean-first-passage-resetting}
    \langle t_f(r, L) \rangle = \int_0^{+\infty} \dd t \; Q_r(0, t) = \tilde{Q}_r(0, 0) = \frac{1}{r} \left(e^{L \sqrt{\frac{r}{D}}} - 1\right) \; .
\end{equation}
A plot of the mean first-passage time as a function of the resetting rate is given in the right panel of Fig.~\ref{fig:resetting-NESS}. As $r \to 0^+$ the mean first-passage diverges which we expect since the mean {\color{blue}first-passage} time for Brownian motion is divergent
\begin{equation}
    \langle t_f(r = 0, L) \rangle = \int_0^{+\infty} \dd t \; Q_0(0, t) = \int_0^{+\infty}\dd t\; {\rm erf}\left(\frac{L}{\sqrt{4 D t}}\right) = +\infty \;.
\end{equation}
On the other hand, as $r \to +\infty$ we get a classical equivalent of the quantum Zeno effect. Resetting happens infinitely frequently and hence the process becomes `stuck' at the origin. In between, these two extremes there is a unique minimum at $r^\star$ given by
\begin{equation}
    r^\star = D \left(\frac{\gamma}{L^2}\right)^2 \;,
\end{equation}
where the constant $\gamma \approx 1.5936\ldots$ is defined via the transcendental equation
\begin{equation}
    \frac{\gamma}{2} = 1 - e^{-\gamma} \;.
\end{equation}
{\color{blue} Although the existence of a minimum was guaranteed from our knowledge of the asymptotes, i.e. the resetting `freezing' when $r \to +\infty$ and the diverging search when $r \to 0^+$, the fact that this minimum is unique and can be obtained analytically is a major result. Notice that the knowledge of the exact optimal $r^\star$ is predicated on the knowledge of the target position $L$ which might seem like cheating. Some works \cite{ER24} have looked into relaxing this assumption and finding the optimal resetting distribution when a prior distribution is known for $L$. However, for most practical applications, the knowledge of the existence of such a minimum, which is not predicated on any knowledge of $L$ is sufficient.} This result is one of the reasons behind the broad inter-disciplinary interest resetting garnered. Searching for targets randomly is generally a terrible strategy, even more so in higher dimensions. Resetting is an easy modification which is also generally simple to implement that can lead to significant speedups. 

\section{Non-interacting dynamics with correlations induced via resetting} \label{sec:ciid-reset}

We have seen in Section~\ref{sec:iid} that the statistics of independent random variables are well understood. From a statistical physics perspective, the random variables will now always represent the positions of randomly moving particles. Hence a collection of independent random variables now represents a gas of noninteracting particles, whose properties are well understood, as we have seen in Section~\ref{sec:iid}. However, in Sections~\ref{subsec:weak} and \ref{subsec:strong} we saw that in the presence of correlations significantly less progress can be made, in particular in the presence of strong correlations. Once again, from a statistical physics perspective, if the random variables are particle positions, correlations between them represent interactions. When these interactions are short-ranged, similar arguments to the weakly correlated variables presented in Section~\ref{subsec:weak} can usually be made. However, when there are strong long-range correlations, there is no general way to approach the problem, and each model has to be studied in a case-by-case basis. Some notable exceptions are the Dyson log-gas \cite{M91, F10,LNV18}, the Generalized Random Energy model \cite{D81,D85} and the one-dimensional jellium model \cite{L61,P62,B63,DKMSS17}.

\vspace{0.2cm}

We saw in Chapter~\ref{ch:ciid} that there exists a family of strongly correlated variables for which we are still able to make general analytical progress, namely conditionally independent identically distributed random variables. The goal of the upcoming chapter is two-fold. We will introduce a series of multiparticle systems whose steady state contains strong long-range correlations and which can nevertheless be described in great analytical detail. Secondly, we will start building some intuition about what could be the general statistical physics equivalent of conditionally independent random variables. Motivated by the similarity between the conditional independent structure - see Eq.~(\ref{eq:def-ciid-compact}) - and the renewal equation - see Eq.~(\ref{eq:resetting-ness}) - our weapon of choice to generate these conditionally independent correlations is resetting.  

\vspace{0.2cm}

Specifically, we consider $N$ independent particles whose positions $X_1(t)$, $\cdots, X_N(t)$ (which we abbreviate as $\vec{X}(t)$) evolve on the line
under some dynamics which we refer to as their ``natural dynamics''. For example, they could be $N$ independent Brownian motions or $N$ independent ballistic particles, etc. For simplicity we consider the evolution in continuous time but this can easily be generalized to discrete time processes such as L\'evy flights. We introduce resetting to the origin by
\begin{equation} \label{eq:reset_dyn}
\vec{X}(t + \dd t) = \begin{dcases}
0, \cdots, 0 &\mbox{with probability~~} r \dd t \\
\, \\
\mbox{each~} X_i \mbox{~evolves independently} &\mbox{with complementary} \\
\mbox{via its natural dynamics} &\mbox{probability~~} 1 - r \dd t \;.
\end{dcases}
\end{equation}
In other words, after every time step $\dd t$, we either reset all the processes simultaneously to the origin with probability $r \dd t$, or we let each process evolve independently, as it would have in the absence of resetting. A cartoon illustrating the trajectories for three Brownian motions evolving under these dynamics is shown in Fig.~\ref{fig:intro-BM-simreset}. Consistently with the notations in the previous Sections, we denote by $p_0(x, t)$ the free-propagator (in the absence of resetting) at time $t$ of this process starting at $x=0$, i.e., the probability density for the process to arrive at $x$ at time $t$, starting at $x=0$ at time $t = 0$. In the absence of resetting the variables $X_1(t), X_2(t), \cdots, X_N(t)$ are independent. Hence, for a given $t$, their joint distribution factorizes  
\begin{equation}
{\rm Prob.}[\vec{X}(t) = \vec{x} \, | \, r = 0] = \prod_{i = 1}^N p_0(x_i, t)\;.
\end{equation}
We can write a renewal equation for the simultaneously resetting process which reads
\begin{equation} \label{eq:general_renewal}
{\rm Prob.}[\vec{X}(t) = \vec{x}] = e^{-r t} \prod_{i = 1}^N p_0(x_i, t)  + r \int_0^{t} \dd\tau \; e^{-r \tau} \prod_{i = 1}^N p_0(x_i, \tau) \;.
\end{equation}
This equation can be understood as follows. There is a possibility that the process never resets in the time interval $[0, t]$. 
The probability of never resetting in this interval is given by $e^{-r t}$ and the propagator $\prod_{i = 1}^N p_0(x_i, t)$ gives us the probability for the free-process to reach $\vec{x}$ at time $t$. This corresponds to the first term of Eq. (\ref{eq:general_renewal}). Otherwise, the process will reset at least once before reaching $\vec{x}$ at time $t$. Suppose it has reset for the last time at time $t - \tau$, it then has to reach $\vec{x}$ from $\vec{0}$ in the time interval $[t - \tau, t]$ while never resetting. The probability of resetting once is given by $r \dd \tau$, the probability of never resetting in the interval $[t - \tau, t]$ is given by $e^{-r\tau}$ and the probability of reaching $\vec{x}$ from $\vec{0}$ in time $\tau$ is given by the free-propagator $\prod_{i = 1}^N p_0(x_i, \tau)$. Multiplying this free propagator by the distribution of $\tau$ and integrating over $\tau$ gives the second term of Eq.~(\ref{eq:general_renewal}). We can see from Eq.~(\ref{eq:general_renewal}) that in the long time limit, $t \gg 1$, the first term drops out and the simultaneously resetting process reaches a non-equilibrium steady state
\begin{equation} \label{eq:general_NESS}
{\rm Prob.}[\vec{X} = \vec{x}]_{\rm NESS} = r \int_0^{+\infty} \dd\tau~ e^{-r \tau} \prod_{i = 1}^N p_0(x_i, \tau) \;,
\end{equation}
where we denoted by $\vec{X} = \vec{X}(t \to +\infty)$ the positions of the particles in the steady state. This steady state is out-of-equilibrium because resetting manifestly breaks detailed balance in the configuration space. 
Comparing Eq.~(\ref{eq:def-ciid-compact}) and Eq.~(\ref{eq:general_NESS}) we see that this system provides an example of {\color{blue}conditionally} independent random variables. Here, we have only $M=1$ conditioning variable $Y_1 = \tau$, which is the time elapsed since the last resetting event before $t$. Since $\tau$ has an exponential distribution, we then have $h(\tau) = r e^{-r \tau}$ with parameter $r$. As discussed in Chapter \ref{ch:ciid}, this joint probability density function does not factorize. Indeed the simultaneous resetting induces strong "all-to-all" correlations in our gas. 

\vspace{0.2cm}

Given the non-equilibrium steady state joint distribution given in Eq. (\ref{eq:general_NESS}), one can investigate various physical observables, such as the average density of the gas
\begin{equation} \label{eq:def-density}
    \rho(x, N | r) = \left\langle \frac{1}{N} \sum_{i = 1}^N \delta(X_i - x) \right\rangle 
\end{equation}
the center of mass
\begin{equation} \label{eq:def-center-mass}
    C_N = \frac{1}{N} \sum_{i = 1}^N X_i
\end{equation}
and the observables presented in Chapter \ref{ch:extreme}. Namely, the order statistics $\max_{i} X_i = M_1 \geq \cdots \geq M_N = \min_i X_i$, the statistics of gaps between particles $d_k = M_k - M_{k+1} \geq 0$ or the full counting statistics $N_L = \#\{X_i \, :\, X_i \in [-L, L]\}$, i.e., the distribution of the number of particles in a given interval. 

\subsection{Diffusive particles with simultaneous resetting \cite{BLMS23,BLMS24}} \label{subsec:BM-simreset}

\begin{figure}
    \centering
    \includegraphics[width=0.7\textwidth]{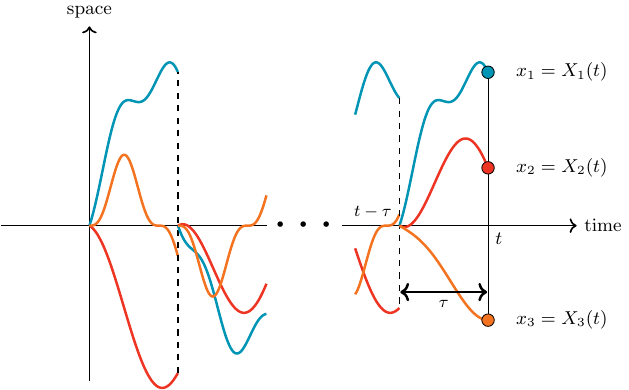}
    \caption{{Schematic trajectories of $N=3$ Brownian motions undergoing simultaneous resetting to the origin at random times. The observation time is marked by $t$ and the time of the last reset before $t$ is marked by $t-\tau$. During the last period $\tau$, the particles evolve independently as free Brownian motions.}}\label{fig:intro-BM-simreset}
\end{figure}

\begin{figure}
    \centering
    \includegraphics[width=0.7\textwidth]{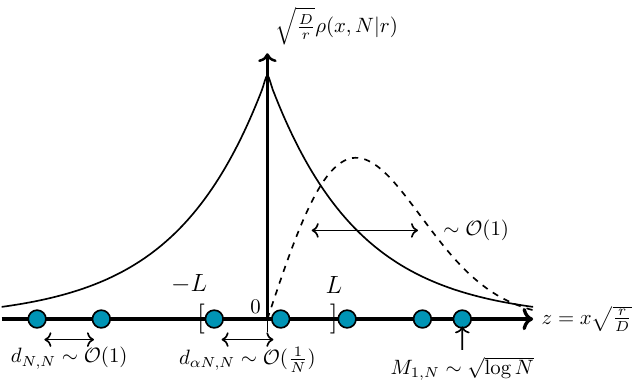}
    \caption{A schematic representation of the position of the particles in the simultaneously resetting Brownian gas and a summary of the main observables. The scaled average density profile, is supported over all the real line and is given in Eq. (\ref{eq:density-A}). Notice the cusp at 0 which is a consequence of the resetting. The distribution of the position $M_{1, N}$ of the rightmost particle is shown schematically by a dashed curve. The distribution is supported over $x \geq 0$ and is given in Eq. (\ref{eq:def-order-f-BM-reset}). The
    inter-particle spacing $d_{k, N}$ is of order one near the edge of the gas and $1/N$ in the bulk.} \label{fig:overiew-resetting-BM}
\end{figure}

In this Section, we consider $N$ independent Brownian particles on a line, all starting at the origin, that are {\it simultaneously} reset to the origin with rate $r$ as defined in Eq. (\ref{eq:reset_dyn}) (this is different from independently resetting Brownian particles studied before \cite{EM11PRL,EM11JPhysA,VAM22}). See Fig. \ref{fig:intro-BM-simreset} for a sketch of the dynamics of this system for $N = 3$. This {\it simultaneous} resetting makes the system strongly correlated, and this correlation persists even in the resulting many-body non-equilibrium steady state at long times. See Fig. \ref{fig:overiew-resetting-BM} for a sketch of the nonequilibrium steady state and the most pertinent observables. 

\vspace{0.2cm}

The propagator $p_0(x, t)$ of a single particle in the absence of resetting, i.e., the probability density to arrive at $x$ at time $t$ starting from $x=0$ is simply diffusive, i.e., 
\begin{equation} \label{free-propagator-A}
p_0(x, t) = \frac{1}{\sqrt{4 \pi D t}} \exp[ -\frac{x^2}{4 D t} ] \;.
\end{equation}
Plugging this propagator into Eq. (\ref{eq:general_NESS}) gives the joint distribution in the non-equilibrium steady state 
\begin{equation} \label{JPDF-stat}
{\rm Prob.}[\vec{X} = \vec{x}]_{\rm NESS} = r \int_0^{+\infty} \frac{\dd \tau}{(4 \pi D \tau)^{N/2}} \exp[ - r \tau - \frac{1}{4 D \tau} \sum_{i = 1}^N x_i^2 ] \;,
\end{equation}
which is manifestly non-factorizable, illustrating the fact that the positions of the particles become correlated in this non-equilibrium steady state. The origin of these correlations can be traced back to the simultaneous resetting of the particles. Given this joint distribution, one can compute our observables of interest which we will now treat one by one.

\subsubsection{The average density}

The average density in the steady state is given by
\begin{equation} \label{eq:density-A}
\rho(x, N \vert r) = r \int_0^{+\infty} \dd\tau\; e^{-r \tau} p_0(x \,|\, \tau) = \frac{1}{2} \sqrt{\frac{r}{D}} e^{-\sqrt{\frac{r}{D}} |x|} \;.
\end{equation}
Thus, even though at fixed $\tau$ the marginal distribution of the conditionally independent variables is Gaussian, once averaged over the conditioning variable $\tau$, the steady-state average density becomes highly non-Gaussian. A plot of this density is shown in Fig. \ref{fig:overiew-resetting-BM}. 

\subsubsection{The center of mass}

We will now look at the center mass defined in Eq. (\ref{eq:def-center-mass}). The two first moments of the distribution $p_0(x, t)$ are clearly finite and are given by
\begin{equation} \label{eq:mv_gaussian}
m(t) = 0 \mbox{~~and~~} \sigma^2(t) = 2 D t \;.
\end{equation}
Since $m(t) = 0$, we have to use Eq. (\ref{eq:scaling_C}) to obtain the statistics of the center of mass. Plugging Eq. (\ref{eq:mv_gaussian}) in Eq. (\ref{eq:scaling_C}) and Eq. (\ref{eq:P_of_Z2}) we immediately obtain {\color{blue} the Laplace distribution}
\begin{equation}
{\rm Prob.}[C_N = c] \underset{N\to\infty}{\longrightarrow} r \int_0^{+\infty} \dd \tau \; \sqrt{\frac{N}{4 \pi D \tau }} e^{- \frac{N c^2}{4 D \tau}} e^{-r \tau} = \frac{1}{2} \sqrt{\frac{r N}{D}} e^{- \sqrt{\frac{r N}{D}} |c|} \;. \label{eq:C2rho}
\end{equation}
\begin{figure}[t]
\centering
\begin{minipage}[b]{0.49\textwidth}
\includegraphics[width=\textwidth]{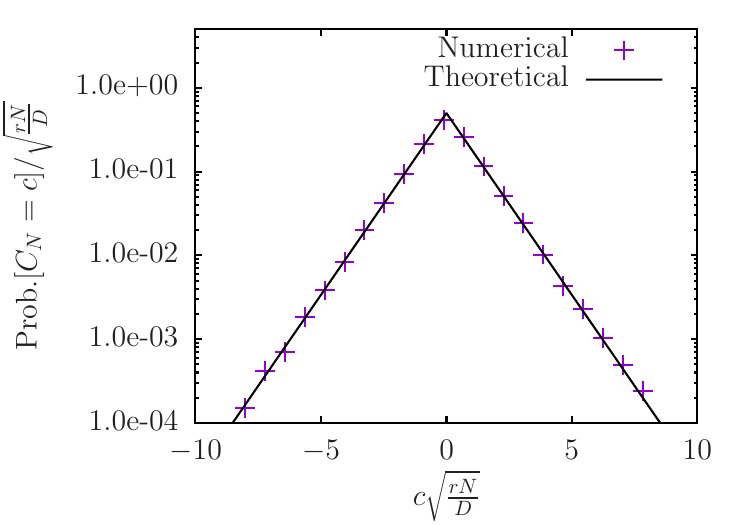}
\end{minipage}
\hfill
\begin{minipage}[b]{0.49\textwidth}
\includegraphics[width=\textwidth]{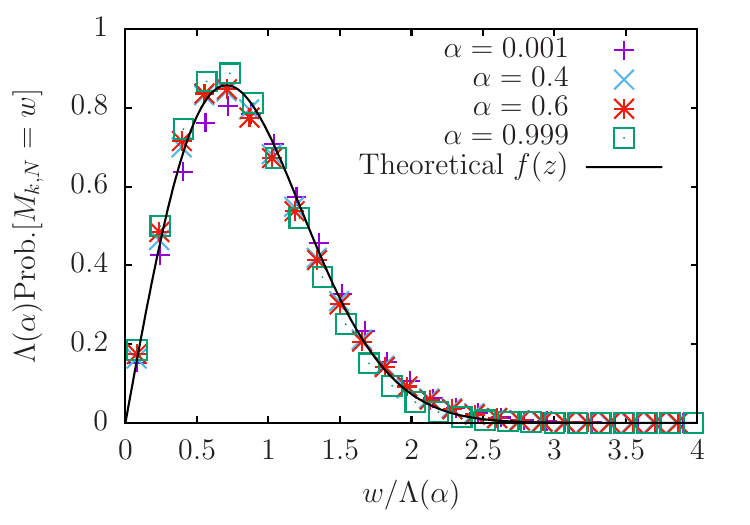}
\end{minipage}
\caption{Plots of the probability density functions of the center of mass (left panel) and the order statistics (right panel) obtained in Eq. \ref{eq:C2rho} and Eq. \ref{eq:res_brownian} respectively. The lines correspond to the theoretical predictions in Eq. \ref{eq:C2rho} -- left panel -- and Eq. \ref{eq:res_brownian} -- right panel. The symbols represent the results of numerical simulations. Different colors and symbols correspond to different values of $\alpha = k/N$, where we used $N=1000$ and $r = D = 1$.} \label{scaling_brownian}
\end{figure}
A plot of this scaling function is given in the left panel of Fig. \ref{scaling_brownian}, where it is also compared to numerical simulations.
Note that this is simply a rescaling of the average density we derived in Eq. (\ref{eq:density-A}). This is because Gaussian variables are stable under summation, which leads to a particular kind of simplification; see Ref. \cite{BLMS24}.

\subsubsection{The order and {\color{blue}extreme-value} statistics}

We now consider the order statistics in the bulk as well as the edge of this gas. The conditional distribution $p_0(x, t)$ in Eq. (\ref{free-propagator-A}) has a Gaussian tail, and hence it clearly belongs to the Gumbel class in Eq. (\ref{eq:gumbel_tail}). Calculating the explicit form of the quantile $q(\alpha, \vec{y}) \equiv q(\alpha, \tau)$ from Eq. (\ref{eq:def_wstar}), we obtain
\begin{equation} \label{w-star-sol-A}
q(\alpha, \tau) = \sqrt{4 D \tau} \, {\rm erfc}^{-1}(2 \alpha)\;,
\end{equation}
where ${\rm erfc}(z) = 2/\sqrt{\pi} \int_z^{\infty} e^{-u^2}\, \dd u$ and ${\rm erfc}^{-1}(z)$ is the associated inverse function. 
As argued in Section \ref{subsec:max-ciid}, the order statistics in the large $N$ limit, both in the bulk and at the edges, can be obtained within the same framework, namely from Eq. (\ref{eq:res_bulk_integral_form}), with the substitutions $\vec{y} \to \tau$ and $h(\vec{y}) = r\,e^{-r\tau}$. Plugging Eq. (\ref{w-star-sol-A}) in Eq.~(\ref{eq:res_bulk_integral_form}) gives
\begin{equation} \label{eq:Mk-delta}
{\rm Prob.}[M_{k, N} = w] \underset{N\to\infty}{\longrightarrow} r \int_0^{+\infty} \dd\tau e^{-r\tau} \delta\left[w -  \sqrt{4  D \tau} \, {\rm erfc}^{-1}(2\alpha) \right] \;.
\end{equation}
Performing this integral we immediately obtain 
\begin{equation} \label{eq:res_brownian}
{\rm Prob.}[M_{k, N} = w] \underset{N\to\infty}{\longrightarrow} \frac{1}{\Lambda(\alpha)} f\left( \frac{w}{\Lambda(\alpha)} \right) \mbox{~~with~~} \Lambda(\alpha) = \sqrt{\frac{4 D}{r}} {\rm erfc}^{-1}(2\alpha) \;,
\end{equation}
where the normalized scaling function $f(z)$ defined on $z > 0$ is given by
\begin{equation} \label{eq:def-order-f-BM-reset}
f(z) = 2 \, z \, e^{-z^2}, \quad z \geq 0 \;,
\end{equation}
In the right panel of Fig. \ref{scaling_brownian}, we compare this analytical prediction to numerical simulations, finding excellent agreement. The {\color{blue}extreme-value} statistics can be obtained by taking the limit $\alpha = k/N \to 0$ of $\Lambda(\alpha)$. Using the known {\color{blue}asymptotes} of the error complementary function we have
\begin{equation}
    {\rm erfc}^{-1}(2\alpha) \underset{\alpha \to 0^+}{\longrightarrow} \sqrt{\log N} \;.
\end{equation}
Hence the maximum (and more generally all the order statistics $M_{k, N}$ for $k \sim \mathcal{O}(1)$) is distributed as 
\begin{equation}
    {\rm Prob.}[M_{1, N} = w] \underset{N\to\infty}{\longrightarrow} \sqrt{\frac{r}{4 D \log N}} \; f\left(w \cdot \sqrt{\frac{r}{4 D \log N}} \right)\;.
\end{equation}

\subsubsection{The gaps}

\begin{figure}
    \centering
    \includegraphics[width=0.5\textwidth]{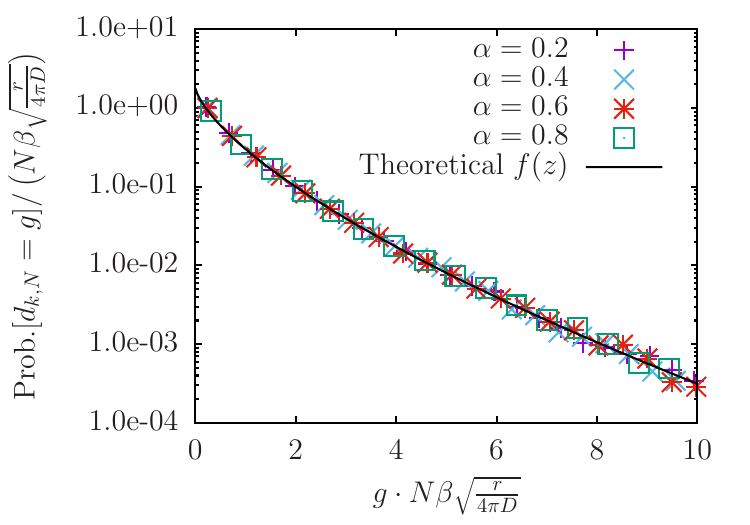}\hfill%
    \includegraphics[width=0.5\textwidth]{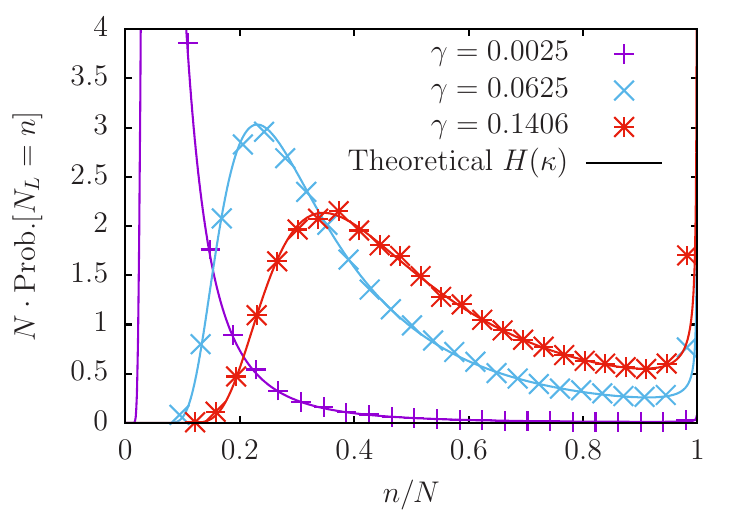}
    \caption{Plots of the probability density functions of the gaps (left panel) and the full counting statistics (right panel). The lines correspond to the theoretical predictions in Eq. \ref{eq:gaps-BM-simreset} -- left panel -- and Eq. \ref{eq:fcs-BM-simreset} -- right panel. The symbols represent the results of numerical simulations. Different colors and symbols correspond to different values of $\alpha = k/N$, where we used $N=1000$ and $r = D = 1$.} \label{fig:gaps-fcs-reset-BM}
\end{figure}

We know from Chapter \ref{ch:ciid}, and specifically from Eq.~(\ref{eq:ciid-gaps-bulk}), that the gap statistics for conditionally independent variables of the Gumbel class are given by
\begin{equation}\label{eq:general-gap-A}
    {\rm Prob.}[d_{k, N} = g] \underset{N\to\infty}{\longrightarrow} \int_0^{+\infty} \dd \tau \; r e^{-r \tau} N p_0(q(\alpha, \tau), \tau) e^{- N p_0(q(\alpha, \tau), \tau) g} \;,
\end{equation} 
both in the bulk ($\alpha \sim \mathcal{O}(1)$) and at the edge ($\alpha \sim \mathcal{O}(1/N)$). Using Eq.~(\ref{w-star-sol-A}) in Eq.~(\ref{free-propagator-A}) we get
\begin{equation} \label{eq:p0-q-A}
    p_0(q(\alpha, \tau), \tau) = \frac{1}{\sqrt{4 \pi D \tau}} e^{-{\rm erfc}^{-1}(2\alpha)^2} \;,
\end{equation}
and for compactness we introduce $\beta = e^{-{\rm erfc}^{-1}(2\alpha)^2}$. Using Eq.~(\ref{eq:p0-q-A}) in Eq.~(\ref{eq:general-gap-A}) we get
\begin{equation} \label{eq:gaps-BM-simreset}
    {\rm Prob.}[d_{k, N} = g] \underset{N\to\infty}{\longrightarrow} N \beta \sqrt{\frac{r}{4 \pi D}} \, F\left(g \cdot N \beta \sqrt{\frac{r}{4 \pi D}}\right) \;,
\end{equation}
where the scaling function $F(z)$ defined for $z \geq 0$ and normalized to unity is given by
\begin{equation}
    F(z) = 2 \int_0^{+\infty} \dd u \; e^{-u^2 - z/u} \;.
\end{equation}
A plot of the scaling function is given in the left panel of Fig. \ref{fig:gaps-fcs-reset-BM}, in perfect agreement with the numerical simulations. Although we cannot compute $F(z)$ explicitly, we can still extract its asymptotic behaviors. The large $z$ {\color{blue}asymptotes} can be obtained via a straightforward saddle point approximation. On the other hand, for the small behavior $z$, we have to separate the integration into two intervals: $[0, z]$ and $[z, +\infty[$. The leading contribution will come from the second interval where we can expand $e^{-z/u}$ in powers of $z$. Overall, the asymptotic behaviors of $F(z)$ are
\begin{equation}
    F(z) \approx \begin{dcases}
        \sqrt{\pi} + 2 z \ln z \;, &\mbox{~~when~~} z \to 0^+\\
        2 \sqrt{\frac{\pi}{3}} \exp[ - 3 \left(\frac{z}{2}\right)^{2/3} ] \;, &\mbox{~~when~~} z \to +\infty \;.
    \end{dcases}
\end{equation}
Thus, the universal scaling function $F(z)$ has rather non-trivial asymptotic behaviors. Its derivatives diverge logarithmically at $z = 0$ and have a stretched exponential tail for large $z$, with a stretching exponent $2/3$. A plot of this scaling function and comparison to the numerical simulation is given in the left panel of Fig. \ref{fig:gaps-fcs-reset-BM}, finding excellent agreement.

\subsubsection{The full counting statistics}

We now turn our attention to the full counting statistics $N_L = \#\{X_i : X_i \in [-L, L]\}$, i.e. the number of particles in a box $[-L, L]$ around the origin. Using the general results for conditionally independent variables in Chapter \ref{ch:ciid}, and specifically Eq.~(\ref{eq:ciid-fcs}), we know that the full counting statistics are given by 
\begin{equation} \label{eq:fcs-dirac-A}
    {\rm Prob.}[N_L = n] \underset{N\to\infty}{\longrightarrow} \int \dd^M \vec{y} \; r e^{-r \tau}\delta\left[n - N \, {\rm erf}\left(\frac{L}{\sqrt{4 D \tau}}\right) \right] \;,
\end{equation}
where we used 
\begin{equation} \label{eq:fcs-mean}
    \int_{-L}^{L} \dd x \; p_0(x, \tau) = {\rm erf}\left(\frac{L}{\sqrt{4 D \tau}}\right) \;.
\end{equation}
The integral in Eq. (\ref{eq:fcs-dirac-A}) can be computed exactly and written in a scaling form
\begin{equation} \label{eq:fcs-BM-simreset}
    {\rm Prob.}[N_L = n] \underset{N\to\infty}{\longrightarrow} \frac{1}{N} H\left(\frac{n}{N}\right) \;,
\end{equation} 
where the scaling function $H(\kappa)$ defined on $0 \leq \kappa \leq 1$ is given by
\begin{equation} \label{eq:H-A}
    H(\kappa) = \frac{\gamma \sqrt{\pi}}{u(\kappa)^3} \exp[ -\frac{\gamma}{u(\kappa)^2} + u(\kappa)^2 ] \;,
\end{equation}
and we introduced 
\begin{equation}
    \gamma = \frac{r L^2}{4 D} \mbox{~~and~~} u(\kappa) = {\rm erf}^{-1}(\kappa) {\color{blue}\;,}
\end{equation}
{\color{blue}where ${\rm erf}^{-1}$ denotes the inverse error function.}
A plot of the scaling function is given in the right panel of Fig. \ref{fig:gaps-fcs-reset-BM} in perfect agreement with numerical simulations. The scaling function $H(\kappa)$ is normalized to unity and has a highly non-trivial shape as can be seen in Fig. \ref{fig:gaps-fcs-reset-BM}. In the right panel of Fig. \ref{fig:gaps-fcs-reset-BM} we compare the analytical prediction of Eq. (\ref{eq:H-A}) with numerical simulation and find excellent agreement. The {\color{blue}asymptotes} of $H(\kappa)$ are given by
\begin{equation}
    H(\kappa) \approx \begin{dcases}
        \frac{8 \gamma}{\pi \kappa^3} \exp( - \frac{4 \gamma}{\pi \kappa^2} ) \;, &\mbox{~~when~~} \kappa \to 0^+\\
        \frac{\gamma \sqrt{\pi}}{(1 - \kappa)|\ln(1 - \kappa)|^{3/2}} \;, &\mbox{~~when~~} \kappa \to 1^- \;.
    \end{dcases}
\end{equation}
The {\color{blue}asymptotes} show that the full counting statistics decay incredibly quickly as $\kappa \to 0^+$ while they diverge extremely fast, barely integrable, as $\kappa \to 1^{-}$. The divergence at $\kappa \to 1^{-}$ comes from the resetting events which concentrate all the particles at the origin, and hence inside of $[-L, L]$, compounded with the one-dimensional Brownian motion which is already recurrent.

\subsection{Ballistic particles with simultaneous resetting \cite{BLMS24}}
\label{subsec:ballistic-simreset}

\begin{figure}
    \centering
    \includegraphics[width=0.7\textwidth]{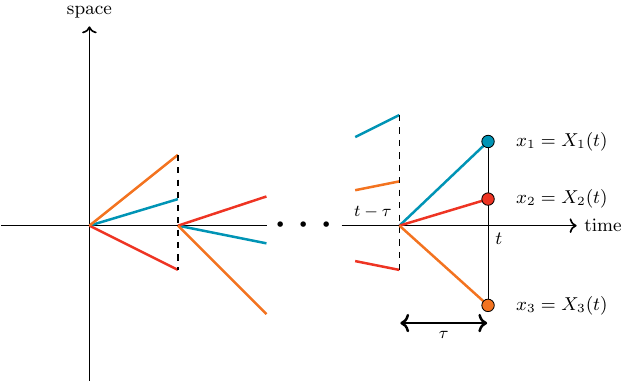}
    \caption{{Schematic trajectories of $N=3$ ballistic particles undergoing simultaneous resetting to the origin at random times. The observation time is marked by $t$ and the time of the last reset before $t$ is marked by $t-\tau$. During the last period $\tau$, the particles evolve independently as free ballistic particles. At each resetting event each particle selects a new velocity at random uniformly in $[-1, 1]$.}}\label{fig:intro-ballistic-simreset}
\end{figure}

\begin{figure}
    \centering
    \includegraphics[width=0.7\textwidth]{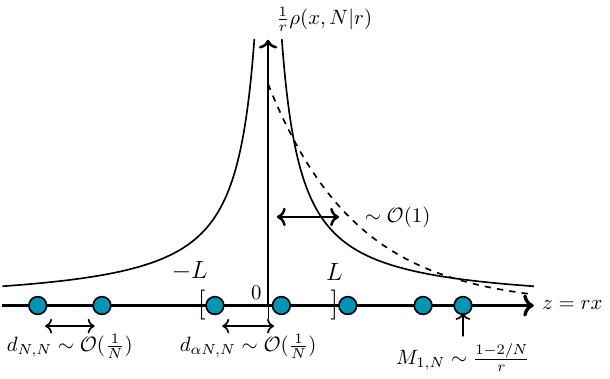}
    \caption{A schematic representation of the position of the particles in the simultaneously resetting ballistic gas and a summary of the main observables. The scaled average density profile, is supported over all the real line and is given in Eq. (\ref{rho_bal1}). Notice the cusp at 0 which is a consequence of the resetting. The exponential distribution of the position $M_{1, N}$ of the rightmost particle is shown schematically by a dashed curve. The distribution is a simple exponential distribution with rate $(1 - 2/N)/r$, i.e. ${\rm Prob.}[\chi] = e^{-\chi}$. The
    inter-particle spacing $d_{k, N}$ is of order $1/N$ both near the edge of the gas and in the bulk.} \label{fig:overiew-resetting-ballistic}
\end{figure}

A physical example belonging to the Weibull class of conditionally independent variables is a gas of $N$ particles undergoing ballistic motions on the line. All particles start at the origin and reset simultaneously to the origin with the rate $r$. At the end of every resetting event, each particle is assigned independently a random velocity $v_i$ drawn from a uniform distribution $n(v)$  
\begin{equation} \label{eq:def_nv}
n(v) = \begin{dcases}
\frac{1}{2} &\mbox{~~if~~} v \in [-1, 1]\\
0 &\mbox{~~otherwise} \;.
\end{dcases}
\end{equation}
In this case, the free propagator of a single particle at time $t$ (in the absence of resetting) is given by 
\begin{equation} \label{free-propagator-C}
p_0(x, t) = \int_{-\infty}^{+\infty} \dd v \, \delta(x - v t) n(v) = \frac{1}{t} n\left( \frac{x}{t} \right) \;.
\end{equation}
An example of the typical trajectories of this system of particles is shown in Fig. \ref{fig:intro-ballistic-simreset}. Plugging in the free propagator from Eq. (\ref{free-propagator-C}) into the general formula for the non-equilibrium steady state given in Eq. (\ref{eq:general_NESS}), we get
\begin{equation}
{\rm Prob.}[\vec{X} = \vec{x}]_{\rm NESS} = r \int_0^{+\infty} \dd\tau \; e^{-r \tau} \frac{1}{\tau^N} \prod_{i = 1}^N n\left( \frac{x_i}{\tau} \right) \;.
\end{equation}
Once again, the particle positions are strongly correlated in the non-equilibr\-ium steady state since the joint distribution does not factorize. We will now compute the same observables as the ones we looked at in the Brownian case.

\subsubsection{The average density}

The average density is given by  
\begin{equation}\label{rho_bal}
\rho(x, N \vert r) = r \int_0^{+\infty} \dd\tau\; e^{-r \tau} p_0(x, \tau) = \frac{r}{2} \int_{r |x|}^{+\infty} \frac{\dd v}{v} e^{-v} = r \, \rho_s\left( r x \right) \;,
\end{equation}
where the normalized scaling function $\rho_s(z)$ defined for $z \in \mathbb{R}$ is given by
\begin{equation}\label{rho_bal1}
\rho_s(z) = \frac{1}{2} \int_{|z|}^{+\infty} \frac{\dd v}{v} e^{-v} = - \frac{1}{2} {\rm Ei}(-|z|) \;.
\end{equation}
Here, ${\rm Ei}(z)$ is the exponential integral function \cite{GR14} and the scaling function $\rho_s(z)$ is plotted in Fig. \ref{fig:overiew-resetting-ballistic}. The {\color{blue}asymptotes} of this scaling function are given by
\begin{equation}
\rho_s(z) \simeq \begin{cases}
- \frac{1}{2} \log z &\mbox{~~for~~} |z| \ll 1\\
e^{-z}/(2 z) &\mbox{~~for~~} |z| \gg 1
\end{cases}\;.
\end{equation}

\subsubsection{The center of mass}

\begin{figure}
\centering
\begin{minipage}[b]{0.48\textwidth}
\centering
\includegraphics[width=\textwidth]{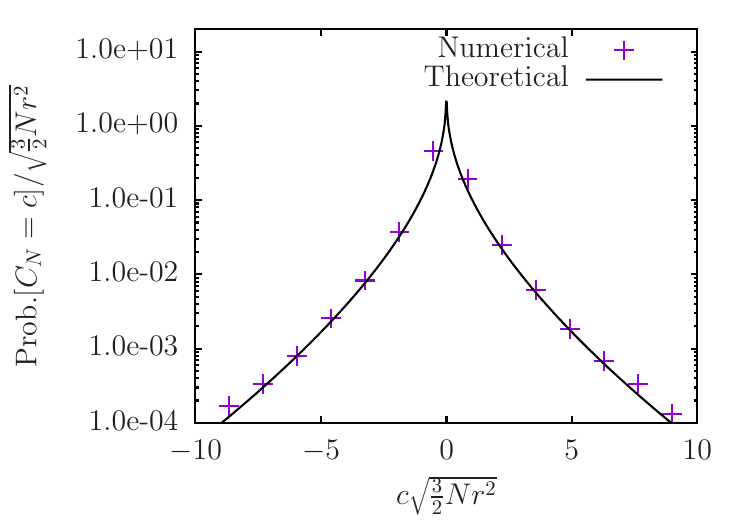}
\end{minipage}\hfill
\begin{minipage}[b]{0.48\textwidth}
\centering
\includegraphics[width=\textwidth]{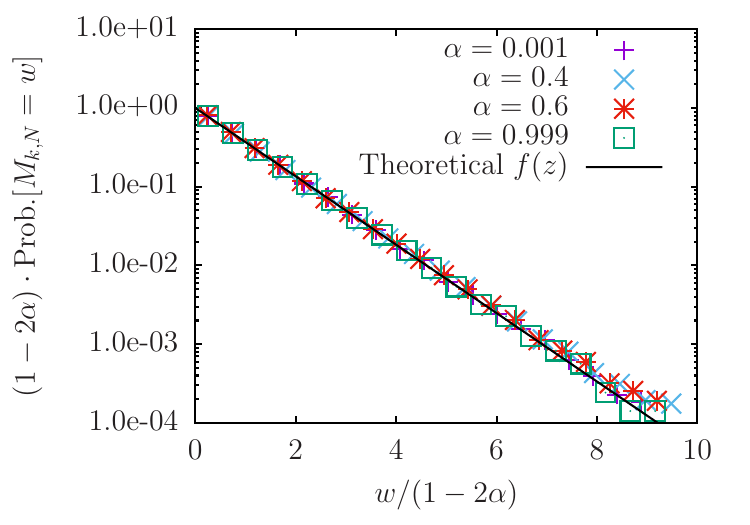}
\end{minipage}
\caption{{\bf Left:} The scaled distribution of the center of mass. The black line denotes the analytical scaling function $\ell(z)$ in Eq. \ref{eq:ell} and the points represent numerical simulations. {\bf Right:} The distribution of the $k$-th maximum $M_{k = \alpha N,N}$ in the bulk, given in Eq. \ref{eq:res_os_ballistic}, is plotted on a log-linear scale. The black line represents the analytical prediction given in Eq. \ref{eq:res_os_ballistic} and the symbols represent simulation results for $N=1000$ and $r=1$, different colors and symbol markers correspond to different values of $\alpha = k/N$.} \label{scaling_ballistic}
\end{figure}

We will now look at the center of mass defined in Eq. (\ref{eq:def-center-mass}). The first and second moments of the distribution $p_0(x, t)$ in this case are also finite and given by
\begin{equation} \label{eq:mv_ballistic}
m(t) = 0 \;,
\end{equation}
and 
\begin{equation}
\sigma^2(t) = \frac{t^2}{3} \;.
\end{equation}
Since $m(t) = 0$ is a constant, independent of $t$, we must use Eq. (\ref{eq:scaling_C}) to obtain the statistics of the center of mass. Plugging Eq. (\ref{eq:mv_ballistic}) in Eq. (\ref{eq:scaling_C}) and Eq. (\ref{eq:P_of_Z2}) we immediately obtain
\begin{equation}
{\rm Prob.}[C_N = c] \underset{N\to\infty}{\longrightarrow} r \int_0^{+\infty} \dd \tau\; \sqrt{\frac{3 N}{2 \pi \tau^2}} \exp[ - \frac{3 N c^2}{2 \tau^2} - r \tau] \;.
\end{equation}
Changing variable to $\nu = r \tau$ we obtain
\begin{equation}
{\rm Prob.}[C_N = c] \underset{N\to\infty}{\longrightarrow} \sqrt{\frac{3}{2} N r^2} \;\; \ell\left( c \sqrt{\frac{3}{2} N r^2}  \right) \;,
\end{equation}
where the normalized scaling function $\ell(z)$ defined for $z \in \mathbb{R}$ is given by
\begin{equation} \label{eq:ell}
\ell(z) = \frac{1}{\sqrt{\pi}} \int_0^{+\infty} \frac{\dd \nu}{\nu} e^{-\nu - z^2/\nu^2} \;.
\end{equation}
While this integral does not admit a simple closed form, we can easily compute the asymptotic behaviors, which are given by
\begin{equation}
\ell(z) \to \begin{cases}
- \frac{1}{\sqrt{\pi}} \log(z) &\mbox{~~for~~} |z| \ll 1\\
\left( \frac{\sqrt{3} |z|^{1/3}}{2^{1/3}} \right)^{-1} \exp( - \frac{3 |z|^{2/3}}{2^{2/3}} ) &\mbox{~~for~~} |z| \gg 1
\end{cases} \;.
\end{equation}
A plot of this normalized scaling function $\ell(z)$ in Eq. (\ref{eq:ell}) is given in the left panel of Fig. \ref{scaling_ballistic}, where it is also compared to the numerical simulations, finding excellent agreement. 

\subsubsection{The order and {\color{blue}extreme-value} statistics}

We now consider the order statistics in the bulk as well as the edge of the gas. As in the Gumbel case, one can obtain the distribution of the $k$-th maximum, both in the bulk and at the edge, using the same framework leading to Eq. (\ref{eq:res_bulk_integral_form}), with the substitutions $\vec{y} \to \tau$ and $h(\vec{y}) = r\,e^{-r\tau}$. We start by computing the explicit form of the quantile $q(\alpha, \vec{y}) \equiv q(\alpha, \tau)$. 
Substituting Eq. (\ref{free-propagator-C}) in Eq. (\ref{eq:def_wstar}) we obtain
\begin{equation} \label{quantile_weibull}
\alpha = \int_{q(\alpha, \tau)}^{+\infty} p_0(x, \tau) \dd x = \int_{q(\alpha, \tau)}^{+\infty} \frac{\dd x}{\tau} n\left( \frac{x}{\tau} \right) = \int_{q(\alpha, \tau)/\tau}^{+\infty} \dd v \, n(v) \;.
\end{equation}
Using the fact that $n(v)$ in Eq.~(\ref{eq:def_nv}) is supported over the finite interval $v \in [-1,+1]$, we need to consider three cases: if $q(\alpha, \tau) \leq -\tau$ then the integral is equal to $1$, if $q(\alpha, \tau) \geq \tau$ then the integral is equal to 0 and otherwise we have
\begin{equation}
\alpha = \frac{\tau - q(\alpha, \tau)}{2\tau} \; .
\end{equation}
Hence we get
\begin{equation} \label{w-star-sol-C}
q(\alpha, \tau) = \tau(1 - 2 \alpha) \quad {\rm where} \quad 0 \leq \alpha \leq 1 \;.
\end{equation}
The free propagator, defined in Eq. (\ref{free-propagator-C}), clearly belongs to the Weibull class in Eq. (\ref{eq:tail_weibull}) with $\mu(\vec{y}) = 1$. 
Therefore, as discussed in Section \ref{subsec:order-ciid}, one can once more replace, for large $N$, the Gaussian distribution in Eq.( \ref{eq:Mk_gauss}) by a 
delta function, leading to Eq. (\ref{eq:res_bulk_integral_form}) for the order statistics in both the bulk and the edges. 
Plugging Eq. (\ref{w-star-sol-C}) into Eq.( \ref{eq:res_bulk_integral_form}) we obtain
\begin{align} 
{\rm Prob.}[M_{k,N} = w] &\underset{N\to\infty}{\longrightarrow} r \int_0^{+\infty} \dd \tau \, e^{- r\tau} \delta\left[w - {\tau} (1 - 2\alpha)\right] \\
&= \frac{r}{|1 - 2\alpha|} \exp\left(-\frac{r |w|}{|1 - 2\alpha|} \right) \Theta(w(1 - 2\alpha)) \;. \label{eq:res_os_ballistic}
\end{align}
Thus the distribution of the $k$-th maximum $M_{k,N}$, both in the bulk as well as at the edges, is given by an exponential distribution, which is   
supported on the positive half-line (resp. negative half-line) for $\alpha < 1/2$ (resp. for $\alpha > 1/2$), as shown in the right panel of Fig. \ref{scaling_ballistic}. Similarly to the Brownian motion case, the {\color{blue}extreme-value} statistics can be obtained by taking the $\alpha \to 0^+$ limit of Eq.~(\ref{eq:res_os_ballistic}) which tells us that the {\color{blue}extreme-value} statistics are given by
\begin{equation}
    {\rm Prob.}[M_{1, N} = w] \underset{N\to\infty}{\longrightarrow} r \exp(- r |w|) \Theta(w) \;.
\end{equation}

\subsubsection{The gap statistics}

\begin{figure}
    \centering
    \includegraphics[width=0.5\textwidth]{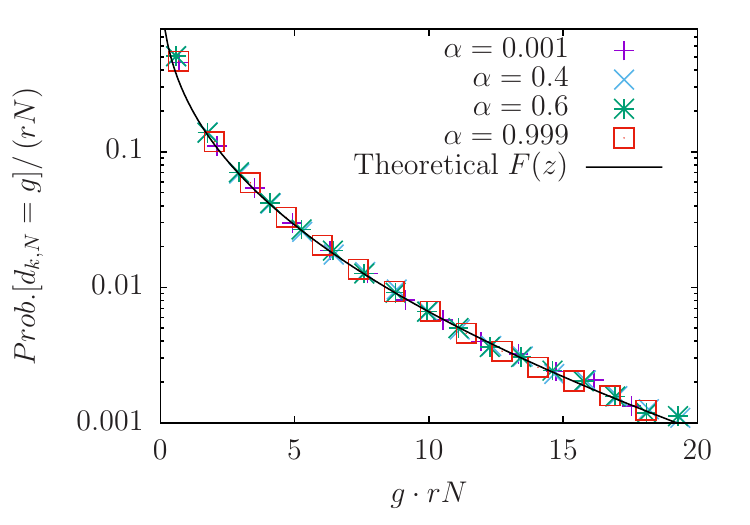}%
    \includegraphics[width=0.5\textwidth]{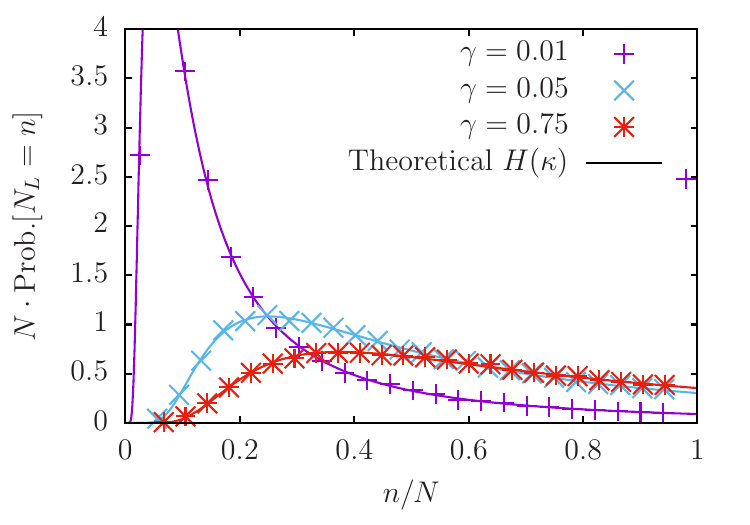}
    \caption{Plots of the probability density functions of the gaps (left panel) and the full counting statistics (right panel) obtained in Eq. (\ref{eq:gaps-B-full}) and Eq. (\ref{eq:fcs-reset-ballistic}) respectively. The black line represents the analytical prediction given in Eq. \ref{eq:res_os_ballistic} and the symbols represent simulation results for $N=1000$ and $r=1$, different colors and symbol markers correspond to different values of $\alpha = k/N$ (left panel) or $\gamma = r L$ (right panel).} \label{fig:gaps-fcs-reset-ballistic}
\end{figure}

We have seen in Chapter \ref{ch:ciid} that the gap distribution for the Weibull class is generically different in the bulk or at the edge. However, in the special case where $\mu = 1$, which is the case for our current model, they are distributed exponentially both in the bulk and at the edge. Using  
\begin{equation} \label{eq:p-q-B}
    p_0(q(\alpha, \tau), \tau) = \frac{1}{\tau} n(1 - 2\alpha) = \frac{1}{2 \tau}
\end{equation}
in Eq.~(\ref{eq:ciid-gaps-bulk}) we get the gap distribution
\begin{equation} \label{eq:gaps-B-full}
    {\rm Prob.}[d_{k,N} = g] \underset{N\to\infty}{\longrightarrow} \int_0^{+\infty}\dd\tau \; r e^{-r \tau} \frac{N}{2 \tau} e^{-\frac{N}{2 \tau} g} = r N K_0(\sqrt{2 r N g}) \;,
\end{equation}
where $K_0(z)$ is the Bessel $K$-function. Eq. (\ref{eq:gaps-B-full}) can be re-written in a scaling form as 
\begin{equation}
    {\rm Prob.}[d_{k, N} = g] \underset{N\to\infty}{\longrightarrow} r N \, F(g \cdot r N ) \;,
\end{equation}
where the normalized scaling function $F(z)$ is given by
\begin{equation}
    F(z) = K_0 (\sqrt{2 z}) \;,
\end{equation}
whose {\color{blue}asymptotes} are also well-known
\begin{equation}
    F(z) \approx \begin{dcases}
        - \frac{1}{2}\log z \;, &\mbox{~~when~~} z \to 0^+ \\
        \sqrt{\frac{\pi}{2}} \frac{e^{-\sqrt{2 z}}}{(2 z)^{1/4}}  \;, &\mbox{~~when~~} z \to +\infty \;.
    \end{dcases}
\end{equation}
Once again, we see a stretched exponential tail as $z \to +\infty$ but with a different exponent compared to the Brownian case. That is, $1/2$ instead of $2/3$, so the gaps are ``more stretched'' in the ballistic case compared to the Brownian case. The small$z$ {\color{blue}asymptotes} are also significantly different, instead of going to a constant with a {\color{blue}nonanalytic} derivative as in the Brownian case, the gap statistics diverge as $z \to 0^+$.

\subsubsection{The full counting statistics}

We now turn our attention to the full counting statistics. The probability of a reset-free particle being inside $[-L, L]$ is given by
\begin{equation}
    \int_{-L}^{L} \dd x \; p_0(x, t) = \min\left(1, \frac{L}{t}\right) \;.
\end{equation}
Using the general conditionally independent result in Eq.~(\ref{eq:ciid-fcs}) we get the distribution of the full counting statistics
\begin{equation}
    {\rm Prob.}[N_L = n] \underset{N\to\infty}{\longrightarrow} r \int_0^{+\infty} \dd \tau \; e^{-r \tau} \delta\left[n - N \min \left(1, \frac{L}{\tau}\right)\right] 
\end{equation}
and the integral can be computed explicitly and written in a scaling form as
\begin{equation} \label{eq:fcs-reset-ballistic}
    {\rm Prob.}[N_L = n] \underset{N\to\infty}{\longrightarrow} \frac{1}{N} H\left( \frac{n}{N} \right)
\end{equation}
where the scaling function $H(\kappa)$ is given by 
\begin{equation}
    H(\kappa) = (1 - e^{- r L}) \delta(\kappa - 1) + \frac{r L e^{-r L / \kappa}}{\kappa^2} \;.
\end{equation}
Notice that $H(\kappa)$ does not diverge as $\kappa \to 1^-$ unlike in the Brownian motion case. The reason for this difference comes from the underlying motion escaping very quickly from the origin, unlike Brownian motion in one dimension, which is recurrent, i.e. it comes back to the origin infinitely many times.

\subsection{Lévy flights with simultaneous resetting \cite{BLMS24}}
\label{subsec:levy-simreset}

\begin{figure}
    \centering
    \includegraphics[width=0.7\textwidth]{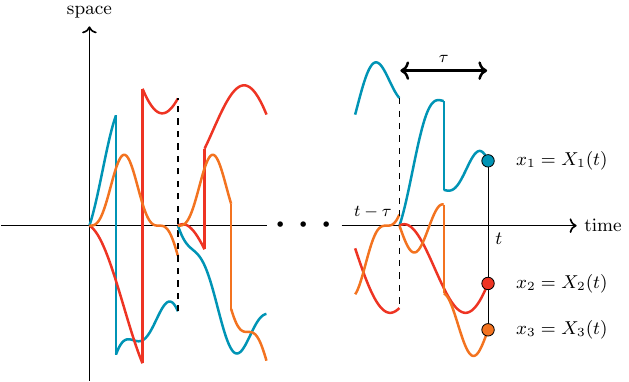}
    \caption{{Schematic trajectories of $N=3$ L\'evy flights undergoing simultaneous resetting to the origin at random times. The observation time is marked by $t$ and the time of the last reset before $t$ is marked by $t-\tau$. During the last period $\tau$, the particles evolve independently as free L\'evy flights.}}\label{fig:intro-levy-simreset}
\end{figure}

\begin{figure}
    \centering
    \includegraphics[width=0.7\textwidth]{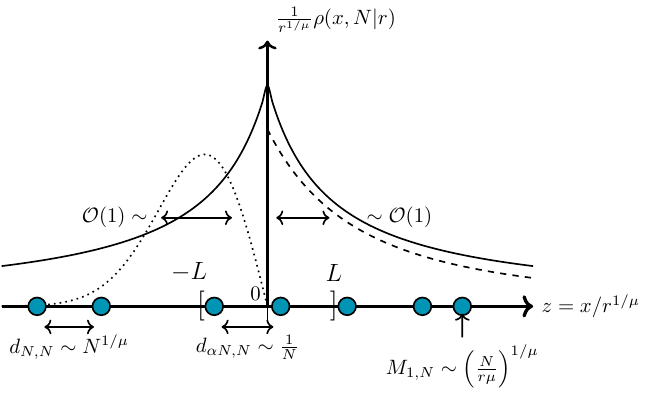}
    \caption{A schematic representation of the position of the particles for the simultaneously resetting Lévy flights and a summary of the main observables. The scaled average density profile, is supported over all the real line and is given in Eq. (\ref{eq:rho_levy}). Notice the cusp at 0 which is a consequence of the resetting. The scaling function in Eq. (\ref{scaling_Smu}) of the extreme statistics $M_{1, N}$ of the rightmost particle is shown schematically by a dashed curve. Compared to the scaling function in Eq. (\ref{f_mu}) of the order statistics $M_{\alpha N, N}$ in the bulk shown by the dotted curve on the left. The inter-particle spacing $d_{k, N}$ is of order $1/(r^{1/\mu} N)$ in the bulk and $(N/r\mu)^{1/\mu}$ at the edge.} \label{fig:overiew-resetting-levy}
\end{figure}

In this section, we consider a gas of $N$ L\'evy flights on the line, all starting at the origin and resetting 
simultaneously to the origin. Usually L\'evy flights (in the absence of resetting) are defined in discrete time where the position of a walker evolves via
\begin{equation} \label{xn}
X_{n} = X_{n-1} + \eta_n \quad, \quad X_0 = 0 \;,
\end{equation}
where $\eta_n$'s are independent identically distributed random noises with a power-law tail, $p(\eta) \sim |\eta|^{-1 - \mu}$ with $0 < \mu < 2$. For large number of steps $n$, one can replace the discrete time $n$ by a continuous variable $t$ and it is well known \cite{BG90} that the free propagator of the L\'evy flight for large $t$ converges to
\begin{equation} \label{free-propagator-B}
p_0(x, t) \approx \frac{1}{t^{1/\mu}} \mathcal{L}_\mu \left( \frac{x}{t^{1/\mu}} \right) \;,
\end{equation}
where $\mathcal{L}_\mu(z)$ is the probability density function of a centered, normalized to unity, symmetric stable function parametrized by $0<\mu<2$, as defined in Eq. (\ref{eq:levy-stable}). A sketch of the trajectories for $N=3$ L\'evy flights is shown in Fig.~\ref{fig:intro-levy-simreset}. Inserting this expression for the free propagator in the general formula for the non-equilibrium steady state in Eq. (\ref{eq:general_NESS}), we get the joint distribution of the positions of the simultaneously resetting L\'evy flights in the non-equilibrium steady state as
\begin{equation}
{\rm Prob.}[\vec{X} = \vec{x}]_{\rm NESS} \approx r \int_{0}^{+\infty} \dd \tau\; e^{-r\tau} \prod_{i = 1}^N \frac{1}{\tau^{1/\mu}} \mathcal{L}_\mu\left(\frac{x_i}{\tau^{1/\mu}}\right)\;.
\end{equation}
Given this joint distribution, one can compute various physical observables, as in the two preceding cases. 

\subsubsection{The average density}

We start with the average density, which is given by
\begin{equation} \label{eq:rho_levy}
\rho(x, N\vert r) \approx r \int_0^{+\infty} \dd \tau\; e^{-r\tau} \frac{1}{\tau^{1/\mu}}\mathcal{L}_\mu\left( \frac{x}{\tau^{1/\mu}} \right) = r^{1/\mu} \rho_\mu(r^{1/\mu} x) \;,
\end{equation}
where the normalized scaling function $\rho_\mu(z)$ is symmetric and is given by
\begin{equation} \label{eq:rho_mu}
\rho_\mu(z) = \mu |z|^{\mu - 1} \int_0^{+\infty} \dd u \; \frac{1}{u^\mu} e^{-(|z|/u)^\mu} \mathcal{L}_\mu(u) \;.
\end{equation}
One can compute the asymptotic behavior of the average density. For large $|z|$, one finds that, for all $0 < \mu < 2$,  
\begin{equation} \label{rho_large_z}
\rho_\mu(z) \approx \frac{1}{2\pi} \frac{1}{|z|^{1 + \mu}} \quad, \quad |z| \to \infty \;.
\end{equation}
Interestingly, the small $z$ behavior is quite different depending on the value of $\mu$. Indeed, we get, as $|z| \to 0$
\begin{equation} \label{rho_small_z}
\rho_{\mu}(z)  \approx \begin{dcases}
\frac{c_1}{|z|^{1 - \mu}} &\mbox{~~when~~} 0 < \mu < 1 \;,\\
\frac{1}{\pi} (- \log |z|) &\mbox{~~when~~} \mu = 1 \;,\\
\frac{1}{\mu \sin(\pi/\mu)} &\mbox{~~when~~} 1 < \mu < 2 \;,
\end{dcases} 
\end{equation}
where we introduced the constant
\begin{equation}
c_1 = \mu \int_0^{+\infty} \frac{\dd v}{v^{\mu}} \mathcal{L}_\mu(v) \;.
\end{equation}
Using $\mathcal{L}_\mu(v) \to O(1)$ as $v \to 0$, we see that the constant $c_1$ is well defined for $\mu < 1$. Thus, the average density diverges as $|z| \to 0$ for $0 < \mu \leq 1$ (but is still integrable), while it approaches a constant as $|z| \to 0$ for $1 < \mu < 2$. 

\subsubsection{The center of mass}

We now turn our attention to the center of mass defined in Eq. (\ref{eq:def-center-mass}). The stable distribution ${\cal L}_\mu(z)$ has a power-law tail for large $z$.
\begin{figure}
\centering
\begin{minipage}[b]{0.32\textwidth}
\centering
\includegraphics[width=\textwidth]{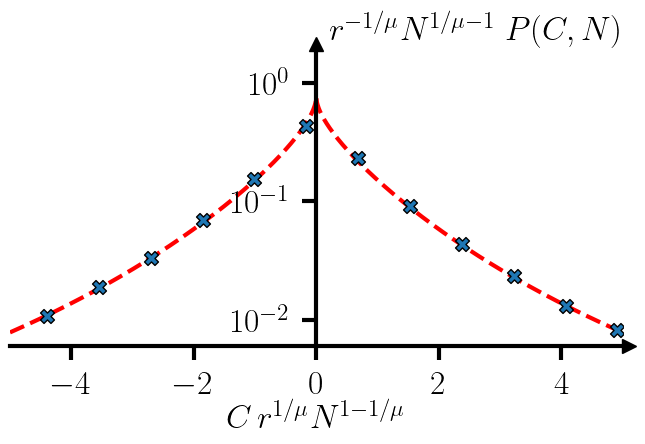}
\end{minipage}
\hfill
\begin{minipage}[b]{0.32\textwidth}
\centering
\includegraphics[width=\textwidth]{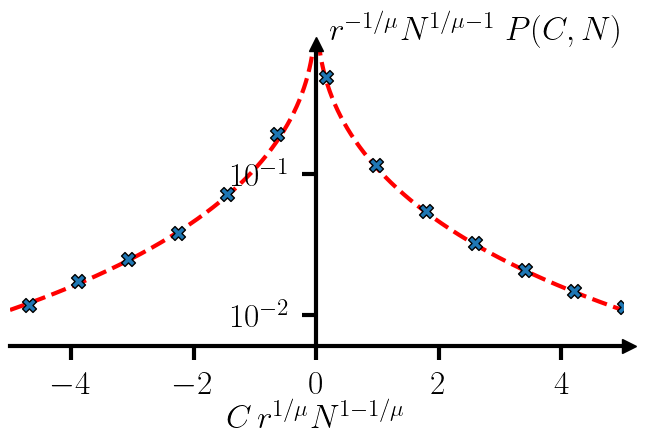}
\end{minipage}
\hfill
\begin{minipage}[b]{0.32\textwidth}
\centering
\includegraphics[width=\textwidth]{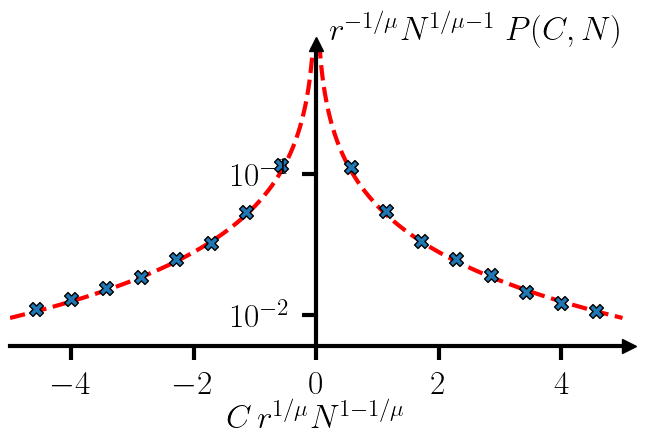}
\end{minipage}
\caption{The p.d.f. of the center of mass $P(C,N)$ in Eq. \ref{eq:res_com_levy} is shown by dashed lines for $N=1000$ and 
for $\mu = 1.5, 1, 0.5$ in the left, middle and right panel respectively. The dots represent numerical simulations.}\label{fig:com_levy}
\end{figure}
Hence, the free propagator in Eq. (\ref{free-propagator-B}) belongs to the Fréchet class of conditionally independent variables. In this case, we can use the general result stated in Eq. (\ref{eq:PC_TF_mu}) and Eq. (\ref{eq:scaling_mu3}). A straightforward computation, with suitable change of variables leads us to
\begin{equation} \label{P_C_mu1}
{\rm Prob.}[C_N = c] \underset{N\to\infty}{\longrightarrow} r N \int_0^{+\infty} \dd \tau \; e^{-r\tau} \frac{1}{(N\,\tau)^{1/\mu}} \mathcal{L}_\mu\left(\frac{c N}{(N\,\tau)^{1/\mu}}\right) \;.
\end{equation}
Making a further change of variable $\nu = N \tau$, one gets  
\begin{equation}\label{P_C_mu2}
{\rm Prob.}[C_N = c] \underset{N\to\infty}{\longrightarrow} N \rho_\mu\left(c N, N \Big| r' = \frac{r}{N} \right) \;,
\end{equation}
where we used the expression for the average density in Eq. (\ref{eq:rho_levy}). The fact that the distribution of the center of mass is related to the rescaled single particle propagator can again be traced back to the fact that L\'evy variables are stable under addition. Finally, from Eqs. (\ref{eq:rho_levy}) and (\ref{P_C_mu2}) we find that the probability density function of the center of mass admits the scaling form for large $N$
\begin{equation}
{\rm Prob.}[C_N = c] \approx N^{1 - 1/\mu} r^{1/\mu} \rho_\mu( N^{1 - 1/\mu} r^{1/\mu} c) \;, \label{eq:res_com_levy}
\end{equation}
where $\rho_\mu(z)$ is a symmetric function defined in Eq. (\ref{eq:rho_mu}). A plot of this scaling function is given in Fig. \ref{fig:com_levy} where it is compared to numerical simulations, showing an excellent agreement. 

\subsubsection{The order statistics (in the bulk)}

Unlike in the two cases (Gumbel and Weibull) discussed before, it turns out that for the Fréchet case, the order statistics in the bulk can not be extrapolated all the way to the edge. Hence, one needs to study separately the statistics of $M_{k,N}$ when $k = {\cal O}(N)$ (bulk) and when $k = {\cal O}(1)$ (edge).

\vspace*{0.2cm}

\begin{figure}[t]
\centering
\begin{minipage}[b]{0.32\textwidth}
\centering
\includegraphics[width=\textwidth]{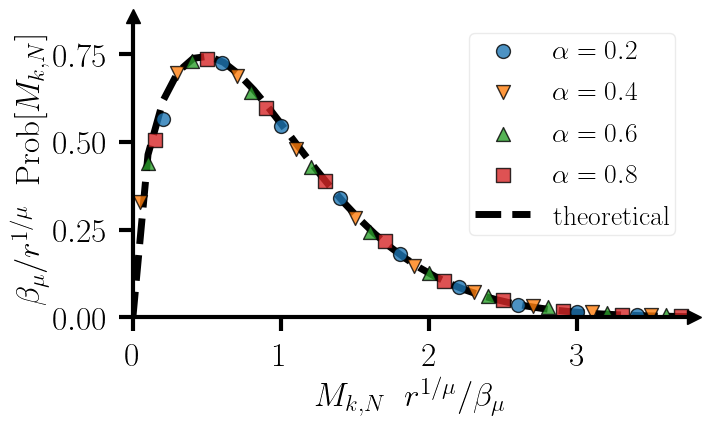}
\end{minipage}
\hfill
\begin{minipage}[b]{0.32\textwidth}
\centering
\includegraphics[width=\textwidth]{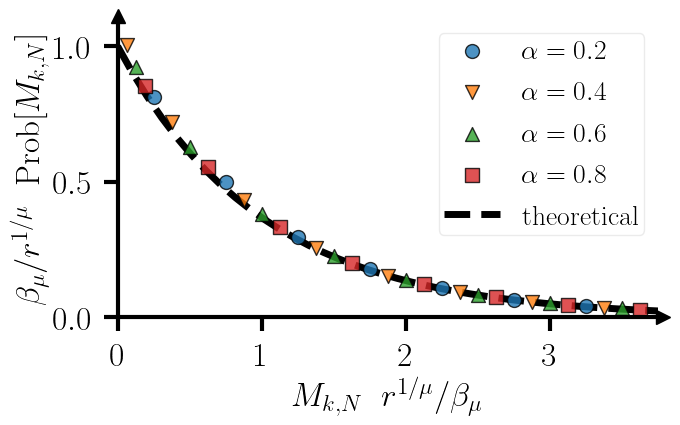}
\end{minipage}
\hfill
\begin{minipage}[b]{0.32\textwidth}
\centering
\includegraphics[width=\textwidth]{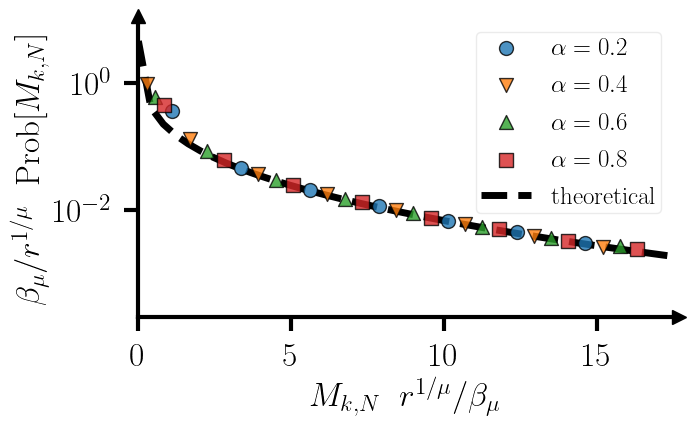}
\end{minipage}
\caption{The probability density function of the $k$-th maximum in the bulk $M_{k=\alpha N,N}$ in Eqs. \ref{eq:res_levy_bulk} and \ref{f_mu} is shown for 
$\mu = 1.5, 1, 0.5$ in the left, middle and right panels respectively. Different colors and symbols correspond to different values of $k$. The dashed black line corresponds to the scaling function in Eq. \ref{eq:res_levy_bulk}.}\label{fig:levy}
\end{figure}

In the bulk, i.e. for $k \sim \mathcal{O}(N)$, we know that the order statistics are given by Eq. (\ref{eq:res_bulk_integral_form}) with $\vec{y}$ replaced by $\tau$ and $h(\vec{y})$ replaced by $r\,e^{-r\tau}$. Hence, we start by computing the explicit form of the quantile $q(\alpha, \vec{y}) \equiv q(\alpha, \tau)$. Replacing Eq. (\ref{free-propagator-B}) in Eq. (\ref{eq:def_wstar}) we obtain
\begin{equation} \label{eq_alpha_mu}
\alpha = \int_{q(\alpha, \tau)}^{+\infty} p_0(x, \tau)\, \dd x = \int_{q(\alpha, \tau)}^{+\infty} \frac{1}{\tau^{1/\mu}}\mathcal{L}_\mu\left(\frac{x}{\tau^{1/\mu}}\right) \dd x \;.
\end{equation}
Changing variable to $z = x/\tau^{1/\mu}$ and denoting by $F_\mu(z) = \int_{-\infty}^z \mathcal{L}_\mu(x) \dd x$ the cumulative distribution function of the stable law, we can invert the relation in Eq. (\ref{eq_alpha_mu}) and express it as
\begin{equation} \label{eq:w-star-levy}
q(\alpha, \tau) = \tau^{1/\mu} F_\mu^{-1}(1 - \alpha) = \tau^{1/\mu} \beta_\mu\;,
\end{equation}
where $F_\mu^{-1}$ is the inverse function of $F_\mu$. Although we do not have a closed form for $\beta_\mu = F_\mu^{-1}(1 - \alpha)$, it is simply a constant which we can numerically compute for any practical purpose. 
Furthermore, from the symmetry of the probability density function ${\cal L}_\mu(z)$ it follows that if $\alpha < 1/2$ then $\beta_\mu > 0$, while if $\alpha > 1/2$ then $\beta_\mu < 0$. Exactly at $\alpha = 1/2$, $\beta_\mu = 0$. Hence, in the bulk of the system, in the large $N$ limit, the order statistics will be given by Eq. (\ref{eq:res_bulk_integral_form}), which can be simplified as
\begin{equation}\label{eq:res_levy_bulk}
{\rm Prob.}[M_{k, N} = w] \underset{N\to\infty}{\longrightarrow} \frac{r^{1/\mu}}{\beta_\mu} f_\mu \left( r^{1/\mu} \frac{w}{\beta_\mu} \right) \;,
\end{equation}
where the normalized scaling function $f_\mu(z)$ defined for $z > 0$ is given by
\begin{equation}\label{f_mu}
f_\mu(z) = \mu \, z^{\mu - 1} e^{- z^\mu} \,\theta(z) \;,
\end{equation}
and is plotted in Fig. \ref{fig:levy} for three different values of $\mu$. For $\mu > 1$, the scaling function $f_\mu(z)$ vanishes as $z \to 0$, while for $\mu < 1$, it diverges as $z \to 0$.  \\ 

\subsubsection{{\color{blue}extreme-value} statistics}

\begin{figure}
\centering
\begin{minipage}[b]{0.32\textwidth}
\centering
\includegraphics[width=\textwidth]{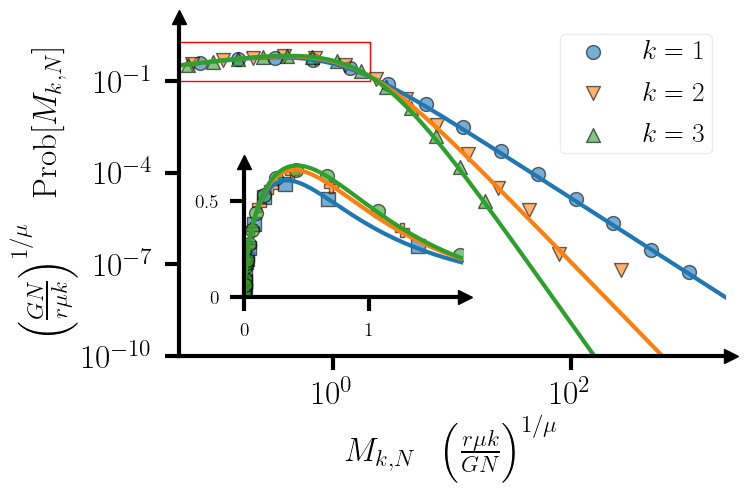}
\end{minipage}
\hfill
\begin{minipage}[b]{0.32\textwidth}
\centering
\includegraphics[width=\textwidth]{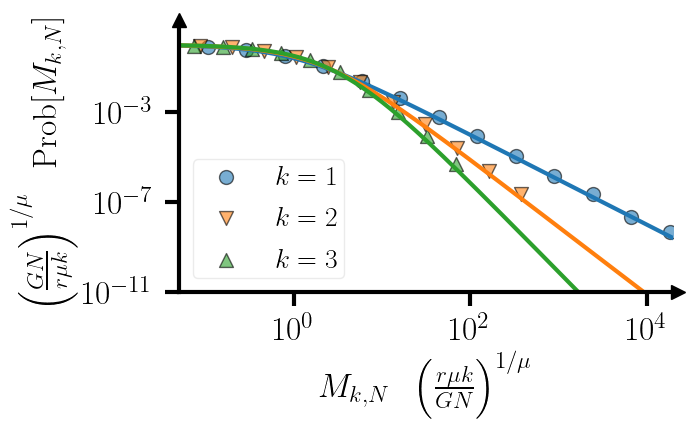}
\end{minipage}
\hfill
\begin{minipage}[b]{0.32\textwidth}
\centering
\includegraphics[width=\textwidth]{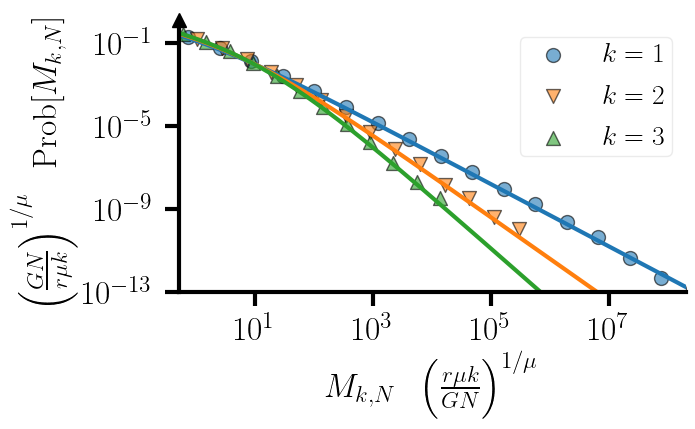}
\end{minipage}
\caption{The probability density functions of the first, second and the third maximum from the right, i.e., $M_{1,N}$, $M_{2,N}$ and $M_{3,N}$, given in Eqs.  \ref{eq:res_levy_edge} and \ref{scaling_Smu} are plotted as solid lines. The three figures correspond to $\mu = 1.5$ (left panel), $\mu = 1$ (middle panel) and $\mu = 0.5$ (right panel). The inset in the left panel shows the behavior close to $z =0$ where it vanishes as $z^{\mu-1} \sim z^{1/2}$. Different colors and symbols correspond to different values of $k$.}\label{fig:levy_edge}
\end{figure}

As argued in the general discussion for the Fréchet class in Section \ref{subsec:order-ciid}, for the order statistics at the edge, we cannot longer use the bulk result in Eq. (\ref{eq:res_bulk_integral_form}). Instead, we need to use 
Eq. (\ref{eq:res:frechet}) which holds for any $k = {\cal O}(1)$. To proceed, let us first evaluate the large $x$ behavior of $p_0(x, t)$ given in Eq. (\ref{free-propagator-B}). Using the large $z$ tail of ${\cal L}_\mu(z)$ we get 
\begin{equation} \label{large_X_mu}
p_0(x, t) \underset{x \gg 1}{\sim} \frac{\tau}{x^{1 + \mu}} \sin( \frac{\pi \mu}{2} ) \frac{\Gamma(\mu + 1)}{\pi} \;.
\end{equation}
By replacing $\vec{y}$ by $t$ and $\mu(\vec{y})$ by $\mu$ in Eq. (\ref{eq:tail_frechet}), we see that Eq. (\ref{large_X_mu}) is a special case of Eq. (\ref{eq:tail_frechet}), namely
\begin{equation}\label{large_X_mu_2}
p_0(x, t) \underset{x \gg 1}{\sim} \frac{G\, \tau}{x^{1 + \mu}}  \;,
\end{equation}
where 
\begin{equation} \label{def_G}
G = \sin( \frac{\pi \mu}{2} ) \frac{\Gamma(\mu + 1)}{\pi} \;.
\end{equation}
Then, using Eq. (\ref{large_X_mu_2}) and Eq. (\ref{eq:def_lambda}) it follows that
\begin{equation} \label{lambda_mu}
\lambda_N(w, \tau) = \frac{G \,\tau \,N}{\mu \, w^\mu} \;.
\end{equation}
Hence, the random variable $\lambda_N(w, \tau)$ is proportional to the random variable $\tau$, which itself is distributed exponentially with rate $r$. This leads to
\begin{equation}
{\rm Prob.}[\lambda_N(w, \tau) \geq v] = \exp(- r\, \frac{\mu v w^\mu}{G N} ) \;.
\end{equation}
Plugging this expression in Eq. (\ref{eq:res:frechet}) we then obtain the cumulative distribution function of $M_{k,N}$, which reads
\begin{align}
{\rm Prob.}[M_{k, N} \leq w] &\approx 1 - \frac{1}{\Gamma(k)} \int_0^{+\infty} \dd v \; v^{k-1} \exp( - r\, \frac{\mu v w^\mu}{G N} - v ) \\
&= S_k\left( \frac{r \mu w^\mu}{G N} \right) \;,
\end{align}
where the scaling function $S_k(z)$ defined for $z \geq 0$ is given by
\begin{equation}
S_k(z) = 1 - \frac{1}{(1 + z)^k} \;.
\end{equation}
The probability density function is thus given by
\begin{equation} \label{eq:res_levy_edge}
{\rm Prob.}[M_{k, N} = w] = \frac{r \mu^2 w^{\mu - 1} }{G N} S_k'\left(\frac{r \mu w^\mu}{G N}\right) = \left(\frac{r \mu k}{G N}\right)^{1/\mu} {\cal S}_{\mu,k}\left[  \left( \frac{r \mu k}{G N} \right)^{1/\mu} w  \right] \;,
\end{equation}
where the normalized scaling function ${\cal S}_{\mu,k}(z)$ reads
\begin{equation}\label{scaling_Smu}
{\cal S}_{\mu, k}(z) = \frac{\mu \, z^{\mu - 1}}{(1 + z^\mu/k)^{1 + k}} \quad, \quad z \geq 0 \;.
\end{equation}
For large $z$, the scaling function decays as a power law ${\cal S}_{\mu, k}(z) \sim z^{-1 - \mu\,k}$, while for small $z$, it behaves as ${\cal S}_{\mu, k}(z) \sim z^{\mu -1}$. This scaling function thus fully characterizes the large $N$ behavior of the order statistics at the edge for simultaneously resetting L\'evy flights. In Fig. \ref{fig:levy_edge}, this scaling function is plotted for three different values of $\mu$ in the three panels and compared to the numerical simulations, showing a good agreement. \\

\subsubsection{The gap statistics (in the bulk)}

\begin{figure}
    \centering
    \includegraphics[width=0.5\textwidth]{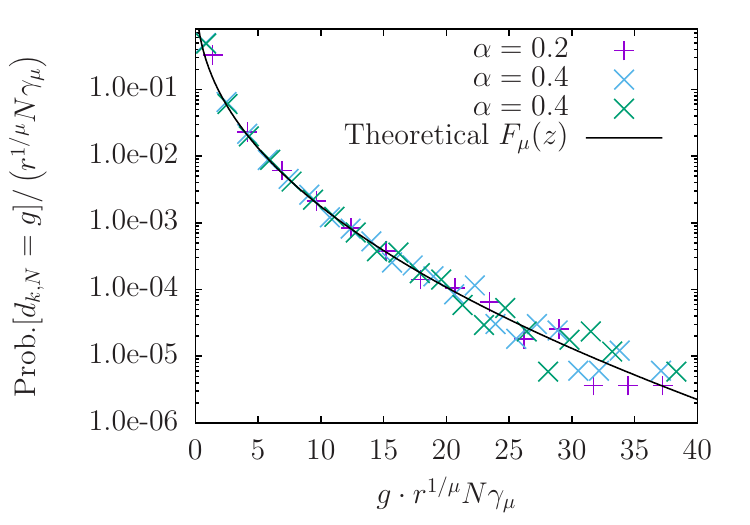}%
    \includegraphics[width=0.5\textwidth]{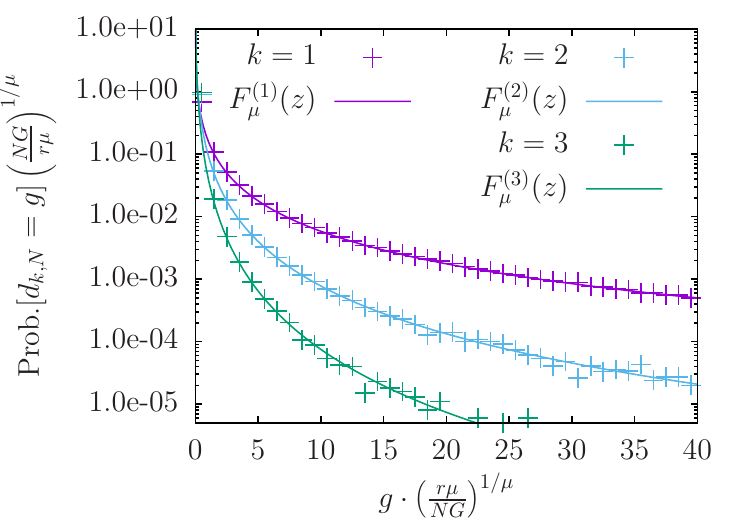}
    \caption{Plots of the probability density functions of the gaps in the bulk (left panel) and at the edge (right panel) obtained in Eq. (\ref{eq:gap-bulk-levy}) and Eq. (\ref{eq:gap-edge-levy}) respectively.} \label{fig:gaps-reset-levy}
\end{figure}

We saw in Chapter \ref{ch:ciid} that for conditionally independent variables of the Fréchet universality class, the gap statistics have a separate behavior in the bulk and at the edge. The behavior in the bulk is considerably simpler and can be obtained from 
\begin{equation} \label{eq:p-q-C}
    p_0(q(\alpha, t), t) = \frac{1}{t^{1/\mu}} \mathcal{L}_\mu\left(\beta_\mu\right) \;,
\end{equation}
where we used Eq.~(\ref{eq:w-star-levy}). Although we cannot simplify $\gamma_\mu = \mathcal{L}_\mu\left(\beta_\mu\right)$, it is simply a constant that can be evaluated numerically for any practical purpose. Using Eq.~(\ref{eq:ciid-gaps-bulk}) we obtain
\begin{equation}
    {\rm Prob.}[d_{k, N} = g] \underset{N\to\infty}{\longrightarrow} \int_0^{+\infty} \dd \tau \; r e^{-r\tau}  \frac{N \gamma_\mu}{\tau^{1/\mu}} e^{- g\frac{N \gamma_\mu}{\tau^{1/\mu}}} \;,
\end{equation}
which can be written in a scaling form
\begin{equation} \label{eq:gap-bulk-levy}
    {\rm Prob.}[d_{k, N} = g] \underset{N\to\infty}{\longrightarrow} r^{1/\mu} N \gamma_\mu \; F_\mu\left(g \cdot r^{1/\mu} N \gamma_\mu \right) \;,
\end{equation}
where the scaling function $F_\mu(z)$ defined for $z \geq 0$ is given by
\begin{equation}
    F_\mu(z) = \mu \int_0^{+\infty} \dd u \; u^{\mu - 2} e^{-u^\mu - z / u} \;.
\end{equation}
Notice that similarly to the order statistics the scaling function for $\mu = 2$ recovers the scaling function of Brownian motion. 

\subsubsection{The gap statistics (at the edge)}

At the edge, we have to use Eq.~(\ref{eq:ciid-gaps-frechet}) instead of Eq.~(\ref{eq:ciid-gaps-bulk}). Notice that $b_N(\vec{y})$ in Eq.~(\ref{eq:ciid-gaps-frechet}) is equivalent to $q(1/N, \tau) \equiv q_N(\tau)$ which can be obtained from the tail of the Lévy distribution
\begin{equation}
    q_N(\tau) = \left( \frac{ N G \tau}{\mu}  \right)^{1/\mu} \;.
\end{equation}
Then, using Eq.~(\ref{eq:ciid-gaps-frechet}) we can write the gap distribution in a scaling form
\begin{equation}
    {\rm Prob.}[d_{k, N} = g] = \left( \frac{r \mu}{N G} \right)^{1/\mu} F_\mu^{(k)}\left[g \left(\frac{r \mu}{N G}\right)^{1/\mu} \right] \;, \label{eq:gap-edge-levy}
\end{equation}
where the scaling function $F_\mu^{(k)}(z)$ is defined for $z \geq 0$ and given by
\begin{equation}
    F_\mu^{(k)}(z) = \mu \int_0^{+\infty} \dd u \; u^{\mu-2} e^{-u^\mu} \frac{\mu^2}{(k-1)!} \int_0^{+\infty} \dd x \; e^{-x^{-\mu}} x^{-\mu-1} \left(x + \frac{z}{u}\right)^{-\mu k - 1} \;.
\end{equation}

\subsubsection{The full counting statistics}

The full counting statistics cannot be computed for the Lévy flights because we have no knowledge of the cumulative distribution of the Lévy stable law that would allow us to compute 
\begin{equation}
    \int_{-L}^{L} \frac{1}{\tau^{1/\mu}} \mathcal{L}_\mu \left(\frac{x}{\tau^{1/\mu}}\right) \;.
\end{equation}

\subsection{Diffusive particles with first-passage resetting} \label{subsec:fpt-simreset}

\begin{figure}
    \centering
    \includegraphics[width=0.7\textwidth]{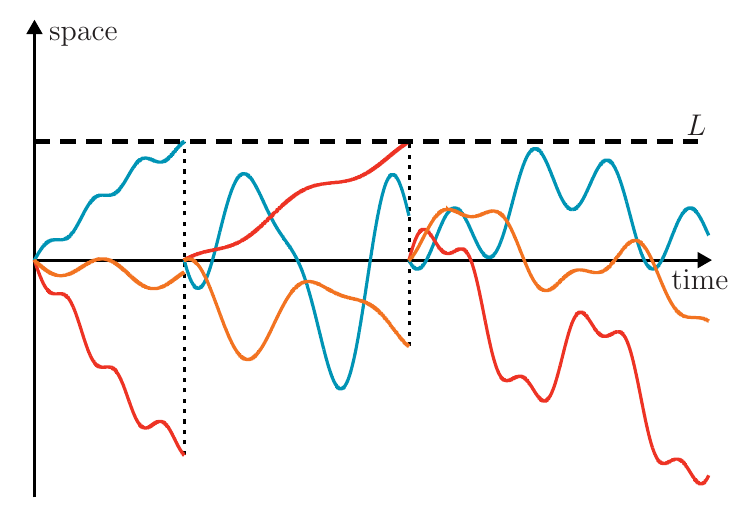}
    \caption{A sketch of the {\color{blue}first-passage} resetting model with $N = 3$ particles and a defect located at $L > 0$. Whenever any of the particles reach the defect all the particles reset to the origin.} \label{fig:first_passage_model}
\end{figure}

\begin{figure}
    \centering
    \includegraphics[width=0.7\textwidth]{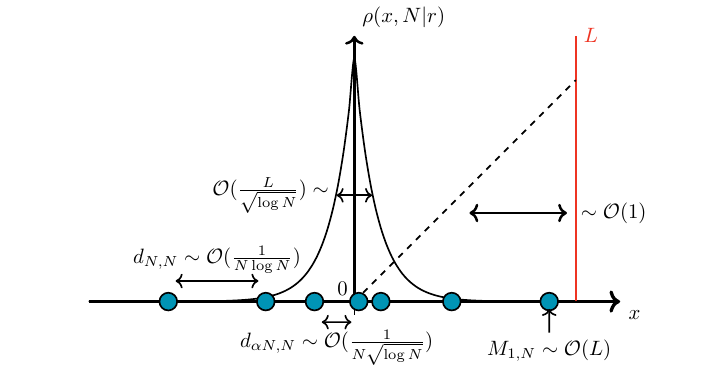}
    \caption{A schematic representation of the position of the particles in the {\color{blue}first-passage} simultaneously resetting gas and a summary of the main observables. The scaled average density profile, is supported over all the real line and is given in Eq. (\ref{eq:rhoS-fpt-reset}). However, the width of the density scales like $L / \sqrt{\log N}$ hence the large $N$ limit is {\color{blue}condensing} the particles close to the origin. The linear distribution of the position $M_{1, N}$ of the rightmost particle is shown by a dashed curve. The distribution is a simple linear function supported on $[0, L]$. The
    inter-particle spacing $d_{k, N}$ is of order $1/(N \sqrt{\log N})$ both near the edge of the gas and in the bulk.} \label{fig:overiew-resetting-fpt}
\end{figure}

We saw in Sections \ref{subsec:BM-simreset} to \ref{subsec:levy-simreset} that a global clock that performs simultaneous resetting events induces strong long-range correlations in our gas. We will now take another approach to create these correlations. Instead of a global clock, the particles will now share an environment with a defect, and the common response to the defect will induce strong long-range correlations between all the particles. Specifically, we consider $N$ particles whose positions we denote by $X_1(t), \cdots, X_N(t)$ and assume that there is a defect in $L > 0$. Whenever {\it any} of the particles touch the defect, they {\it all} reset to the origin. This simple model could be used to model a variety of systems with a global failure mode, such as $N$ countries' electrical consumption on the same grid with a blackout level, or $N$ stresses along a fault in a stick-slip system with a critical stress level, a close reminder of the famous Hébraud-Lequeux model \cite{HL98}. 

\vspace{0.2cm}

The single particle case $N = 1$, has a very rich history. It was first introduced by Gerstein and Mandelbrot \cite{GM64} as a neuronal activity model and has since become one of the fundamental theoretical models in neuroscience for analytical computations \cite{T88, SG13}. An equivalent of what the Ising model is for us statistical physicists. In the language of theoretical neuroscience \cite{GM64}
\begin{itemize}
    \item the position $X_1(t)$ represents the voltage of the somatic and dendritic membrane of the neuron
    \item the origin is the resting potential of the neuron
    \item the defect $L > 0$ is an architectural property of the neuron
    \item the voltage, $X_1(t)$, performs a Brownian motion because the neuron receives infinitely many infinitely small excitatory and inhibitory post synaptic potentials. Refer to Section \ref{subsec:strong} to see how such a discrete random walk tends to a Brownian motion.
    \item Whenever the voltage reaches $L > 0$ the neuron `fires', i.e. produces an action potential, the voltage subsequently relaxes instantly to its base resting potential, i.e. the origin.
\end{itemize}
In our language, $X_1(t)$ performs a Brownian motion with an absorbing boundary condition at $L > 0$, and whenever it reaches the boundary the process restarts from the origin. We have already seen in Section~\ref{subsec:strong} in Eq.~(\ref{eq:constrained-propagator}) that the probability $p_{\leq L}(x, t)$ for a diffusing particle to reach $x$ at time $t$ while staying below $L$ is given by
\begin{equation} \label{eq:free-first-passage-reset-propagator}
    p_{\leq L}(x, t) = \frac{1}{\sqrt{4 \pi D t}} \left[ e^{-\frac{x^2}{4 D t}} - e^{-\frac{(x - 2 z)^2}{4 D t}} \right] \;,
\end{equation}
and the survival probability $Q(L, t)$ characterizing the probability with which the particle will stay below $L$ up to time $t$ was also given in Eq.~(\ref{eq:cumulative-BM}) and reads
\begin{equation}
    Q(L, t) = \int_{-\infty}^{L} \dd x \; p_{\leq L}(x, t) = {\rm erf}\left( \frac{L}{\sqrt{4 D t}} \right) \;.
\end{equation} 
We also introduce $F_1(L, t)$ characterizing the probability that a single particle reaches the defect at $L$ for the first time at time $t$. This first-passage probability can be obtained directly from the survival probability noticing
\begin{equation} \label{eq:first-passage-prob-simreset}
    F_1(L, t) = - \pdv{Q(L, t)}{t} = \frac{L}{\sqrt{4 \pi D t^3}} e^{-\frac{L^2}{4 D t}} \;,
\end{equation}
which can be understood as follows. The trajectories that reach $L$ for the first time at time $t$ correspond exactly to the trajectories with `die' at time $t$. This first-passage probability $F_1(L, t)$ is the main quantity of interest from a neuroscience perspective because it will determine the interval between successive neuron spikes, which is the typical observable obtained experimentally. However, we can immediately see from Eq.~(\ref{eq:first-passage-prob-simreset}) that the mean first-passage time diverges, which does not seem very physical given that neurons fire repeatedly. Gerstein and Mandelbrot resolved this {\color{blue}incongruence} by introducing a positive drift \cite{GM64} that makes the mean first-passage time finite. However, this fundamentally changes the assumptions made by the model; the existence of a positive drift is predicated on the neuron receiving more excitatory than inhibitory signals. Although this assumption may be valid in some scenarios, it is clearly quite constraining. We will see shortly that another approach is to consider that there is not a single voltage regulating the activation of the neuron, but a collection of $N > 2$ potentials and whenever any of them reach the defect, the neuron fires and relaxes all the voltages to the resting potential. Gerstein and Mandelbrot mentioned \cite{GM64} such a generalization and attempted computations for $N = 2$ but saw no significant difference with their previous model, but stopped too early. In fact, significantly different behaviors occur when $N > 2$ rather than when $N \leq 2$. Most strikingly, the system reaches a non-equilibrium steady state at long times, and the mean first-passage time becomes finite when $N > 2$.

\vspace{0.2cm}

The dynamics of the $N$ particle system are described by
\begin{equation} \label{eq:dynamics}
    X_i(t + \dd t)  = \begin{cases}
    0 \mbox{~~if any particle goes above~} L \\
    X_i(t) + \sqrt{2 D \dd t} \, \eta_i(t) \mbox{~~otherwise,}
    \end{cases}
\end{equation}
where $D > 0$ is the diffusion constant and $\eta_i(t)$ is a zero-mean Gaussian white noise with a correlator $\langle \eta_i(t) \eta_j(t') \rangle = \delta_{ij} \delta(t - t')$. The joint probability density function ${\rm Prob.}[\vec{X}(t) = \vec{x}]$ of finding the $N$ particles at $\vec{x}$ at time $t$ given that they started at $\vec{x} = \vec{0}$ at time $t = 0$ and resetted to $\vec{0}$ each time one of them crossed $L$ can be written as
\begin{equation} \label{eq:renewal-fpt-resetting}
    {\rm Prob.}[\vec{X}(t) = \vec{x}] = \prod_{i = 1}^N p_{\leq L}(x_i, t) + \int_0^t \dd t' \; F_N(L, t') {\rm Prob.}[\vec{X}(t - t') = \vec{x}] \;,
\end{equation}
where $F_N(L, t)$ is the {\color{blue}first-passage} probability of the $N$ particle system, i.e. the probability that all particles stay below $L$ up to time $t$ and a particle reaches $L$ for the first time at time $t$. This equation can be understood as follows. There are two possibilities for the particles to reach $\vec{x}$ at time $t$. Either they do so in such a way that none of them ever cross the barrier, which corresponds to the first term, in which case they are all diffusing independently and hence we can split the joint probability density function. Otherwise, at least one of them must have crossed the barrier at least once, denote the time of the first crossing by $t'$. The probability that a particle hits the barrier at time $t'$ is given by the {\color{blue}first-passage} probability $F_N(L, t')$ and the particles are then reset to $\vec{0}$. Hence, they must now reach $\vec{x}$ at time $t$ from $\vec{0}$ at time $t'$ and since the problem is time-translationally invariant this is equivalent to ${\rm Prob.}[\vec{X}(t - t') = \vec{x}]$. Finally, we must integrate over all possible {\color{blue}first-passage} times $t'$. In between resetting events the particles are independent, hence the {\color{blue}first-passage} probability of the $N$ particle system can be obtained
\begin{equation} \label{eq:survival-first-relation}
    F_N(L, t) = - \pdv{}{t} \left(Q(L, t)^N\right) = N F_1(L, t) Q(L, t)^{N-1} \;.
\end{equation}
The convolution structure of Eq. (\ref{eq:renewal-fpt-resetting}) lends itself to a Laplace transform. Taking the Laplace transform of Eq. (\ref{eq:renewal-fpt-resetting}) we obtain
\begin{equation} \label{eq:laplace-renewal-1}
    \int_0^{+\infty} \dd t \; e^{-s t} {\rm Prob.}[\vec{X}(t) = \vec{x}] = \frac{\int_0^{+\infty} \dd t \; e^{-s t} \prod_{i = 1}^N p_{\leq L}(x_i, t)}{1 - \tilde{F}_N(L, s)} \;.
\end{equation}
Using Eq. (\ref{eq:survival-first-relation}) we can simplify the expression for the Laplace transform of the {\color{blue}first-passage} probability 
\begin{equation} \label{eq:laplace-FS-relation}
    \tilde{F}_N(L, s) = 1 - s \, \int_0^{+\infty} \dd t \; e^{-s t} Q(L, t)^N  \;,
\end{equation}
allowing us to re-write Eq. (\ref{eq:laplace-renewal-1}) as
\begin{equation} \label{eq:laplace-renewal-2}
    s \, \int_0^{+\infty} \dd t \; e^{-s t} {\rm Prob.}[\vec{X}(t) = \vec{x}] = \frac{\int_0^{+\infty} \dd t\; e^{-s t} \, \prod_{i = 1}^N p_{\leq L}(x_i, s)}{\int_0^{+\infty} \dd t\; e^{-s t} \, {\rm erf}\left( \frac{L}{\sqrt{4 D t}} \right)^N} \;.
\end{equation}
From Eq. (\ref{eq:laplace-renewal-2}) we can see that the model admits a steady state for $N > 2$ of which the joint probability density function can be obtained from the $s \to 0^+$ limit of the Laplace transform which is
\begin{equation} \label{eq:full-jpdf-NESS}
    {\rm Prob.}[\vec{X} = \vec{x}]_{\rm NESS} = \left[\int_0^{+\infty} \dd t \; {\rm erf}\left( \frac{L}{\sqrt{4 D t}} \right)^N\right]^{-1} \int_0^{+\infty} \dd t\; \prod_{i = 1}^N p_{\leq L}(x_i, t) \;.
\end{equation}
Taking a crude Laplace saddle-point approximation of Eq.~(\ref{eq:full-jpdf-NESS}) shows that the integral is dominated for $t^{-1/2} \sim \sqrt{\log N}$. Therefore, making the appropriate change of variable $\frac{L}{\sqrt{4 D t}} = u \sqrt{\log N}$ and taking the large-$N$ limit we obtain the asymptotic behavior of the joint probability density function 
\begin{equation}\label{eq:ciid-prob-fpt}
    {\rm Prob.}[\vec{X} = \vec{x}] \underset{N\to\infty}{\longrightarrow} \int_1^{+\infty} \dd u \; \frac{2}{u^3} \prod_{i = 1}^N \frac{u}{L} \sqrt{\frac{\log N}{\pi}} \exp[ - \frac{u^2 \log N x_i^2}{L^2} ]
\end{equation}
The natural length scale of the system is $\frac{L}{\sqrt{\log N}}$, the multi-particle correlations are creating a condensation of particles closer and closer to the origin as $N$ increases. We recognize in Eq.~(\ref{eq:ciid-prob-fpt}) the conditionally independent form we are now familiar with. Consistently with the notations in the previous Sections we can introduce
\begin{equation} \label{eq:p0-h-fpt}
    p_0(x, u) = \frac{u}{L} \sqrt{\frac{\log N}{\pi}} e^{- \frac{u^2 \log N x_i^2}{L^2}} \mbox{~~and~~} h(u) = \frac{2}{u^3} \Theta(u - 1) \;, 
\end{equation} 
to rewrite Eq.~(\ref{eq:ciid-prob-fpt}) in the usual conditional independent form
\begin{equation} \label{eq:ciid-fpt-reset}
    {\rm Prob.}[\vec{X} = \vec{x}] \underset{N\to\infty}{\longrightarrow} \int_0^{+\infty} \dd u \; h(u) \prod_{i = 1}^N p_0(x_i, u) \;.
\end{equation}
While we may have been able to guess that the particles should be conditionally independent, the simple form in Eq.~(\ref{eq:p0-h-fpt}) and Eq.~(\ref{eq:ciid-fpt-reset}) would be impossible to guess. Even {\it a posteriori}, the interpretation of this conditional variable $u$ is not obvious. However, armed with the conditional form in Eq.~(\ref{eq:ciid-prob-fpt}) we are ready to compute all of our usual observables to gain some insight on the behavior of this gas. 

\subsubsection{The average density}

\begin{figure}
    \centering
    \includegraphics[width=0.5\textwidth]{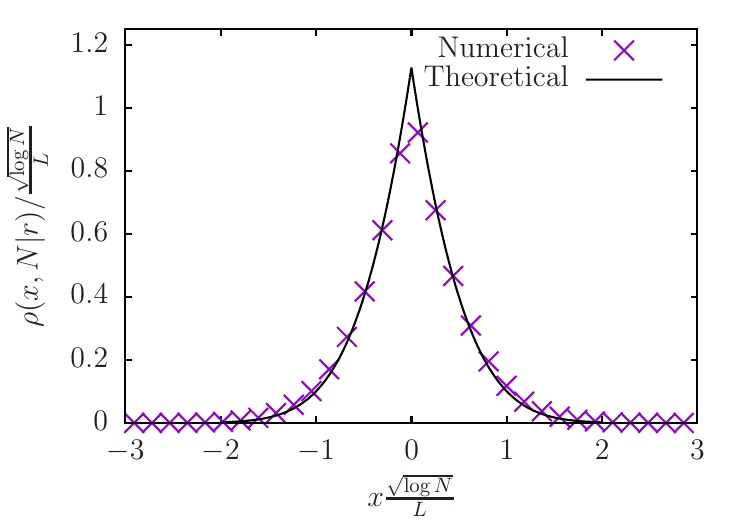}%
    \includegraphics[width=0.5\textwidth]{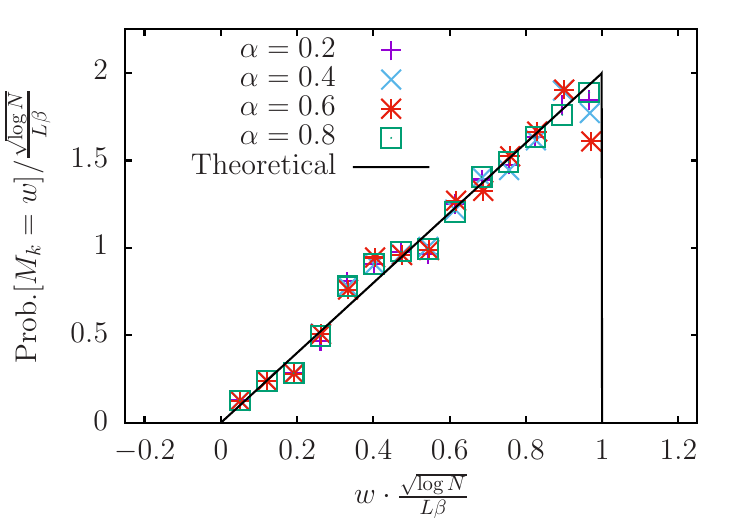}
    \caption{{\bf Left:} The scaled average density of particles in the non-equilibrium steady state. The black line denotes the analytical scaling function $\rho_S(z)$ in Eq. \ref{eq:rhoS-fpt-reset} and the points represent numerical simulations. {\bf Right:} The distribution of the $k$-th maximum $M_{k = \alpha N,N}$, given in Eq. \ref{eq:order_res_fpr}. The black line represents the analytical prediction given in Eq. \ref{eq:order_res_fpr} and the symbols represent simulation results for $N=100\,000$ and $L=1$, different colors and symbol markers correspond to different values of $\alpha = k/N$.} \label{fig:density_order_fpr}
\end{figure}

The average density $\rho(x, N | r)$ can be obtained directly from Eq.~(\ref{eq:ciid-fpt-reset}) 
\begin{equation}
    \rho(x, N | r) \approx \int_{0}^{+\infty} \dd u\; h(u) p_0(z, u) \;,
\end{equation}
yielding
\begin{equation}
    \rho(x, N | r) \approx \frac{\sqrt{\log N}}{L} \; \rho_S\left(x \frac{\sqrt{\log N}}{L} \right) \;,
\end{equation}
where the normalized scaling function $\rho_S(z)$ is defined for any $z \in \mathbb{R}$ and is given by
\begin{equation} \label{eq:rhoS-fpt-reset}
    \rho_S(z) = \int_1^{+\infty} \dd u \; \frac{2}{u^3} \frac{u}{\sqrt{\pi}} e^{-u^2 z^2} = \frac{2}{\sqrt{\pi}} e^{-z^2} - 2 |z| {\rm erfc}|z| \;.
\end{equation}
A comparison of this theoretical {\color{blue}prediction} with numerical experiments is given in the left panel of Fig. \ref{fig:density_order_fpr} finding excellent agreement.

\subsubsection{Order and {\color{blue}extreme-value} statistics}

We now consider the order statistics in the bulk and edge of the gas. Clearly $p_0(z, u)$ belongs to the Gumbel class. Hence applying the results derived in Chapter \ref{ch:ciid} we know that the behavior of the order statistics are encoded in the $\alpha$-quantile of the conditional distribution $p_0(z, u)$ which is given by
\begin{equation}
    q(\alpha, u) = \frac{L}{\sqrt{\log N}} \frac{1}{u}{\rm erfc}^{-1}(2\alpha) \;,
\end{equation}
and we introduce the usual constant
\begin{equation}
    \beta = {\rm erfc}^{-1}(2\alpha) \;.
\end{equation}
Applying the universal conditionally independent result in Eq. (\ref{eq:res_bulk_integral_form}) we obtain
\begin{equation} \label{eq:order_res_fpr}
    {\rm Prob.}[M_k = w] \underset{N\to\infty}{\longrightarrow} \frac{\sqrt{\log N}}{L \beta} \; f\left( w \frac{\sqrt{\log N}}{L \beta}\right) \;,
\end{equation}
where the scaling function $f(z)$ is defined on $[0, 1]$ and is given by
\begin{equation}
    f(z) = 2 z \;.
\end{equation}
Since we are in the Gumbel class, to obtain the {\color{blue}extreme-value} statistics it suffices to take the small $\alpha = k/N$ asymptotic of $\beta$ which can obtained from the known asymptotic of the error complementary function
\begin{equation}
    \beta \sim \sqrt{\log N}\;, \mbox{~~when~~} k \sim \mathcal{O}(1) \mbox{~~and~~} N \to +\infty\;.
\end{equation}
Hence, the {\color{blue}extreme-value} statistics are distributed as
\begin{equation}
    {\rm Prob.}[M_k = w] = \frac{1}{L} f\left(\frac{w}{L} \right)\;.
\end{equation}
A comparison of this {\color{blue}theoretical} prediction with numerical simulations is given in the right panel of Fig. \ref{fig:density_order_fpr} finding excellent agreement.

\subsubsection{The gap statistics}

\begin{figure}
    \centering
    \includegraphics[width=0.5\textwidth]{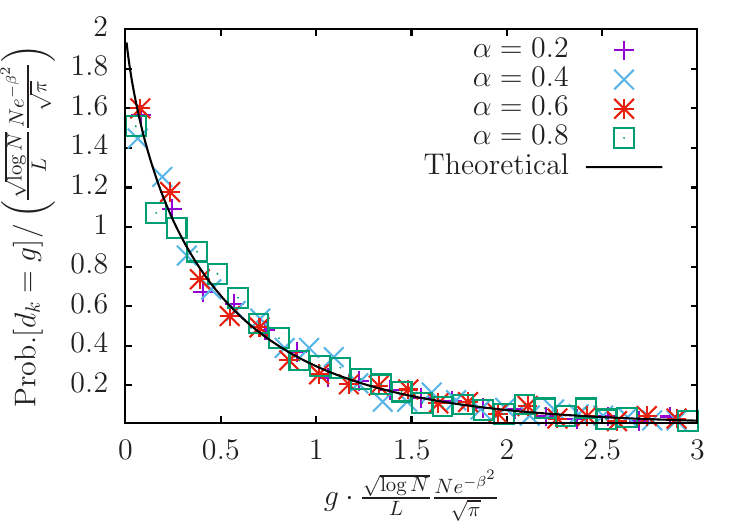}%
    \includegraphics[width=0.5\textwidth]{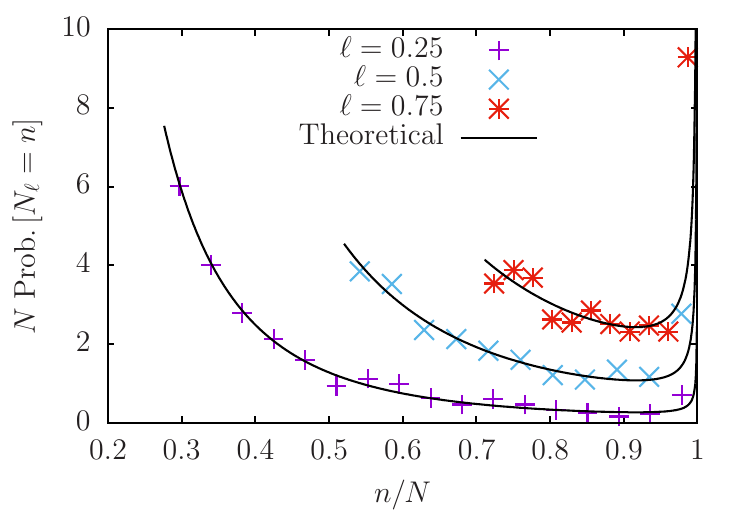}
    \caption{Plots of the probability density functions of the gaps (left panel) and the full counting statistics (right panel) obtained in Eq. (\ref{eq:gaps_fpr}) and Eq. (\ref{eq:fcs_fpr}) respectively. The black line represents the analytical prediction given in Eq. (\ref{eq:gaps_fpr}) and Eq. (\ref{eq:fcs_fpr}) respectively and the symbols represent simulation results for $N=100\,000$ and $L=1$, different colors and symbol markers correspond to different values of $\alpha = k/N$ (left panel) or $\ell$ (right panel).} \label{fig:gaps-fcs-reset-fp}
\end{figure}

Since we are in the Gumbel conditionally independent class we know that the gap statistics are universally described by Eq.~(\ref{eq:ciid-gaps-bulk}) and in our system
\begin{equation}
    p_0(q(\alpha, u), u) = \frac{u}{L} \sqrt{\frac{\log N}{\pi}} e^{-{\rm erfc}^{-1}(2\alpha)^2} = \frac{u \beta}{L}\sqrt{\frac{\log N}{\pi}}  \;,
\end{equation}
where we used the usual constant $\beta = {\rm erfc}^{-1}(2\alpha)$. Then applying Eq.~(\ref{eq:ciid-gaps-bulk}) we obtain
\begin{equation} 
    {\rm Prob.}[d_{k, N} = g] \approx N \beta \sqrt{\frac{\log N}{\pi}} \; F\left(g \cdot N \beta \sqrt{\frac{\log N}{\pi}}\right) \;,
\end{equation}
where the scaling function $F(z)$ is defined for $z \geq 0$, given by
\begin{equation} \label{eq:gaps_fpr}
    F(z) = 2 e^{-z} + 2 z E_i(-z) \;,
\end{equation}
and normalized to unity, where $E_i(z)$ is the exponential integral function. Using the known {\color{blue}asymptotes} of $E_i(z)$ we directly obtain the {\color{blue}asymptotes} of the gaps
\begin{equation}
    F(z) \sim \begin{dcases}
    2 + z (\log z + \gamma_E - 1) \;, \mbox{~~when~~} z \to 0^+ \\
    2 e^{-z} / z \;, \mbox{~~when~~} z \to +\infty\;,
    \end{dcases}
\end{equation}
where $\gamma_E \approx 0.57721\ldots$ is Euler's gamma constant. A plot of this scaling function is given in Fig. \ref{fig:gaps-fcs-reset-fp} and compared to numerical simulations finding excellent agreement.

\subsubsection{The full counting statistics}

We now look at the full counting statistics, but since we know that the natural length scale of the system is $L / \sqrt{\log N}$ we consider the full counting statistics $N_\ell$ in an interval $[- \ell L /\sqrt{\log N}, \ell L / \sqrt{\log N}]$ such that $\ell$ is dimensionless and the intervals is scaled with the natural length of the non-equilibrium steady state. For our propagator
\begin{equation}
    \int_{- \ell L /\sqrt{\log N}}^{\ell L / \sqrt{\log N}} \dd x \; p_0(x, u) = {\rm erf}(u \ell) \;.
\end{equation}
and using Eq. (\ref{eq:ciid-fcs}) we obtain immediately
\begin{equation} \label{eq:fcs_fpr}
    {\rm Prob.}[N_\ell = n] = \frac{1}{N} H\left( \frac{n}{N} \right)\;,
\end{equation}
where the scaling function $H(\kappa)$ is defined on ${\rm erf}(\ell) < \kappa < 1$ and given by
\begin{equation} \label{eq:def-H-fcs-reset}
    H(\kappa) = \frac{\ell^2 \sqrt{\pi}}{{\rm erf}^{-1}(\kappa)^3} e^{{\rm erf}^{-1}(\kappa)^2} \;,
\end{equation}
and normalized to unity. A plot of this scaling function is given in Fig. \ref{fig:gaps-fcs-reset-fp} and compared to numerical simulations finding excellent agreement.

\chapter{Non-interacting diffusive particles in a switching harmonic trap \cite{BKMS24}} \label{ch:ou-switch}
\section{Experimental considerations} \label{sec:experimental-OU}

Although the theory on resetting stochastic systems rapidly expanded, experiments have had a harder time keeping up. This is due to a couple of assumptions which are made when considering resetting systems which are typically hard to reproduce experimentally:
\begin{enumerate}
    \item {\bf Resetting events are instantaneous} This is almost always an approximation. To our knowledge, the only case where it is realistic to assume that the resetting event is instantaneous is when the resetting is the result of a quantum measurement that instantaneously collapses the wavefunction \cite{MSM18, PCML21, SVH23, KM23}. However, in any other scenario the particle teleporting is not realistic. Some novel experimental techniques, such as the Engineered Swift Equilibration technique \cite{MPGOTC16, GPP20,GORKTMGM19,PGOTP19,CBGOTPC18}, had to be developed to remedy this issue and come as close as possible to instantaneous resetting. Otherwise, the relaxation time it takes for the system to reset cannot be neglected.
    \item {\bf Resetting to a Dirac delta distribution.} Obtaining a perfect Dirac delta distribution experimentally is often unrealistic. Most likely we can reset to a Gaussian with a finite but very small width around the origin. However, even such a simple modification generally leads to very different behaviors \cite{BBPMC20}. 
\end{enumerate}
We describe one possible experimental setup \cite{BBPMC20} of resetting that has previously been used to confirm the theoretical results of the mean {\color{blue}first-passage} time. A small silica microsphere of radius $\sim 1 \mu {\rm m}$ is immersed in a fluid chamber containing pure water, which is the particle in our theoretical model. A near-infrared laser of wavelength $1064 {\rm nm}$ is focused into the chamber through an oil immersion objective creating an optical trap that traps the particle in the $(x, y)$ plane inside of a harmonic potential of stiffness $\mu > 0$. The stiffness can be controlled by modulating the laser. To obtain single one-dimensional trajectories, we exploit the fact that for diffusion the $x$ and $y$ coordinates are independent. Hence we consider the $x$ and $y$ projections of the two-dimensional trajectory as two independent samples of one-dimensional walks. 

\vspace{0.2cm}

\begin{figure} 
    \centering
    \includegraphics[width=0.7\textwidth]{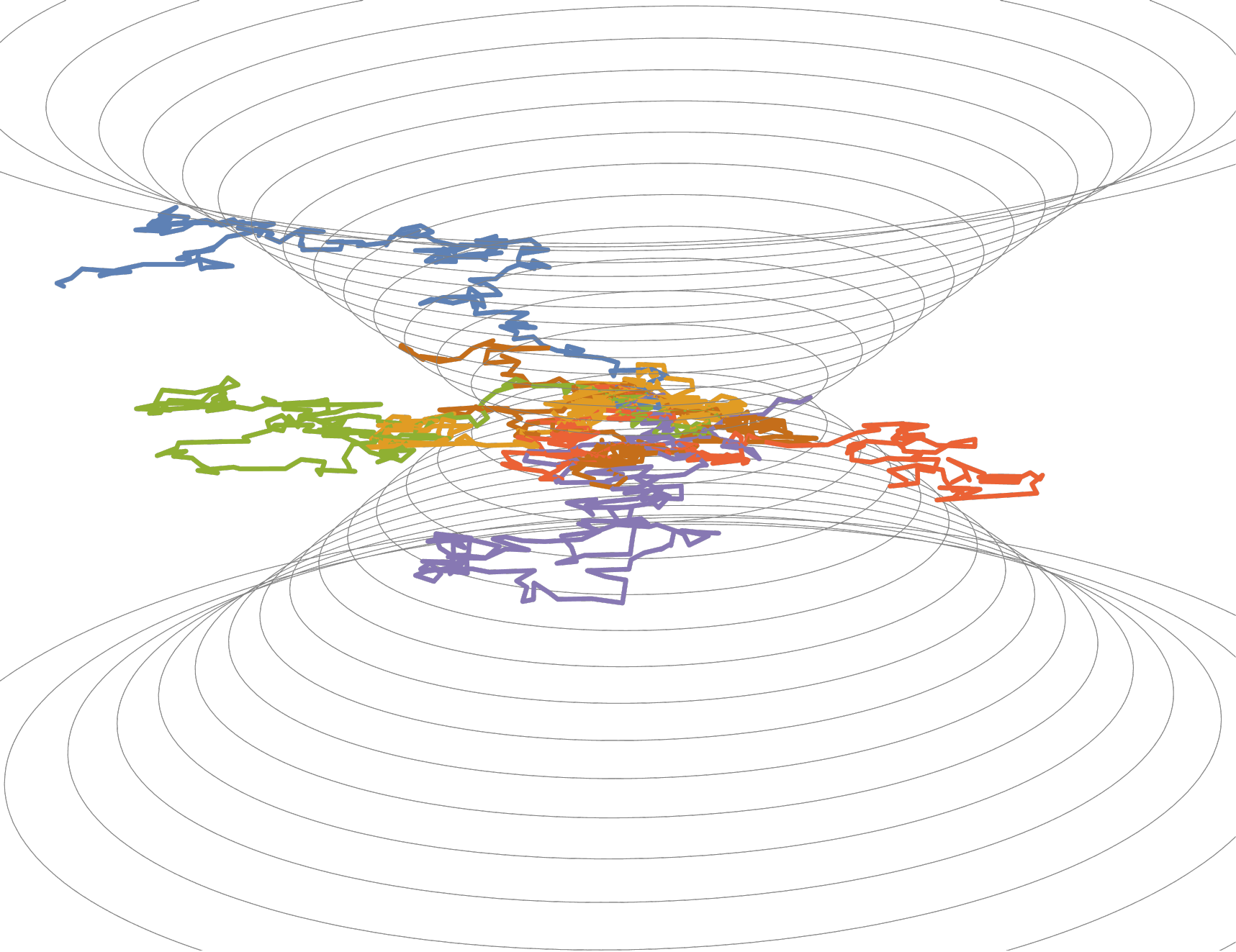}
    \caption{A sketch of the optical trap experimental setup. The potential of the experimental trap is sketched using the black {\color{blue}contour} lines. The optical trap results in a harmonic potential confining the trajectories of the particles to stay close to the origin (the center of the trap). At random Poissonianly distributed times, the stiffness of the trap is increased as much as possible effectively resetting the particle as close to the origin as possible, the stiffness is then placed back to its default value. Trajectories from different resetting events are plotted in different colors.} \label{fig:experimental_trap}
\end{figure}

To perform resetting the stiffness of the trap is increased as much as possible to $\mu_{\rm max}$ which leads to the particle relaxing to a tight Gaussian around the origin with a variance $\propto 1/\sqrt{\mu_{\rm max}}$. As discussed above, both assumptions of theoretical resetting are violated. The resetting is not instantaneous, since there is a finite relaxation time to the trapped Gaussian and we are not resetting to a Dirac delta at the origin. Modification of the theoretical model can be made \cite{BBPMC20} to reconcile some of the predictions with the experiments. Nevertheless, the behavior of the experimental results presents interesting new physics which cannot be captured by the ideal resetting model. 

\vspace{0.2cm}

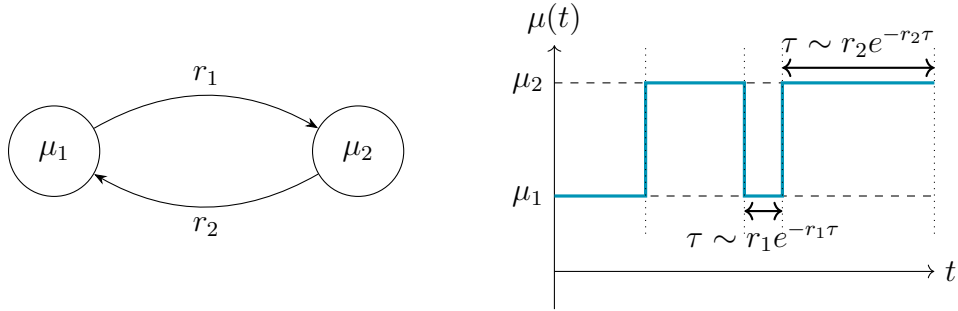
\begin{figure}
    \centering
    \begin{minipage}{0.5\textwidth}
        \begin{tikzpicture}[>=Stealth, node distance=4cm]
            \node[circle, draw, minimum size=1.2cm] (mu1) at (0,0) {$\mu_1$};
            \node[circle, draw, minimum size=1.2cm] (mu2) at (4,0) {$\mu_2$};
          
            \draw[->, bend left=30] (mu1) to node[above] {$r_1$} (mu2);
            \draw[->, bend left=30] (mu2) to node[below] {$r_2$} (mu1);
          \end{tikzpicture}
    \end{minipage}\hfill %
    \begin{minipage}{0.5\textwidth}
        \begin{tikzpicture}[scale=1.0]

            \draw[->] (0,-0.5) -- (5,-0.5) node[right] {$t$};
            \draw[->] (0,-1.0) -- (0,2.5) node[above] {$\mu(t)$};
          
            \draw[dashed] (0,0.5) -- (5,0.5);
            \node[left] at (0,0.5) {$\mu_1$};
          
            \draw[dashed] (0,2) -- (5,2);
            \node[left] at (0,2) {$\mu_2$};
          
            \draw[very thick, C4]
              (0,0.5) -- (1.2,0.5) 
              -- (1.2,2) -- (2.5,2)
              -- (2.5,0.5) -- (3.0,0.5)
              -- (3.0,2) -- (5.0,2);
          
            \foreach \x in {1.2,2.5,3.0,5.0} {
              \draw[dotted] (\x,0) -- (\x,2.5);
            }

            \draw[<->, thick] (3.0, 2.2) -- node[above] {$\tau \sim r_2 e^{-r_2 \tau}$} (5.0, 2.2);
            \draw[<->, thick] (2.5, 0.3) -- node[below] {$\tau \sim r_1 e^{-r_1 \tau}$} (3.0, 0.3);
          \end{tikzpicture}
    \end{minipage}
    \caption{The stiffness of the confining potential is a two state system with values $\mu_1 > \mu_2 > 0$. The stiffness switches from one value to the other at random Poissonianly distributed time intervals. The switching rates from one state to the other are not necessarily symmetric. The stiffness switches from $\mu_1$ to $\mu_2$ with rate $r_1$ and from $\mu_2$ to $\mu_1$ with rate $r_2$.} \label{fig:switching-potential}
\end{figure}

Inspired by this experimental protocol, we studied the following theoretical model, which is an exact reproduction of the experimental realization. Instead of implementing stochastic resetting, we consider diffusing particles $X_1(t), \cdots, X_N(t)$ in a harmonic trap of stiffness $\mu(t)$ where $\mu(t)$ switches between two values $\mu_1 > \mu_2$ at random Poissonianly distributed intervals. The stiffness switches from $\mu_1$ to $\mu_2$ with rate $r_1$ and from $\mu_2$ to $\mu_1$ with rate $r_2$. A sketch of this two-state system is provided in Fig.~\ref{fig:switching-potential}. In the limits $\mu_1 \to \infty$, $\mu_2 \to 0$ and $r_1 \to \infty$, this general protocol reduces to the standard model of diffusion under stochastic resetting to the origin. The limit $\mu_1 \to \infty$ and $\mu_2 \to 0$ ensures the resetting of a diffusing particle to the origin, while the limit $r_1 \to \infty$ guarantees that once it is reset to the origin, it immediately restarts, thus achieving the instantaneous resetting. For a {\it single} particle undergoing this switching intermittent potential, the resulting position distribution in the non-equilibrium steady state has been studied only recently~\cite{MBMS20,GPKP20,SDN21,XZMD22,GP22,ACB22,MBM22}. Our goal now is to study $N$ independent particles undergoing this switching intermittent protocol. We will show that the switching dynamics between two stiffnesses of the trap drives the system into a non-equilibrium steady state with strong collective correlations that emerge purely out of the dynamics. Thus, the emergence of strong correlations without direct interaction is a robust phenomenon and is not just an artifact of instantaneous resetting. 

\section{Switching \OU process} \label{sec:model-switching-OU}

We consider $N$ independent Brownian particles on a line, all starting at the origin which feel a potential that switches between
$V_1(x) = \mu_1 x^2/2$ and $V_2(x) = \mu_2 x^2/2$,  
with Poissonian rate $r_1$ (from $\mu_1$ to $\mu_2$) and rate $r_2$ (from $\mu_2$ to $\mu_1$). Hence, the duration $\tau$ of the time intervals between successive switches is distributed via $h(\tau) = r_i e^{-r_i\tau}$, where $r_i$ is $r_1$ or $r_2$. Moreover, the intervals are statistically independent. In each phase the positions $\{ x_i\}$ evolve as independent Ornstein-Uhlenbeck processes~\cite{UO30} 
\begin{equation} \label{eq:sde-ou-reset}
\dv{x_i}{t} = - \mu_k x_i + \sqrt{2 D} \eta_i(t) \;,
\end{equation}
where $\mu_k = \mu_1$ or $\mu_2$ depending on the phase, $D$ is the diffusion constant and $\eta_i(t)$ is a zero-mean Gaussian white noise with a correlator $\langle \eta_i(t) \eta_j(t') \rangle = \delta_{ij} \delta(t - t')$. In order to write the Fokker-Planck equation associated with the stochastic differential equation in Eq. (\ref{eq:sde-ou-reset}) it is easier to introduce two joint probability densities $p_1(\vec{x}, t)$ and $p_2(\vec{x}, t)$ characterizing the probability of the particles being located at $\vec{x}$ at time $t$ and that the system is in phase 1 or 2 respectively. The indiscriminate joint probability, regardless of the potential, is then given by
\begin{equation} \label{eq:tot-p-p1-p2}
    p(\vec{x}, t) = p_1(\vec{x}, t) + p_2(\vec{x}, t) \;.
\end{equation}
We can now express the Fokker-Planck equation of the process defined in Eq. (\ref{eq:sde-ou-reset}) as
\begin{align}
    \pdv{p_1(\vec{x}, t)}{t} &= \sum\limits_{i = 1}^N \left[D \pdv[2]{p_1(\vec{x}, t)}{x_i} + \mu_1 \pdv{}{x_i} \left(x_i p_1(\vec{x}, t) \right) \right] - r_1 p_1(\vec{x}, t) + r_2 p_2(\vec{x}, t) \label{fp-p1}\\
    \pdv{p_2(\vec{x}, t)}{t} &= \sum\limits_{i = 1}^N \left[ D \pdv[2]{p_2(\vec{x}, t)}{x_i} + \mu_2 \pdv{}{x_i} \left( x_i p_2(\vec{x}, t) \right) \right] - r_2 p_2(\vec{x}, t) + r_1 p_1(\vec{x}, t) \label{fp-p2}
\end{align}
with the initial conditions
\begin{equation} \label{init-fp}
p_1(\vec{x}, t = 0) = \frac{1}{2}\delta(\vec{x}) \mbox{~~and~~} p_2(\vec{x}, t = 0) = \frac{1}{2}\delta(\vec{x}) \;,
\end{equation}
where we assumed that, initially, both phases occur equally likely. The first terms on the right hand side of Eqs. (\ref{fp-p1}) and (\ref{fp-p2}) represent diffusion and advection in a harmonic potential, while the last two terms represent loss and gain due to the switching between potentials, with rates $r_1$ and $r_2$ respectively.  

\vspace{0.2cm}

To solve this pair of Fokker-Planck equations, it is convenient to work in the Fourier space where we define 
\begin{equation} \label{eq:def-fourier-p1-p2}
    \hat{p}_1(\vec{k}, t) = \int_{-\infty}^{+\infty} \dd \vec{x} \; e^{i \vec{k} \cdot \vec{x}} p_1(\vec{x}, t) \mbox{~~and~~} \hat{p}_2(\vec{k}, t) = \int_{-\infty}^{+\infty} \dd \vec{x} \; e^{i \vec{k} \cdot \vec{x}} p_2(\vec{x}, t) \;.
\end{equation}
In the steady state, setting $\partial_t \hat{p}_1 = \partial_t \hat{p}_2 = 0$, Eqs. (\ref{fp-p1}) and (\ref{fp-p2}) and taking the Fourier transform reduces to
\begin{align}
\left(D \sum\limits_{i = 1}^N k_i^2 + r_1\right) \hat{p}_1(\vec{k}) + \mu_1 \sum\limits_{i = 1}^N k_i \pdv{\hat{p}_1(\vec{k})}{k_i} &= r_2 \hat{p}_2(\vec{k}) \label{fourier-fp-p1}\\
\left(D \sum\limits_{i = 1}^N k_i^2 + r_2\right) \hat{p}_2(\vec{k}) + \mu_2 \sum_{i = 1}^N k_i \pdv{\hat{p}_2(\vec{k})}{k_i} &= r_1 \hat{p}_1(\vec{k}, t) \label{fourier-fp-p2}\;,
\end{align}
where we dropped the explicit time dependence $\hat p_1(\vec{k}) \equiv \hat p_1(\vec{k}, t \to +\infty)$ since we are in the steady state. Notice that Eqs (\ref{fourier-fp-p1}-\ref{fourier-fp-p2}) are spherically symmetric. It is therefore much easier to move to hyperspherical coordinates where $k = \sqrt{\sum_{i = 1}^N k_i^2}$ is the distance to the origin and $\theta_i$, for $i = 1, \cdots, N-1$ are the different angular coordinates. Then Eqs. (\ref{fourier-fp-p1}-\ref{fourier-fp-p2}) simplify to
\begin{align}
\left[(D k^2 + r_1) + \mu_1 k \partial_k \right]\hat{p}_1(k) &= r_2 \hat{p}_2(k) \label{spherical-fourier-fp-p1}\\
\left[(D k^2 + r_2) + \mu_2 k \partial_k\right] \hat{p}_2(k) &= r_1 \hat{p}_1(k) \label{spherical-fourier-fp-p2} \;.
\end{align}
Notice that by permuting the indices $1 \leftrightarrow 2$ in Eq. (\ref{spherical-fourier-fp-p1}) leads to Eq. (\ref{spherical-fourier-fp-p2}). Hence, we can solve only for $\hat{p}_1(k, t)$ and the solution for $\hat{p}_2(k, t)$ will follow by permuting the indices. Notice that when setting $k = 0$ in Eq. (\ref{spherical-fourier-fp-p1}) we obtain
\begin{equation} \label{eq:init-p1-p2}
    {\color{blue}r_1 \hat p_1(0) = r_2 \hat p_2(0) \;,}
\end{equation}
and from Eq. (\ref{eq:tot-p-p1-p2}) and the normalization condition 
\begin{equation} \label{eq:normalization-condition}
    1 = \int \dd^N \vec{x} \; p(\vec{x}, t) \;,
\end{equation}
we must have
\begin{equation} \label{eq:fourier-normalization-condition}
    {\color{blue} 1 = \hat p_1 (0) + \hat p_2(0) \;.}
\end{equation}
Putting Eq.~(\ref{eq:init-p1-p2}) and Eq.~(\ref{eq:fourier-normalization-condition}) together we obtain the initial conditions for the coupled ordinary differential equations in Eq.~(\ref{spherical-fourier-fp-p1}) and Eq.~(\ref{spherical-fourier-fp-p2}) 
\begin{equation} \label{bc_Fourier}
    {\color{blue}\hat p_1(0) = \frac{r_2}{r_1 + r_2} \mbox{~~and~~} \hat p_2(0) = \frac{r_1}{r_1 + r_2} \;.}
\end{equation}

\vspace{0.2cm}

To solve the coupled first-order ordinary differential equations in Eqs. (\ref{spherical-fourier-fp-p1}-\ref{spherical-fourier-fp-p2}) we apply the differential operator a second time to obtain two uncoupled second order ordinary differential equations
\begin{equation}
    \left[ \left(D k^2 + r_2 \right) + \mu_2 k \dv{}{k} \right]\left[ \left(D k^2 + r_1 \right) + \mu_1 k \dv{}{k} \right] \hat p_1(k) = r_1 r_2 \hat{p}_1(k) \label{ode-p1} \;,
\end{equation}
as stated previously the equivalent ordinary differential equation for $\hat p_2(k)$ can be obtained by permuting the indices $1 \leftrightarrow 2$. Solving (which can be done with Mathematica) the ordinary differential equation in Eq. (\ref{ode-p1}) yields the most general solution
\begin{align} \label{M-U-res}
\hat{p}_1(k) =  e^{-\frac{D k^2}{2 \mu_1}} \;\Bigg[ &A_1 M\left( R_1; 1 + R_1 + R_2; - \frac{D k^2 (\mu_1 - \mu_2)}{2 \mu_1 \mu_2} \right) \\
+ &B_1 U\left( R_1; 1 + R_1 + R_2; - \frac{D k^2 (\mu_1 - \mu_2)}{2 \mu_1 \mu_2} \right) \Bigg]\;,
\end{align}
where $A_1, B_1$ are arbitrary constants and we denote
\begin{equation} \label{R1-R2}
R_1 = \frac{r_1}{2 \mu_1} \mbox{~~and~~} R_2 = \frac{r_2}{2 \mu_2} \;.
\end{equation}
Here $M(a;b;z)$ is the Kummer function defined by the power series \cite{NIST}
\begin{equation} \label{Kummer}
M(a;b;z) = 1 + \frac{a}{b}z + \frac{a(a+1)}{b(b+1)} \frac{z^2}{2!} + \cdots \;. 
\end{equation}
and $U(a;b;z)$ is the confluent hypergeometric $U$ function. To fix the constants $A_1$ and $B_1$ we will use the boundary conditions in Eq. (\ref{bc_Fourier}). We use small-argument {\color{blue}asymptotes} of the hypergeometric functions, namely $M(a;b;z) \to 1$ and $U(a;b;z) \sim z^{1-b}$ as $z \to 0$. Taking the $k\to 0$ limit in Eq. (\ref{M-U-res}), one sees that the second term diverges as $k^{-2(R_1+R_2)}$ as $k \to 0$. However, from Eq. (\ref{bc_Fourier}), we see that $\hat p_1(k=0) = {r_2}/{(r_1+r_2)}$. Hence, we must have $B_1 = 0$. Taking the limit $k \to 0$ in Eq. (\ref{M-U-res}) then fixes $A_1 = r_2/(r_1+r_2)$. Similarly, the solution for $\hat p_2(k)$ can be written down by exchanging the indices $1$ and $2$. This gives
\begin{equation} \label{tilde-p1}
\hat{p}_1(k) = \frac{r_2}{r_1 + r_2} e^{-\frac{D k^2}{2 \mu_1}} M \left( R_1; 1 + R_1 + R_2; - \frac{D k^2 (\mu_1 - \mu_2)}{2 \mu_1 \mu_2} \right)
\end{equation}
and
\begin{equation} \label{tilde-p2}
\hat{p}_2(k) = \frac{r_1}{r_1 + r_2} e^{-\frac{D k^2}{2 \mu_2}} M \left( R_2; 1 + R_1 + R_2; - \frac{D k^2 (\mu_2 - \mu_1)}{2 \mu_1 \mu_2} \right) \;.
\end{equation}
Fortunately, there is a very nice integral representation of the Kummer's function which reads~\cite{NIST}
\begin{equation} \label{M-integral}
M(a; b; z) = \frac{\Gamma(b)}{\Gamma(a)\Gamma(b - a)} \int_0^1 \dd u \; e^{z u} u^{a - 1} (1 - u)^{b - a - 1} \;,
\end{equation}
which allows us to invert the Fourier transforms in Eqs. (\ref{tilde-p1}-\ref{tilde-p2}). Using Eq. (\ref{M-integral}) we can re-express Eq. (\ref{tilde-p1}) as
\begin{equation} \label{tilde-p1-int}
\hat{p}_1(k) = \frac{r_2}{r_1 + r_2} \frac{\Gamma(1 + R_1 + R_2)}{\Gamma(R_1) \Gamma(1 + R_2)} \int_0^1 \dd u \; u^{R_1 - 1} (1 - u)^{R_2} e^{-k^2\left(\frac{D(u \mu_1 + (1 - u)\mu_2)}{2 \mu_1 \mu_2}\right)} \;.
\end{equation}
Under this form, one can now easily invert the Fourier transform and we can obtain the real space representation of the joint probability density functions $p_1(\vec{x})$ and $p_2(\vec{x})$. Using Eq.~(\ref{eq:tot-p-p1-p2}) we obtain the total joint probability density function which can be written compactly as
\begin{equation} \label{jpdf}
p(\vec{x}, t \to+\infty) \equiv p(\vec x) = \int_0^1 {\rm d}u\, h(u) \prod_{i=1}^N p_0(x_i, u) \;,
\end{equation}
where 
\begin{equation} \label{h_u_sm}
h(u) = \frac{c\,r_H}{4} u^{R_1 - 1} (1 - u)^{R_2 - 1} \left[ \frac{1 - u}{\mu_1} + \frac{u}{\mu_2} \right] 
\end{equation}
with $c = \Gamma(R_1 + R_2 + 1) / ( \Gamma(R_1 + 1) \Gamma(R_2 + 1) )$ and $r_H = 2 \, r_1 r_2/(r_1+r_2)$ and 
\begin{equation} \label{eq:p0-OU}
    p_0(x, u) = \frac{1}{\sqrt{2 \pi V(u)}} \; e^{- \frac{x^2}{2 V(u)}} \;,
\end{equation}
where 
\begin{equation}\label{def_V_of_u}
    V(u) = D\,\left( \frac{u}{\mu_2} + \frac{1 - u}{\mu_1} \right) \;.
\end{equation}
One can check that $h(u)$ is normalized to unity, i.e.,
\begin{equation} \label{norm_h}
\int_0^1 {\rm d}u\, h(u) = 1 \;.
\end{equation}
Since $h(u) \geq 0$ and is normalized to unity, it can be interpreted as a probability density function of a random variable $u$. Thus one can interpret Eq.~(\ref{jpdf}) as the joint distribution of $N$ independent identically distributed Gaussian variables with zero mean and a common variance $V(u)$ parametrized by $u$, which itself is a random variable distributed via the probability density function $h(u)$.

\vspace{0.2cm}

The conditionally independent form in Eq.~(\ref{jpdf}) is an incredible, but welcome, surprise not only from an analytical standpoint but physically also. Now that we have such a conditionally independent form we can use our previous universal results from Chapter \ref{ch:ciid} to obtain the behavior of all of our previously studied observables. This conditional independence was not built `{\color{blue}artificially}' via resetting events leading to a renewal equation, it arises naturally from the internal dynamics of the system. A priori, nothing indicated or suggested the existence of this conditional independence. In order to understand the origin behind this phenomenon we have to clarify the physical {\color{blue}interpretation} of this latent variable $u$. To do so we need to stare at Eq.~(\ref{def_V_of_u}) for a second. If the particle was entirely in phase $2$, its stationary distribution would be a Gaussian (the Gibbs state) with a variance $V(u=1) = D/\mu_2$. In contrast, if it was in phase $1$, it will again be a Gaussian with a variance $V(u = 0) = D/\mu_1$. Hence from the formula for $V(u)$ in Eq.~(\ref{def_V_of_u}), we see that $0 \leq u \leq 1$ seems to be interpolating between the steady states of phase 1 and phase 2. In fact, we will prove in Section~\ref{sec:kesten} through an alternative derivation that $u$ (resp. $1-u$) represents the fraction of time the system spends in phase 2 (resp. phase 1).

{\color{blue}
\section{Experimental setup~\cite{BCKMPS25}} \label{sec:experimental-setup}

\begin{figure}
    \centering
    \includegraphics[width=0.7\textwidth]{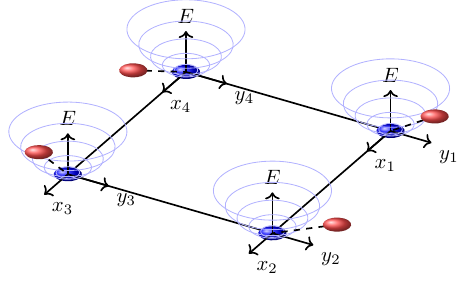}
    \caption{{\color{blue}A sketch of the experimental setup. Four colloidal particles a placed at the vertices of a square. Using a laser we create four harmonic traps confining the particles close to their respective vertex. At random Poissonianly distributed times we switch the stiffness of the trap by modulating the intensity of the laser. Using a camera we track the position of each particle in the plane $x_i, y_i$. Since for diffusion in a harmonic trap the $x$ and $y$ coordinates are uncorrelated we consider each coordinate $\{x_1, y_1, \cdots, x_4, y_4\}$ as a separate one-dimensional realization of our process, i.e. 8 one-dimensional particles.}}
    \label{fig:sketch-experiment}
\end{figure}

Following these theoretical considerations we have collaborated with Pr. Ciliberto from ENS Lyon to design an experimental setup corresponding to this theoretical model. We immersed four colloidal particles in water and trapped each particle separately using an optical tweezer. The stiffness of the traps can be changed by modulating the intensity of the laser used for the optical tweezer. A sketch of the experimental setup is given in Fig.~\ref{fig:sketch-experiment}. Going to the large-$N$ limit was experimentally unfeasible, however the analytical derivations in the following sections can be adapted for finite $N = 8$. In the experimental setup the colloidal particles are not independent, there are hydrodynamic interactions between all the particles. Nonetheless, as can be seen in Fig.~\ref{fig:exp-ou}, the theoretical predictions from our conditionally independent theoretical model perfectly match the experimental data. This means that the correlations generated dynamically via the fluctuating environment are so strong that they overpower any existing underlying hydrodynamic correlations in the experimental system. Proving once again the importance of studying these environmental fluctuations and how strong the correlations they induce can be.

\begin{figure}
    \centering
    \includegraphics[width=\textwidth]{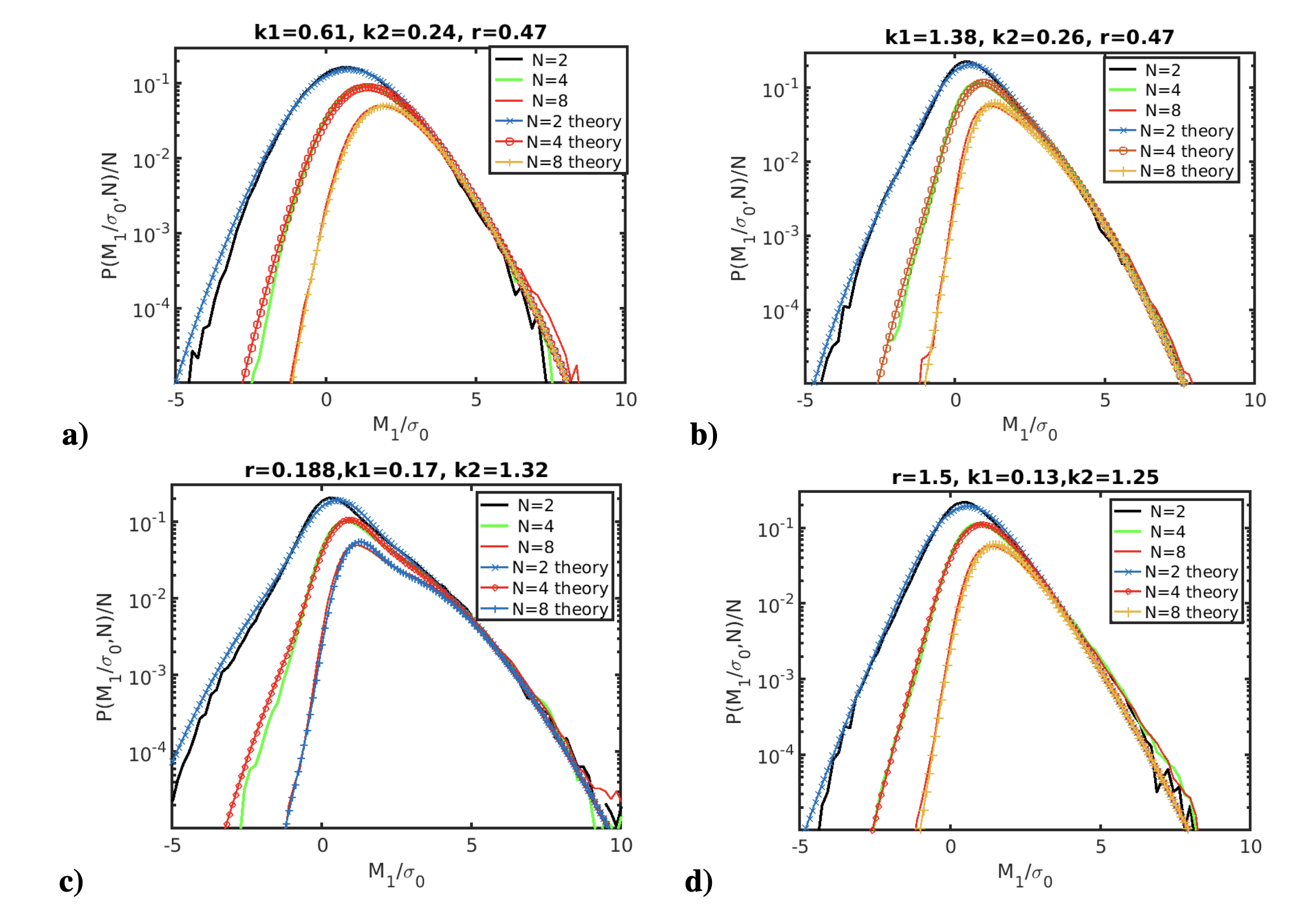}
    \caption{{\color{blue}A recall of Fig.~\ref{fig:main-res-exp-ou}. The experimental measurements for the distribution of the maximum $M_{1, N}$ for 2,4 and 8 particles are compared
    with the theoretical predictions obtained which follow from the analytical treatment in Section~\ref{sec:ou-order}. Each panel corresponds to different resetting rates $r$, and stiffnesses $k_1, k_2$ and each line corresponds to different number of {\color{blue} particles} $N$, continuous lines are theoretical predictions and marked lines are the experimental results.}} \label{fig:exp-ou}
\end{figure}

}

\vspace{0.2cm}

\section{Order statistics} \label{sec:ou-order}

\begin{figure}
    \centering
    \includegraphics[width=0.32\textwidth]{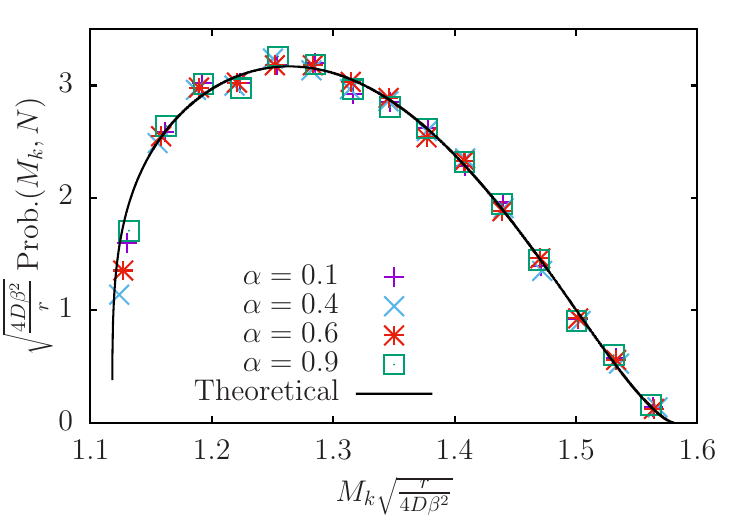} 
    \hfill%
    \includegraphics[width=0.32\textwidth]{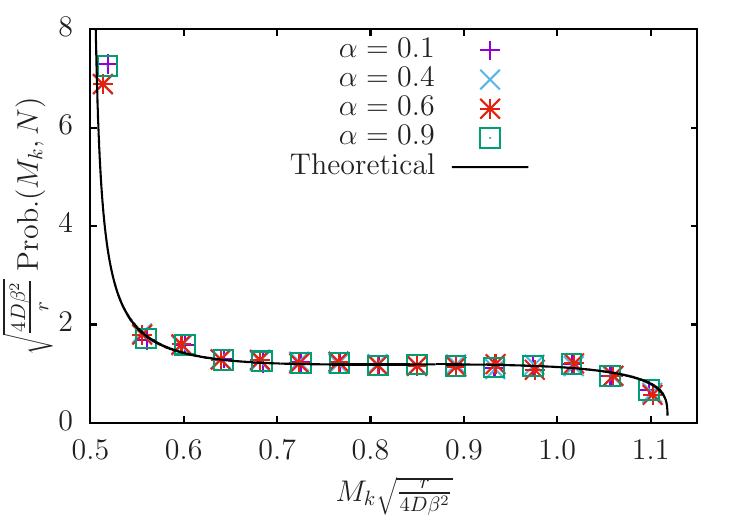} 
    \hfill%
    \includegraphics[width=0.32\textwidth]{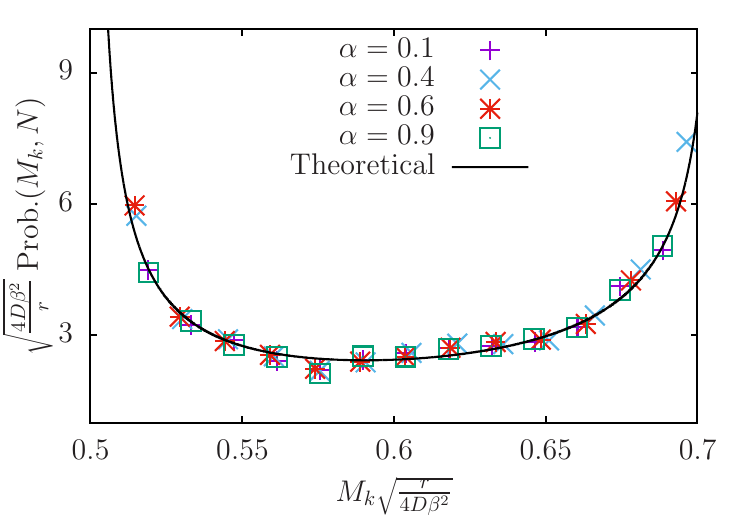} 
    \caption{Scaling collapse of the distribution of the $k$-th maximum as in Eq.~(\ref{Mk}) for different values of $\alpha = k/N$ and different values of the parameters. We set $r_1 = r_2 = 1$, $D=1$, $N = 10^{6}$ and vary $\mu_1$ and $\mu_2$. From left to right we used respectively $\mu_1 = 0.4, \, \mu_2 = 0.2$ then $\mu_1 = 2, \, \mu_2 = 0.4$ and finally $\mu_1 = 2,\, \mu_2 = 1$. The dashed black line corresponds to the theoretical prediction and the symbols are the numerical results. Different colors correspond to different values of $\alpha$. The numerical results were obtained by sampling $10^{5}$ examples directly from the \NESS distribution given in Eq. (\ref{jpdf}).} \label{fig:Mk-OU}
\end{figure}

We will now look at the usual observables we use to characterize this gas of particles, starting from the order statistics. As seen in Chapter \ref{ch:ciid}, the first step to compute the order statistics is compute the $\alpha$-quantile 
\begin{equation} \label{q}
q(\alpha, u) =  \sqrt{2\,V(u)} \, \; {\rm erfc}^{-1}\, \left(2\alpha\right) \;,
\end{equation}
where ${\rm erfc}^{-1}(z)$ is the inverse of the complementary error function ${\rm erfc}(z) = (2/\sqrt{\pi}) \int_z^{+\infty} e^{-u^2} \dd u$.
In terms of the $\alpha$-quantile, the probability density function of $M_k$ can then be expressed as (see Eq. (\ref{eq:res_bulk_integral_form}))
\begin{equation} \label{mk-delta}
{\rm Prob.}[M_k = w] \underset{N \to \infty}{\longrightarrow} \int_0^1 \dd u \; h(u) \delta( q(\alpha, u) - w ) \;.
\end{equation}
For compactness, let us denote 
\begin{equation} \label{def_beta}
\beta = {\rm erfc}^{-1}(2\alpha) \;.
\end{equation}
Substituting Eq. (\ref{q}) in Eq. (\ref{mk-delta}) we obtain
\begin{equation} \label{Mk-explicit}
{\rm Prob.}[M_k = w] \underset{N \to \infty}{\longrightarrow}  \int_0^1 \dd u \; h(u) \; \delta\left( \beta \sqrt{\frac{2 D}{\mu_1 \mu_2} (u \mu_1 + (1 - u) \mu_2)} - w\right) \;.
\end{equation}
This integral can be performed leading to the scaling form
\begin{equation} \label{Mk}
{\rm Prob.}[M_k = w] \underset{N \to \infty}{\longrightarrow}  \sqrt{\frac{r_H}{4 D \beta^2}} \; f\left(w \sqrt{\frac{r_H}{4 D \beta^2}}\right) \;,
\end{equation}
where $r_H$ is the harmonic mean of the switching rates, defined as $r_H = \frac{2}{\frac{1}{r_1} + \frac{1}{r_2}}$. The scaling function $f(z)$ is supported over the finite interval $\sqrt{R_{H,1}} < z < \sqrt{R_{H,2}}$ and is given explicitly by
\begin{equation} \label{f-z}
f(z) = c \frac{R_{H,1}^{R_1 - 1} R_{H,2}^{R_2 - 1}}{(R_{H,2} - R_{H,1})^{R_1 + R_2 - 1}}  |z|^3 \left(1 - \frac{z^2}{R_{H, 2}}\right)^{R_2 - 1} \left(\frac{z^2}{R_{H, 1}} - 1 \right)^{R_1 - 1}   \;,
\end{equation}
where
\begin{equation}
R_{H, 1} = \frac{r_H}{2 \mu_1} \mbox{~~and~~} R_{H, 2} = \frac{r_H}{2 \mu_2} \;.
\end{equation}
One can easily verify the normalization 
\begin{equation} \label{norm}
\int_{\sqrt{R_{H,1}}}^{\sqrt{R_{H,2}}} \dd z \, f(z) = 1 \;.
\end{equation}
The fact that the scaling function for the $k$-th maximum is supported over a finite interval is rather unusual since in most known examples \cite{MPS20}, the associated scaling function of $M_k$ has an infinite (or semi-infinite) support. Moreover, the shape of this scaling function can be tuned by varying the parameters $r_1, r_2, \mu_1$ and $\mu_2$. For instance, from Eq. (\ref{f-z}), if $R_1 >1$ and $R_2>1$, the scaling function $f(z)$ vanishes at both edges of the support (see left panel of Fig. \ref{fig:Mk-OU}). If $R_1<1$ and $R_2>1$, the scaling function diverges at the lower edge but vanishes at the upper edge (see middle panel of Fig. \ref{fig:Mk-OU}). Similarly, if $R_1<1$ and $R_2<1$, the scaling function diverges at both edges (see right panel of Fig. \ref{fig:Mk-OU}). 

\vspace{0.2cm}

Since the scaling function $f(z)$, given in Eq. (\ref{f-z}), is independent {of $\alpha$}, it also holds for $M_k$ when $k =O(1)$, i.e., $\alpha = O(1/N)$. The only difference is in the scale factor in Eq. (\ref{Mk}). Indeed, by setting $\alpha = k/N$, with $k=O(1)$, 
one finds from Eq.~(\ref{def_beta}) to leading order for large $N$ 
\begin{equation} \label{asympt_beta}
\beta = {\rm erfc}^{-1}[2 \alpha] \approx \sqrt{ \ln N} \;,
\end{equation}
independent of $k$. Hence, for all $k=O(1)$, we have
\begin{equation} \label{M1}
{\rm Prob.}[M_k = w] = \sqrt{\frac{r_H}{4 D \ln N}} \; f\left(w \sqrt{\frac{r_H}{4 D \ln N}} \right) \;,
\end{equation}
where the scaling function $f(z)$ is given in Eq. (\ref{f-z}). Thus, the scaling function $f(z)$ is universal, i.e., independent of the order $k$, either in the bulk or at the edges, and the {\color{blue}extreme-value} statistics scale as $\sqrt{4 D \ln N / r_H}$.

\section{Gap statistics}

We now turn our attention to the statistics of the gaps $d_k = M_{k} - M_{k+1}$. Once again, we will exploit the conditionally independent structure of the joint distribution in Eq. (\ref{jpdf}) and follow the general procedure outlined in Chapter \ref{ch:ciid}, from Eq.~(\ref{eq:ciid-gaps-bulk}) we know that the gap distribution will be given by
\begin{equation} \label{dk-orig}
{\rm Prob.}[d_k = g] \underset{N\to\infty}{\longrightarrow} \int_0^{1} \dd u \; h(u) \; N p[q(\alpha, u) | u] e^{- N p[q(\alpha, u)] g} \;.
\end{equation}
Using Eqs. (\ref{h_u_sm}-\ref{eq:p0-OU}) and Eq. (\ref{q}) we can re-write Eq. (\ref{dk-orig}) in a scaling form
\begin{equation} \label{dk}
{\rm Prob.}[d_k = g] \underset{N\to\infty}{\longrightarrow} N\,\sqrt{\frac{\mu_H}{4 \pi D}} \,e^{-\beta^2}\, 
F\left( \sqrt{\frac{\mu_H}{4 \pi D}} e^{-\beta^2} N\, g \right) \;,
\end{equation}
where $\mu_H = 2 \mu_1 \mu_2/(\mu_1+\mu_2)$, the constant $\beta$ is given in Eq. (\ref{def_beta}) and the scaling function $F(z)$, supported on $z \geq  0$, is given by
\begin{align} 
F(z) = c \frac{r_H}{4} \int_0^1 \dd u\; &u^{R_1-1} (1 - u)^{R_2-1} \left[ \frac{1 - u}{\mu_1} + \frac{u}{\mu_2} \right] \; \\
&\times \sqrt{\frac{\mu_1 + \mu_2}{u \mu_1 + (1 - u)\mu_2}} \exp( - z \sqrt{\frac{\mu_1 + \mu_2}{u \mu_1 + (1 - u)\mu_2}} ) \;. \label{G}
\end{align}
One can check that $F(z)$ is normalized to 1, i.e., $\int_0^\infty \dd z \, F(z) = 1$. While we could not compute this integral explicitly, the asymptotic behavior of $F(z)$ can be easily extracted from Eq. (\ref{G}).  

\vspace{0.2cm}

\noindent{\it The limit $z \to 0$.} In this limit, expanding $e^{-s z} \sim 1 - s z$, we get
\begin{equation} \label{G-small-z}
F(z) \approx  B_1 - B_2 z \;,
\end{equation}
where the constants $B_1$ and $B_2$ are given by
\begin{equation} \label{A1}
B_1 = \frac{c\, r_H}{4} \int_0^{1}\dd u\; u^{R_1 - 1} (1 - u)^{R_2 - 1} \left[\frac{1-u}{\mu_1} + \frac{u}{\mu_2}\right] \sqrt{\frac{\mu_1 + \mu_2}{u \mu_1 + (1 - u) \mu_2}} \;,
\end{equation}
and 
\begin{align} \label{A2}
B_2 &= \frac{c\, r_H}{4} \int_0^{1}\dd u\; u^{R_1 - 1} (1 - u)^{R_2 - 1} \left[ \frac{1 - u}{\mu_1} + \frac{u}{\mu_2} \right] \frac{\mu_1 + \mu_2}{u \mu_1 + (1 - u) \mu_2} \\
&= \frac{r_H}{\mu_H} \frac{r_2 \mu_1 + r_1 \mu_2}{r_1 r_2}\;.
\end{align}

\vspace{0.2cm}

\begin{figure*}
\centering
\includegraphics[width=0.32\textwidth]{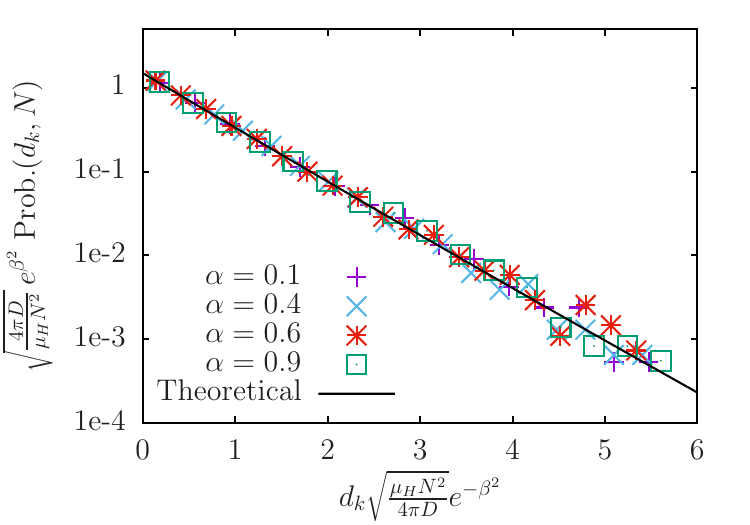} 
\hfill
\includegraphics[width=0.32\textwidth]{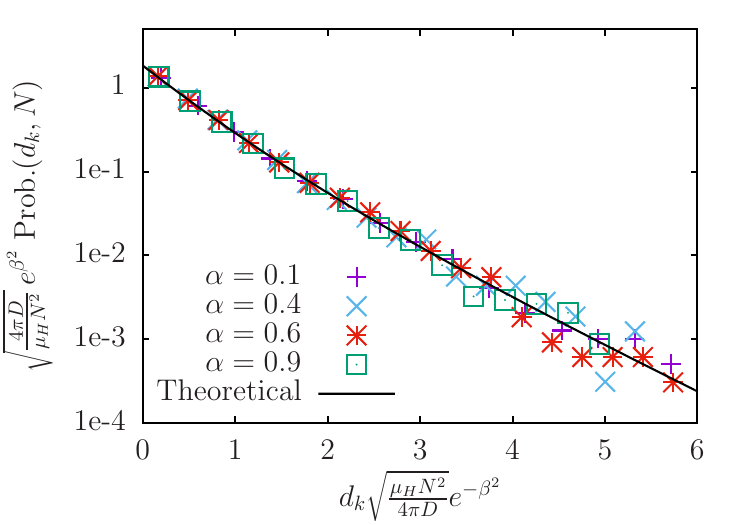} 
\hfill
\includegraphics[width=0.32\textwidth]{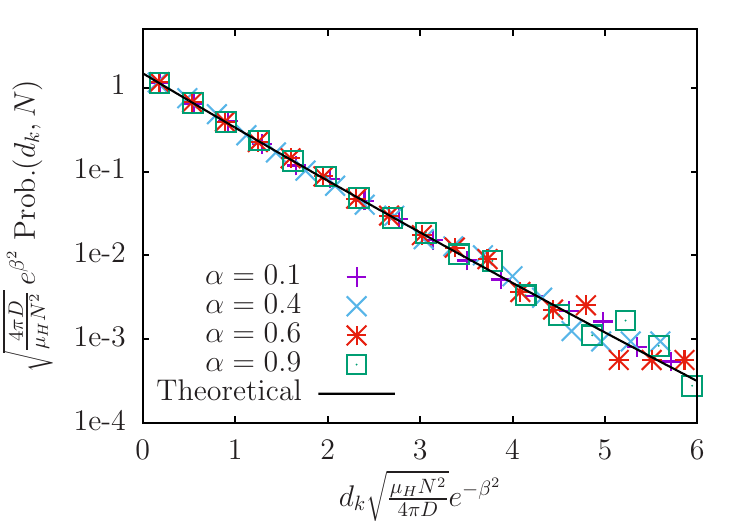} 
\caption{Scaling collapse of the distribution of the $k$-th gap as in Eq.~(\ref{dk}) for different values of $\alpha = k/N$ and different values of the parameters. We set $r_1 = r_2 = 1$, $D=1$, $N = 10^{6}$ and vary $\mu_1$ and $\mu_2$. From left to right we used respectively $\mu_1 = 0.4, \, \mu_2 = 0.2$ then $\mu_1 = 2, \, \mu_2 = 0.4$ and finally $\mu_1 = 2,\, \mu_2 = 1$. The dashed black line corresponds to the theoretical prediction given in Eq. (\ref{G}) and the symbols are the numerical results. Different colors correspond to different values of $\alpha$. The numerical results were obtained by sampling $10^{5}$ examples directly from the NESS distribution given in Eq. (\ref{jpdf}).} \label{fig:gap}
\end{figure*}

\noindent{\it The limit $z \to \infty$.} Since we have set $\mu_1 > \mu_2$, we find that the function 
\begin{equation} \label{def_phi2}
\phi(u) = \sqrt{\frac{\mu_1 + \mu_2}{u \mu_1 + (1 - u)\mu_2}} 
\end{equation}
that appears inside the argument of the exponential in Eq. (\ref{G}) is a monotonically decreasing function of $u$ for $u \in [0,1]$. Consequently, the dominant contribution to the integral for large $z$ comes from the vicinity of $u=1$. Expanding $\phi(u)$ near $u=1$, we get for large $z$
\begin{equation}
F(z) \approx \frac{c\, r_H}{4} \,e^{-z \phi(1)}\, \phi(1) \, \int_0^1 \dd u \; u^{R_1 - 1}(1 - u)^{R_2 -1} \left[ \frac{1 - u}{\mu_1} + \frac{u}{\mu_2} \right]  e^{- z (u-1)\phi'(1)} \;.
\end{equation}
Using Eq. (\ref{def_phi2}) gives 
\begin{align}
F(z) \approx &\frac{c\, r_H}{4} \sqrt{\frac{\mu_1+\mu_2}{\mu_1}}e^{-z \sqrt{\frac{\mu_1 + \mu_2}{\mu_1}}} \\
&\int_0^1 \dd u \; u^{R_1 - 1} (1 - u)^{R_2 - 1} \left[ \frac{1 - u}{\mu_1} + \frac{u}{\mu_2} \right] e^{-z (1 - u) \frac{\sqrt{\mu_1 + \mu_2}}{2\mu_1^{3/2}}(\mu_1 - \mu_2) } \;.
\end{align}
Changing variable to $v = (1 - u) z$ we get, for large $z$,
\begin{equation}
F(z) \approx \frac{c \, r_H}{4} z^{-R_2} \frac{1}{\mu_2} \sqrt{\frac{\mu_1 + \mu_2}{\mu_1}} e^{-z\sqrt{\frac{\mu_1 + \mu_2}{\mu_1}}} \int_0^{+\infty} \dd v \; v^{R_2 - 1} e^{-v \frac{\sqrt{\mu_1 + \mu_2}}{2 \mu_1^{3/2}} (\mu_1 - \mu_2)} \;.
\end{equation}
This integral over $v$ can be done explicitly, leading to 
\begin{equation} \label{G-large-z}
F(z) \approx B \, \frac{e^{-z \sqrt{\frac{\mu_1 + \mu_2}{\mu_1}}}}{z^{R_2}} \;,
\end{equation}
where 
\begin{equation}
    B = \frac{\Gamma(1 + R_1 + R_2)}{\Gamma(1 + R_1)} \frac{r_H}{r_2} \left(\frac{\mu_1}{\mu_1 - \mu_2}\right)^{R_2} \left(\sqrt{\frac{4 \mu_1}{\mu_1 + \mu_2}}\right)^{R_2 - 1}  \;.
\end{equation}
To summarize, the {\color{blue}asymptotes} of $F(z)$ are given by
\begin{equation} \label{F-tails}
F(z) \longrightarrow \begin{cases}
B_1 - B_2 z \;  &\mbox{~~for~~} z \ll 1\\
B \, z^{-R_2} e^{-z \sqrt{\frac{\mu_1 + \mu_2}{\mu_1}}}  &\mbox{~~for~~} z \gg 1
\end{cases} \;.
\end{equation}
In Fig. \ref{fig:gap}, we compare this analytical scaling function $F(z)$ in Eq. (\ref{G}) with numerical simulations, showing an excellent agreement. 

\section{Full counting statistics}

\begin{figure*}
\centering
\includegraphics[width=0.32\textwidth]{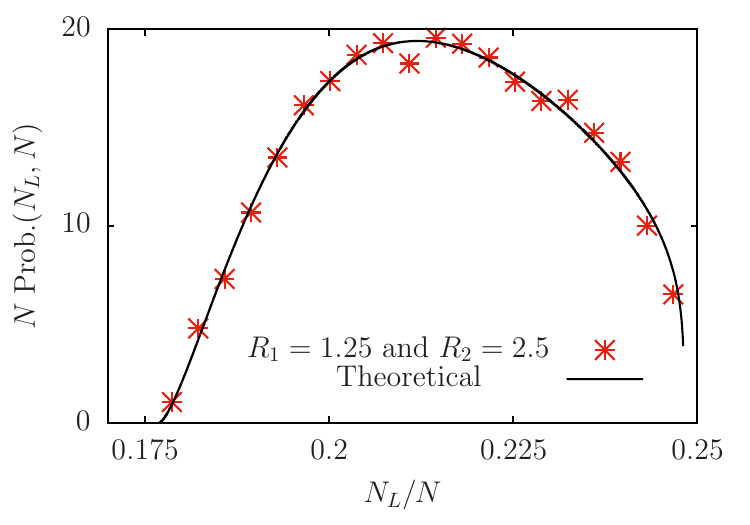} 
\hfill
\includegraphics[width=0.32\textwidth]{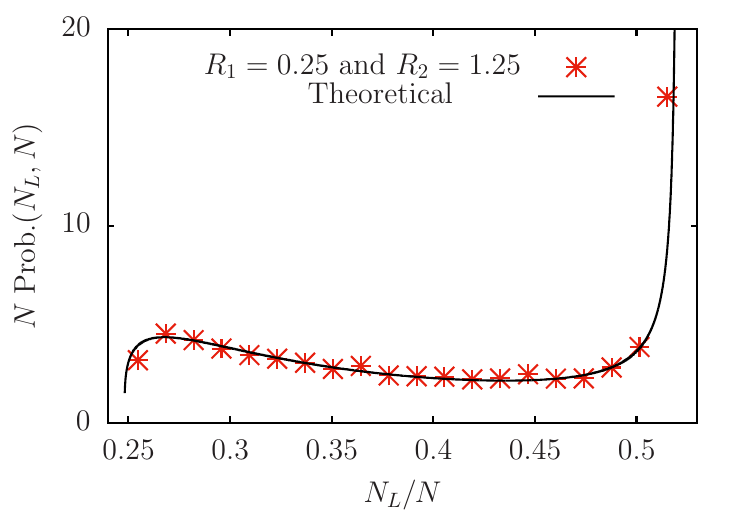} 
\hfill
\includegraphics[width=0.32\textwidth]{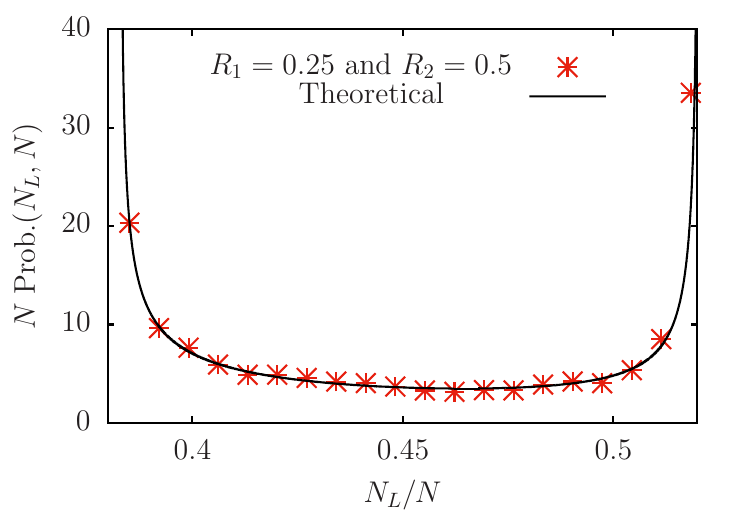} 
\caption{Scaling collapse of the distribution of the number of particles $N_L$ in $[-L,L]$ as in Eq.~(\ref{NL}) for different values of the parameters $R_1$ and $R_2$. We set $L=0.5, r_1 = r_2 = 1$, $D=1$, $N = 10^{6}$ and vary $\mu_1$ and $\mu_2$. From left to right we used respectively $\mu_1 = 0.4, \, \mu_2 = 0.2$ then $\mu_1 = 2, \, \mu_2 = 0.4$ and finally $\mu_1 = 2,\, \mu_2 = 1$. The dashed black line corresponds to the theoretical prediction given in Eq. (\ref{H}) and the symbols are the numerical results. The numerical results were obtained by sampling $10^{5}$ examples directly from the NESS distribution given in Eq. (\ref{jpdf}).} \label{fig_fcs}
\end{figure*}

Finally, we compute the full counting statistics, that is, the distribution of the number $N_L$ of particles inside the interval $[-L, L]$ around the origin. Exploiting again the conditionally independent structure of the joint probability density function in Eq.~(\ref{jpdf}) and applying the general results derived in Chapter~\ref{ch:ciid} we get for the probability distribution of $N_L$
\begin{equation}
{\rm Prob.}[N_L = n] \underset{N\to\infty}{\longrightarrow} \frac{1}{N} \int_0^1 \dd u\; h(u) \; \delta\left[ \frac{n}{N} - \int_{-L}^L \dd x \; p_0(x, u) \right] \;.
\end{equation}
where $h(u)$ and $p_0(x, u)$ are given in Eq. (\ref{h_u_sm}) and Eq.~(\ref{eq:p0-OU}) respectively. Using Eqs. (\ref{h_u_sm}-\ref{eq:p0-OU}) we can express the distribution in a scaling form as
\begin{equation} \label{NL}
{\rm Prob.}[N_L = n] \underset{N\to\infty}{\longrightarrow} \frac{1}{N} H\left( \frac{n}{N} \right) \;,
\end{equation}
where the scaling function $H(z)$ is supported from ${\rm erf}(\sqrt{\gamma/R_{H, 2}})$ all the way to ${\rm erf}(\sqrt{\gamma/R_{H, 1}})$. Here we have denoted 
\begin{equation} \label{gamma}
\gamma= \frac{r_H L^2}{4 D} \;.
\end{equation}
The scaling function $H(z)$ is given explicitly by
\begin{align}
H(z) = &\frac{c \, \, \gamma^2}{(R_{H, 2} - R_{H, 1})^{R_1 + R_2 - 1}} \frac{\sqrt{\pi}}{2} e^{u(z)^2} \frac{1}{u(z)^5} \\
&\quad \times \left( \frac{\gamma}{u(z)^2} - R_{H, 1} \right)^{R_1 - 1} \left(R_{H, 2} - \frac{\gamma}{u(z)^2}\right)^{R_2 - 1} \;, \label{H}
\end{align}
where {$u(z) = {\rm erf}^{-1}(z)$ is} the inverse error function. In contrast to the results obtained in Section \ref{subsec:BM-simreset}, we can see that for this system the full counting statistics have a richer variety of behaviors with a finite support contained in $[0, 1]$ and possible divergences at the edges of the support. 
We can check that this scaling function $H(z)$ is normalized to unity on its support ${\rm erf}(\sqrt{\gamma/R_{H, 2}}) < z < {\rm erf}(\sqrt{\gamma/R_{H, 1}})$. In Fig. \ref{fig_fcs}, we compare this analytical scaling function $H(z)$ in Eq. (\ref{H}) with the numerically obtained scaling function, and find an excellent agreement. 

\vspace{0.2cm}

We remark on an interesting fact. Since $N_L \in [0, N]$, the scaling variable $z=N_L/N$ has an allowed range $z \in [0,1]$. However, we find that, in the limit $N \to \infty$, the scaling function $H(z)$ is supported over a smaller interval $z \in [{\rm erf}(\sqrt{\gamma/R_{H, 2}}), {\rm erf}(\sqrt{\gamma/R_{H, 1}})] \subset [0,1]$. Hence, the probability of having $N_L < N {\rm erf}(\sqrt{\gamma/R_{H, 2}})$ or $N_L > N {\rm erf}(\sqrt{\gamma/R_{H, 1}})$ is vanishingly small in the large $N$ limit. It would be interesting to investigate the leading large $N$ behavior of this vanishing probability outside this shorter range. 

\section{Kesten variables approach to reveal the magic behind the conditional independence} \label{sec:kesten}

The conditionally independent form in Eq.~(\ref{jpdf}) is nice because it allows us to compute all our desired observables, but what is its physical {\color{blue}interpretation}? To understand the origin of this conditional {\color{blue}form} it is instructive to study the form of $p_0(x, \tau)$ in Eq.~(\ref{eq:p0-OU}). Notice that $p_0(x, \tau)$ is a simple gaussian with variance $V(u)$ given in Eq.~(\ref{def_V_of_u}) which we remind for convenience
\begin{equation}\label{def_V_of_u_2}
    V(u) = D\,\left( \frac{u}{\mu_2} + \frac{1 - u}{\mu_1} \right) \;.
\end{equation}
Secondly, from Eq.~(\ref{jpdf}) we know that this hidden latent variable $u$ belongs to $[0, 1]$ and dimensional analysis also quickly reveals $u$ to be adimensional. Let's study the edge behaviors of the variance
\begin{equation} \label{eq:edge-V-of-u}
    V(u = 0) = \frac{D}{\mu_1} \mbox{~~and~~} V(u = 1) = \frac{D}{\mu_2} \;.
\end{equation}
From Eq.~(\ref{eq:edge-V-of-u}) we recognize that setting $u = 0$ recovers the steady state of the \OU process with potential~$\mu_1$ (see Eq.~(\ref{eq:OU-propagator})), respectively $u = 1$ leads to the steady state of the \OU process with potential~$\mu_2$. From these observations we postulate that $u$ represents a the fraction of time that the system spends in the steady state of phase 2 (hence $1 - u$ is the fraction of time in phase 1). This {\color{blue}interpretation} explains all the behaviors above: $u$ being in $[0, 1]$, $u$ being adimensional and the edge behavior of the variance in Eq.~(\ref{eq:edge-V-of-u}). Amazingly, an alternative approach based on Kesten allows us to prove that this intuition is correct.

\vspace{0.2cm}

As mentioned in Section~\ref{sec:model-switching-OU}, we are considering $N$ particles at positions $\vec{X} = X_1, \cdots, X_N$ which diffuse independently within a potential {that} switches from $V_1(x) = \frac{1}{2} \mu_1 x^2$ and $V_2(x) = \frac{1}{2} \mu_2 x^2$ with Poissonian rates $r_1$ and $r_2$, i.e. with rate $r_1$ the system will switch from $V_1(x)$ to $V_2(x)$ and respectively with rate $r_2$ it will switch back from $V_2(x)$ to $V_1(x)$, see Fig.~\ref{fig:switching-potential}. Hence, the duration $\tau$ of the time intervals {between} successive switches is distributed as
\begin{equation} \label{p-tau}
{\rm Prob.}[\tau] = r_i e^{- r_i \tau} \;,
\end{equation}
where $r_i$ is $r_1$ or $r_2$ according to which potential is on during the interval. {Furthermore, all intervals are distributed independently from each other. We denote by $\{\tau_i\}$ all the successive intervals}. Without loss of generality assume that $V_1(x)$ {is} on during the odd intervals $\{\tau_1, \tau_3, \cdots\}$ and respectively $V_2(x)$ {is} on during the even intervals $\{\tau_2, \tau_4, \cdots\}$. During the odd intervals the equation of motion is
\begin{equation} \label{dynamic1}
\dv{X_i}{\tau} = -\mu_1 X_i + \sqrt{2 D} \eta_i(\tau) \;,
\end{equation}
and respectively during the even intervals the equation of motion is
\begin{equation} \label{dynamic2}
\dv{X_i}{\tau} = -\mu_2 X_i + \sqrt{2 D} \eta_i(\tau) \;,
\end{equation}
where $\eta_i(\tau)$ is a Gaussian white noise such that
\begin{equation} \label{eta}
\langle \eta_i(\tau) \rangle = 0 \mbox{~~and~~} \langle \eta_i(\tau) \eta_j(\tau') \rangle = \delta(\tau - \tau')\delta_{ij} \;.
\end{equation}
We saw in Section~\ref{sec:model-switching-OU} -- see Eq.~(\ref{spherical-fourier-fp-p1}) -- that making use of the spherical symmetry the problem reduces to a one-dimensional problem on 
\begin{equation}
    x = \sqrt{\sum_{i = 1}^N x_i^2} \;.
\end{equation}
For simplicity, we will restrict ourselves below to the one-particle case, but the multi-particle case follows the same derivation.

\vspace{0.2cm}

For a fixed choice of random intervals $\{\tau_i\}$ the process $X(\tau)$ is a Gaussian process since {Eqs. (\ref{dynamic1}) and (\ref{dynamic2})} are linear evolution equations. Hence the probability distribution ${\rm Prob.}[X(\tau) = x | \{\tau_i\}] = p(x, \tau | \{\tau_i\})$ of the system being at $x$ at time $\tau$ knowing the random intervals $\{\tau_i\}$ is given by
\begin{equation} \label{p-x-tau}
p(x, \tau | \{\tau_i\} ) = \frac{1}{\sqrt{2 \pi \; V(\tau, \{\tau_i\})}} \exp( - \frac{x^2}{2 \; V(\tau, \{\tau_i\})}) \;,
\end{equation}
where $V(\tau, \{\tau_i\})$ is the variance at time $\tau$ given the $\{\tau_i\}$'s. In the long time limit {the system reaches a steady state}. Hence when $\tau \to +\infty$ then $V(\tau, \{\tau_i\}) \to V(\{\tau_i\})$. Therefore the distribution $p(x | \{\tau_i\})$ in the steady state is given by
\begin{equation} \label{p-x}
p(x | \{\tau_i\} ) = \frac{1}{\sqrt{2 \pi\; V(\{\tau_i\})}} \exp( - \frac{x^2}{2 \; V(\{\tau_i\})}) \;.
\end{equation}
If we now average over all possible realizations of the $\{\tau_i\}$'s it will give us the stationary distribution $p^{\rm st}(x)$,
\begin{equation} \label{p-st}
p^{\rm st}(x) = \int_0^{+\infty} \dd V \; {\rm Prob.}[V] \frac{1}{\sqrt{2 \pi  V}} \exp( - \frac{x^2}{2  V})\;,
\end{equation}
{where $V$ stands for $V(\{\tau_i\})$ averaged over the interval lengths $\tau_i$'s.} We recognize here a form similar to Eq. (\ref{eq:p0-OU}). The goal now is to find the distribution ${\rm Prob.}[V]$ which is induced by the random variables $\{\tau_1, \tau_2, \cdots\}$, this should allow us to recover Eq. (\ref{jpdf}). To do so we will proceed recursively. {Let $X_n$ denote the position of the particle $X(\tau)$ 
at the end of the $n$-th interval, i.e., $X_n= X\left(\tau= \sum_{i=1}^n \tau_i\right)$.}
Furthermore, we denote by $V_n = \langle x_n^2 \rangle$ the variance at the end of the $n$-th interval. If $n$ is odd, then the potential $V_1(x)$ was on during that interval. Hence 
\begin{equation} \label{xodd-xeven}
X_{2n + 1} = X_{2n} e^{-\mu_1 \tau_{2n + 1}} + \sqrt{2 D} \; e^{-\mu_1 \tau_{2n + 1}} \int_0^{\tau_{2n + 1}} \eta(\tau) e^{\mu_1 \tau} \dd \tau \;,
\end{equation} 
and consequently
\begin{equation} \label{Vodd-Veven}
V_{2n + 1} = V_{2n} e^{-2 \mu_1 \tau_{2n + 1}} + \frac{D}{\mu_1}\left(1 - e^{-2 \mu_1 \tau_{2n + 1}}\right)
\end{equation}
Inversely, if $n$ is even then the potential $V_2(x)$ was on during that interval. Hence
\begin{equation} \label{xeven-xodd}
X_{2n} = X_{2n-1} e^{-\mu_2 \tau_{2n}} + \sqrt{2 D} \; e^{-\mu_2 \tau_{2n}} \int_0^{\tau_{2n}} \eta(\tau) e^{\mu_2 \tau} \dd \tau \;,
\end{equation}
and consequently
\begin{equation} \label{Veven-Vodd}
V_{2n} = V_{2n-1} e^{-2 \mu_2 \tau_{2n}} + \frac{D}{\mu_2}\left(1 - e^{-2 \mu_2 \tau_{2n}}\right) \;.
\end{equation}
Here $\tau_n$'s are random variables drawn from two exponential distributions{\color{blue}, either $r_1 e^{-r_1 \tau}$ or  $r_2 e^{-r_2 \tau}$} alternatively. Thus the recursion relations satisfied by the $V_n$'s involve random variables. Such linear recursion relations with random coefficients are known as ``Kesten recursion relations'' and they have been studied in different contexts such as in probability theory and disordered systems~\cite{K73,KKS75,DH83,KS84,CLNP85,G91,BDMZ13,GBL21,GMS23}.   

\vspace{0.2cm}

From (\ref{Vodd-Veven}) and (\ref{Veven-Vodd}), we see that $V_n$ fluctuates between $D/\mu_1$ and $D/\mu_2$ in the large $n$ limit. It is then convenient to define {\color{blue} the Kesten} variable 
\begin{equation} \label{def_un}
u_n = \frac{V_n - \frac{D}{\mu_1}}{\frac{D}{\mu_2} - \frac{D}{\mu_1}} \;.
\end{equation}
Therefore $u_n$ lies in the interval $[0,1]$ in the $n \to \infty$ {limit} (we recall that $\mu_1>\mu_2$).  Applying the reparametrization in Eq. (\ref{def_un}) to Eqs. (\ref{Vodd-Veven}) and (\ref{Veven-Vodd}) we get the recursion relations
\begin{equation} \label{t-rec}
u_{2n + 1} = u_{2n} e^{-2 \mu_1 \tau_{2n + 1}} \mbox{~~and~~} 1 - u_{2n} = (1 - u_{2n - 1}) e^{-2 \mu_2 \tau_n} \;.
\end{equation} 
It is convenient to define $z_n = e^{-2 \mu_i \tau_n}$ where $\mu_i = \mu_1$ if $n$ is odd and $\mu_i = \mu_2$ otherwise. From Eq. (\ref{p-tau}) we get
\begin{equation} \label{z}
\begin{cases} {\rm Prob.}[z_{2n + 1} = z] &= R_1 z^{R_1 - 1} \\
{\rm Prob.}[z_n = z] &= R_2 z^{R_2 - 1}
\end{cases}
\;, \qquad {\rm with} \; 0 \leq z \leq 1 \;,
\end{equation}
where $R_i = r_i/(2 \mu_i)$. In the $n \to +\infty$ limit we expect to reach a steady state. Hence we expect that $u_{2n + 1} \to u_{\rm odd}$ and $u_{2n} \to u_{\rm even}$ as $n \to +\infty$, and from Eqs. (\ref{t-rec}) and (\ref{z}) we know that
\begin{equation} \label{t-st-coupled}
u_{\rm odd} = u_{\rm even} z_1 \mbox{~~and~~} 1 - u_{\rm even} = (1 - u_{\rm odd}) z_2 \;.
\end{equation}
Let $P_{\rm even}(u)$ and $P_{\rm odd}(u)$ denote respectively the stationary distribution of $u_{\rm even}$ and $u_{\rm odd}$ in the limit $n \to \infty$. Then, from Eq. (\ref{t-st-coupled}) we have 
\begin{align}
P_{\rm odd}(u) &= \int_0^1 \dd u' \int_0^1 \dd z \; P_{\rm even}(u') R_1 z^{R_1 - 1} \delta(u - u' z) \\
&= \int_u^1 \dd z\; P_{\rm even}\left(\frac{u}{z}\right) R_1 z^{R_1 - 1} \;.
\end{align}
Making a change of variable to $y = u/z$ we get
\begin{equation}
P_{\rm odd}(u) = R_1 u^{R_1 - 1} \int_u^{1} \frac{P_{\rm even}(y)}{y^{R_1}} \dd y\;.
\end{equation}
Taking a derivative with respect to $u$, one gets
\begin{equation} \label{ode-odd}
\dv{}{u} \left[ \frac{1}{u^{R_1 - 1}} P_{\rm odd}(u) \right] = - \frac{R_1}{u^{R_1}} P_{\rm even}(u) \;.
\end{equation}
A similar derivation for $P_{\rm even}(u)$ leads to
\begin{equation} \label{ode-even}
\dv{}{(1 - u)} \left[ \frac{1}{(1 - u)^{R_2 -1}} P_{\rm even}(1 - u) \right] = \frac{-R_2}{(1 - u)^{R_2}} P_{\rm odd}(1 - u) \;.
\end{equation}
Given the ODEs in Eqs. (\ref{ode-odd}) and (\ref{ode-even}) one can easily check that the solutions are given by 
\begin{equation} \label{eq:Podd-Peven-sol}
P_{\rm odd}(u) = {c\, R_1 \, u^{R_1 - 1} (1 - u)^{R_2} }\mbox{~~and~~} P_{\rm even}(u) = {c \, R_2 \, u^{R_1} (1 - u)^{R_2 - 1}} \;,
\end{equation}
where $c$ is an arbitrary constant, yet to be fixed. To fix this constant, we proceed as follows. From Eq. (\ref{bc_Fourier}), we know that, in the stationary state, the potential itself is in phase $1$ (with stiffness $\mu_1$) with probability $r_2/(r_1+r_2)$ and is in phase $2$ (with stiffness $\mu_2$) with the complementary probability $r_1/(r_1+r_2)$. Therefore, the probability density function $P_{\rm odd}(u)$ will occur with probability $r_2/(r_1+r_2)$ and $P_{\rm even}(u)$ will occur with probability $r_1/(r_1+r_2)$. Hence the full probability density function of the random variable $u$ in the stationary state is given by
\begin{equation} 
h(u) = \frac{r_2 P_{\rm odd}(u) + r_1 P_{\rm even}(u)}{r_1 + r_2}  \;,
\end{equation}
replacing with the forms found in Eq.~(\ref{eq:Podd-Peven-sol}) we obtain
\begin{equation}\label{ht-prelim}
    h(u) = \frac{{c\,R_1\, R_2}}{r_1 + r_2} u^{R_1-1} (1 - u)^{R_2-1} \left[ \frac{1 - u}{R_2} r_1 + \frac{ u }{R_1} r_2 \right] \;.
\end{equation}
The normalization condition $\int_0^{1} \dd u \; h(u) = 1$ then fixes the constant $c$ to be
\begin{equation} \label{c}
{c = \frac{\Gamma(R_1 + R_2 + 1)}{\Gamma(R_1+1) \Gamma(R_2+1)}} \;.
\end{equation}
Placing Eq. (\ref{c}) back in Eq. (\ref{ht-prelim}) we obtain
\begin{equation} \label{h}
h(u) = \frac{r_1 r_2}{2(r_1 + r_2)} \frac{\Gamma(R_1 + R_2 + 1)}{\Gamma(R_1 + 1) \Gamma(R_2 + 1)} \; u^{R_1 - 1} (1 - u)^{R_2 - 1} \left[\frac{u}{\mu_2} +\frac{1 - u}{\mu_1}\right] \;,
\end{equation}
recovering exactly Eq. (\ref{h_u_sm}). This Kesten approach thus shows clearly that the random variable $u$ has the physical interpretation of the fraction of time the particle spends in phase 2.

\section{Generalization to negative stiffness}

Although we have assumed that $\mu_1 > \mu_2 > 0$ we could, in theory, consider $\mu_1 > 0 > \mu_2$ and $0 > \mu_1 > \mu_2$. When both $\mu_1$ and $\mu_2$ are negative the system will obviously very quickly diverge and will never admit a steady state. However, when $\mu_1 > 0 > \mu_2$ it may be possible to reach a steady state. Physically, the system with $\mu_2 < 0$ is completely different to the $\mu_2 > 0$ case. Indeed, when $\mu_2 > 0$ the particle is always confined around the origin with varying stiffnesses whereas when $\mu_2 < 0$ the particle alternates between being propelled exponentially fast far from the origin and then being brought back exponentially fast close to the origin. The solution derived in Eq.~(\ref{M-U-res}) is valid for any value $\mu_1, \mu_2$ even if $\mu_2 < 0$. From now on we will set $\mu_2 = -|\mu_2| < 0$. To set the constants $A_1$ and $B_1$ in Eq. (\ref{M-U-res}) we need to look at its {\color{blue}asymptotes}. For large $z$, $M(a, b, z) \sim \Gamma(b)/\Gamma(a) \; z^{a-b} e^z$ hence the first term in Eq. (\ref{M-U-res}) behaves as
\begin{equation} \label{eq:large-k-M}
    A_1 M\left( R_1, 1 + R_1 + |R_2|, - \frac{D k^2 (\mu_1 + |\mu_2|)}{2 \mu_1 |\mu_2|} \right) \sim A_1 k^{\frac{r}{|\mu_2|} - 2} e^{\frac{D k^2}{2 |\mu_2|}} 
\end{equation}
when $|k| \gg 1$. Eq. (\ref{eq:large-k-M}) is clearly divergent for large $k$, which forces us to set $A_1 = 0$. Hence the general solution when $\mu_2 < 0$ is 
\begin{equation}
    \hat{p}_1(k) = A_2 e^{- \frac{D k^2}{2 \mu_1}} U\left(R_1, 1 + R_1 - |R_2|, - \frac{D k^2 (\mu_1 + |\mu_2|)}{2 \mu_1 |\mu_2|}\right) \;. \label{eq:general-solution-U}
\end{equation}
To fix the constant $A_2$ we have to look at the $k \to 0$ limit and use the normalization conditions in Eqs. (\ref{bc_Fourier}). The small $z$ {\color{blue}asymptotes} of $U(a, b, z)$ depend on whether $b < 1$, $b = 1$ or $b > 1$. If $b \geq 1$ then $U(a, b, z)$ is divergent when $z \to 0^+$, which is incompatible with Eq. (\ref{bc_Fourier}). Hence for a steady state to exist we must have
\begin{equation} \label{eq:ss-condition}
    |R_2| > R_1 \Leftrightarrow \frac{|\mu_2|}{\mu_1} < \frac{r_2}{r_1} \;.
\end{equation}
From now on we place ourselves in the case were there exists a steady state, i.e. we suppose that Eq. (\ref{eq:ss-condition}) holds. The small $z$ asymptotic of $U(a, b, z)$ when $b < 1$ is $U(a, b, z) \sim \Gamma(1 -b) / \Gamma(1 + a - b)$. Hence 
\begin{equation} \label{eq:small-k-P1}
    \hat{p}_1(k) \stackrel{k \to 0}{\longrightarrow} A_2 \frac{\Gamma(|R_2| - R_1)}{\Gamma(|R_2|)} \;.
\end{equation}
Using Eq. (\ref{bc_Fourier}) we can set
\begin{equation} \label{eq:A2}
    A_2 = \frac{r_2}{r_1 + r_2} \frac{\Gamma(|R_2|)}{\Gamma(|R_2| - R_1)} \;.
\end{equation}
We can obtain the expression of $\hat{p}_2(k)$ by using Eq. (\ref{eq:general-solution-U}) in Eq. (\ref{spherical-fourier-fp-p2}). We can further simplify the expressions of $\hat{p}_1(k)$ and $\hat{p}_2(k)$ by using the integral representation of Tricomi's function
\begin{equation} \label{eq:U-integral-form}
    U(a, b, z) = \frac{1}{\Gamma(a)} \int_0^{+\infty} e^{-z t} t^{a - 1} (1 + t)^{b - a - 1} \dd t \;,
\end{equation}
which is valid for $a > 0$. Doing so allows us to re-write $\hat{p}(k) = \hat{p}_1(k) + \hat{p}_2(k)$ as 
\begin{align}
    \hat{p}(k) = &\frac{r_H}{4} \frac{\Gamma(|R_2|)}{\Gamma(R_1 + 1) \Gamma(|R_2| - R_1)} \nonumber\\
    &\quad \times \int_0^{+\infty} \dd t \; \frac{t^{R_1 - 1}}{(1 + t)^{|R_2| + 1}} \left[ \frac{t}{|\mu_2|} + \frac{1 + t}{\mu_1} \right] e^{-k^2 / (2 V_-(t))}  \;, \label{eq:FT-solution}
\end{align}
where $r_H = 2 r_1 r_2 / (r_1 + r_2)$ is the harmonic mean of the switching rates and 
\begin{equation} \label{eq:def-V}
    V_-(t) = D \left[ \frac{t}{|\mu_2|} + \frac{1 + t}{\mu_1} \right] \;,
\end{equation}
is a variance which appears naturally in the problem. For compactness we will denote the fraction of $\Gamma$-functions by
\begin{equation} \label{eq:def-c-gamma}
    c_\Gamma = \frac{\Gamma(|R_2|)}{\Gamma(R_1 + 1) \Gamma(|R_2| - R_1)} 
\end{equation}
from now on. The form in Eq. (\ref{eq:FT-solution}) can be inverted back from Fourier space easily because it is simply as mixture of Gaussian distributions. Therefore the general solution for $\mu_2 < 0$ is given by
\begin{align} \label{eq:real-N-solution}
    {\rm Prob.}[\vec{X} = \vec{x}] = \frac{c_\Gamma r_H}{4} \int_0^{+\infty} \dd t \; \frac{t^{R_1 - 1}}{(1 + t)^{|R_2| + 1}} \left[ \frac{t}{|\mu_2|} + \frac{1 + t}{\mu_1} \right] \prod_{i = 1}^N \frac{e^{-x_i^2 / (2 V_-(t))}}{\sqrt{2 \pi V_-(t)}}  \;.
\end{align}
The joint probability density function in Eq. (\ref{eq:real-N-solution}) has the form of conditionally independent identically distributed random variables which can be made more explicit by re-writing Eq. (\ref{eq:real-N-solution}) as
\begin{equation} \label{eq:ciid-structure}
    {\rm Prob.}[\vec{X} = \vec{x}] = \int_0^{+\infty} \dd t \; h(t) \prod_{i = 1}^N p_0(x_i | t) \;,
\end{equation}
where the distributions $p_0(x | t)$ and $h(t)$ are given by
\begin{equation} \label{eq:p-def}
    p_0(x | t) = \frac{1}{\sqrt{2 \pi V_-(t)}} e^{-\frac{x^2}{2 V(t)}}
\end{equation}
and 
\begin{equation} \label{eq:h-def-OU}
    h(t) = \frac{c_\Gamma r_H}{4} \frac{t^{R_1 - 1}}{(1 + t)^{|R_2| + 1}} \left[\frac{t}{|\mu_2|} + \frac{1 + t}{\mu_1}  \right] \;.
\end{equation}

\subsection{The average density}

The average density can be obtained easily by exploiting the conditionally independent form and is given by
\begin{equation}
    \rho(x) = \int_0^{+\infty} \dd t \; h(t) p_0(x | t)\;. 
\end{equation}
Although we cannot compute $\rho(x)$ explicitly, we can look at its {\color{blue}asymptotes}. The small $x$ asymptotic is obtained by expanding the exponential to second order and gives
\begin{equation}
    \rho(x) \simeq 1 - B x^2 \mbox{~~when~~} |x| \ll 1 \;,
\end{equation}
where $B$ is a constant. For large $x$, the integral will be dominated by large values of $t$, specifically for $t \sim x^2$. Hence with the appropriate change of variable $t = x^2 s$ we can extract the asymptotic of $\rho(x)$ for large $x$ which is
\begin{align}
    \rho(x) \simeq &\frac{1}{x^{2 |R_2| - 2 R_1 + 1}} \times \nonumber \\
    &\frac{c_\Gamma \Gamma(\frac{1}{2} - R_1 + |R_2|)}{4} \sqrt{\frac{r_H}{\mu_H \pi D}} \left( \frac{4 D}{\mu_H} \right)^{\frac{1}{2} - R_1 + |R_2|} \;,
\end{align}
when $|x| \gg 1$ and $\mu_H = 2 \mu_1 |\mu_2| / (\mu_1 + |\mu_2|)$.

\subsection{Order and {\color{blue}extreme-value} statistics}

Similar calculations as those carried out for the $\mu_2 > 0$ case can be repeated for the $\mu_2 < 0$ case. We start with the study of the order statistics, i.e. the behavior of the sorted positions $M_1 \geq M_2 \geq \cdots \geq M_N$ of the particles. To do so, we use the general results derived in Chapter~\ref{ch:ciid}. We are going to study the distribution of $M_k$ for any $k = \alpha N$ in the large $N$ limit. The first step consists in computing the $\alpha$-quantiles $q(\alpha, t)$ which is given by
\begin{equation} \label{eq:q-result}
    q(\alpha, t) = \sqrt{2 V_-(t)}\;  {\rm erfc}^{-1}(2 \alpha) = \beta \sqrt{2 V_-(t)}\;,
\end{equation}
where ${\rm erfc}^{-1}(z)$ is the inverse error complementary function. Results in Eq. (\ref{eq:res_bulk_integral_form}) in Chapter~\ref{ch:ciid} tell us that
\begin{equation} \label{eq:Mk-int}
    {\rm Prob.}[M_k = w] \underset{N\to\infty}{\longrightarrow} \int_0^{+\infty} \dd t \; h(t) \delta(q(\alpha, t) - w) \;.
\end{equation}
Using Eq. (\ref{eq:q-result}) and Eq. (\ref{eq:h-def-OU}) we can compute this integral explicitly and re-write the density in a scaling form
\begin{equation} \label{eq:Mk-scaling}
    {\rm Prob.}[M_k = w] \underset{N\to\infty}{\longrightarrow} \sqrt{\frac{\mu_H}{4 D \beta^2}} f\left( w \sqrt{\frac{\mu_H}{4 D \beta^2}} \right) \;,
\end{equation}
where the normalized scaling function $f(z)$ defined for $z > \sqrt{\hat{\mu}_2}$ is given by
\begin{equation} \label{eq:def-f}
    f(z) = \frac{c_\Gamma r_H}{2} z^3 \frac{\left(z^2 - \hat{\mu}_2\right)^{R_1 - 1}}{\left(z^2 + \hat{\mu}_1\right)^{|R_2| + 1}} \;,
\end{equation}
where $\hat{\mu}_1 = \mu_1/(\mu_1 + |\mu_2|)$ and $\hat{\mu_2} = |\mu_2| / (\mu_1 + |\mu_2|)$. The asymptotic of $f(z)$ are
\begin{equation} \label{eq:f-asymp}
    f(z) \to \begin{dcases}
    \frac{c_\Gamma r_H}{\mu_H} \hat{\mu}_2^{3/2} \left(z^2 - \hat{\mu}_2  \right)^{R_1 - 1} &\mbox{~~for~~} (z - \hat{\mu}_2) \ll 1\\
    \frac{c_\Gamma r_H}{\mu_H} \frac{1}{z^{2|R_2| - 2 R_1 - 1}} &\mbox{~~for~~} z \gg 1 \;.
    \end{dcases}
\end{equation}
To obtain the behavior of the maximum, i.e. $M_1$, it suffices to take the asymptotic behavior of $\beta = {\rm erfc}^{-1}(2 \alpha) \simeq \sqrt{\ln N}$.

\subsection{Gap statistics}

Similarly, using Eq.~(\ref{eq:ciid-gaps-bulk}), the gap statistics i.e. the distribution of $d_k = M_k - M_{k+1} > 0$ for $k = \alpha N$ are given by
\begin{equation} \label{eq:dk-int}
    {\rm Prob.}[d_k = g] \underset{N\to\infty}{\longrightarrow} \int_0^{t} \dd t \; h(t) N p(q(\alpha, t) | t) e^{- N p(q(\alpha, t) | t) g} \;.
\end{equation}
Using Eq. (\ref{eq:h-def-OU}) and Eq. (\ref{eq:q-result}) we can rewrite Eq. (\ref{eq:dk-int}) in a scaling form
\begin{equation}
    {\rm Prob.}[d_k = g] \underset{N\to\infty}{\longrightarrow} N e^{-\beta^2} \sqrt{\frac{\mu_H}{4 \pi D}} \; F\left(N e^{-\beta^2} g \sqrt{\frac{\mu_H}{4 \pi D}}\right) \;,
\end{equation}
where the normalized scaling function $F(z)$ defined for $z > 0$ is given by
\begin{align}
    F(z) = \frac{c_\Gamma r_H}{4} \int_0^{+\infty} \dd t \; \frac{t^{R_1 - 1}}{(1 + t)^{|R_2| + 1}} \left[ \frac{t}{|\mu_2|} + \frac{1 + t}{\mu_1}  \right] \times \nonumber \\
    \sqrt{\frac{\mu_1 + |\mu_2|}{t \mu_1 + (1 + t)|\mu_2|}} \exp[ -z \sqrt{\frac{\mu_1 + |\mu_2|}{t \mu_1 + (1 + t)|\mu_2|}} ] \;. \label{eq:def-F}
\end{align}
Although we cannot compute the integral in Eq. (\ref{eq:def-F}) explicitly we can still obtain its {\color{blue}asymptotes}. For small $z$ we can expand the exponential and
\begin{equation}
    F(z) \simeq B_1 - z B_2 \mbox{~~when~~} z \ll 1 \;.
\end{equation}
For large $z$ the integral is dominated by large values of $t \sim z^2$ so making the change of variable $t = z^2 s$ we can obtain the asymptotic
\begin{equation}
    F(z) \simeq \frac{c_\Gamma r_H}{2} \left[ \frac{1}{|\mu_2|} + \frac{1}{\mu_1} \right] \frac{\Gamma(1 - 2 R_1 + 2 |R_2|)}{z^{2|R_2| - 2 R_1 + 1}} \;,
\end{equation}
when $z \gg 1$.

\chapter{{\color{blue}Logarithmically} repelling diffusive particles with simultaneous resetting \cite{BMS25}} \label{ch:dyson}
\section{A quick history of random matrix theory}

A random matrix is a matrix whose entries are filled randomly with a chosen probability distribution. Random matrix theory has a rich history dating back to the early 1900s in the mathematical literature \cite{W28}. A couple {\color{blue} of decades} later random matrix theory became extremely popular in physics because of its very successful modeling of the spectra of heavy nuclei \cite{M91}. According to quantum mechanics, the energy levels of a system are described by the eigenvalues of its Hamiltonian. After a choice of basis, the Hamiltonian can usually be expressed as a (possibly very large) matrix $H$. If we can solve the eigenvalue equation
\begin{equation}
    H \ket{\psi_i} = \lambda_i \ket{\psi_i} \;.
\end{equation} 
and get the eigenvalues and eigenvectors of the system we can usually obtain all the behavior of all typical physical observables from them. However, for heavy nuclei not only do we not know the explicit form of $H$, even if we did, it would be an {\color{blue}untractable} large matrix whose exact diagonalization would be computationally unrealistic. However, as statistical physicists, we know that when a very large system has a very complicated microscopic description, we might be able to obtain general results by abstracting the microscopic details to random noise. The idea is then simple; if we cannot describe $H$ we are going to assume that $H$ is random, only making sure that it respects the necessary symmetries pertinent to the problem we are trying to describe. The construction of a statistical theory of energy levels was the program set forth by Wigner and Dyson \cite{M91}. In Dyson's words: ``such theory will not predict the detailed sequence of levels in any one nucleus, but it will describe the general appearance and the degree of irregularity of the level structure that is expected to occur in any nucleus which is too complicated to be
understood in detail.''

\vspace{0.2cm}

Measurements of energy spectra were already available, obtained through scattering experiments, and some experimental statistical properties were puzzling, namely the so-called `level spacings'. When looking at a given energy $E$ of the spectrum and studying levels close to it, i.e. levels within $[E, E + \delta E]$ experimental data showed that the density of the levels is nearly constant in this interval and the fluctuations in the precise positions of the levels seem not to depend on the nucleus or the excitation energy. This might lead us to believe that energy levels are independently distributed random variables. Although this seemed to be true for levels with different spin, parity, or quantum numbers, levels with the same quantum numbers were clearly strongly correlated. The most obvious indication of such correlation was the fact that it is extremely rare to find levels close together; this is called the `level repulsion'.

\vspace{0.2cm}

\begin{figure}
    \centering
    \includegraphics[width=0.7\textwidth]{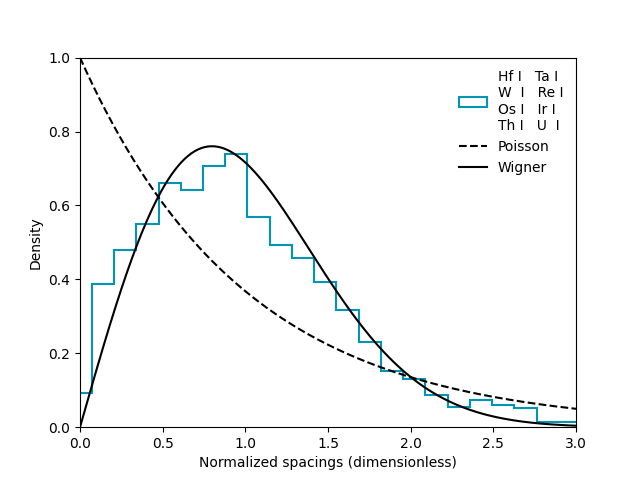}
    \caption{Experimental data (obtained from the Atomic Spectral Database \cite{ASD}) for the level spacings of Hafnium (Hf I), Tantalum (Ta I), Tungsten (W I), Rhenium (Re I), Osmium (Os I), Iridium (Ir I), Thorium (Th I) and Uranium (U I) compared to the independent hypothesis (Poisson) or the Wigner surmise for the gaussian orthogonal ensemble ($\beta = 1$). The spacings of each element were computed from the odd levels grouped by angular momentum. The energy levels were also `unfolded', meaning that they were locally renormalized in such a way that they would be uniformly distributed.} \label{fig:usual_atomic}
\end{figure}

\subsubsection{Wigner's surmise}

Although unconventional, for consistency with the rest of the text, let $M_1 \geq M_2 \geq \cdots \geq M_N$ denote the different ordered energy levels of the spectra and let $d_k = M_k - M_{k + 1}$ be the spacings. The experimental observations suggest that the distribution of $d_k$ does not depend on $k$ hence we generally consider the dimensionless relative spacings $s_k = d_k / S$ where 
\begin{equation}
    S = \frac{1}{N - 1} \sum_{k = 1}^{N - 1} d_k \;,
\end{equation}
is the average spacing. The assumption is then that the relative spacings $s_i$ are different samples of the same underlying spacing distribution $p(s)$, and the goal is to characterize $p(s)$. If the energy levels were independently distributed from one another then they would be described by a Poisson point process and hence we would obtain
\begin{equation}
    p(s) = p_{\rm poisson}(s) = e^{-s} \;.
\end{equation}
However, this is incompatible with the experimental results which showed that energy levels with the same spin, parity, and angular momenta are almost never close to each other, as can be seen in Fig. \ref{fig:usual_atomic}.

\vspace{0.2cm}

Instead of assuming that the levels are independent, we assume that they are the eigenvalues of a random Hamiltonian $H $ filled with Gaussian values. The general study of these random matrices is the object of random-matrix theory. However, to get us started we will follow in Wigner's footsteps and compute exactly the statistics of the gap $d = |x_1 - x_2|$ for a 2 by 2 matrix and conjecture that it will be a good approximation for the mean level spacing for matrices of any size. Numerical results have shown that this is indeed the case~\cite{M91}. We consider a 2 by 2 matrix
\begin{equation} \label{eq:def-2x2-X}
    X = \begin{pmatrix}
    y_1 & y_3 \\
    y_3 & y_2
    \end{pmatrix} \;,
\end{equation}
such that $y_1, y_2$ are Gaussian random variables with variance $\sigma^2/2$ and $y_3$ is a Gaussian random variable with variance $\sigma^2$. The eigenvalues of $X$, defined in Eq. (\ref{eq:def-2x2-X}) are given by
\begin{equation} \label{eq:eigenvalues-2x2}
x_{1,2} = \frac{1}{2} \left[ y_1 + y_2 \pm \sqrt{(y_1 - y_2)^2 + 4 y_3^2} \right] \;,
\end{equation}
and hence the gap is given by
\begin{equation} \label{eq:explicit-2x2-gap}
d = \sqrt{(y_1 - y_2)^2 + 4 y_3^2} \;.
\end{equation}
Changing variables from $y_1, y_2, y_3$ to $d, \theta, \psi$ where
\begin{equation} \label{eq:spherical-variables}
\begin{dcases}
y_1 - y_2 = d \cos \theta \\
2 y_3 = d \sin \theta \\
y_1 + y_2 = \psi
\end{dcases}
\end{equation}
and integrating out the extra degrees of freedom in $\theta$ and $\psi$ we obtain the probability density function of the gap $d$, defined in Eq. (\ref{eq:explicit-2x2-gap}), which is given by
\begin{equation} \label{eq:beta-1-Wigner}
{\rm Prob.}[d] = \frac{d \,\sigma^2}{2} \exp[ - \frac{d^2 \sigma^2}{4}]
\end{equation}
and hence the dimensionless relative gap $s = d / \langle d \rangle$ is distributed as
\begin{equation}
    p(s) = p_{\rm Wigner}(s) = \frac{\pi}{2} s \exp[ - \frac{\pi}{4} s^2 ] \;,
\end{equation}
which as can be seen in Fig. (\ref{fig:usual_atomic}) is a much better description of the measured spacing in scattering experiments.

\vspace{0.2cm}

The reason why random matrix theory is such an effective model is due to the fact that the eigenvalues of a random matrix repel each other with a logarithmic pairwise potential, the so-called `eigenvalue repulsion' or `level spacing' \cite{W51,M91,F10}. However, thanks to a special determinental structure the statistical properties of the eigenvalues can be computed exactly \cite{M91,F10,LNV18}. This is what makes random matrix theory such a powerful framework for studying systems with strongly correlated~variables~\cite{W28,W51,M91,F10,LNV18,PB20,MPS20,MSbook}. It is one of the rare models with strong long-range repulsive interactions, which can nonetheless be described in great detail analytically. Applications of random matrix theory are not limited to nuclear physics. The spectral statistics of the eigenvalues of Gaussian random matrices appear also in quantum chaos, mesoscopic transport, finance, or information theory \cite{G58,DN04,SM14,MPS20}.

\subsubsection{Dyson Log Gas}

We will now introduce some of the main results from random matrix theory that are relevant to us. This is not intended to be a broad overview of random matrix theory; for a more complete and detailed description, see \cite{M91, F10,LNV18}. Let us consider a real symmetric $N \times N$ matrix $X$ whose entries $X_{j,k}$'s perform independent \OU processes in some fictitious time $t$, i.e.
\begin{equation} \label{eq:dynamics-GUE}
\dv{X_{j, k}(t)}{t} = - \mu X_{j, k}(t) + \sqrt{(1 + \delta_{jk}) D } \; \eta_{jk}(t)
\end{equation}
where 
\begin{equation}\label{eq:eta-mu-def}
\langle \eta_{jk}(t) \eta_{mn}(t') \rangle = \delta_{jm} \delta_{kn} \delta(t - t') \;.
\end{equation}
We assume that the process starts at the origin $X_{j,k}(t=0)=0$. Thus, $X_{j,k}$ can be thought of as the position of a particle labeled by $(j,k)$ on the real line, performing Brownian motion in the presence of a confining potential $\mu X_{j,k}^2/2$, which we have already solved in Section \ref{subsec:weak}. The position distribution at any time $t$ is a simple Gaussian
\begin{equation} \label{eq:OU-propagator-sigma}
{\rm Prob.}[X_{j,k}(t)] = \frac{1}{\sqrt{\pi \sigma^2(t) (1 + \delta_{jk})}} \exp( - \frac{X_{j,k}(t)^2}{\sigma^2(t) (1 + \delta_{jk})} ) \;,
\end{equation} 
where 
\begin{equation} \label{eq:def-sigma}
\sigma^2(t) = \frac{D (1 - e^{-2 \mu t})}{\mu} \;.
\end{equation}
The distribution of the full matrix is obtained by taking the product of the distributions of all its independent components as given in Eq. (\ref{eq:OU-propagator-sigma}), which yields
\begin{equation} \label{eq:matrix-propagator}
{\rm Prob.}[X(t) = X] = \left( \frac{1}{2 \pi \sigma^2(t)} \right)^{N/2} \left(\frac{1}{\pi \sigma^2(t)} \right)^{N(N-1)/4} \exp( - \frac{\Tr[X^2]}{2 \sigma^2(t)} ) \;.
\end{equation}
The prefactors take the form above because the degrees of freedom of the matrix are $N$ diagonal elements of variance $\sigma^2(t)$ and $N(N-1)/2$ off-diagonal elements of variance $\sigma^2(t)/2$. In the long-time limit, the variance $\sigma^2(t)$ reaches a stationary value $\sigma^2(t \to +\infty) = \frac{D}{\mu}$ and hence the process in Eq. (\ref{eq:matrix-propagator}) reaches an equilibrium steady state
\begin{equation} \label{eq:matrix-ss}
{\rm Prob.}[X(t \to \infty) = X] = \frac{1}{2^{N/2}} \left( \frac{\mu}{\pi D}\right)^{\frac{N(N+1)}{4}} \exp( - \frac{\mu}{2D} \Tr[X^2]  ) \;.
\end{equation}
The $N$ real eigenvalues associated to this Gaussian matrix have the joint distribution~\cite{M91,F10} 
\begin{equation} \label{P_joint_Xeq_GOE}
{\rm Prob.}[x_1, \cdots x_N] \propto e^{- \frac{\mu}{2D} \sum_{i=1}^N x_i^2} \prod_{i<j} |x_i - x_j| \;,
\end{equation}
where the proportionality constant can be fixed from the normalization condition $\int dx_1 \dots \int dx_N {\rm Prob.}[x_1, \cdots x_N] =1$. 

\vspace{0.2cm}

\begin{figure}
    \centering
    \includegraphics[width=0.7\textwidth]{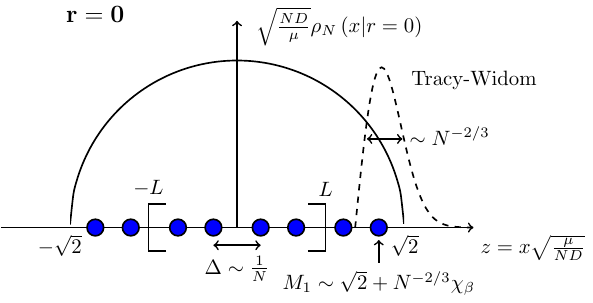}
    \caption{A schematic representation of the positions of the particles in the log-gas. 
    The scaled average density profile is supported over $z \in [-\sqrt{2},+\sqrt{2}]$ and has a semi-circular shape. The distribution of the position $x_{\max}$ of the rightmost particle is shown schematically by a dashed curve. The distribution is peaked around $\sqrt{2}$ with a width of order $O(N^{-2/3})$ given by the Tracy-Widom distribution. The inter-particle spacing $\Delta$ is of order $O(1/N)$ in scaled units.} \label{fig:free-dyson-gas}
\end{figure}

Following exactly the same method as the one detailed above, one can similarly compute the distribution of the matrix $X(t)$ at any time $t$ where the entries are complex or quaternionic. The nature of the matrix $X(t)$ is usually encoded via Dyson's index $\beta = 1, 2 $ or $4$ for real, complex or quaternionic entries, respectively. Skipping details, one finds that for any $\beta=  1, 2, 4$, the distribution of the matrix remains Gaussian at any time $t$ and is given by
\begin{equation} \label{eq:matrix-propagator_beta}
{\rm Prob.}[X(t) = X] \propto \exp( - \beta \, \frac{\Tr[X^\dagger X]}{2 \sigma^2(t)} ) \;,
\end{equation}
where we have introduced the factor $\beta$ inside the exponential for later convenience. Taking the limit $t \to \infty$ in Eq. (\ref{eq:matrix-propagator_beta}), the stationary distribution of the matrix for $\beta = 1, 2, 4$ reads
\begin{equation}
{\rm Prob.}[X(t \to \infty) = X] \propto \exp( - \frac{\beta \mu}{2D} \Tr[X^\dagger X]  ) 
\end{equation}
and the associated joint distribution of eigenvalues is given by \cite{F10}
\begin{equation} \label{P_joint_Xeq_Gbeta}
{\rm Prob.} [x_1, \cdots, x_N] \propto e^{- \frac{\beta \mu}{2D} \sum_{i=1}^N x_i^2} \prod_{i<j} |x_i - x_j|^\beta \propto e^{-\beta E[\{ x_i\}]} \;.
\end{equation}
Here the energy $E[\{ x_i\}]$ associated with 
the positions $\{x_i\}$'s of the gas is given by
\begin{equation} \label{energy}
E[\{ x_i\}] = \frac{\mu}{2D} \sum_{i=1}^N x_i^2 - \frac{1}{2} \sum_{i \neq j} \ln |x_i-x_j| \;.
\end{equation}
Thus, in the stationary state, the eigenvalues form a gas of $N$ particles which is at thermal equilibrium with $\beta$ playing the role of the
inverse temperature. The energy of the gas in Eq. (\ref{energy}) has two components: the first term corresponds to an external harmonic
potential $\mu/(2 D) x_i^2$, while the second term represents pairwise logarithmic repulsion between particles. Hence why this gas is referred to
as the ``Dyson log gas''. 

\vspace*{0.2cm}

An alternative way to arrive at the same stationary state of the eigenvalues, for $\beta = 1,2,4$, is to consider an overdamped Langevin equation~\cite{D62,D62b} 
\begin{equation} \label{Langevin}
\dv{x_i}{t} = - \mu x_i + D \sum_{j (\neq i)} \frac{1}{x_i - x_j} + \sqrt{\frac{2 D}{\beta}} \, \eta_i(t) \;,
\end{equation}
where $\eta_i(t)$ is a Gaussian white noise with $\langle \eta_i(t) \rangle = 0$ and $\langle \eta_i(t) \eta_j(t') \rangle = \delta_{ij} \delta(t - t')$. We assume that the Langevin equations in Eq. (\ref{Langevin}) start from an initial condition where all the particles are localized close to the origin, e.g., over an interval $[-\epsilon,+\epsilon]$ with 
\begin{equation} \label{IC}
x_i(0) = \epsilon\left(-1 + \frac{2i}{N} \right) \;, \; i =1,2, \cdots, N \;.
\end{equation}
Eventually, we will take the limit $\epsilon \to 0$. Thus, the eigenvalues can be interpreted as the positions of $N$ particles on a line, each performing an \OU process in the presence of a pairwise repulsive force. Consistently with the notation in the other sections, let ${\rm Prob.}[x_1, x_2, \cdots, x_N,t] \equiv p_0(\vec{x},t)$ denote the joint distribution of the positions of the particles at time $t$, that is, the propagator at time $t$ (the reason for the subscript 0 will be apparent in a second). Its time evolution is governed by the Fokker-Planck equation associated to the Langevin equation (\ref{Langevin}) and it reads
\begin{equation} \label{eq:FP-DBM_text}
\pdv{p_0(\vec{x}, t)}{t} = \frac{D}{\beta} \sum_{i = 1}^N \pdv[2]{p_0(\vec{x}, t)}{x_i} - \sum_{i = 1}^N \pdv{}{x_i} \left[\left(-\mu x_i + D \sum_{j \neq i} \frac{1}{x_i - x_j} \right) p_0(\vec{x}, t) \right] \;.
\end{equation}
Setting $\partial{p_0(\vec{x}, t)}/{\partial t} = 0$ as $t \to \infty$, it is easy to check that the stationary joint distribution ${\rm Prob.}[x_1, x_2, \cdots, x_N] = \lim_{t \to \infty} {\rm Prob.}[x_1, x_2, \cdots, x_N,t]$ is indeed given by the Gibbs state in the equation. (\ref{P_joint_Xeq_Gbeta}). Note that for $\beta = 1,2,4$ the positions $x_i$'s can be interpreted as the eigenvalues of an underlying Gaussian matrix. However, the Langevin equation (\ref{Langevin}) is well defined for any $\beta > 0$. But for $\beta \neq 1,2,4$, they do not have any interpretation as the eigenvalues of an underlying Gaussian matrix. We refer to this Langevin process in Eq. (\ref{Langevin}) as the~$\beta$-Dyson Brownian Motion. 

\subsubsection{Resetting Dyson Log Gas}

\begin{figure}
    \centering
    \includegraphics[width=0.7\textwidth]{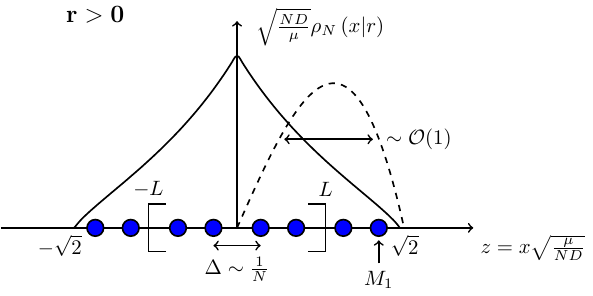}
    \caption{A schematic representation of the positions of the particles for the resetting Dyson log-gas. The scaled average density profile is supported over $z \in [-\sqrt{2},+\sqrt{2}]$. But instead of the semi-circular shape in Fig. \ref{fig:free-dyson-gas} it now has a cusp at $z=0$, where the simultaneous resetting takes place. The distribution of the 
    position $x_{\max}$ of the rightmost particle is shown schematically by a dashed curve. In contrast with the maximum in the absence of resetting (as seen in Fig. \ref{fig:free-dyson-gas}) where the distribution is peaked around $\sqrt{2}$ with a width of order $O(N^{-2/3})$ (the Tracy-Widom distribution), the distribution is now supported over $[0,\sqrt{2}]$. The inter-particle spacing $\Delta$ is of order $O(1/N)$ in scaled units in both models.} \label{fig:reset-dyson-gas}
\end{figure}

We can introduce stochastic resetting in DBM in two alternative ways: 
\begin{itemize}

\item[(i)] For $\beta=1,2,4$, we can start from a $N \times N$ Gaussian matrix whose elements perform independent Ornstein-Uhlenbeck processes, as in Eq. (\ref{eq:dynamics-GUE}), starting from $X_{j,k}(t=0)=0$, but with a constant rate $r$, they are reset {\it simultaneously} to the origin. This is equivalent to evolving the matrix entries via Eq. (\ref{eq:dynamics-GUE}) up to a certain random time $\tau$ drawn from an exponential distribution $h(\tau) = r\, e^{- r \tau}$ and then resetting them simultaneously to the initial condition close to the origin as in Eq.~(\ref{IC}), followed by a restart of the dynamics. In other words, in a small time interval $\dd t$, with probability $1 - r \dd t$ the entries evolve via independent Ornstein-Uhlenbeck processes, while with complementary probability $r \dd t$ one resets all matrix entries to their initial value simultaneously. At any instant of time, one can diagonalize this matrix and observe the evolution of the eigenvalues. Between any two consecutive resettings, the eigenvalues will evolve via the Dyson-Brownian motion in Eq. (\ref{Langevin}) and then get reset to their initial conditions. We will refer to this process of eigenvalues as a resetting Dyson Brownian motion with $\beta = 1,2,4$.

\item[(ii)] For general $\beta > 0$, we can think of a set of particles with positions $x_i$'s that evolve via the $\beta$-Dyson Brownian motion in Eq. (\ref{Langevin}) and then simultaneously reset to their initial conditions with rate $r$. For a general $\beta>0$, the $x_i$'s do not have the interpretation of the eigenvalues, but nevertheless they represent a system of harmonically confined particles in one dimension with pairwise repulsion and stochastic resetting. This is a wider class of models parameterized by $\beta > 0$ and for the special cases $\beta = 1,2,4$ it has the interpretation of evolving eigenvalues with stochastic resetting. 
For general $\beta$, we will refer to this process as the $\beta$-Resetting Dyson Brownian motion. 

\end{itemize}

\vspace{0.2cm}

In this Section, we will focus on the $\beta$-Resetting Dyson Brownian motion. Our principal goal is to derive the joint distribution of the positions of the particles in the stationary state of this $\beta$-Resetting Dyson Brownian motion. Consistent with previous notation, let us define this joint distribution at time $t$ by $p_r(\vec{x}, t) = {\rm Prob.}[x_1,\cdots, x_N,t \vert r]$, where $r$ denotes the resetting rate. Using a renewal approach, it is possible to express this joint distribution in terms of the propagator $p_0(\vec{x}, t) = {\rm Prob.}[x_1,\cdots, x_N,t \vert r=0]$ of the reset-free $\beta$-Dyson Brownian motion evolving via Eq. (\ref{Langevin}) given in the previous Section. The usual resetting renewal equation which we have seen multiple times still holds
\begin{equation} \label{renew_x_t}
p_r(\vec{x}, t) = e^{-rt}\, p_0(\vec{x}, t) + r \int_0^t \dd \tau \, e^{-r \tau}\, p_0(\vec{x}, t) \;.
\end{equation}
In the long time limit $t \to \infty$, the first term on the right hand side of Eq.~(\ref{renew_x_t}) drops out and one gets the stationary joint distribution of the positions   
\begin{equation} \label{renew_x}
p_r(\vec{x}) \equiv p_r(\vec{x}, t \to +\infty) = r \int_0^\infty \dd \tau \, e^{-r \tau}\, p_0(\vec{x}, \tau) \;. 
\end{equation}
However, unlike in the previously studied models in Sections \ref{subsec:BM-simreset}-\ref{subsec:levy-simreset} the particles of the underlying reset-free process are not independent. Hence
\begin{equation}
    p_0(\vec{x}, t) \neq \prod_{i = 1}^N p_0(x_i, t) \;,
\end{equation}
and the non-equilibrium steady state in Eq.~(\ref{renew_x}) cannot be described using conditionally independent identically distributed random variables. Thus, to compute the joint distribution of $\beta$-\RDBM in \NESS we need the propagator $p_0(\vec{x}, t)$ of the reset-free $\beta$-\DBM {\it at all times $\tau$}, and not just at late times. Hence, we need the solution of the Fokker-Planck equation (\ref{eq:FP-DBM_text}) at all times $\tau$. Interestingly, this exact solution for all $\tau$ and all $\beta >0$ can be explicitly written down as
\begin{align} 
p_0(\vec{x}, \tau) = \frac{1}{Z_N(\beta)}
&\frac{1}{\sigma(\tau)^{N + \beta N(N-1)/2}} \nonumber \\
&\times \exp[ -\frac{\beta}{2}\left(\frac{1}{\sigma^2(\tau)} \sum_{i = 1}^N x_i^2 - \sum_{i \neq j} \ln |x_i - x_j| \right)] \;, \label{propag_DBM_intro}
\end{align}
where $\sigma^2(\tau) = {D (1 - e^{-2 \mu \tau})}/{\mu}$ is already defined in Eq. (\ref{eq:def-sigma}) and $Z_N(\beta)$ is a normalization constant that can be computed explicitly. Note that while for $\beta = 1,2,4$ this exact propagator of the \DBM at any time $\tau$ was well known by exploiting its connection to the rotationally invariant Gaussian matrices (see e.g. \cite{F10}), for general $\beta$ we have not seen this result in the literature. 
However, it turns out that the form in Eq.~(\ref{propag_DBM_intro}) is valid for any $\beta > 0$ not only $\beta = 1,2,4$. This can be checked by substituting the solution in Eq.~(\ref{propag_DBM_intro}) in the Fokker-Planck equation in Eq.~(\ref{eq:FP-DBM_text}), see the Appendix of Ref. \cite{BMS25} for details. Substituting Eq. (\ref{propag_DBM_intro}) in Eq. (\ref{renew_x}), we get the exact joint distribution of the $\beta$-\RDBM process in its NESS
\begin{align} 
p_r(\vec{x}, t) = \frac{1}{Z_N(\beta)} \int_0^{+\infty} r \dd\tau\; e^{-r\tau} \frac{1}{\sigma(\tau)^{N + \beta N(N-1)/2}}\\
\times \exp[ -\frac{\beta}{2}\left(\frac{1}{\sigma^2(\tau)} \sum_{i = 1}^N x_i^2 - \sum_{i \neq j} \ln |x_i - x_j| \right)] \;.\label{eq:eigenvalues-jpdf}
\end{align}
We can rewrite this joint distribution as 
\begin{equation}\label{eq:eigenvalues-jpdf-E}
p_r(\vec{x}, t) = \frac{1}{Z_N(\beta)} \int_0^{+\infty} r \dd \tau \;  \frac{e^{-r \tau}}{\sigma(\tau)^{N + \beta N(N-1)/2}} \exp[ - \beta E\left[ \vec{x} ; \tau \right] ] \;,
\end{equation}
where $E(\vec{x}; \tau)$ can be interpreted as the energy of a Coulomb gas parametrized by $\tau$ 
\begin{equation} \label{eq:def-Z-E}
E[\vec{x}; \tau] = \frac{1}{2 \sigma^2(\tau)} \sum_{i = 1}^N x_i^2 - \frac{1}{2} \sum_{i \neq j} \ln |x_i - x_j| \;.
\end{equation}
Here the $\tau$ dependence arises from the factor $\sigma^2(\tau) = D(1-e^{-2 \mu \tau})/\mu$. Eqs. (\ref{eq:eigenvalues-jpdf-E}) and (\ref{eq:def-Z-E}) form the starting point of our analysis. Note that in the limit $r \to 0$, we can rescale $r \tau = u$ in Eq. (\ref{eq:eigenvalues-jpdf}) and one recovers the standard log-gas with the joint distribution of eigenvalues 
\begin{equation} \label{GOE}
p_0(\vec{x}) = \frac{1}{Z_N(\beta)} \left(\frac{\mu}{D}\right)^{N + \beta\frac{N(N+1)}{2}} e^{-\beta E[\{x_i\}]} \;,
\end{equation} 
where the energy $E[\{x_i\}]$ of the log-gas is given by Eq. (\ref{energy}).

\vspace*{0.3cm}
Note again that for the three special values $\beta = 1,2, 4$ the stationary distribution of the positions of the particles can be interpreted as the eigenvalues of a rotationally invariant $N \times N$ Gaussian matrix $X$ whose entries are distributed via
\begin{align} 
{\rm Prob.}&[X   | r] \propto \int_0^{+\infty} r \dd \tau \; \frac{e^{- r \tau} }{\sigma_\beta(\tau)^{N + \beta N(N-1)/2}} \exp( - \beta\,\frac{\Tr[ X^\dagger X ]}{2 \sigma^2(\tau)}) \;.\label{eq:eigenvalues-jpdf-beta}
\end{align}
where $\sigma^2(\tau)  = D (1 - e^{-2 \mu \tau})/\mu$, as defined in Eq. (\ref{eq:def-sigma}). This probability measure does not factorize into independent Gaussian for each entry. This is due to the simultaneous resetting of all the values, which induces strong correlations between all the components. However, Eq. (\ref{eq:eigenvalues-jpdf-beta}) shows that the distribution is still spherically symmetric, i.e., rotationally invariant, since it is only a function of the trace of the matrix. 
Such deformed Gaussian random matrix ensembles have appeared before in the literature under the name ``superstatistics'' where the variance of the matrix entries is considered as a random variable with some distribution \cite{AM05,BCP08,AAV09,B09, RB14,HW04,DGKJS16}. However, in these papers there was no ``microscopic'' dynamics involved that led to such deformed ensembles and the distribution of the variance was put in by hand. In contrast, the deformed Gaussian ensemble in Eq. (\ref{eq:eigenvalues-jpdf-beta}) for the three special values $\beta = 1,2,4$ appears naturally via the underlying stochastic resetting. Moreover, our $\beta$-\RDBM model for the evolution of $x_i$'s is more general and valid for arbitrary $\beta >0$ and not restricted to $\beta = 1,2,4$. 

\subsubsection{Connection to Resetting Vicious Brownian Walkers}

One can show that in the special $\beta =2$ and $\mu=0$ (untrapped particles) of our \RDBM model, there is a close connection with another model, namely vicious Brownian motions with simultaneous resetting. \VBM refers to a system of $N$ Brownian motions conditioned not to intersect each other (the word ``vicious'' refers to the fact that if any pair of particles intersect each other, they kill each other). Consider first the case $r=0$, $\mu = 0$ and $\beta = 2$. In this case, it is well known that the propagators of the \DBM and the \VBM are related to each other via  
(see e.g. \cite{GMS21,RS11})
\begin{equation} \label{eq:dbm-vicious-link}
{\rm Prob.}^{\rm DBM}_{\beta = 2, \mu=0}[\vec{x}, t | \vec{x_0}, 0] = \frac{\prod_{i < j} (x_i - x_j) }{\prod_{i < j} (x_{0i} - x_{0j}) } {\rm Prob.}^{\rm VBM}[\vec{x}, t | \vec{x_0}, 0] \;,
\end{equation}
{where ${\rm Prob.}_{\beta = 2, \mu=0}^{\rm DBM}[\vec{x}, t | \vec{x_0}, t_0]$ denotes the probability for a \DBM with $\beta = 2$ and $\mu=0$, 
starting at $\vec{x_0}$ at time $t_0$, to reach $\vec{x}$ at time $t$ (and equivalently for the VBM).}
In the presence of resetting Eq. (\ref{renew_x_t}) allows us to express the \RDBM propagator in the \NESS as 
\begin{equation} \label{eq:resetting-free-link}
{\rm Prob.}^{\rm RDBM}[\vec{x} | r, \vec{x}_{\rm reset} = \vec{0}] = r \int_0^{+\infty} \dd \tau \; e^{-r \tau} {\rm Prob.}^{\rm DBM}[\vec{x}, \tau | \vec{0}, 0] \;,
\end{equation}
where ${\rm Prob.}^{\rm RDBM}[\vec{x} | r, \vec{x}_{\rm reset} = \vec{0}] $ denotes the probability for a \RDBM to be at $\vec{x}$ in the \NESS while resetting to $\vec{0}$ with rate $r$.
From Eq. (\ref{eq:dbm-vicious-link}) we see that when $\vec{x_0} = \vec{0}$ the right hand side seems undefined. In reality, the divergence of the denominator will be regularized by the normalization constant in the \VBM propagator. Hence, taking $\vec{x_0} = \epsilon \vec{a}$, where $a_{i} = -1 + 2 i /N$, we can rewrite Eq. (\ref{eq:resetting-free-link}) as
\begin{align} 
{\rm Prob.}^{\rm RDBM}_{\beta = 2, \mu=0}&[\vec{x} | r, \vec{x}_{\rm reset} = \vec{0}] = \nonumber \\
&\lim_{\epsilon \to 0^+} \frac{r}{\epsilon} \frac{\prod_{i < j} (x_i - x_j)}{ \prod_{i < j} (a_i - a_j) } \int_0^{+\infty} \dd \tau \; e^{-r \tau} {\rm Prob.}^{\rm VBM}[\vec{x}, \tau | \epsilon \vec{a}, 0] \;. \label{eq:RDBM-VBM-renewal}
\end{align}
From the same renewal arguments we know that the integral is the propagator of Resetting Vicious Brownian Motions which do not reset to the origin but to $\epsilon \vec{a}$, hence
\begin{align} 
{\rm Prob.}^{\rm RDBM}_{\beta = 2,\mu=0}&[\vec{x} | r, \vec{x}_{\rm reset} = \vec{0}] = \nonumber \\
&\lim_{\epsilon \to 0^+} \frac{1}{\epsilon} \frac{\prod_{i < j} (x_i - x_j)}{ \prod_{i < j} (a_i - a_j) } {\rm Prob.}^{\rm RVBM}[\vec{x} | r, \vec{x}_{\rm reset} = \epsilon \vec{a}] \;.\label{eq:RDBM-RVBM-link}
\end{align}
Therefore, in the following sections, all results derived for the un-trapped \RDBM, i.e. $\mu = 0$, can be directly used to obtain the behavior of a gas of $N$ resetting \VBM by setting $\beta = 2$ and applying the appropriate weight defined in Eq. (\ref{eq:RDBM-RVBM-link}).

\vspace{0.2cm}

\begin{figure}[t]
    \centering
    \includegraphics[width=0.48\textwidth]{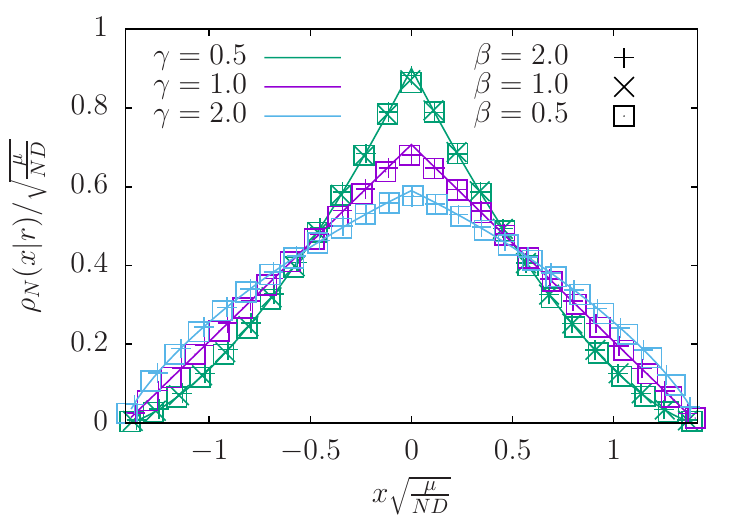} \hfill
    \includegraphics[width=0.48\textwidth]{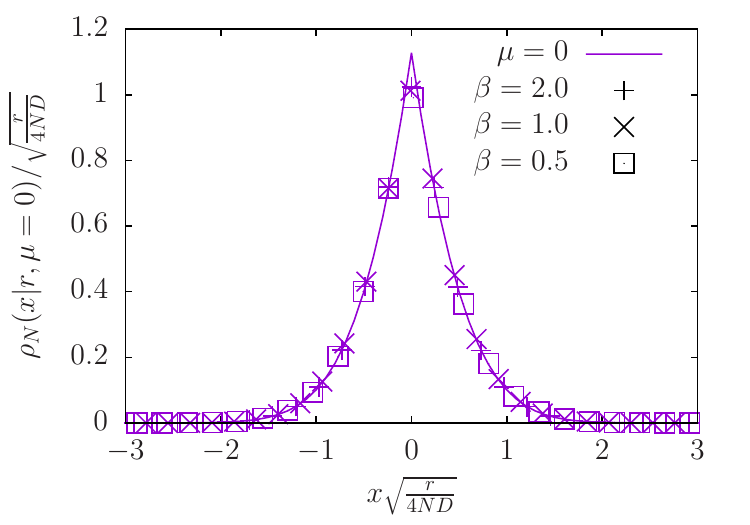}
    \caption{{\bf Left.} Plot of the scaling function $\rho_S(z, \gamma)$ given in Eq. (\ref{eq:def-rhoS}) vs the rescaled length $z=x \sqrt{\mu/(ND)}$, describing the average density profile of the \RDBM in the \NESS. The lines are the analytical prediction given by Eq. (\ref{eq:def-rhoS}) and the points are the results of numerical direct sampling simulations. Different colors correspond to different values of $\gamma = \mu/r$ (0.5, 1 and 2 for green, purple and blue respectively) and different symbols correspond to different values of $\beta =  0.5, 1, 2$. {\bf Right.} Plot of the scaling function $f_0(y)$ defined in Eq. (\ref{eq:def-f0}) vs the rescaled length $y = x \sqrt{r/(4ND)}$, describing the density of the untrapped ($\mu = 0$) \RDBM in the \NESS. The line corresponds to the theoretical prediction given in Eq. (\ref{eq:def-f0}) and the points are the results of numerical direct sampling simulations. Different symbols correspond to different values of $\beta = 0.5, 1, 2$.}\label{fig:density}
\end{figure}

\section{The average density}

Although we do not have access to a conditionally independent form that would allow us to directly obtain our usual observables we can still exploit the renewal structure in Eq.~(\ref{renew_x}) and our knowledge of the reset-free Dyson log gas. We start by studying the average spectral density in the \NESS, which is defined as
\begin{equation} \label{eq:def-rho}
\rho_N(x | r) = \left\langle \frac{1}{N} \sum_{i = 1}^N \delta(x - x_i) \right\rangle_{\{x_i\} \sim p_r(\vec{x})} \;,
\end{equation}
where the subscript outside the expectation denotes that the $x_i$'s are drawn from the joint distribution $p_r(\vec{x})$ in Eqs. \eqref{eq:eigenvalues-jpdf-E} and \eqref{eq:def-Z-E}. We then get
\begin{align} 
\rho_N(x | r) &= r \int_0^{+\infty} \dd \tau \; e^{-r \tau} \left\langle \frac{1}{N} \sum_{i = 1}^N \delta(x - x_i) \right\rangle_{\{x_i\} \sim p_r(\vec{x})} \nonumber \\
&= r \int_0^{+\infty} \dd \tau \; e^{-r \tau} \rho_N(x | r = 0, \tau) \;, \label{eq:renewal-density}
\end{align}
where $\rho_N(x | r = 0, \tau)$ is the average density of particles in the $\beta$-\DBM at time $\tau$ without resetting. The latter is well known to be of the Wigner semi-circular form in the large $N$ limit and is independent of $\beta>0$~\cite{M91, PB20}. Using this result and the explicit expression $\sigma^2(\tau) = D(1-e^{-2\mu \tau})/\mu$, we get
{
\begin{equation} \label{eq:free-density}
\rho_N(x | r = 0, \tau) \simeq \sqrt{\frac{2 \mu}{N D (1 - e^{- 2 \mu \tau })\pi^2}\left(1 - \frac{\mu x^2}{ 2 N D ( 1 - e^{-2 \mu \tau} ) }\right)} \;.
\end{equation}
}
Substituting this result in Eq. (\ref{eq:renewal-density}) we obtain the large $N$ scaling form
\begin{equation}\label{eq:density-scaling}
\rho_N(x | r) \simeq \sqrt{\frac{\mu}{N D}} \; \rho_S\left(x \sqrt{\frac{\mu}{N D}}, \frac{\mu}{r}\right) \;,
\end{equation}
where the normalized scaling function $\rho_S(z, \gamma)$ supported over $z \in [-\sqrt{2}, \sqrt{2}]$ and for $\gamma > 0$ is given by
\begin{equation} \label{eq:def-rhoS}
\rho_S(z, \gamma) = \frac{1}{2 \gamma} \int_{\log\frac{2}{2 - z^2}}^{+\infty} \dd u \; e^{- u/(2 \gamma)} \sqrt{\frac{2}{\pi^2 ( 1 - e^{-u} )}\left( 1 - \frac{z^2}{2 (1 - e^{-u})} \right)} \;.
\end{equation}
This integral can be performed explicitly using hypergeometric $\,_2F_1$ functions from which we can easily extract the asymptotic behaviors
\begin{equation} \label{eq:main-f-asymp}
    \rho_S(z, \gamma) \approx \begin{dcases}
    A_0 - \frac{|z|}{2 \gamma} &\mbox{~~when~~} |z| \to 0 \\
    A_{+} (\sqrt{2} - |z|)^{(1 + 1/\gamma)/2} &\mbox{~~when~~} |z| \to \sqrt{2}
    \end{dcases}\;,
\end{equation}
where the amplitudes $A_0$ and $A_+$ are given by
\begin{equation} \label{A0Ap}
A_0 = \sqrt{\frac{2}{\pi}} \frac{\Gamma\left(1+\frac{1}{2\gamma}\right)}{\Gamma \left(\frac{1}{2}+\frac{1}{2 \gamma}\right)}  \quad, \quad A_+ = \frac{2^{(1+\gamma)/(4 \gamma)}}{\sqrt{2\pi}} \frac{\Gamma\left( 1 + \frac{1}{2 \gamma}\right)}{\Gamma\left(\frac{3}{2} + \frac{1}{2 \gamma} \right)} \;.
\end{equation}
This scaling function is shown in Fig.~\ref{fig:density} for different values of $\gamma$ and is in perfect agreement with the numerical simulations. From the asymptotic behaviors of the scaling function it is clear that the shape of this average density profile is drastically different from that of the Wigner semi-circular law. First it has a cusp as $z \to 0$, which results from this simultaneous resetting of the particles to $x=0$: this is quite different from the smooth quadratic behavior in the semi-circular form valid for $r=0$. Secondly, when $|z|$ approaches the edge of the support $z = \pm \sqrt{2}$, the spectral density vanishes with an exponent $(1+1/\gamma)/2$ larger than $1/2$ valid for $r=0$. In the limit when $r \to 0$, i.e. $\gamma = \mu/r \to \infty$, our spectral density reduces to the Wigner semi-circular law. To see this, we note that the integral in Eq. (\ref{eq:def-rhoS}) is dominated by values of $u \gtrsim 2 \gamma$ for which $e^{-u} \ll 1$, recovering the usual \RMT~semi-circle law
\begin{align} 
\rho_S(z, \gamma) &\stackrel{\gamma \to +\infty}{\longrightarrow} \sqrt{\frac{2}{\pi^2}\left( 1 - \frac{z^2}{2}\right)}\; \frac{1}{2 \gamma} \int_{\log \frac{2}{2 - z^2}}^{+\infty} e^{-u/(2 \gamma)} \, \dd u \nonumber \\
&= \sqrt{\frac{2}{\pi^2}\left( 1 - \frac{z^2}{2}\right)^{1 + 1/\gamma}} \stackrel{\gamma \to +\infty}{\longrightarrow} \sqrt{\frac{2}{\pi^2}\left( 1 - \frac{z^2}{2}\right)} \;. \label{eq:f-inf}
\end{align}

\vspace{0.2cm}

\noindent{\bf The limit $\mu \to 0$.} The competition between the long-range attraction from resetting and the long-range repulsion from the logarithmic interaction balances out even in the absence of a confining potential. Hence, it is possible to reach a new \NESS with Dyson's logarithmic repulsion combined with resetting. To obtain this limit, we consider the scaling form of the spectral density in Eq. (\ref{eq:density-scaling}). When $\mu \to 0$, the ratio $\gamma = \mu/r \to 0$ and also the scaled distance $z = x \sqrt{\mu/(N\,D)} \to 0$. Since we want to find the density at a fixed position $x$, we need to take the simultaneous limits $\gamma \to 0$, $z \to 0$, but with the ratio $z^2 / \gamma = x^2 r / (N D)$ fixed. Taking these limits in Eq. (\ref{eq:density-scaling}) and Eq. (\ref{eq:def-rhoS}) we get
\begin{equation} \label{eq:density0}
\rho_N(x | r) \stackrel{\mu \to 0}{\longrightarrow} \sqrt{\frac{r}{4 N D}} \rho_{S, 0} \left( x \sqrt{\frac{r}{4 N D}} \right) \;,
\end{equation}
where 
\begin{equation} \label{eq:def-f0}
\rho_{S, 0}(y) = \frac{2}{\pi} \int_{y^2}^{+\infty} \dd v \; e^{-v} \sqrt{\frac{v - y^2}{v^2}} = \frac{2 e^{-y^2}}{\sqrt{\pi}} - 2 |y| \, {\rm erfc} |y| \;,
\end{equation}
is the normalized scaling function which is now defined on the unbounded support $y \in \mathbb{R}$. It is plotted in the right panel of Fig. \ref{fig:density} and is in perfect agreement with the numerical simulations. The {\color{blue}asymptotes} of this scaling function $\rho_{S, 0}(y)$ are given by
\begin{equation} \label{eq:f0-asymptotics}
\rho_{S, 0}(y) \approx \begin{dcases}
\frac{2}{\sqrt{\pi}} - 2 |y| &\mbox{~~as~~} |y| \to 0\\
\frac{e^{-y^2}}{y^2 \sqrt{\pi}} &\mbox{~~as~~} |y| \to \infty
\end{dcases} \;.
\end{equation}
In the absence of a confining trap, the density is now defined over the full real line instead of a bounded region around the origin, but the resetting forms a cusp at the origin, concentrating the measure there, which allows the gas to reach a \NESS with an unbounded support. 

\section{{\color{blue}Extreme-value} statistics} \label{sec:model-evs}

\begin{figure}[t]
\centering
\includegraphics[width=0.48\textwidth]{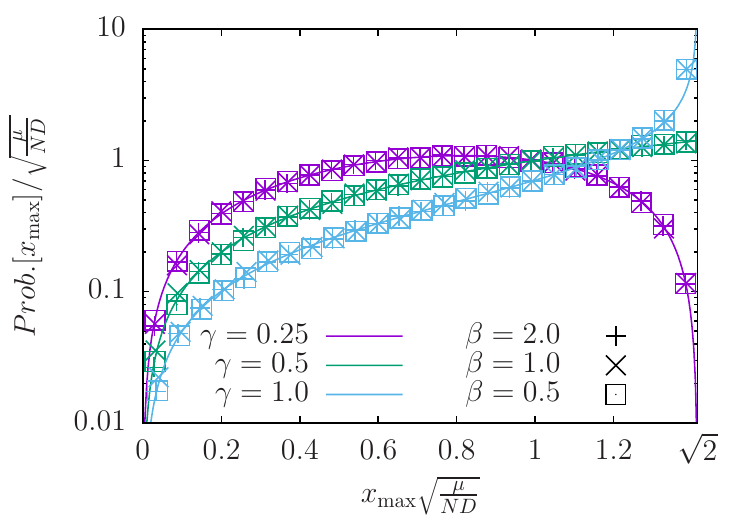}\hfill
\includegraphics[width=0.48\textwidth]{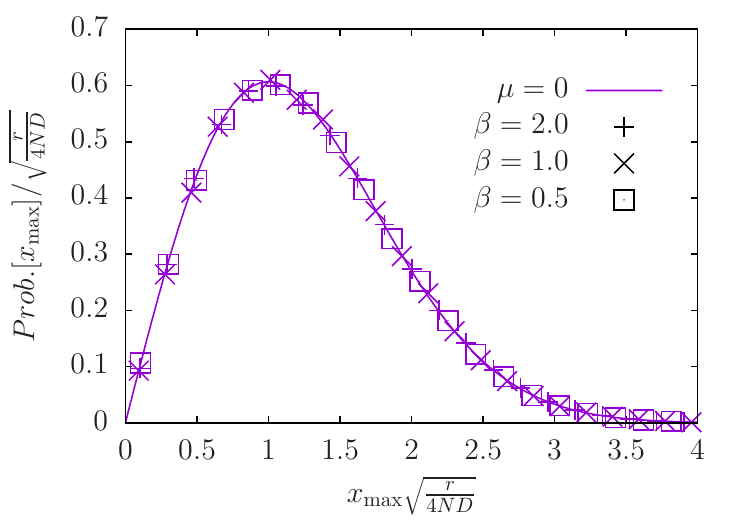}
\caption{{\bf Left.} Plot of the scaling function $g(z, \gamma)$ defined in Eq. (\ref{eq:g-def}) vs the rescaled maximum $z =x_{\max} \sqrt{\mu/(ND)}$, describing the probability distribution function of $x_{\max}$ representing the position of the rightmost particle of the \RDBM in its \NESS. The lines are the analytical prediction given in Eq. (\ref{eq:g-def}) and the points are the results of numerical direct sampling simulations using the tri-diagonal matrix diagonalizations \cite{DE02}. Different colors correspond to different values of $\gamma$ (0.25, 0.5 and 1 for purple, green and blue) and different symbols corresponds to different values of $\beta = 0.5,1,2$. {\bf Right.} Plot of the scaling function $f_0(y)$ defined in Eq. (\ref{eq:def-g0}) vs the rescaled maximum $y=x_{\max}\sqrt{r/(4 ND)}$ describing the probability density of the rightmost particle's position of the untrapped ($\mu = 0$) \RDBM in its \NESS. The line is the analytical prediction given by Eq. (\ref{eq:def-g0}) and the crosses are the results of numerical direct sampling simulations using the tri-diagonal matrix diagonalizations \cite{DE02}. Different symbols correspond to different values of $\beta=0.5,1,2$.}\label{fig:max}
\end{figure}

We now turn our attention to the \EVS, i.e., to the study of the distribution of the position $x_{\rm max} = \max \{x_1, \cdots, x_N\}$ of the rightmost particle in the \RDBM. In the absence of resetting $r=0$, it is well known that $x_{\max}$, appropriately centered and scaled in the limit of large $N$, is distributed via the 
Tracy-Widom law \cite{F10,TW94,TW96}. More precisely, $x_{\max}$, as a random variable, can be expressed, for large $N$, as  
\begin{equation}\label{eq:free-max}
x_{\rm max}(\tau, r = 0) \approx \sqrt{\frac{D (1 - e^{-2 \mu \tau})}{\mu}} \left[ \sqrt{2 N} + \frac{1}{\sqrt{2}} N^{-1/6} \chi_\beta\right] \;,
\end{equation}
where $\chi_\beta$ is an $\mathcal{O}(1)$ random variable whose law is given by the Tracy-Widom distribution of index $\beta$. Consequently the scaled maximum can be expressed as 
\begin{equation} \label{scaled_xmax}
\sqrt{\frac{\mu}{N\, D}} x_{\max}(\tau, r= 0)\approx \sqrt{(1 - e^{-2 \mu \tau})} \left[ \sqrt{2} + \frac{1}{\sqrt{2}} N^{-2/3}\, \chi_\beta\right] \;.
\end{equation}
Since the width of the distribution of $x_{\max}$ scales as $N^{-1/6}$ in Eq. (\ref{eq:free-max}) (or equivalently as $O(N^{-2/3})$ for the scaled maximum in Eq. (\ref{scaled_xmax})), it follows that, for large $N$, the distribution gets more and more concentrated around the mean $\sqrt{2 N \sigma^2(\tau)}$. When the simultaneous resetting $r$ is switched on, using Eqs. \eqref{eq:eigenvalues-jpdf-E} and \eqref{eq:def-Z-E}, 
the probability distribution can be written as  
\begin{equation} \label{eq:full-renewal-evs}
{\rm Prob.}[x_{\rm max} = x | r] = r \int_0^{+\infty} \dd \tau \; e^{-r \tau}\; {\rm Prob.}[x_{\rm max}(\tau,r=0) = x] \;,
\end{equation}
where the random variable $x_{\rm max}(\tau, r = 0)$ is distributed via Eq. (\ref{eq:free-max}). In the large $N$ limit, it turns out that the dominant contribution to this integral comes from the deterministic part in Eq. (\ref{eq:free-max}). The stochastic part containing the Tracy-Widom variable gives only subleading contributions for all $r>0$. Hence one can approximate Eq. (\ref{eq:full-renewal-evs}) by
\begin{equation} \label{eq:full-renewal-evs-delta}
{\rm Prob.}[x_{\rm max} | r] \underset{N\to\infty}{\longrightarrow} r \int_0^{+\infty} \dd \tau \; e^{-r \tau}\; \delta\left[x_{\rm max} - \sqrt{\frac{2 N D(1 - e^{-2\mu \tau})}{\mu}}\right] \;.
\end{equation} 
This integral can be done explicitly and has a scaling form
\begin{equation}\label{eq:max-scaling}
{\rm Prob.}[x_{\rm max} = x | r] \underset{N\to\infty}{\longrightarrow} \sqrt{\frac{\mu}{N D}} \; f\left( x \sqrt{\frac{\mu}{N D}}  , \frac{\mu}{r} \right) \;,
\end{equation}
where the normalized scaling function $f(z, \gamma)$ is supported over $z \in [0, \sqrt{2}]$ (strictly for $\gamma > 0$) by
\begin{equation} \label{eq:g-def}
f(z, \gamma) = \frac{1}{2\gamma} z \left(1 - \frac{z^2}{2}\right)^{\frac{1}{2\gamma} - 1} \quad, \quad {\rm with} \quad \gamma = \frac{\mu}{r} \;. 
\end{equation}
This scaling function is shown in Fig. \ref{fig:max} and is in perfect agreement with numerical simulations. Notice that when $\gamma \to +\infty$ the distribution concentrates around the upper limit $z=\sqrt{2}$ of the support recovering the deterministic part of Eq. (\ref{eq:free-max}). However, for any $\gamma >0$, the maximal eigenvalue has a nontrivial distribution parametrized by $\gamma$. This distribution diverges as $z \to \sqrt{2}$ for $\gamma>1/2$, while it vanishes  as $z \to \sqrt{2}$ for $\gamma <1/2$, thus displaying a ``shape transition'' at $\gamma=1/2$. Once again, this is extremely different from the behavior of the {\color{blue}extreme-value} statistics for the Dyson log gas. In Dyson log gas, the rightmost particle is always at the right edge of the support of the density, i.e., around the scaled distance $z = \sqrt{2}$, with small fluctuations of order $O(N^{-2/3})$. The centered and the scaled distribution is described by the Tracy-Widom law. On the other hand, in our gas, the 
rightmost particle can be anywhere between $z=0$ and $z= \sqrt{2}$ and its distribution has a width of order $O(1)$ in the large $N$ limit. This distribution may either diverge or vanish as $z \to \sqrt{2}$, depending on whether $\gamma > 1/2$ or $\gamma<1/2$. Thus, as one switches on resetting, the whole distribution of $x_{\max}$ is shifted from the upper edge $\sqrt{2}$ towards the trap center $z=0$, in stark contrast with the $r=0$ case. 
This is a clear indication of the strong attractive correlations generated by resetting. Indeed, in Dyson log gas the strong repulsive correlations ensure that the particles will always try to spread as much as possible within the trap, therefore pushing the rightmost particle to the right edge of the support. In our gas, the strong attractive correlations balance the repulsive ones and hence allow for the rightmost particle to not necessarily concentrate near the upper edge $z=\sqrt{2}$ of the support. 

\vspace*{0.3cm}
\noindent{\bf The limit $\mu \to 0$.} As was done previously for the density, to obtain the result for the untrapped gas, i.e. $\mu \to 0$, we have to take $\gamma \to 0$ and $z \to 0$ while keeping $z^2 / \gamma = x^2 r  / (N D)$ fixed. Taking these limits in Eq. (\ref{eq:max-scaling}) and Eq. (\ref{eq:g-def}) we obtain in the large $N$ limit
\begin{equation} \label{eq:evs-mu0}
{\rm Prob.}[x_{\rm max} = x | r] \stackrel{\mu \to 0}{\longrightarrow} \sqrt{\frac{r}{4 N D}} \; f_0\left(x \sqrt{\frac{r}{4 N D}} \right) 
\end{equation}
where
\begin{equation} \label{eq:def-g0}
f_0\left(y \right) = 2 \, y \, e^{-y^2} \;,
\end{equation}
is the normalized scaling function defined on $y > 0$. Interestingly, the same scaling function $f_0(y)$ appears in the problem of $N$ simultaneously resetting noninteracting Brownian motions which we have seen in Section \ref{subsec:BM-simreset}. In that problem it we showed that the distribution of $x_{\max}$ exhibits a scaling form 
 \begin{equation} \label{eq:evs_inde}
{\rm Prob.}[x_{\rm max} = x | r] \underset{N\to\infty}{\longrightarrow} \sqrt{\frac{r}{4  D \ln N}} \; g_0\left(x \sqrt{\frac{r}{4 D \ln N}} \right) \;,
\end{equation}
where $f_0(y)$ is given in (\ref{eq:def-g0}). Hence, both problems exhibit exactly the same scaling function, but the $N$-dependence of the
scale of $x_{\max}$ is different. In the former case, it scales as $\sqrt{N}$, while in the latter case it scales as $\sqrt{\ln N}$. The reason behind the appearance
of the same scaling function $f_0(y)$ can be qualitatively understood as follows. In \RDBM with $\mu = 0$, since the
particles are now untrapped, diffusion will make the semicircle density spread like $\sim \sqrt{t}$. Logarithmic repulsion does not change the diffusive spread but just renormalizes $\sqrt{t}$ to $\sqrt{N \, t}$. Thus, the deterministic part of the maximum behaves as $\sqrt{N\,t}$, which gives rise to the scaling function $f_0(y)$ as shown above. In the case of non-interacting Brownian particles with simultaneous resetting, the deterministic part of $x_{\max}$ behaves as $\sqrt{(\ln N)\,t}$. Although the fluctuating parts are very different in the two problems, the deterministic part scales as $\sqrt{t}$ in both problems, although with different dependent prefactors $N$. When reset is turned on, the dominant contribution to the distribution of $x_{\max}$ comes only from the deterministic part,   
while the stochastic part contributes to subleading behaviors. Since both problems share the same deterministic time-dependent evolution, the leading large $N$ behavior in the presence of resetting is exactly the same in both problems, up to the $N$ dependent rescaling factor.  

\section{Gap statistics} \label{sec:model-gap}

We now turn our attention to the spacings in the gas. The simplest statistical description of the spacing in \RMT is given by Wigner's surmise, which we have described previously. We will follow a similar procedure here in the presence of resetting $r>0$. Our starting point then is the joint distribution of eigenvalues in Eq. (\ref{eq:eigenvalues-jpdf}) which reads, for $N=2$, 
\begin{align} 
{\rm Prob.}&[x_1, x_2 | r ] = \nonumber \\
\times & A_\beta \int_0^{+\infty} r \dd\tau\; e^{-r\tau} \frac{1}{\sigma(\tau)^{\beta + 2}} \exp[ -\frac{\beta}{2\sigma^2(\tau)} (x_1^2 + x_2^2) ] \, |x_1-x_2|^\beta\;, \label{eq:eigenvalues-jpdf_N2}
\end{align}
where $\sigma^2(\tau) = D(1-e^{-2\mu \tau})/\mu$ is given in Eq. (\ref{eq:def-sigma}) and the normalization constant $A_\beta$ is given by
\begin{equation} \label{Ab}
A_\beta = \frac{2}{\sqrt{\pi}}\left(\frac{\sqrt{\beta}}{2} \right)^{\beta + 2} \frac{1}{\Gamma\left( \frac{1+\beta}{2}\right)} \;.
\end{equation} 
The distribution of the spacing $\Delta = |x_1 - x_2|$ is then given by
\begin{equation} \label{p_gap1}
{\rm Prob.}[\Delta = s \vert r] = \int {\rm Prob.}[x_1, x_2 | r ] \; \delta \left[ |x_1-x_2| = s\right] \; \dd x_1 \dd x_2 \;,
\end{equation}
where ${\rm Prob.}[x_1, x_2 | r ]$ is given in Eq. (\ref{eq:eigenvalues-jpdf_N2}). This double integral can be easily done by making the change of variables $x_1+x_2 = y_1$ and $x_1 - x_2 = y_2$. Performing these integrals we get  
{
\begin{equation} \label{eq:renewal-s}
{\rm Prob.}[\Delta = s | r] = \frac{r \sqrt{\beta} }{2^\beta \Gamma\left(\frac{1 + \beta}{2} \right)} \left(s \sqrt{\beta} \right)^\beta  \int_0^{+\infty} \dd \tau \; \frac{e^{-r \tau}}{\sigma(\tau)^{1 + \beta}}  \exp[ - \frac{s^2}{4} \frac{\beta}{\sigma^2(\tau)} ]  \;.
\end{equation}
}
This can be written in a scaling form as
{
\begin{equation} \label{eq:s-scaling}
{\rm Prob.}[\Delta =s \vert r] = \sqrt{\frac{\mu \beta}{D}} F_\beta\left( s \sqrt{\frac{\mu \beta}{D}}, \frac{\mu}{r}  \right) \;,
\end{equation}
}
where 
\begin{align} 
F_\beta(u, \gamma) &= \nonumber \\
&\frac{u^\beta}{2^\beta \Gamma\left(\frac{1 + \beta}{2} \right)} \int_0^{+\infty} \dd T \; e^{-T} \left(\frac{1}{1 - e^{-2\gamma T}}\right)^{\frac{1+\beta}{2}} \exp[ - \frac{u^2}{4(1 - e^{-2\gamma T})}] \;. \label{eq:h-def}
\end{align}
Making the change of variable $v = 1 - e^{-2 \gamma T}$ we can compute this integral explicitly in terms of hypergeometric $\,_1 F_1$ functions, which allow us to easily extract its {\color{blue}asymptotic} behavior
\begin{equation} \label{eq:main-h-asymptotics-small-z}
    F_\beta(u, \gamma) \underset{u \to 0}{\approx} \begin{dcases}
    \left( \frac{\Gamma\left(\frac{1 - \beta}{2}\right) \Gamma\left(1 + \frac{1}{2\gamma}\right) }{2^\beta \Gamma\left(\frac{1 + \beta}{2}\right) \Gamma\left( \frac{1}{2\gamma} + \frac{1 - \beta}{2} \right)} \right)\,u^\beta & \mbox{~~for~~} \beta < 1 \\
     - \frac{u \log u}{2 \gamma} & \mbox{~~for~~} \beta = 1 \\
    \frac{u}{2 \gamma (\beta - 1)} &\mbox{~~for~~} \beta > 1
    \end{dcases} \;.
\end{equation}
On the other hand, for large $u$ and any $\beta>0$, the scaling function behaves as
\begin{equation} \label{eq:main-h-asymptotics-large-z}
F_\beta(u, \gamma) \underset{u \to \infty}\approx \frac{\Gamma\left(1 + \frac{1}{2 \gamma}\right)}{\Gamma\left(\frac{1 + \beta}{2}\right)} \left(\frac{u}{2}\right)^{\beta - 1/\gamma} e^{-u^2/4} \;.
\end{equation} 

\vspace{0.2cm}

\begin{figure}[t]
\centering
\includegraphics[width=0.32\textwidth]{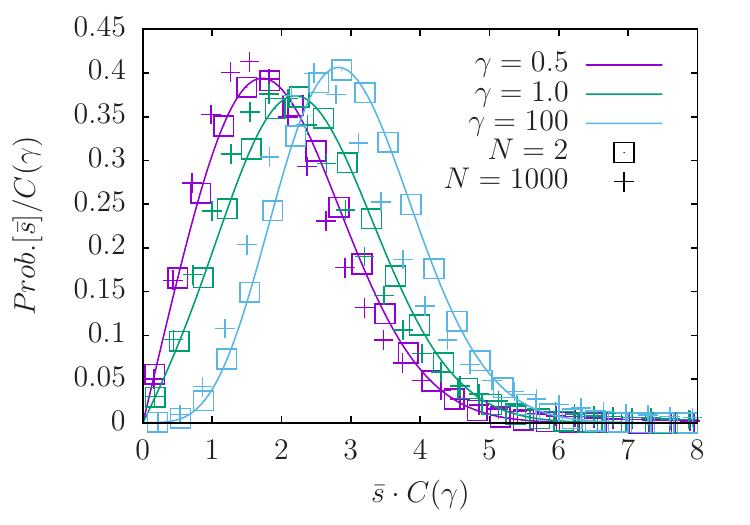}
\hfill
\includegraphics[width=0.32\textwidth]{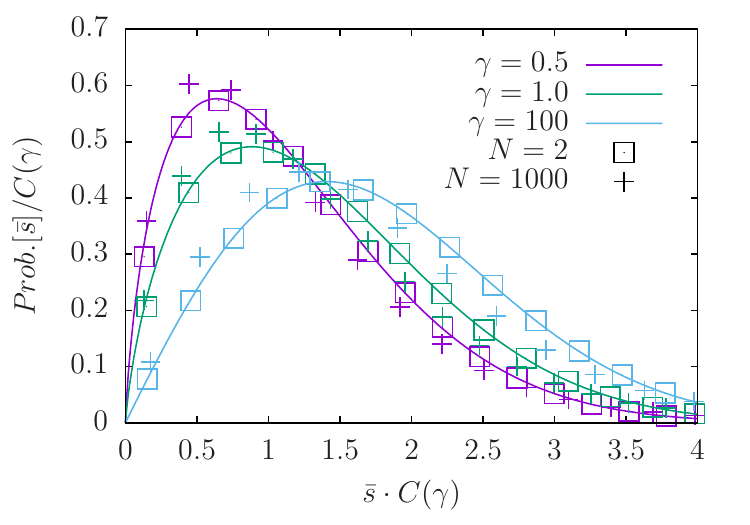}
\hfill
\includegraphics[width=0.32\textwidth]{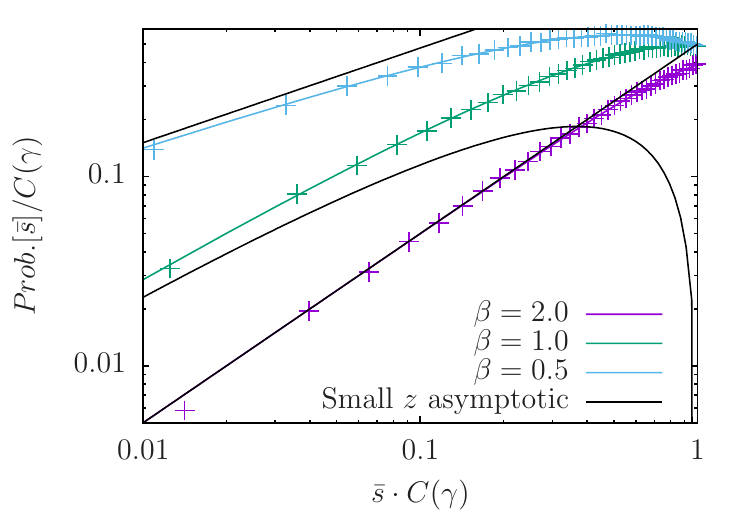}
\caption{{\bf Left panel:} Plot of ${\rm Prob.}[\bar{s}\vert r]$ in Eq. \eqref{eq:s-mean-C} as a function of the scaled distance $\bar{s}\,C(\gamma)$ with $C(\gamma)$ given in Eq. (\ref{eq:def-Cgamma}) and the scaling function $F_{\beta}(u, \gamma)$ given in Eq. (\ref{eq:h-def}) describing the probability density of the gap of the $\beta$-\RDBM for $\beta=4$ in its \NESS with $N=2$. The lines correspond to the analytical prediction given by Eq. (\ref{eq:h-def}) and the points are the results from numerical direct sampling simulations for $N=2$ and for $N=1000$. {\bf Middle panel:} the same as the left panel but with $\beta = 1$. The curves in
the left and middle panels confirm that the Wigner surmise is a good approximation even in the presence of resetting. Different colors correspond to different values of $\gamma$ (0.5, 1 and 100 for purple, green and blue) and different symbols correspond to $N=2$ and $N=1000$. {\bf Right panel:} 
it represents  a zoom of the small-$z$ {\color{blue}asymptotes} given in Eq. (\ref{eq:main-h-asymptotics-small-z}), where we fixed $\gamma = 1$ and the different colors correspond to $\beta = 0.5,1,2$.}\label{fig:gaps}
\end{figure}

\begin{figure}
    \centering
    \includegraphics[width=0.7\textwidth]{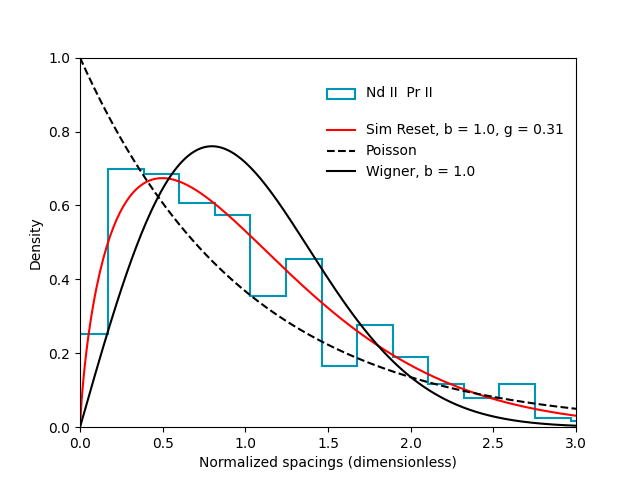}
    \caption{Experimental data (obtained from the Atomic Spectral Database \cite{ASD}) for the level spacings of the ion of Neodymium (Nd II) and the ion of Praseodymium (Pr II) compared to the independent hypothesis (Poisson) or the Wigner surmise for the gaussian orthogonal ensemble ($\beta = 1$) and the gap statistics for a simultaneously resetting log-gas with $\beta = 1$ and $\gamma = 0.31$.} \label{fig:atomic-unusual}
\end{figure}

From Eq. (\ref{eq:s-scaling}) and Eq. (\ref{eq:h-def}) we can obtain the average gap explicitly, yielding
{
\begin{equation} \label{eq:s-mean}
\langle s \rangle = \sqrt{\frac{\pi D}{\mu \beta}} \frac{\Gamma\left(1 + \frac{\beta}{2}\right)\Gamma\left( 1 + \frac{1}{2\gamma} \right)}{\Gamma\left(\frac{1 + \beta}{2}\right) \Gamma\left( \frac{3}{2} + \frac{1}{2 \gamma} \right)}
\end{equation}
}
Hence the distribution of the scaled spacing $\bar{s} = s / \langle s \rangle$ is given by
\begin{equation} \label{eq:s-mean-C}
{\rm Prob.}[\bar{s}\vert r] = C(\gamma) \, F_\beta\left( \bar{s} \, C(\gamma), \, \frac{\mu}{r} \right) \;,
\end{equation}
where 
\begin{equation}\label{eq:def-Cgamma}
C(\gamma) = \sqrt{\pi} \frac{\Gamma\left(1 + \frac{\beta}{2}\right)\Gamma\left( 1 + \frac{1}{2\gamma} \right)}{\Gamma\left(\frac{1 + \beta}{2}\right) \Gamma\left( \frac{3}{2} + \frac{1}{2 \gamma} \right)} \;,
\end{equation}
and the scaling function $F_\beta(u,\gamma)$ is given in Eq. (\ref{eq:h-def}). Once again, for some special values of $\gamma$ the scaling function simplifies. Indeed for $\gamma = 1/2$ we have
\begin{equation} \label{eq:h-gamma-half}
F_\beta(u, 1/2) = \frac{u}{2} \frac{\Gamma\left(\frac{\beta - 1}{2}, \frac{u^2}{4}\right)}{\Gamma\left( \frac{1 + \beta}{2} \right)} \;,
\end{equation}
where $\Gamma(a, z)$ is the incomplete $\Gamma$-function. For $\gamma = 1$ and $\beta = 1$ we get
\begin{equation} \label{eq:h-gamma-1}
F_{\beta = 1}(u, 1) = \frac{u}{4} e^{-u^2/8} \, K_0(u^2 / 8) \;,
\end{equation}
where $K_0(z)$ is the modified Bessel function of index $0$. Furthermore, for $r \to 0$, i.e., for $\gamma \to +\infty$, we recover the original form of Wigner's surmise
\begin{equation} \label{eq:h-gamma-inf}
F_\beta(u, +\infty) = \frac{1}{2^\beta \Gamma\left( \frac{1 + \beta}{2} \right)} u^\beta e^{- u^2/4} \; .
\end{equation}
The scaling function in Eq. (\ref{eq:s-mean-C}) is plotted in Fig. \ref{fig:gaps} and is in perfect agreement with numerical simulations when $N = 2$. As is the case in random matrix theory, that is, for $r=0$, the $N = 2$ result remains a good approximation of the scaled spacing distribution for $N>2$. As can be seen in Fig. \ref{fig:gaps}, as one increases $r$, that is, decreases $\gamma$, the peak of the distributions moves towards a smaller value of $\bar{s}$, indicating that the typical gap decreases with increasing $r$. This is consistent with the physical expectation that resetting generates effective attractive correlations between particles in the \NESS and thus decreases the typical spacing between particles. This can also be seen from the small $u$ {\color{blue}asymptotes} in Eq. (\ref{eq:main-h-asymptotics-small-z}). As $u \to 0$, the gap distribution vanishes as $\sim u^\beta$ for $\beta < 1$, as $-u\log u$ for $\beta = 1$ and linearly for $\beta > 1$. This should be compared with the case $r=0$ \cite{M91,F10}, where the distribution disappears as $u^\beta$ as $u \to 0$ for all $\beta>0$. Hence, for $r>0$, the small gaps have a larger probability of occurring, reflecting the attractive correlations generated by resetting. Furthermore, we can tune this competition between attraction and repulsion by playing with the $\gamma$ parameter. A large $\gamma$ leads to weaker attractive correlation whereas a smaller $\gamma$ corresponds to stronger attractive correlations. {\color{blue}Using this extra expressive power we can fit better some real-world behaviors. As a toy example we show how this gap function can fit the atomic level spacings of the ion of Neodymium (Nd II) and the ion of Praseodymium (Pr II) better than the standard Wigner surmise in Fig. \ref{fig:atomic-unusual}.}

\vspace*{0.3cm}
\noindent{\bf The limit $\mu \to 0$.} As in the previous cases, when we take the limit $\mu \to 0$ to study the gap distribution for the untrapped gas, we see from Eq. (\ref{eq:s-scaling}) that we have to simultaneously take the limits $\gamma \to 0$, $u \to 0$ while keeping the ratio 
$u^2 / \gamma =  r s^2 \beta / D$ fixed in Eqs. (\ref{eq:s-scaling}) and (\ref{eq:h-def}). This yields
{
\begin{equation} \label{eq:gap-mu0}
{\rm Prob.}[\Delta = s | r] = \sqrt{\frac{r \beta}{2 D}} \; F_0\left( s \sqrt{\frac{r \beta}{2 D}} \right) \;,
\end{equation}
}
where
\begin{equation} \label{eq:h0}
F_0(y) =  \frac{y^{\frac{1 + \beta}{2}}}{2^{\frac{\beta - 1}{2}} \Gamma\left( \frac{1+\beta}{2} \right)} K_{\frac{\beta - 1}{2}}\left( y \right) \;,
\end{equation}
where $K_\nu(z)$ is the modified Bessel function of index $\nu$.  

\vspace*{0.3cm}
Similarly, in the untrapped limit $\mu \to 0$, the scaled PDF of the level-spacing, in units of mean level-spacing $\langle s \rangle$, is given by
\begin{equation} \label{eq:mean-gap-mu0}
{\rm Prob.}[\bar{s} \vert r] = \sqrt{\pi} \frac{\Gamma\left(1 + \frac{\beta}{2}\right)}{\Gamma\left( \frac{1 + \beta}{2} \right)} \, F_0\left( \sqrt{\pi} \frac{\Gamma\left(1 + \frac{\beta}{2}\right)}{\Gamma\left( \frac{1 + \beta}{2} \right)} y \right) \;.
\end{equation}
The asymptotic behavior of the scaling function $h_0(y)$ as $y \to 0$ is different for $\beta = 1$ and $\beta = 2,4$. One finds 
\begin{equation} \label{eq:h0-asymptotics-small-z}
F_0(y)  \underset{y \to 0}{\approx} \begin{dcases}
\left( \frac{\Gamma\left(\frac{1 - \beta}{2}\right)}{2^\beta \Gamma\left( \frac{1 + \beta}{2} \right)} \right)\,y^\beta \quad &{\rm for} \quad \beta < 1\\
- y \log y & {\rm for}\quad \beta = 1 \\
\frac{y}{\beta - 1} &{\rm for} \quad \beta >1 
\end{dcases} \;.
\end{equation}
The large-$y$ asymptotic behavior is given by
\begin{equation} \label{eq:h0-asymptotics-large-z}
F_0(y) \underset{y \to \infty}{\approx}  \sqrt{\frac{\pi}{2^\beta}} \frac{1}{\Gamma\left( \frac{1 + \beta}{2} \right)} y^{\beta/2} e^{-y} \;.
\end{equation}
Comparing Eq. (\ref{eq:h0-asymptotics-small-z}) and Eq. (\ref{eq:h0-asymptotics-large-z}) with Eq. (\ref{eq:main-h-asymptotics-small-z}) and Eq. (\ref{eq:main-h-asymptotics-large-z}) we see that untrapping the gas has not changed the small $y$ asymptotic behavior (apart from a prefactor) but the large $y$ {\color{blue}asymptotes} are significantly different: the Gaussian tail for $\mu > 0$ changes to a much slower exponential tail when $\mu \to 0$. Thus the right tail is significantly stretched in the absence of a confining potential. Physically this is what we would expect, since the gas is now un-trapped it can reach configurations with possibly much larger gaps.

\section{Full counting statistics} \label{sec:model-fcs}

In this section, we compute the distribution of $N_L$ that denotes the number of particles in the  interval $[-L, L]$ in the NESS induced by a nonzero resetting rate $r>0$. Clearly $N_L$ is a random variable that fluctuates from one sample to another sample of the gas. This random variable can be expressed as 
\begin{equation} \label{indic}
N_L = \sum_{i=1}^N {\mathbb I}_{x_i \in [-L,L]} \;,
\end{equation}
where ${\mathbb I}_{x_i \in [-L,L]}$ is an indicator function, which takes the value ${\mathbb I}_{x_i \in [-L,L]} = 1$ if the $i$-th particle is inside the interval $[-L,L]$ and ${\mathbb I}_{x_i \in [-L,L]} = 0$ otherwise. Therefore the probability distribution of $N_L$ is given by
\begin{equation} \label{FCS_1}
{\rm Prob.}[N_L = M \vert r] = \int \dd x_1 \cdots \dd x_N \delta \left[ M - \sum_{i=1}^N {\mathbb I}_{x_i \in [-L,L]} \right] {\rm Prob.}[x_1, \cdots, x_N \vert r]
\end{equation}
where $ {\rm Prob.}[x_1, \cdots, x_N \vert r]$ is given in Eq. (\ref{eq:eigenvalues-jpdf}). Interchanging the integrals over $\tau$ and the $x_i$'s one can re-write Eq.~(\ref{FCS_1}) as
\begin{align} \label{FCS_2}
{\rm Prob.}&[N_L = M \vert r] = r\int_0^\infty \dd \tau  e^{-r \tau} \int \dd x_1 \cdots \int \dd x_N \delta\left[M -  \sum_{i=1}^N {\mathbb I}_{x_i \in [-L,L]}\right]\nonumber \\
\times &\frac{1}{Z_N(\beta)} \frac{1}{\sigma(\tau)^{N + \beta N(N-1)/2}} \exp[ -\frac{\beta}{2}\left(\frac{1}{\sigma^2(\tau)} \sum_{i = 1}^N x_i^2 - \sum_{i \neq j} \ln |x_i - x_j| \right)] \nonumber \\
&\qquad\qquad\quad=   r\int_0^\infty \dd \tau  \,e^{-r \tau} \, {\rm Prob.}[N_L(\tau) = M ; \tau] \;,
\end{align}  
where we have identified the integrals over the $x_i$'s as the probability distribution ${\rm Prob.}[N_L(\tau) = M ; \tau]$ of the number of particles $N_L(\tau)$ in the interval $[-L,L]$ in the Coulomb gas with energy in Eq. (\ref{eq:def-Z-E}) parametrized by $\tau$ that appears inside the variance $\sigma^2(\tau) = D(1-e^{-2 \mu \tau})/\mu$. This full counting statistics ${\rm Prob.}[N_L(\tau) = M ; \tau]$ of the Coulomb gas has been studied in the random matrix theory literature~\cite{DM63,CL95,FS95,MMSV14,CLM15,MMSV16}. In particular, the distribution has a peak around its mean $\langle N_L(\tau)\rangle = O(N)$ and it fluctuates around this mean over a scale $O(\sqrt{\ln N})$. In other words, the random variable $N_L(\tau)$ in the Coulomb gas with parameter $\tau$ can be expressed, for large $N$, as 
\begin{equation} \label{FCS_3}
N_L(\tau) \approx \langle N_L(\tau)\rangle + \sqrt{\ln N} \, W \;,
\end{equation}
where $W$ is an $N$-independent, but $\ell$-dependent, random variable with magnitude of order $O(1)$. This is because the log-gas is very rigid: in fact, this is an example of hyperuniformity~\cite{T18} where the fluctuations scale less than linearly as $N \to \infty$. The mean value $ \langle N_L(\tau)\rangle$ can be computed explicitly limit as
\begin{equation} \label{eq:def-FCS}
\langle N_L(\tau) \rangle = N \int_{-L}^{L} \dd x \; \rho_N(x | r = 0, \tau) \;,
\end{equation}
where the average density $\rho_N(x | r = 0, \tau)$ of the Coulomb gas with parameter $\tau$ is given explicitly (for large $N$) by the Wigner semi-circular form 
in Eq. (\ref{eq:free-density}). It is convenient to first introduce the dimensionless length $\ell$ as
\begin{equation}\label{eq:def-ell}
\ell = L \sqrt{\frac{\mu}{N D}} \;.
\end{equation}
Since, for large $N$, the gas is confined on the interval $z \in [-\sqrt{2},+\sqrt{2}]$, we will henceforth consider only $\ell \in [0, \sqrt{2}]$. For $\ell > \sqrt{2}$, $N_L(\tau)$ saturates to $N$ with probability one and there are no fluctuations. Therefore FCS have a nontrivial behavior only for $\ell < \sqrt{2}$. 

\begin{figure}[t]
    \centering
    \includegraphics[width = 0.55 \linewidth]{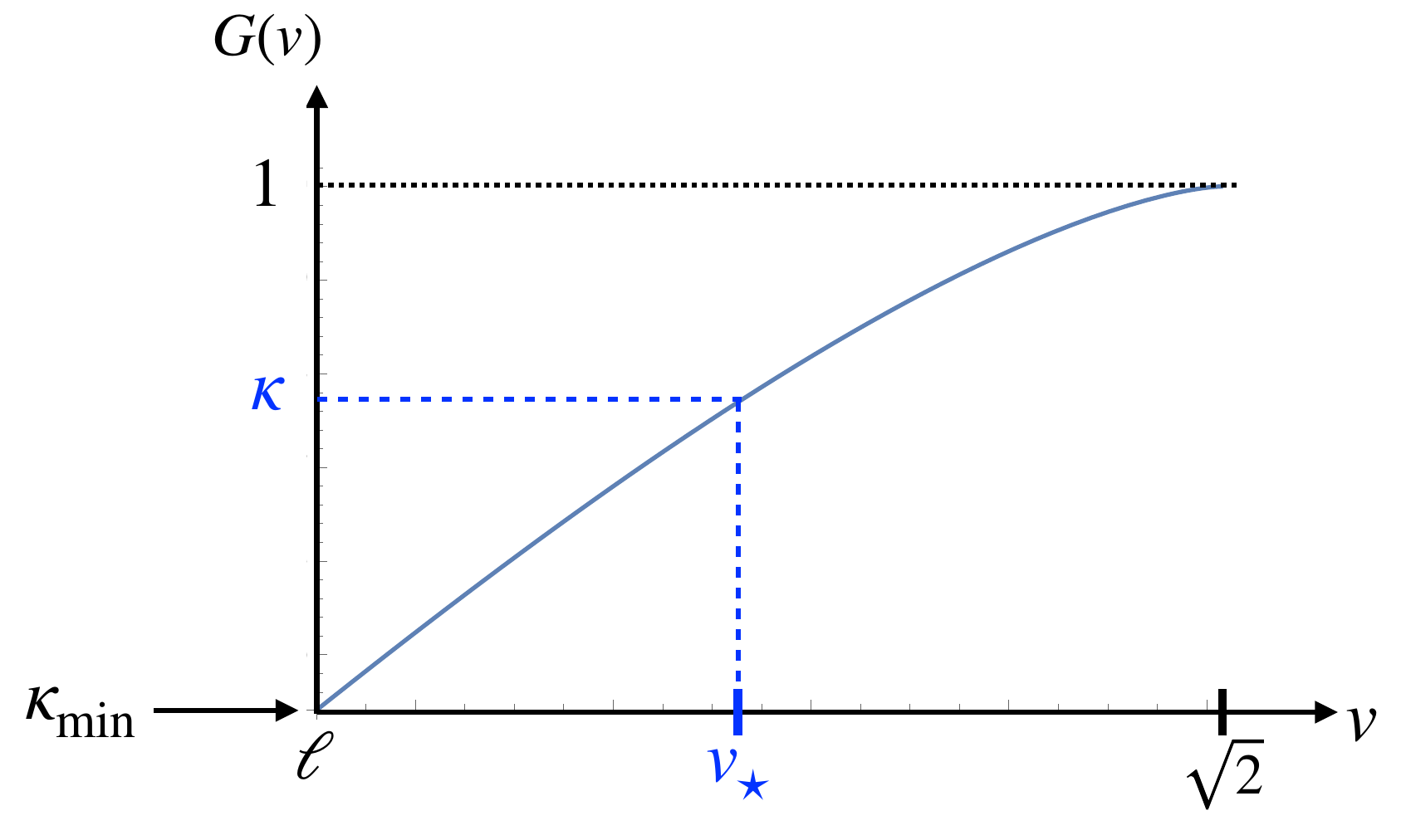}
    \caption{Plot of the function $G(v)$ given in Eq. (\ref{eq:def_G}). For a given $\kappa$, one can then read off $v_\star$ such that $G(v_\star) = \kappa$ as stated in Eq. (\ref{eq:v-eq}). The lowest allowed value $\kappa_{\min}$ is given in Eq. (\ref{eq:def-zmin}).}\label{Fig_G}
\end{figure}

\begin{figure}[t]
\centering
\includegraphics[width=0.32\textwidth]{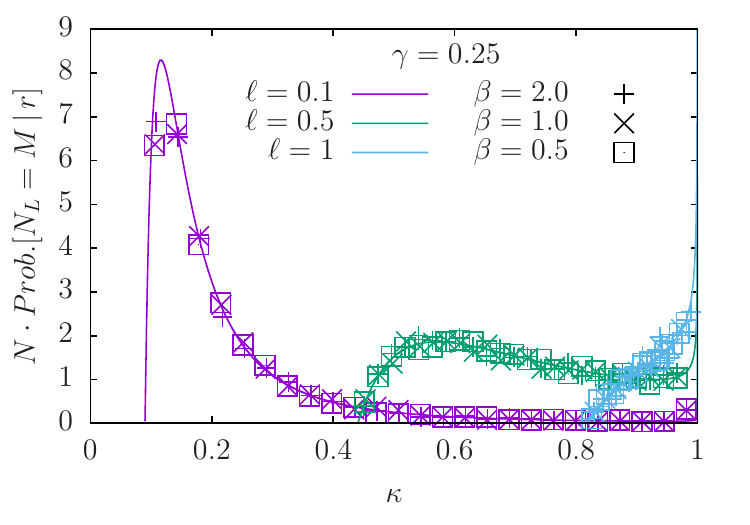}
\hfill
\includegraphics[width=0.32\textwidth]{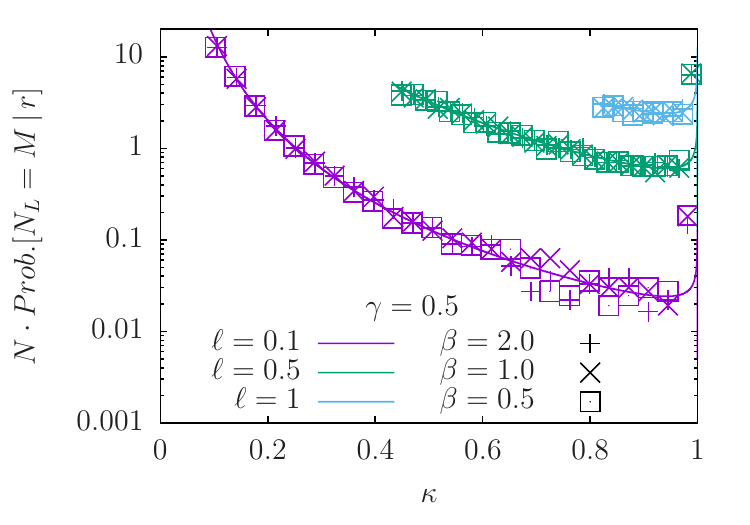}
\hfill
\includegraphics[width=0.32\textwidth]{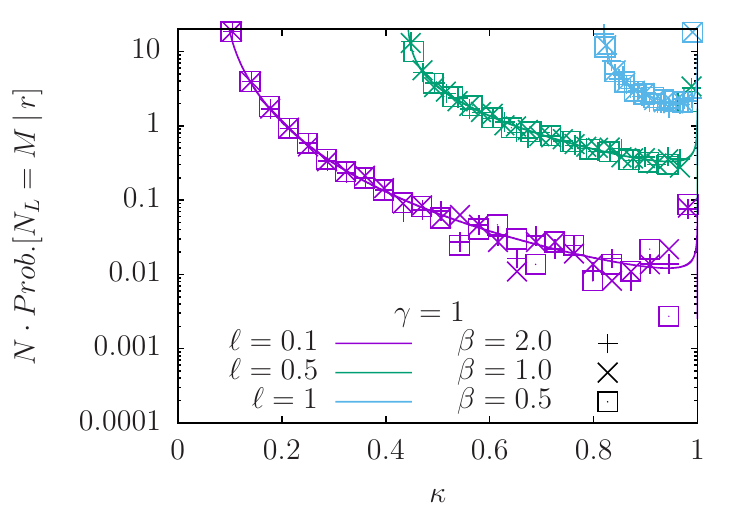}
\caption{Plot of the scaling function $q(\kappa, \ell, \gamma)$ vs $\kappa$ given in Eq. (\ref{eq:d-def-2}) describing the full counting statistics, i.e., the probability density function of the number of particles in the interval $[-L, L]$. In this plot, we only show the smooth part of the scaling function in (\ref{eq:d-def-2}), i.e., without the delta function. 
The lines are the analytical predictions given by Eq. (\ref{eq:d-def-2}) while the points are obtained from numerical direct sampling simulations. Different colors correspond to different values of $\ell$ (0.1, 0.5 and 1 for purple, green and blue) and different symbols correspond to different values of $\beta = 0.5,1,2$. The different panels correspond to different values of $\gamma = 1/4, 1/2, 1$ from left to right.}\label{fig:fcs}
\end{figure}

\vspace*{0.3cm}
Noting that the semicircular density in Eq. (\ref{eq:free-density}) has a finite support over $x \in [-\sqrt{2N\sigma^2(\tau)}, + \sqrt{2N \sigma^2(\tau)}]$, where $\sigma^2(\tau) = D(1-e^{-2\mu \tau})/\mu$, the integral in (\ref{eq:def-FCS}) can be expressed in a compact form as
\begin{equation} \label{NL_1}
\frac{\langle N_L(\tau)\rangle}{N} \approx G\left( v = \frac{\ell}{\sqrt{1-e^{-2 \mu \tau}}}\right) \;,
\end{equation}
where the scaling variable $v \geq \ell$ necessarily and the function $G(v)$ is given by (see also Fig. \ref{Fig_G} for a plot of this function)
\begin{equation} \label{eq:def_G}
G(v) =
\begin{cases} 
&\frac{1}{\pi} v \sqrt{2-v^2} + \frac{2}{\pi} {\rm ArcTan}\left( \frac{v}{\sqrt{2-v^2}}\right) \quad, \quad \ell \leq v < \sqrt{2} \\
& \\
& 1 \quad, \quad \hspace*{5.1cm} v > \sqrt{2} \;.
\end{cases}
\end{equation}
We have thus determined the first term in Eq. (\ref{FCS_3}) (which is ``deterministic'') explicitly for large $N$. The fluctuating second term (of order $O(\sqrt{\ln N})$) is much smaller compared to this ``deterministic'' term (of order $O(N)$). Hence to leading order for large $N$ we neglect the fluctuating term and express the distribution of $N_L(\tau)$ by
\begin{equation} \label{delta_NL}
{\rm Prob.}[N_L(\tau) = M ; \tau] \approx \delta\left( M - \langle N_L(\tau)\rangle \right) \;.
\end{equation}
Substituting this into Eq. (\ref{FCS_2}) we get, for large $N$,
\begin{equation} \label{FCS_4}
{\rm Prob.}[N_L = M \vert r] \approx r\int_0^\infty \dd \tau  \,e^{-r \tau} \, \delta\left( M - \langle N_L(\tau)\rangle \right) \;.
\end{equation}
We now substitute the explicit expression for $\langle N_L(\tau) \rangle$ in Eq. (\ref{NL_1}) into (\ref{FCS_4}) to write
\begin{align} 
{\rm Prob.}&[N_L = M | r] = \nonumber\\
\times &\frac{r}{N} \int_{I_1} \dd \tau \; e^{-r \tau} \delta\left(\frac{M}{N} - 1\right) + \frac{r}{N} \int_{I_2} \dd \tau \; e^{-r \tau} \delta\left(\frac{M}{N} - \frac{1}{N}\langle N_L(\tau) \rangle\right)  \;, \label{eq:2-interval-split}
\end{align}
where the integration domains are $I_1 = \left[0,-\ln(1-\ell^2/2)/(2 \mu)\right]$ and $I_2 = \left[-\ln(1-\ell^2/2)/(2 \mu),+\infty\right)$. Substituting Eq. (\ref{eq:def_G}) we can express the distribution as 
\begin{equation} \label{eq:fcs-scaling}
{\rm Prob.}[N_L =M | r] \approx \frac{1}{N} \, H\left( \frac{M}{N}, L \sqrt{\frac{\mu}{N D}}, \mu/r \right) \;,
\end{equation}
where the scaling function $H(\kappa, \ell, \gamma)$ has support over $\kappa \in [\kappa_{\rm min}, 1]$ and is given by
\begin{align} 
H(\kappa, \ell, \gamma) = \;&\delta(\kappa - 1) \left[1 - \left(1 - \frac{\ell^2}{2}\right)^{\frac{1}{2 \gamma}}\right] \nonumber \\
&\qquad+ \frac{\ell^2}{\gamma} \int_{\ell}^{\sqrt{2}} \frac{\dd v}{v^3} \left(1 - \frac{\ell^2}{v^2} \right)^{\frac{1}{2\gamma}-1} \delta\left( \kappa - G(v)\right)\;, \label{eq:d-def}
\end{align}
and
\begin{equation}\label{eq:def-zmin}
\kappa_{\rm min} = G(\ell) =\frac{1}{\pi} \ell \sqrt{2 - \ell^2} + \frac{2}{\pi} {\rm ArcTan}\left(\frac{\ell}{\sqrt{2 - \ell^2}}\right) \;.
\end{equation}
The lower limit $\kappa_{\rm min}$ comes from the limit $v \to \ell$ of the integrand in Eq. (\ref{eq:d-def}). To perform this integral, we look for the root $v_{\star} \equiv v_{\star}(\kappa)$ of the equation (see Fig. \ref{Fig_G})
\begin{equation} \label{eq:v-eq}
\kappa = G(v) \;,
\end{equation}
in terms of which the scaling function $H(\kappa, \ell, \gamma)$ in Eq. (\ref{eq:d-def}) can be expressed as
\begin{align} \label{eq:d-def-2}
H(\kappa, \ell, \gamma) = \delta(\kappa - 1) \left[1 - \left(1 - \frac{\ell^2}{2}\right)^{\frac{1}{2 \gamma}}\right] + \frac{\pi \ell^2}{2 \sqrt{2} \, \gamma \, v_{\star}^3} \frac{1}{\sqrt{1-\frac{v_{\star}^2}{2}}} \left(1 - \frac{\ell^2}{v_{\star}^2} \right)^{\frac{1}{2 \gamma}-1} \;,
\end{align}
where we recall that $\ell < v_{\star}(\kappa) < \sqrt{2}$. Note that this scaling function $q(\kappa, \ell, \gamma)$ is normalized to unity, i.e., $\int_{\kappa_{\min}}^1 \dd \kappa \, H(\kappa, \ell, \gamma) = 1$ and is independent of $\beta$. One can easily work out the asymptotic behaviors of $H(\kappa, \ell, \gamma)$ as $\kappa$ approaches the two edges $\kappa = \kappa_{\min}$ and $\kappa = 1$. Skipping details, one can show that they are given by 
\begin{eqnarray} \label{eq:main-left-asymptotic-d}
    H(\kappa, \ell, \gamma) \approx
    \begin{cases}
    \dfrac{1}{2\gamma} \left( \dfrac{\pi}{\ell\sqrt{2 - \ell^2}} \right)^{\frac{1}{2\gamma}} (\kappa-\kappa_{\min})^{\frac{1}{2\gamma} - 1} \quad, &\kappa \to \kappa_{\rm min} \;,\\
    \\
    \Bigg\{\delta(\kappa - 1) \left[1 - \left(1 - \frac{\ell^2}{2}\right)^{\frac{1}{2 \gamma}}\right] \\
    \, & \kappa \to  1 \;.\\
    \quad + \dfrac{\pi^{2/3}}{4 \gamma \, 6^{1/3}} \ell^2 \left(1 - \frac{\ell^2}{2}\right)^{\frac{1}{2 \gamma} - 1} \left({1 - \kappa}\right)^{-1/3} \Bigg\}\quad, \\
    \end{cases} 
    \end{eqnarray}
We note in particular that the behavior of the scaling function 
$H(\kappa, \ell, \gamma)$ changes at $\gamma = 1/2$. For $\gamma < 1/2$ it vanishes as $\kappa \to \kappa_{\min}$, while it diverges for $\gamma > 1/2$.  
A plot of this scaling function is given in Fig. \ref{fig:fcs} and is in perfect agreement with numerical simulations. 

\chapter{Resetting for search processes} \label{ch:search}
\section{Resetting as an efficient search process} \label{sec:intro-search}

We have previously seen in Section \ref{sec:intro-resetting} that resetting a Brownian walker at a constant rate improves its mean {\color{blue}first-passage} time to an arbitrary target. The natural question is under which conditions is resetting a useful strategy? One can easily imagine that if we have a ballistic searcher pointing straight towards the target, resetting can only be a hindrance. So, where do we draw the line? The usual quantity to look at to determine whether resetting is useful is the mean {\color{blue}first-passage} time $\langle t_f(r, L) \rangle$ that we previously introduced in Section \ref{sec:intro-resetting}. Although some recent arguments were made \cite{GVPPM23} indicating that in some cases one might need to consider higher moments of the {\color{blue}first-passage} time, such as the variance. For our purposes, we will stick to the mean {\color{blue}first-passage} time. Secondly, we will restrict ourselves to Poissonian resetting with rate $r$ since this was the only resetting protocol considered in this thesis. Entirely separate questions can be asked when modifying the resetting protocol such as: if the waiting time is not Poissonian, then what is the best distribution for the resetting time \cite{PKE16, ER24}? or if we keep a Poissonian resetting but make the rate spatially dependent $r \equiv r(x)$ what is the best resetting profile $r(x)$ that optimizes the search \cite{GVPPM23,EMS20, GJ22}? All of these are interesting questions, but out of the scope of this thesis. We refer to Ref. \cite{EMS20, GJ22} for a more general introduction to the subject. 

\vspace{0.2cm}

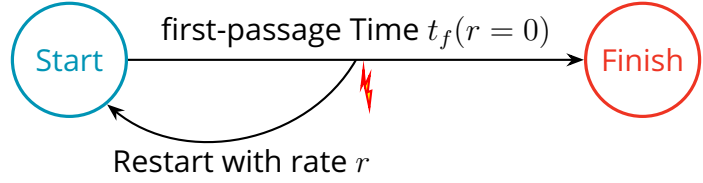
\begin{figure}
    \centering
    \begin{tikzpicture}[node distance=4cm, >=Stealth, thick]
        \node[circle, draw=C4, very thick, minimum size=1.5cm, text=C4] (start) {Start};
        \node[circle, draw=C2, very thick, minimum size=1.5cm, right=6cm of start, text=C2] (finish) {Finish};

        \path (start) -- (finish) coordinate[midway] (mid);

        \draw[->, thick] (start) -- (finish) node[midway, above, text=black] {{\color{blue}first-passage} Time $t_f(r = 0)$};

        \draw[->, thick, bend left=50] (mid) to node[midway, below, text=black, align=center] {Restart with rate $r$} (start);

        \begin{scope}[shift={(mid)}, xshift=5pt, yshift=-18pt, scale=0.1, rotate=20]
            \fill[yellow] (0,0) -- (1,2) -- (0.5,2) -- (1.5,4) -- (0.5,3) -- (1,5) -- (0,3) -- (0.5,3.5) -- cycle;
            \draw[red, thick] (0,0) -- (1,2) -- (0.5,2) -- (1.5,4) -- (0.5,3) -- (1,5) -- (0,3) -- (0.5,3.5) -- cycle;
        \end{scope}

    \end{tikzpicture}
    \caption{A sketch of how resetting influences a general search process. Any search process start from a given starting configuration, represented by the blue node on the left, and tries to reach a target configuration, represented by the red node on the right. According to the internal dynamics of the search process the un-interrupted search process will terminate after some {\color{blue}first-passage} time $t_f$ characterizing the time it took to reach the target configuration for the first time. In the presence of resetting the search may be interrupted (represented by the red lightning bolt) and the process restarts from its initial configuration. This happens with Poissonian rate $r$. } \label{fig:resetting-general-process}
\end{figure}

In this Section, we will introduce some known results from the resetting literature \cite{S16} which set a clear criterion for when can Poissonian resetting expedite a search process. Although we focus only on Poissonian resetting we do not impose any constraints on the search process{\color{blue}, i.e. it could be Brownian, ballistic, Lévy flights or any other stochastic process one may imagine}. Consider {\color{blue} this} arbitrary search process starting from a configuration $\vec{x}_S$ and trying to reach a target configuration $\vec{x}_T$. The search process is characterized by the {\color{blue}first-passage} time $t_f(\vec{x}_S, \vec{x}_T, r = 0)$ which denotes the time it takes for the process to reach $\vec{x}_T$ for the first time, having started from $\vec{x}_S$ at time $t = 0$, in the absence of resetting. We will keep $\vec{x}_S$ and $\vec{x}_T$ fixed and hence they won't play a role in the following, so to lighten the notation we will write $t_f(\vec{x}_S, \vec{x}_T, r) \equiv t_f(r)$. As we have seen previously (see Section \ref{sec:intro-resetting}), to study the {\color{blue}first-passage} time it is useful to introduce the survival probability
\begin{equation}
    Q_r(t) = {\rm Prob.}[t_f(r) > t] \;,
\end{equation}
which represents the probability that the search process has not terminated before time $t$. Then the {\color{blue}first-passage} time is related to the survival probability via
\begin{equation}
    {\rm Prob.}[t_f(r) = t] = - \dv{Q_r(t)}{t} \;,
\end{equation}
since the searches which terminate at $t$ correspond exactly to the searches which `die' at $t$ and hence the mean {\color{blue}first-passage} time is given by
\begin{equation} \label{eq:mfpt-2-survival}
    \langle t_f(r) \rangle = \int_0^{+\infty} \dd t \; t\left(-\dv{Q_r(t)}{t}\right) = \int_0^{+\infty} \dd t\; Q_r(t) \;.
\end{equation}
The survival probability also admits a renewal form
\begin{equation} \label{eq:survival-renewal}
    Q_r(t) = e^{-r t} Q_0(t) + r \int_0^{t} \dd \tau \; e^{-r \tau} Q_0(\tau) Q_r(t - \tau) \;,
\end{equation}
which can be interpreted in the following manner. Either the search does not terminate without doing any resetting, or at least one resetting happened. Then we denote the last resetting event by $t - \tau$ and the search as to fail in the presence of resetting during $[0, t - \tau]$ and it has to fail in the absence of resetting during $[t- \tau, t]$. Taking the Laplace transform of Eq.~(\ref{eq:survival-renewal}) we obtain
\begin{equation} \label{eq:laplace-survival}
    \tilde Q_r(s) = \frac{\tilde Q_0(s + r)}{1 - r \tilde Q_0(s + r)} \;.
\end{equation}
Then using Eq.~(\ref{eq:mfpt-2-survival}) we know the mean {\color{blue}first-passage} time is given by
\begin{equation} \label{eq:mean-mfpt}
    \langle t_f(r) \rangle = \int_0^{+\infty} \dd t \; Q_r(t) = \tilde Q_r(s = 0) = \frac{\tilde Q_0(r)}{1 - r \tilde{Q_0}(r)} \;.
\end{equation}
We can obtain the second moment in a similar fashion by noticing
\begin{equation} \label{eq:mfpt-second-moment}
    \langle t_f^2(r) \rangle = \int_0^{+\infty} t^2 \left(- \dv{Q_r(t)}{t}\right) = 2 \int_0^{+\infty} \dd t\; t Q_r(t) = -\tilde Q_r'(s = 0^+)\;.
\end{equation}
Using Eq.~(\ref{eq:laplace-survival}) we obtain
\begin{equation}
    \langle t_f^2(r) \rangle = 2 \frac{\tilde Q_0'(r)}{(1 - r \tilde Q_0(r))^2}
\end{equation}
and hence the variance of the mean {\color{blue}first-passage} time is given by
\begin{equation} \label{eq:variance-mfpt}
    \sigma^2(r) = \langle t_f^2(r) \rangle - \langle t_f(r) \rangle^2 = \frac{-2 \tilde Q_0'(r) - \tilde Q_0^2(r)}{(1 - r \tilde Q_0(r))^2} \;.
\end{equation}
For our purposes, we define the optimal resetting rate $r^\star$ as the one which minimizes the mean {\color{blue}first-passage} time, we do not take into consideration higher moments as in Ref. \cite{GVPPM23}. Hence $r^\star$ is defined via
\begin{equation}
    \dv{\langle t_f(r)\rangle}{r} \Big|_{r^\star} = 0 \;. 
\end{equation}
Using Eq.~(\ref{eq:laplace-survival}) we obtain that $r^\star$ is defined through the implicit equation
\begin{equation} \label{eq:rstar-relationship}
    -\tilde Q_0'(r^\star) = \tilde Q_0(r^\star)^2 \;.
\end{equation}
Placing Eq.~(\ref{eq:rstar-relationship}) back in Eq.~(\ref{eq:variance-mfpt}) we obtain
\begin{equation}
    \sigma^2(r^\star) = \frac{\tilde Q_0^2(r)}{(1 - r \tilde Q_0(r))^2} = \langle t_f(r^\star) \rangle^2  \;,
\end{equation}
hence the coefficient of variation of the mean {\color{blue}first-passage} time at the optimal resetting rate is unity
\begin{equation}
    {\rm CV}_{t_f(r^\star)} = \frac{\sigma(r^\star)}{\langle t_f(r^\star) \rangle} = 1 \;.
\end{equation}
To know whether resetting is useful we have to check whether introducing resetting lowers the mean {\color{blue}first-passage} time, namely
\begin{equation}
    \dv{\langle t_f(r) \rangle}{r} \Big|_{r = 0^+} \stackrel{?}{<} 0 \;.
\end{equation}
Using the expression of the mean {\color{blue}first-passage} time in Eq.~(\ref{eq:mean-mfpt}) we can reduce the above condition to 
\begin{equation} \label{eq:survival-condition}
    \tilde Q_0'(0) + \tilde Q_0^2 (0) \stackrel{?}{<} 0 \;.
\end{equation}
and the coefficient of variation at $r = 0$ is given by
\begin{equation}
    {\rm CV}_{t_f(r = 0)} = \sqrt{-2 \frac{\tilde Q_0'(0)}{\tilde Q_0(0)^2} - 1} \;.
\end{equation}
Hence we know that for resetting to be useful the reset-free process must have a coefficient of variation larger than 1, i.e. if 
\begin{equation} \label{eq:CV-condition}
    {\rm CV}_{t_f(r = 0)} > 1
\end{equation}
then resetting will improve the search process since the condition in Eq.~(\ref{eq:CV-condition}) is equivalent to the one in Eq.~(\ref{eq:survival-condition}). The condition in Eq.~(\ref{eq:CV-condition}) has a nice physical interpretation. Resetting is useful only if the fluctuations of the {\color{blue}first-passage} time are larger than its mean. This is because resetting will allow the process to discard spurious searches with a very high {\color{blue}first-passage} time in favor of the much shorter paths where the fluctuation acts in our favor. Resetting utilizes these fluctuations of the {\color{blue}first-passage} time to improve the search. However, if the fluctuations are small compared to the mean {\color{blue}first-passage} time, the possible gains of choosing shorter search paths are outweighed by the possibility of killing `good enough' search paths. Hence resetting is not useful anymore. Going back to the original example presented at the beginning of the chapter, a ballistic particle pointing straight towards the target has a fixed mean {\color{blue}first-passage} time and zero variance. Hence this is the worst possible system for resetting, there is no way in which resetting could possibly improve the search since the variance of the reset-free search is null.

\section{Critical numbers of walkers for diffusive search processes with resetting}

This issue of the existence of an optimal $r^*$ has not been addressed
so far, to the best of our knowledge, when the search for the target is
carried out by a team of $N$ searchers with stochastic
resetting. The purpose of this Section is to study the optimal $r^*$ as
a function of $N$ in a simple model of $N$ diffusive searchers on a
line undergoing stochastic resetting at a constant rate $r$. To be
precise, we will consider
$N$ diffusing particles on a line each
with diffusion constant $D$ and starting from the origin,
with the target fixed at $L$. Since the search process is symmetric
with respect to the sign of $L$, we consider only $L>0$ without loss of generality.
For resetting, we will follow
two {distinct} protocols. 

\begin{itemize}

\item {\bf Protocol A.} 
In this protocol, {each one of the $N$ particles} diffuses and resets to the origin
{\it independently} with rate $r$~\cite{EM11PRL}.  
The positions of the particles are thus
{\it uncorrelated} at all times. For a typical schematic representation of the
trajectories, see Fig. 1 (left panel).   

\item {\bf Protocol B.} Here each {one of the $N$ particles} diffuses independently,
but they {all} reset {\it simultaneously} to the origin with rate $r$. This is the model studied in Chapter \ref{ch:sim-reset} in Section \ref{subsec:BM-simreset}. 
This simultaneous resetting makes the particle positions {\it correlated}
at all times $t$. See Fig. 1 (right panel) for typical trajectories under
protocol B.

\end{itemize}

For $N=1$, the two protocols coincide, but they are different for $N>1$.
In protocol $A$, the particles do not interact at all times.
This protocol was first studied in~\cite{EM11PRL} with the initial positions of the
searchers distributed uniformly
with density $\rho$ (that is, the limit $N\to \infty$) on one
side of the target at the origin and the authors
calculated exactly the Laplace transform of the survival probability of the target up to time $t$. From this Laplace transform, the exact asymptotic behavior of survival probability at late times was extracted.
In a recent work~\cite{VAM22}, the two-time correlation function of the
maximum displacement of the particles $N$ (without a target) was studied
numerically. However, \MFPT to a fixed $N>1$ target has not
been studied. Protocol B is the one we have previously described in Section \ref{subsec:BM-simreset}. We showed that in the absence of a target, the system approaches at long times a many-body non-equilibrium stationary state with strong correlations between the positions
of the particles and computed several observables describing the gas of particles. However, \MFPT to a finite target $N>1$ has not been computed for protocol B either.

In this Section, we compute analytically the \MFPT to the target by
$N$ Brownian searchers for both resetting protocols A and B defined
above. For the optimal reset rate $r^*$, we find a rather interesting
and somewhat surprising result for both protocols. 
We show that \MFPT, as a function of the reset rate $r$, exhibits
a unique minimum at $r=r^\star$. However, the optimal value $r^*$ is strictly
positive, that is, resetting is beneficial only for $N\le 7$ in protocol A and
$N\le 6$ in protocol B. When $N\ge 8$ in protocol A or $N\ge 7$ in protocol B,
the optimal reset rate becomes $r^\star=0$. In these cases, \MFPT is a monotonically
increasing function of $r$
with a minimum at $r=0$, which implies that resetting will only increase
the mean search time and therefore is detrimental to the search process. 
To understand the
origin of these two magic numbers $N=7$ and $N=6$ in the two protocols, 
it is convenient to continue analytically our general formula for integer $N$
to real $N$. Following the analytic continuation, we show that
the actual transitions take place, respectively, at $N_c=7.3264773\ldots$
(for protocol A) and $N_c=6.3555864\ldots$ (for protocol B) which
turn out to be the unique roots of two different transcendental equations.

\begin{figure}
\centering
\begin{minipage}[b]{0.49\textwidth}
\centering
\includegraphics[width = \textwidth]{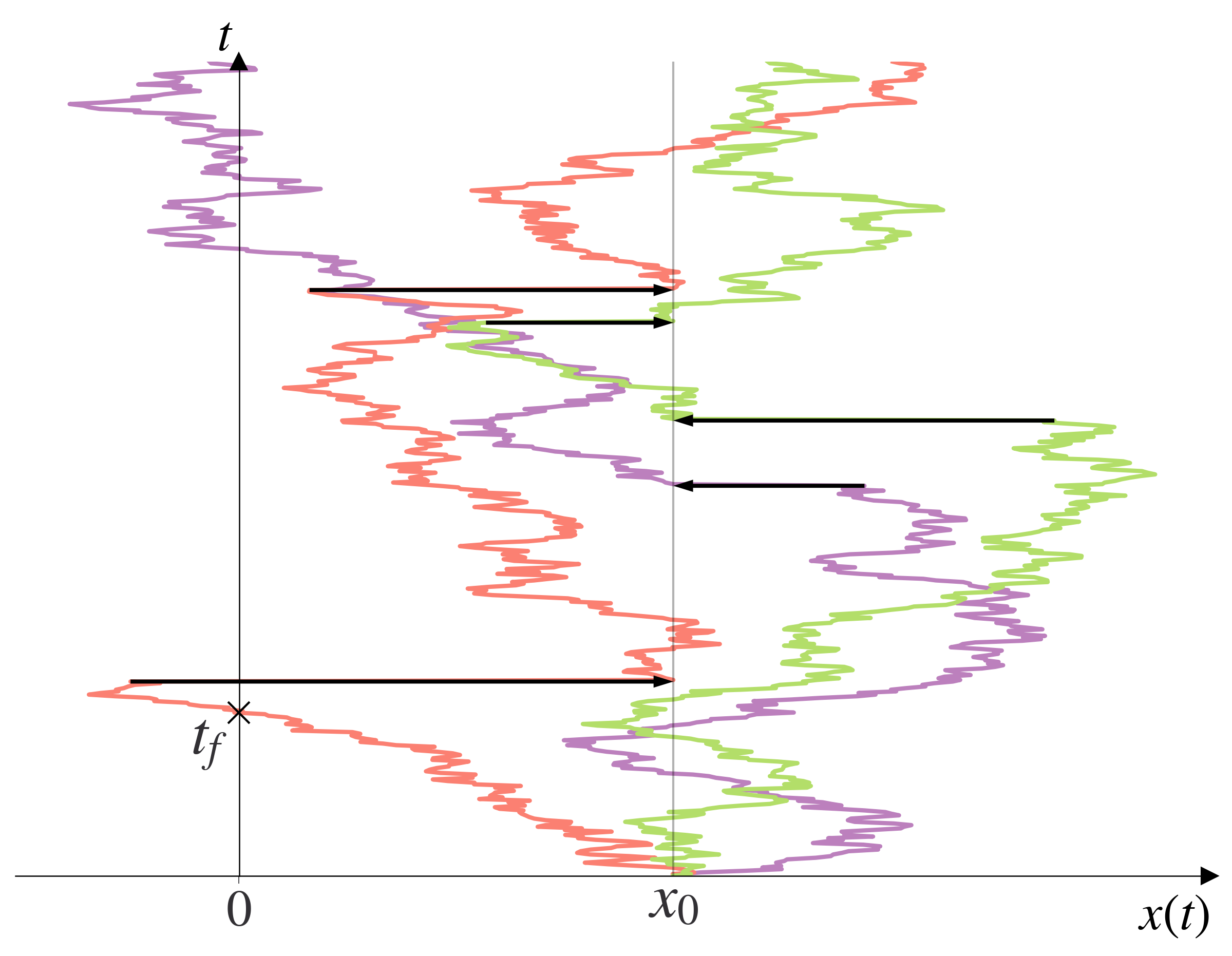}
\end{minipage}\hfill
\begin{minipage}[b]{0.49\textwidth}
\centering
\includegraphics[width = \textwidth]{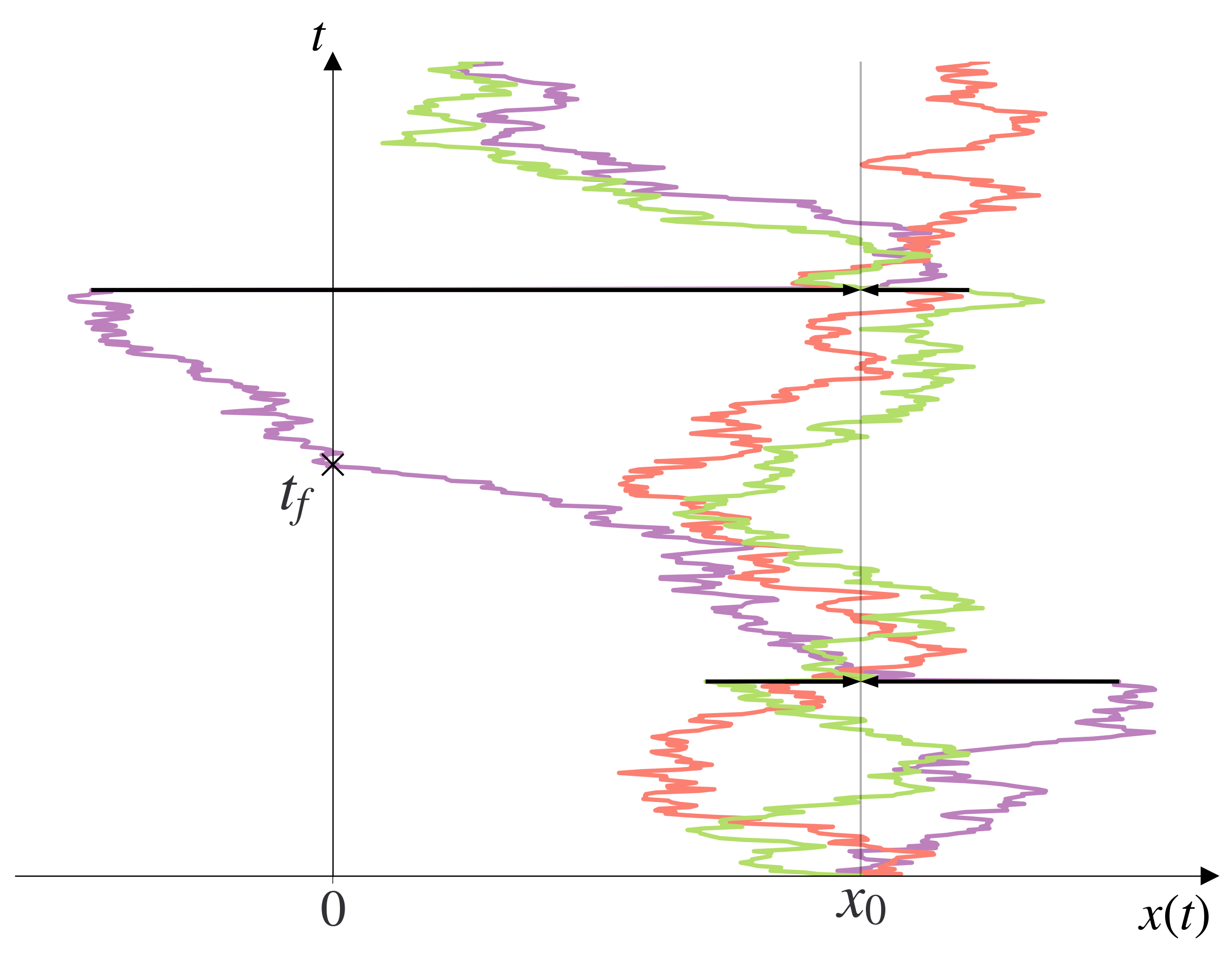}
\end{minipage}
\caption{Typical trajectories for $N = 3$ one-dimensional 
random walkers undergoing {\it independent} 
resetting (protocol A) in the left panel 
and {\it simultaneous} resetting (protocol B) in the right panel. 
Different colors correspond to different walkers and the resetting events 
are shown with full black arrows. {All the} walkers start at $x_0 > 0$ and 
reset to $x_0$ and $t_f$ denotes the first-passage time of the walkers to 
the target located at the origin $x = 0$.} 
\label{fig:sketch}
\end{figure}

\subsection{Mean first-passage time}
\label{sec:MFPT}

We consider $N$ Brownian particles that start at the origin at $t=0$
and undergo stochastic resetting with rate $r$ following either
protocol A or B defined above. We consider a stationary target
at $L > 0$. Whenever any of the $N$ walkers reaches the target,
the search is terminated. We denote by $t_f(r, L, N)$ the first-passage time
to the origin by this $N$-particle process (see Fig. \ref{fig:sketch}). 
Clearly $t_f$ is a random variable, and we will denote the \MFPT by
$\langle t_f(r, L, N)^{(A)} \rangle$ for protocol A and $\langle 
t_f(r, L, N)^{(B)}\rangle$ for protocol B. As was done in Section \ref{sec:intro-search}, in order to compute $\langle 
t_f(r, L, N)^{(A/B)}\rangle$ it is useful to 
consider the cumulative distribution of $t_f$
\begin{equation}
Q_{r, L, N}^{(A/B)}(t)= {\rm Prob.}\left[t_f(r, L, N)^{(A/B)}\ge t\right]\, ,
\label{surv.1}
\end{equation}
known as the survival probability, i.e., the probability
that none of the walkers have reached the target up to time $t$. Identically to Eq.~(\ref{eq:mfpt-2-survival}) the \MFPT can be expressed as
\begin{equation} 
\label{eq:def_Tx}
\langle t_f^{(A/B)}(r, L, N) \rangle = 
\int_0^{+\infty} t \left(-\dv{Q_{r, L, N}^{(A/B)}(t)}{t}\right) \dd t 
= \int_0^{+\infty} Q_{r,L,N}^{(A/B)}(t) \dd t \;,
\end{equation}
We will now treat protocols A and B separately.

\subsection{Protocol A}
\label{subsection:A}

\begin{figure}
\centering
\begin{minipage}[b]{0.49\textwidth}
\centering
\includegraphics[width = \textwidth]{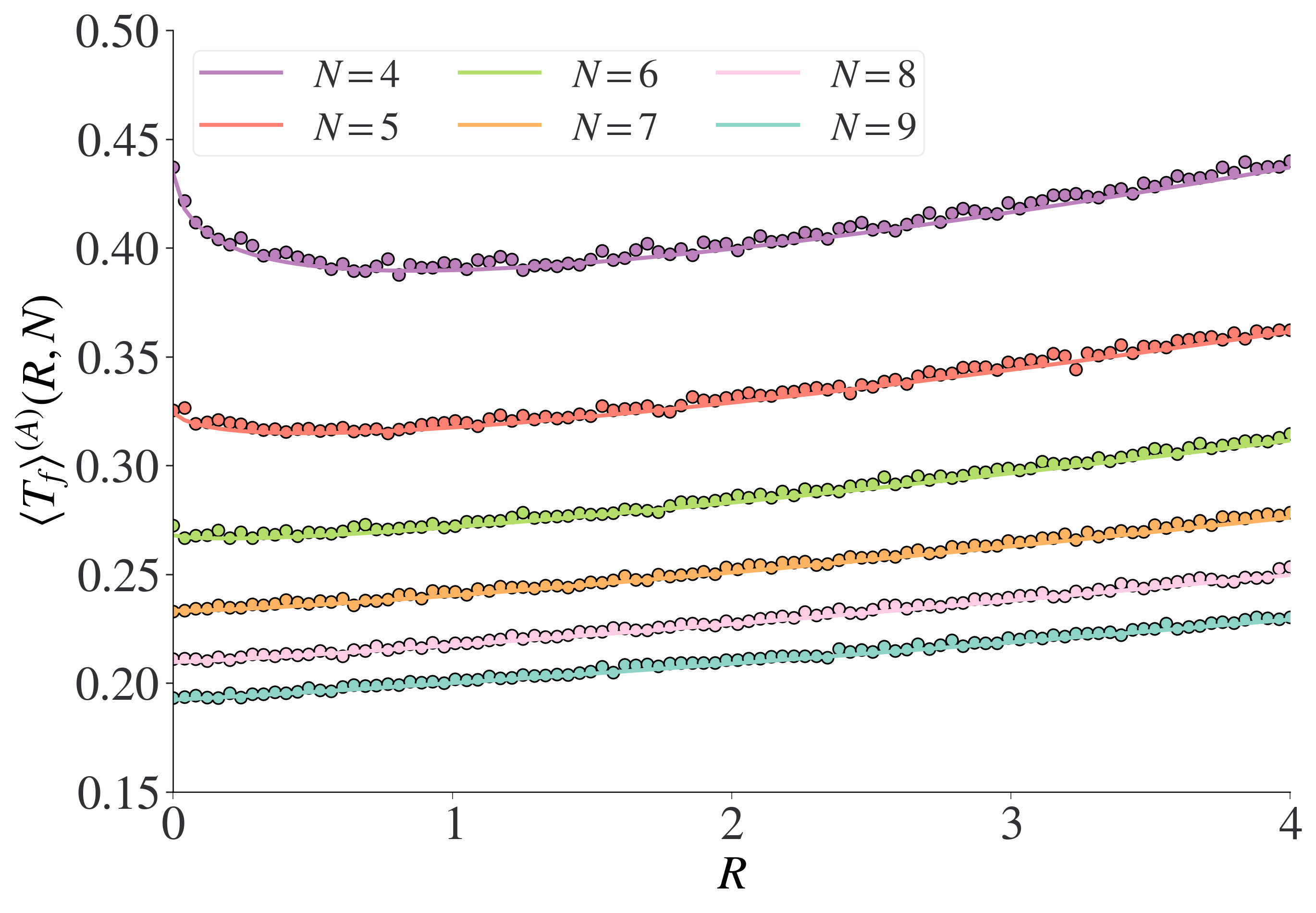}
\end{minipage}\hfill
\begin{minipage}[b]{0.49\textwidth}
\centering
\includegraphics[width = \textwidth]{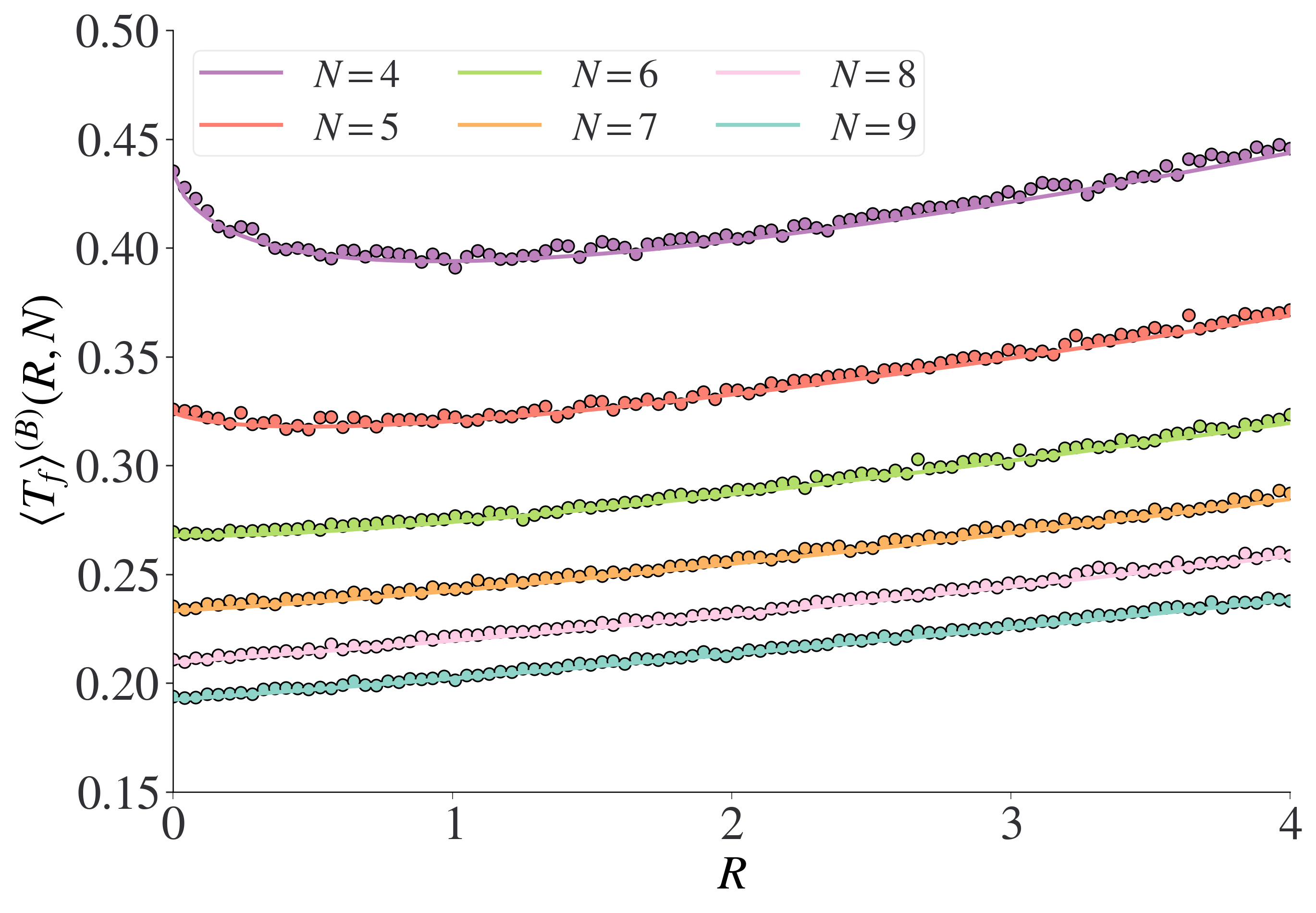}
\end{minipage}
\caption{
Comparison of theoretical and {numerical Langevin} results for the mean 
first-passage time as a function of the resetting rate for
protocol A (left panel) and protocol B 
(right panel). The solid lines {with varying markers and colors} correspond to the theoretical results 
given in Eq. (\ref{eq:res_mfpt_A}) (left panel) and Eqs. 
(\ref{eq:res_mfpt_B}) and (\ref{S2.5}) (right panel). 
The quantity $q(R,T)$ in Eq. (\ref{eq:res_mfpt_A}) is computed
by evaluating the Bromwich integral in Eq. (\ref{S1.8}) numerically.
The {crosses} represent the results 
from {numerical Langevin} simulations with $10^5$ samples. The different 
colors {and markers} correspond to different values of $N$, where $N$ goes from 4 to 9 
from top to bottom. Notice that in both panels we can observe the 
gradual disappearance of the minimum at $R^\star > 0$.} 
\label{fig:mfpt}
\end{figure}

In protocol A we have $N$ independent copies of a one-dimensional 
resetting random walker (see the left panel of Fig. \ref{fig:sketch}). These walkers are independent at all times $t$. 
Hence
\begin{equation} \label{eq:N21}
Q_{r,L,N}^{(A)}(t) = \left[Q_{r,L,1}(t)\right]^N \equiv \left[ Q_{r, L}(t) \right]^N \;,
\end{equation}
where $Q_{r, L}(t)$ is the survival probability of a single walker 
in the presence of resetting with an absorbing boundary condition at $L > 0$, starting at the origin at $t = 0$. From Section \ref{sec:intro-resetting} - specifically Eq.~(\ref{eq:survival-resetfree} and Eq.~(\ref{eq:laplace-reset-survival}) - we can relate the survival probability to its reset free counterpart 
\begin{equation} \label{eq:survival-resetfree-search}
    Q_{0, L}(x, t) = {\rm erf}\left( \frac{L - x}{\sqrt{4 D t}} \right) \;.
\end{equation}
from which we obtain the Laplace transform of the survival probability which is given by
\begin{equation} \label{eq:survival-resetting-search}
    \int_0^{+\infty} \dd t \; e^{-s t} Q_{r, L}(t) = \frac{1 - e^{-L\sqrt{\frac{r + s}{D}}}}{s + r e^{-L\sqrt{\frac{r + s}{D}}}} \;.
\end{equation}
For simplicity, from now on, we re-write all the variables in terms of 
their dimensionless counterparts, i.e.,
\begin{equation}
\label{S1.5}
S = \frac{L^2}{D} s, \quad\, R = \frac{L^2}{D} r, 
\quad\, T = \frac{D}{L^2} t \;.
\end{equation}
Rewriting Eq.~(\ref{eq:survival-resetting-search}) in terms of dimensionless variables,
\begin{equation}
\label{S1.7}
\tilde{Q}_{r, L}(s) = \frac{L^2}{D}\, \frac{1 - 
e^{-\sqrt{S + R}}}{\left[S + R e^{-\sqrt{S + R}}\right]} \;.
\end{equation}
Inverting this Laplace transform formally one gets
\begin{equation}
\label{S1.8}
Q_{r, L}(t) = \int_\Gamma \frac{\dd S}{2\pi i} \; e^{S\, T} 
\frac{1 - e^{-\sqrt{S + R}}}{S + R \,e^{-\sqrt{S + R}}}\equiv q(R, T)\;,
\end{equation}
where $\Gamma$ denotes the Bromwich contour in the complex $S$ plane.
Plugging this result in Eq. (\ref{eq:N21}) and then using 
Eq.~(\ref{eq:def_Tx}) we get
the dimensionless \MFPT
\begin{equation}
\langle T_f^{(A)}(R, N) \rangle = \frac{D}{L^2} 
\langle t_f^{(A)}(r, L, N) \rangle = \int_0^{+\infty} 
\left[q(R, T)\right]^N\, \dd T \;. 
\label{eq:res_mfpt_A}
\end{equation}
We inverted the Laplace transform in Eq. (\ref{S1.8}) numerically
and then evaluated the integral in Eq. (\ref{eq:res_mfpt_A}).
In the right panel of Fig.~\ref{fig:mfpt} we compare this theoretical prediction
with {numerical Langevin} simulation results by plotting 
$\langle T_f^{(A)}(R, N) \rangle$
as a function of $R$, for different values of $N$.
We find excellent agreement.
Physically it is clear that as $R \to +\infty$ we 
expect the \MFPT $\langle T_f^{(A)}(R, N) \rangle$ to diverge since the 
system constantly resets and thus never explores the space. This can be seen 
by noting that $q(R, T) \to 1$ as $R \to +\infty$ in Eq. (\ref{S1.8})
and hence the integral 
of the \MFPT in Eq. (\ref{eq:res_mfpt_A}) diverges.
Let us now investigate the opposite limit $R\to 0$. 
If the \MFPT decreases at small $R$ 
then it is likely 
that there is a certain $R^\star > 0$ where the
curve becomes a global minimum, before starting to increase again
and finally diverging as $R\to \infty$ (see Fig. \ref{fig:mfpt}). 
However if the \MFPT increases for small $R$, then clearly $R^\star = 0$,
provided the \MFPT increases monotonically with increasing $R$ as it
happens to be the case (see Fig. \ref{fig:mfpt}). 
Thus, the existence of a minimum $R^\star>0$ can then be investigated
by analyzing the small $R$ behavior of $\langle T_f^{(A)}(R, N) \rangle$. 

\vspace{0.2cm}

The small $R$ asymptotic
behavior of $\langle T_f \rangle^{(A)}(R, N)$ depends on the value of
$N$. It can be analyzed using Eqs. (\ref{eq:res_mfpt_A})
and (\ref{S1.8}) and even though the search process makes sense only
for integer $N$, our analytical result in Eqs. (\ref{S1.8}) and (\ref{eq:res_mfpt_A}) can be analytically continued to real $N$.
Hence, from now on we will consider $N$ real in this sense.
It turns out that
if $N \leq 2$ then $\langle T_f \rangle^{(A)}(R, N)$ diverges
as $R \to 0$, if $2<N \leq 4$ then $\langle T_f \rangle^{(A)}(0, N)$
is finite but the slope of $\langle T_f 
\rangle^{(A)}(R, N)$ at $R \to 0$ is negatively divergent and
finally if $N > 4$ both the \MFPT and its derivative are finite at $R=0$. 
Let us summarize here the leading small $R$ behavior 
of the \MFPT for different values of $N$:
\begin{equation}
\langle T_f^{(A)}(R, N) \rangle 
{\underset{R \to 0}{\sim}}
\begin{dcases}
        \frac{\Gamma\left(1 - N/2\right)}{\pi^{N/2} N^{1 - N/2}}
\frac{1}{R^{1 - N/2}} \quad &\mbox{ if } N < 2 \;,\\
        -\frac{1}{\pi} \ln R \quad &\mbox{ if } N = 2 \;,\\
        {C^{(A)}_N} -
\frac{2\, \Gamma(2-N/2)}{(N-2)\, \pi^{N/2}}\, N^{N/2-1}\, R^{N/2-1}
\quad &\mbox{ if } N \in ]2, 4[ \;, \\
       {C^{(A)}_N}+ \frac{4}{\pi^2}
R \ln R \quad  &\mbox{ if } N = 4 \;,\\
        {C^{(A)}_N} + R \int_0^{+\infty}
\left[q(0, T)\right]^{N-1} \pdv{q(R, T)}{R} \Big|_{R = 0} \dd
T &\mbox{ if } N > 4 \;,
\end{dcases} 
\label{summary_A}
\end{equation}
where for any $N > 2$ {the constant $C^{(A)}_N$ is given by}
\begin{equation}
{C_N^{(A)}} = \langle T_f^{(A)}(0, N) \rangle = \int_0^{+\infty}
\erf\left(\frac{1}{\sqrt{4 T}}\right)^N \dd T \;.
\label{value_A}
\end{equation}

For $N\le 4$, the small $R$ behavior
of the \MFPT above, 
combined with the divergence of $R\to \infty$,
indicates the existence of a finite $R^\star > 0$ for all $N\le 4$. 
However, for $N > 4$ one has to find the condition
for a non-zero $R^\star > 0$.
For $N > 4$ both the \MFPT and its first derivative with respect
to $R$ are finite and the sign of the derivative can be positive or negative, depending on $N$. In fact, by taking the derivative
of Eq. (\ref{eq:res_mfpt_A}) and setting $R=0$ one gets
\begin{equation} \label{eq:AN4}
\pdv{\langle T_f^{(A)}(R, N) \rangle}{R} \Big|_{R = 0} = 
N\, \int_{0}^{+\infty} \left[q(0, T)\right]^{N-1} 
\pdv{q(R, T)}{R} \Big|_{R = 0} \dd T \;,
\end{equation}
where, using Eq. (\ref{eq:survival-resetfree-search}), one has 
\begin{equation}
q(0,T)={\rm erf}\left(\frac{1}{\sqrt{4\, T}}\right)\, .
\label{q0T.1}
\end{equation}
Taking the derivative of Eq. (\ref{S1.8}) with respect to $R$ and setting
$R=0$ gives
\begin{equation}
\pdv{q(R, T)}{R} \Big|_{R = 0}= \int_\Gamma \frac{\dd S}{2\pi i} \; e^{S\, T}\,
\left[\frac{1}{2\,S^{3/2}}\, e^{-\sqrt{S}}-\frac{1}{S^2}\, \left(
e^{-\sqrt{S}}- e^{-2\, \sqrt{S}}\right)\right]\, .
\label{der_lap.1}
\end{equation}
This Laplace inversion can be explicitly done to give
\begin{align}
\pdv{q(R, T)}{R} \Big|_{R = 0} = (T+1)\, \text{erf}\left(\frac{1}{ 
\sqrt{4\,T}}\right)&-(T+2)\, \text{erf}\left(\frac{1}{\sqrt{T}}\right)\nonumber\\
&+
\frac{2\,\sqrt{T}}{\sqrt{\pi}}\, 
\left( e^{-\frac{1}{4\, T}}- e^{-\frac{1}{T}}\right) +1 \;.
\label{S1.10}
\end{align}
Plugging Eqs. (\ref{q0T.1}) and (\ref{S1.10}) in Eq. (\ref{eq:AN4})
gives us the derivative of the \MFPT at $R=0$ in terms of a single
integral, which unfortunately is not easy to evaluate
explicitly. However, it can be easily evaluated numerically for all
$N>4$ using Mathematica (see Fig. \ref{fig:dT}).
As $N$ increases beyond $4$, the derivative at $R=0$ in Eq. (\ref{eq:AN4})
increases, initially negative, as can be seen in Fig. (\ref{fig:dT}).
As long as this derivative at $R=0$ is negative, we have a nonzero $R^*>0$.
When the derivative changes sign and becomes positive, we have $R^*=0$.
Using a dichotomous algorithm, we find that this sign change occurs at
$N_c =7.3264773 \cdots$. This is our main result in this section.
It says that resetting in protocol A is beneficial for a
team of $N$ searchers as long as $N<N_c$. When $N>N_c$, resetting
increases the search time and is therefore no longer a useful strategy.

\subsection{Protocol B}
\label{subsection:B}

In protocol B the {\it simultaneous} resetting (see the right panel of Fig. \ref{fig:sketch}) induces strong long range 
correlations between the walkers as we have seen in Chapter \ref{ch:sim-reset}. Hence, the system is 
not simply $N$ independent copies of a single resetting random walker. 
However, since the resetting happens {\it simultaneously} we have a new 
renewal equation for the $N$-particle stochastic process
\begin{equation}
\label{S2.1}
Q_{r, L, N}^{(B)}(t) = e^{-r t}\, Q_{0, L, N}^{(B)}(t) + 
r \int_0^{+\infty} \dd \tau \; e^{-r \tau}\, Q_{0, L, N}^{(B)}(\tau)\, 
Q_{r, L, N}^{(B)}(t - \tau) \;.
\end{equation}
This renewal equation has the usual interpretation that we have seen throughout this manuscript. Either the process survives up to time $t$ in the absence of resetting, leading to the first term. Or the process resets at least once, then the last reset time is denoted by $t - \tau$ and the process must survive up to $t - \tau$ in the presence of resetting and survive from $t-\tau$ to $t$ in the absence of resetting. For $r = 0$, the walkers become 
independent and hence
\begin{equation} \label{eq:Br0}
Q_{0, L, N}^{(B)}(t) = \left[Q_{0, L}(t)\right]^N = 
\left[\erf\left( \frac{L}{\sqrt{4 D t}} \right)\right]^N \;.
\end{equation}
Taking the Laplace transform of 
Eq. (\ref{S2.1}) we obtain
\begin{equation}
\tilde{Q}_{r, L, N}^{(B)}(s) = \frac{\tilde{Q}_{0, L, N}^{(B)}(s + r)}{1 - r \tilde{Q}_{0,L,N}^{(B)}(s + r)} \;.
\label{S2.2}
\end{equation}
Finally using Eq. (\ref{eq:def_Tx}) we can express the \MFPT as
\begin{equation}
\label{S2.3}
\langle t_f^{(B)}(r, L, N) \rangle = \tilde{Q}_{r, L, N}^{(B)}(s = 0) = 
\frac{\tilde{Q}_{0, L, N}^{(B)}(r)}{1 - 
r \tilde{Q}_{0, L, N}^{(B)}(r)} \;.
\end{equation}
Inserting Eq. (\ref{eq:Br0}) in Eq. (\ref{S2.3})
we then get an explicit formula
\begin{equation}
\label{S2.4}
\langle t_f^{(B)}(r, L, N) \rangle = \frac{\int_0^{+\infty} \dd 
t\, e^{-r t}\, \left[\erf\left(\frac{L}{\sqrt{4 D t}} 
\right)\right]^N}{1 - r\, \int_0^{+\infty} \dd t\, e^{-r t}\, 
\left[\erf\left(\frac{L}{\sqrt{4 D t}} \right)\right]^N} \;.
\end{equation}
Once again we appropriately re-scale the variables to make them 
dimensionless by setting $T = \frac{D}{L^2} t$ and $R = 
\frac{L^2}{D} r$ and obtain the simpler expression
\begin{align}
\langle T_f^{(B)}(R, N) \rangle &= \frac{D}{L^2} \langle 
t_f^{(B)}(r, L, N) \rangle \\
&= \frac{\int_0^{+\infty} \dd T\, 
e^{- R T}\, \left[\erf\left(\frac{1}{\sqrt{4 T}} \right)\right]^N}{1 - 
R \int_0^{+\infty} \dd T\, e^{-R T}\, \left[\erf\left(\frac{1}{\sqrt{4 T}} 
\right)\right]^N} \\
&= \frac{h(R, N)}{1 - R \, h(R, N)} \;, 
\label{eq:res_mfpt_B}
\end{align}
where for simplicity we introduced the function
\begin{equation}
\label{S2.5}
h(R, N) = \int_0^{+\infty} \dd T\, e^{- R T}\, \left[\erf\left(\frac{1}{\sqrt{4 T}} 
\right)\right]^N \;.
\end{equation}
We {verify} this theoretical result
by comparing it to {numerical} Langevin simulations as shown in the right panel of Fig. \ref{fig:mfpt}. As in the
case of protocol A, we infer the existence or not of a nonzero
optimal $R^*>0$ by analyzing the small $R$ behavior of Eq. (\ref{eq:res_mfpt_B}). 
{For a detailed derivation of the small $R$ behavior for different $N$ refer to the Appendix of Ref. \cite{BFHMS24}.} Here we summarize these results:
\begin{equation}
\hspace*{-0.4cm}\langle T_f^{(B)}(R, N) \rangle \stackrel{R \ll 1}{\sim}
\begin{dcases}
        \frac{\Gamma\left(1 - N/2\right)}{\pi^{N/2}}
\frac{1}{R^{1 - N/2}} \quad &\mbox{ if } N < 2\\
        -\frac{1}{\pi} \ln R \quad &\mbox{ if } N = 2\\
       C_N^{(B)} -
\frac{\Gamma\left(2 - N/2\right)(2 - N/2)}{\pi^{N/2} }
R^{N/2 - 1} \quad &\mbox{ if } 2 < N < 4\\
       C_N^{(B)}+ \frac{1}{\pi^2}\,
R\, \ln R \quad  &\mbox{ if } N = 4\\
       C_N^{(B)} +  \Bigg\{ \left[
\int_0^{+\infty} \left[\erf\left( \frac{1}{\sqrt{4 T}} \right)\right]^N\,
\dd T \right]^2 \\
\qquad\quad - \int_0^{+\infty} T\,\left[\erf\left( \frac{1}{\sqrt{4 T}}
\right)\right]^N\, \dd T  \Bigg\}\, R \quad &\mbox{ if } N > 4 \;,
\end{dcases}  \nonumber \\
\label{summary_B}
\end{equation}
where for any $N > 2$, $C_N^{(B)}$ is a constant given by  
\begin{equation}
C_N^{(B)} = \langle T_f^{(B)}(0, N) \rangle = \langle T_f^{(A)}(0, N) \rangle =
\int_0^{+\infty} \left[\erf\left(\frac{1}{\sqrt{4 T}}\right)\right]^N \dd T \;.
\label{value_B}
\end{equation}

\begin{figure}
\centering
\includegraphics[width = 0.6\textwidth]{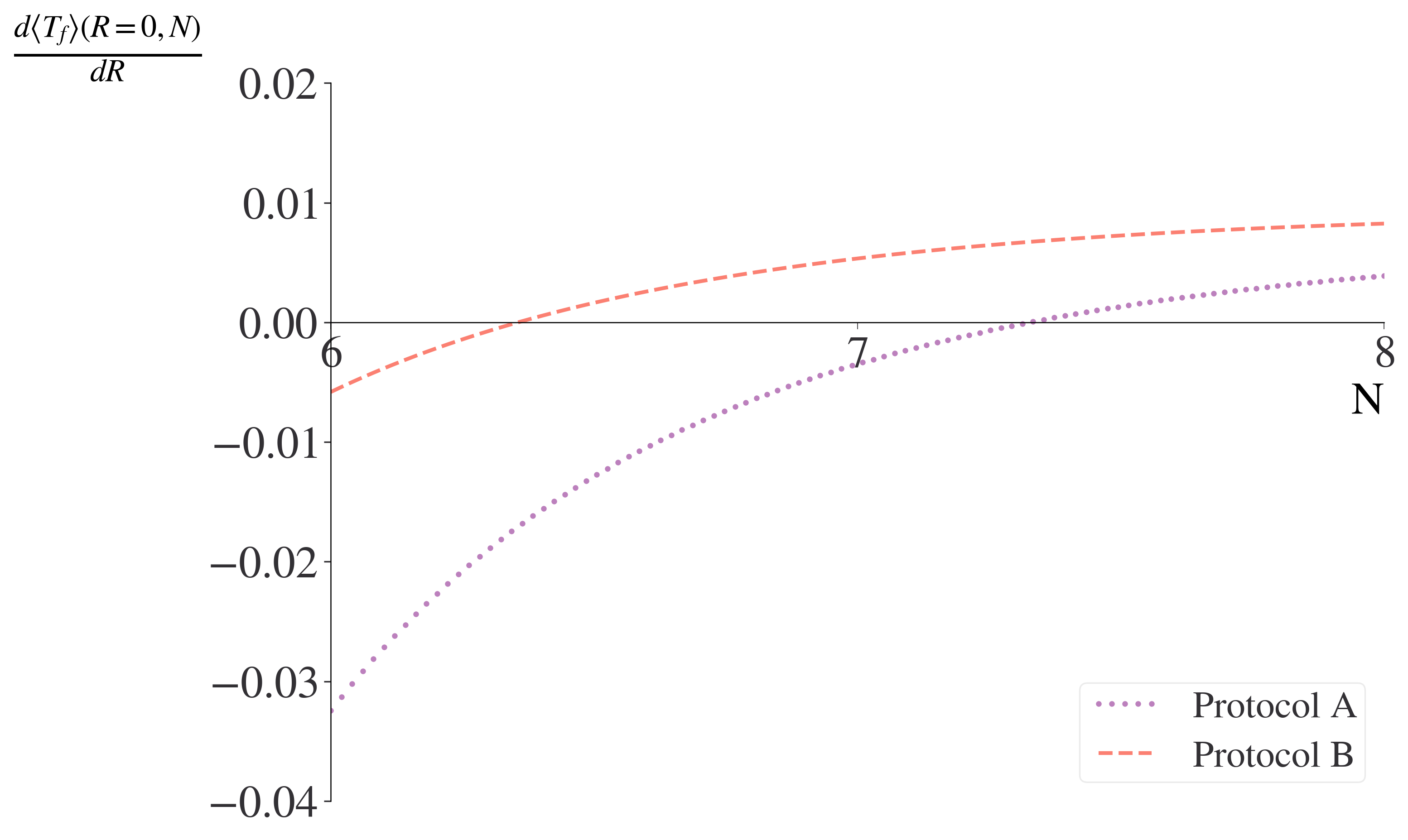}
\caption{The derivative of the \MFPT at $R=0$ in Eq. (\ref{eq:AN4}) 
(protocol A) and Eq. (\ref{eq:res_dT_B}) (protocol B), plotted
as a function of $N$ for $6<N<8$. The derivatives change sign
respectively at $N_c=7.3264773\ldots$
(protocol A) and $N_c=6.3555864\ldots$ (protocol~B).}
\centering
\label{fig:dT}
\end{figure}

As in the case of protocol A, it is clear from the small $R$ behavior
that there is an optimal $R^*>0$ for all $N\le 4$. For
$N>4$, both $h(R, N)$ and its first derivative converge when $R \to 0$. Then the derivative of the \MFPT, for $N>4$, 
is given by
\begin{align}
&\pdv{\langle T_f^{(B)}(R, N) \rangle}{R} \Big|_{R = 0} = 
\left[h(0, N)\right]^2 + 
\partial_R h(R, N) \Big|_{R = 0} \\
&= \left[\int_0^{+\infty} 
\left[\erf\left( \frac{1}{\sqrt{4 T}} \right)\right]^N\, \dd T \right]^2 - 
\int_0^{+\infty} T\, \left[\erf\left( \frac{1}{\sqrt{4 T}} \right)\right]^N
\dd T \;. 
\label{eq:res_dT_B}
\end{align}
The existence of a finite $R^\star > 0$ is uniquely determined by the 
sign of the above expression. If it is negative, then there exists a
finite $R^\star > 0$. However, if it is positive, then $R^\star = 0$ and
the resetting hinders the search process. 
Once again, the integrals in Eq. (\ref{eq:res_dT_B}) can be
easily evaluated using Mathematica
(see Fig. \ref{fig:dT}) and we find that the
critical value of $N$ is defined as the value for which the derivative of the \MFPT at $R = 0$ 
changes sign is given by $N_c = 6.3555864\ldots$. Thus, for protocol B, resetting benefits the search process
as long as $N<N_c$, but delays the search process for $N>N_c$. \\\\
\noindent The value of $N_c$ is slightly lower in protocol B than in protocol A. It is not easy to guess these values of $N_c$ a priori from simple physical arguments. However, we can understand why the value of $N_c$ for protocol B is lower than that of protocol A from the following argument. As was shown in Chapter \ref{ch:sim-reset} in {protocol} B the simultaneous resetting induces an effective attraction between all the particles. The simultaneous resetting {pulls the particles together}. Hence, Protocol B not only restricts the movement of the particles to stay closer to the origin but also creates coordination between the particles, further constraining the movement of the ensemble of particles. Therefore, the constraints of protocol B are stronger than {those} of {protocol} A leading to resetting becoming inefficient faster for protocol B which, in turn, leads to a lower value of $N_c$.

\chapter{Engineering faster search processes}\label{ch:rescaling}
Although previous chapters examined how stochastic resetting can influence search efficiency—often reducing the mean first-passage time and inducing nontrivial steady states—resetting's broader utility is perhaps best appreciated through its real-world implementations \cite{EMS20}. Across fields as varied as web algorithms, molecular biology, and optimization, resetting mechanisms are used, sometimes explicitly, sometimes emergently, to overcome the inefficiencies inherent to purely diffusive or deterministic dynamics.

\vspace{0.2cm}

A striking example comes from cellular biology, in which DNA-binding proteins searching for specific target sites employ intermittent dynamics, alternating between one-dimensional sliding along DNA and three-dimensional excursions \cite{BWH81, CBM04,GMKC18,C19}. These 3D unbinding events serve as resets, allowing the searcher to bypass slow, exhaustive scanning of unpromising regions. Related patterns emerge in animal foraging \cite{BC09, VDLRS11}, where long-range relocations or returns to a home base act as stochastic resets in response to local depletion. In computational settings, restarting is a widely used strategy in randomized algorithms, particularly in difficult optimization or decision problems. Techniques such as simulated annealing or SAT solvers frequently benefit from systematic restarts, which prevent entrapment in poor performing regions of the solution space. Even in chemical physics, resetting-inspired strategies emerge \cite{EMS20}. 

\vspace{0.2cm}

These examples demonstrate that resetting is not merely an abstract theoretical tool but a widely effective mechanism for managing exploration, avoiding dead ends, and accelerating convergence. Building on this insight, the present chapter introduces a novel variant: resetting by rescaling, in which the position of a diffusing particle is rescaled by a fixed factor $a \in ]-1, 1[$ rather than reset to a constant location. This protocol maintains analytical tractability while offering new dynamical features. Most importantly, we will see that in certain regimes - particularly when $a < 0$ — it can obtain mean first-passage times lower than classical resetting.


\section{Resetting by rescaling: exact results for a diffusing particle in one-dimension}

In this original toy model, the walker's position is always reset to the origin. However, this model can be analytically solved even when the reset position is not fixed after every reset, but rather is drawn independently from a distribution after each reset \cite{EM11JPhysA}. In fact, in the experimental setups used in Refs. \cite{BBPMC20,FBPCM21}, the resetting position typically corresponds to the Boltzmann distribution of the particle confined in a potential $U(x)$. However, the two paradigms discussed above, namely the existence of a \NESS and a finite \MFPT remain true when the resetting position is drawn from a distribution. In these models, the position after the resetting is uncorrelated with the position before reset, i.e., the post-resetting position has no memory of the pre-resetting position. A simple model that retains some dependence on the pre-resetting position has recently been studied under the name of  ``backtrack resetting'' where the usual stochastically resetting diffusion evolution rule is modified to~\cite{TRR22,P22}
\begin{equation}\label{eq:summary-model-dynamics}
X(t + \dd t) = \begin{cases}
a \, X(t) &\mbox{~~with probability~~} r \dd t \;, \\
X(t) + \sqrt{2 D\, \dd t} \, \eta(t)\, & \mbox{~~with probability~~} 1 - r \dd t \;,
\end{cases}
\end{equation}
where $0 \leq a \leq 1$ is a backtracking parameter and $\eta(t)$ is zero-mean $\langle \eta(t) \rangle = 0$ delta-correlated $\langle \eta(t) \eta(t') \rangle = \delta(t - t')$ Gaussian white noise. This model represents for example the situation where $X(t)$ denotes the population of a habitat at time $t$, a fraction of which gets wiped out after a catastrophic event that occurs at random times distributed via a Poisson distribution with rate $r$ \cite{TRR22}. This model for $a>0$ sometimes goes by the name of ``partial resetting'' \cite{TRR22,P22,BCHPM23,OG24}. For $a=0$, this model reduces to the standard model of diffusion with stochastic resetting studied in Section \ref{sec:intro-resetting}. For $a=1$, clearly the post and pre-resetting positions are the same (thus effectively there is no resetting) and the particle simply diffuses. 
For $a > 1$ the particle eventually escapes to $\pm \infty$. Thus there is no stationary state for $a \geq 1$ and the \NESS exists only for $0\leq a<1$. 
In this range of $a$, the stationary position distribution has been computed in several recent papers \cite{TRR22,P22,BCHPM23,OG24,H23}. However, to the best of our knowledge, no result for the \MFPT exists for general $a > 0$. In all these studies on the ``partial resetting'' model above, the parameter $a$ was considered to be nonnegative $a \geq 0$. However, in principle, one can study resetting for $a<0$ also \cite{H23}. A negative value of $a$ means a partial resetting coupled with a reflection around the origin. It is clear that for $a \leq -1$, there is no stationary state and the stationary state exists only in the range $-1<a \leq 0$ \cite{H23}. However, the stationary position distribution for $-1<a<0$ is not known explicitly. Furthermore, the \MFPT has also not been studied for $-1<a<0$. To summarize, for this partial resetting model with parameters $-1<a<+1$, the position distribution does become stationary at late times and is known explicitly only in the range $0 \leq a < 1$. The \MFPT is essentially unknown in the full range $-1< a < + 1$, except for $a=0$. One interesting open question is whether this additional parameter $a$ can reduce the \MFPT compared the standard $a=0$ model of stochastic resetting. 

\vspace{0.2cm}

In this Section, we revisit this problem and compute analytically the stationary position distribution in the full range $-1<a<+1$. Our results for $0<a<+1$ coincide with the known results, while the result for $-1<a<0$ is new. Furthermore, we compute exactly the \MFPT in the range $0\leq a < 1$ and show that it increases as $a$ increases from $0$, for any fixed $r$. This indicates that a positive value of the fraction $a$ is not beneficial for the search for a target. For $-1<a<0$, we show that the underlying non-local differential equation for the \MFPT has a fundamentally different structure compared to the case $0 \leq a <1$. Here we succeeded in finding an exact formula for the \MFPT, but only up to an unknown constant that can, however, be easily inferred from numerical simulations. With this numerically determined constant we then have an exact formula for the \MFPT. This solution shows that the \MFPT actually decreases when $a$ decreases from $0$, for fixed $r$. This means a negative value of the parameter $-1<a<0$, i.e., when the partial resetting gets coupled with a reflection, is more optimal than simple resetting ($a=0$). We also performed numerical simulations and found perfect agreement with our theoretical predictions. 

\vspace{0.2cm}

In Subsection \ref{subsec:NESS}, we compute the exact stationary position distribution in the full range $-1<a<1$. In Section \ref{sec:MFPT}, we compute the exact \MFPT first for $0 \leq a <1$ and then for $-1<a<0$.

\begin{figure}
\centering
\includegraphics[width = 0.3\textwidth]{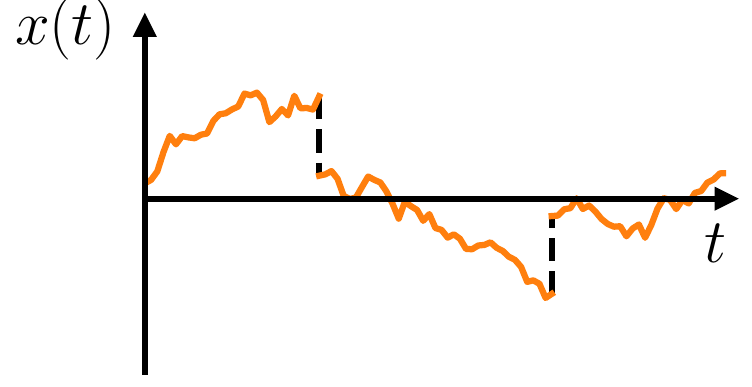}
\hfill
\includegraphics[width = 0.3\textwidth]{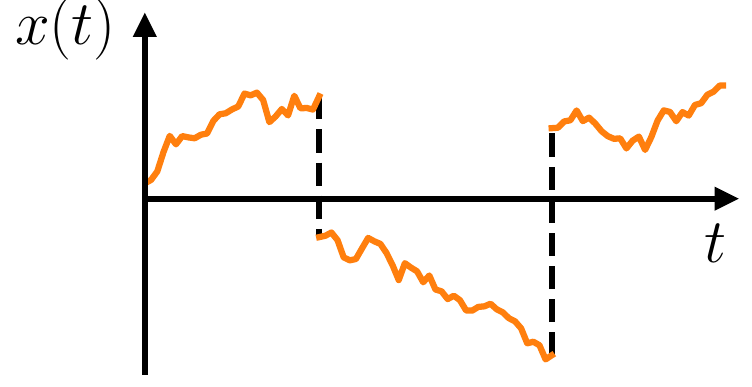}
\hfill
\includegraphics[width = 0.3\textwidth]{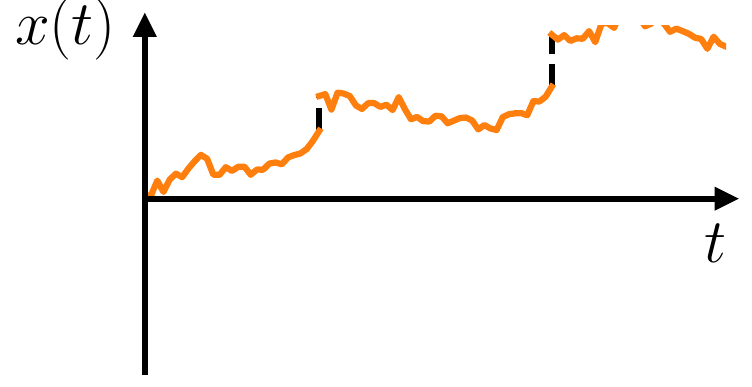}
\caption{Schematic trajectories of a rescaling random walk, as defined in Eq. (\ref{eq:summary-model-dynamics}) for $0 < a < 1$ (left panel), $-1 < a < 0$ (middle panel) and $a > 1$ (right panel). The orange part of the curves represents the purely diffusive part of the trajectory, while the dashed vertical black lines correspond to long-range rescaling (resetting) moves. As shown in the text, the position distribution of the walker becomes stationary only for $-1<a<1$, while it remains non-stationary at all times for $|a| \geq 1$.} \label{fig:model}
\end{figure}

\subsection{Non-equilibrium steady state} \label{subsec:NESS}

We consider the process $X(t)$ evolving via the Langevin Eq. (\ref{eq:summary-model-dynamics}). Let $p_{r, a}(x, t) = {\rm Prob.}[X(t) = x]$ denote the probability density at which the particle arrives at $x$ at time $t$, starting from the origin $x = 0$ at time $t = 0$. We can obtain the Fokker-Planck equation of $p_{r, a}(x, t)$ from the forward Master equation of Eq.~(\ref{eq:summary-model-dynamics}), i.e. in a time step $\dd t$ the probability density will evolve as
\begin{equation} \label{eq:discrete-balance}
  p_{r, a}(x, t + \dd t) = (1 - r \dd t)
  \Big\langle p_{r, a}(x - \sqrt{2 D \, \dd t}\,\eta(t),t)\Big\rangle_\eta +
  \frac{r \, \dd t}{|a|}  \; p_{r, a}\left( \frac{x}{a}, t \right) \;,
\end{equation}
where the notation $\langle \cdots \rangle_\eta$ means an average over the instantaneous noise $\eta(t)$, which is distributed via a Gaussian with zero mean and unit variance. In Eq. (\ref{eq:discrete-balance}), the first term denotes the diffusive movement in time $\dd t$, which occurs with probability $1-r\,\dd t$ (see Eq. (\ref{eq:summary-model-dynamics})). If the particle has to reach $x$ at time $t + \dd t$, it must have been at $x -  \sqrt{2 D \, \dd t}\,\eta(t)$ at time $t$ and it must be averaged over all possible values of $\eta(t)$. The second term denotes the reset that restores the particle from $x/a$ to $x$ in time $\dd t$. This resetting move with probability $r\, \dd t$ (see Eq. (\ref{eq:summary-model-dynamics})). The factor $1/|a|$ is a Jacobian factor associated with the probability density $p_{r, a}\left( \frac{x}{a}, t \right)$ in $x$. In the limit $\dd t \to 0$, we expand the first term in a Taylor series in $\sqrt{\dd t}$. Keeping terms up to order $O(\dd t)$ and taking the limit $\dd t \to 0$ one arrives at the Fokker-Planck equation
\begin{equation} \label{eq:Fokker-Planck}
\pdv{p_{r, a}(x, t)}{t} = D \pdv[2]{p_{r, a}(x, t)}{x} - r p_{r, a}(x, t) + \frac{r}{|a|} p_{r, a}\left( \frac{x}{a}, t \right) \;.
\end{equation}
This equation is valid for all $a$ and, for $a\geq 0$, it was already derived in Refs.~\cite{TRR22,P22,BCHPM23,H23}. By integrating it over $x$ from $-\infty$ to $+\infty$, one clearly sees that Eq.~(\ref{eq:Fokker-Planck}) conserves the normalization
\begin{equation} \label{eq:NESS-normalization}
 \int_{-\infty}^\infty P_{r,a}(x,t)\, \dd x  = 1 \;.
 \end{equation} 
 
 \begin{figure}
\centering
\includegraphics[width=0.6\textwidth]{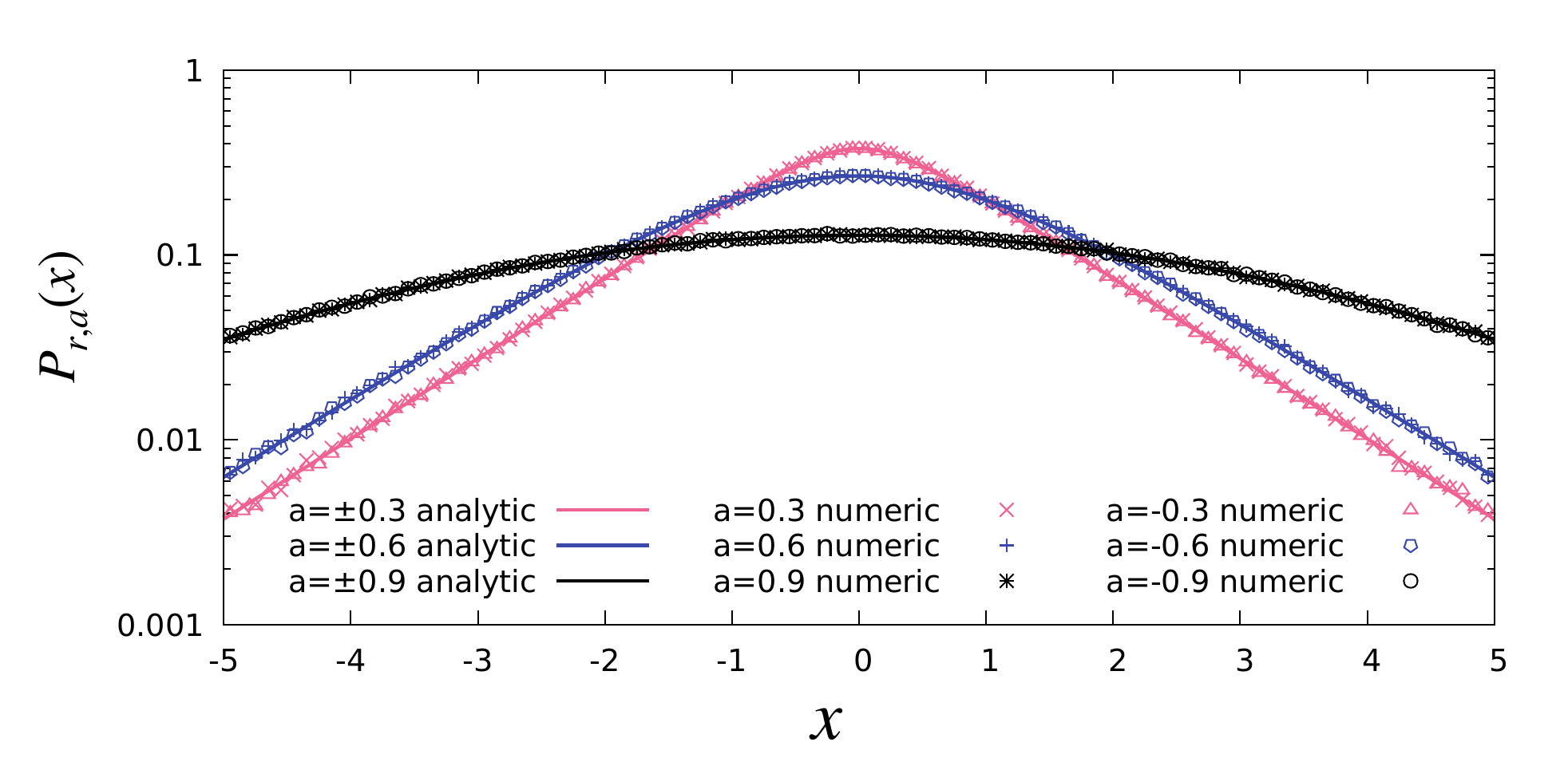}
\caption{Plot of the steady-state probability density $P_{r,a}(x)$ vs $x$ for different values of the parameter $a \in (-1,1)$ and for fixed $r=1$ and $D=1$. 
The analytical result in Eq. (\ref{eq:full-NESS}) is in perfect agreement with the results obtained from numerical simulations.}\label{fig:NESS}
\end{figure}
As mentioned previously, the \NESS exists only in the range $|a| < 1$. Assuming that the \NESS exists, one can obtain it by setting the left hand side of Eq. \eqref{eq:Fokker-Planck} to zero. This gives
\begin{equation} \label{eq:SS-FP}
0 = D \dv[2]{p_{r, a}(x)}{x} - r p_{r, a}(x) + \frac{r}{|a|} p_{r, a}\left( \frac{x}{a}\right) \quad, \quad |a| < 1 \;.
\end{equation}
Even though this is an ordinary differential equation, it is {\it nonlocal} in $x$ and hence is nontrivial to solve. This type of nonlocal equations have appeared in other contexts before, e.g., in the modeling of cell growth \cite{HW89}, in the growth of clusters in a generalized Eden model on a tree \cite{DM06} and also  
in the context of the discrete Ornstein-Uhlenbeck processes~\cite{Lar04,MK07}, etc. These works suggest to look for a solution of Eq.~(\ref{eq:SS-FP}) in the form 
\begin{equation} \label{eq:NESS-Ansatz}
p_{r, a}(x) = \sum_{n = 0}^{+\infty} c_n e^{-\frac{1}{|a|^n} \sqrt{\frac{r}{D}} |x| } \;,
\end{equation}
where $c_n$ are constants. Substituting this in Eq. (\ref{eq:SS-FP}) we obtain
\begin{equation} \label{eq:SS-Ansatz-ODE}
0 = \sum_{n = 0}^{+\infty} \left( \frac{r}{|a|^{2n}} c_{n} - r c_n + \frac{r}{|a|} c_{n - 1} \right) e^{-\frac{1}{|a|^n} \sqrt{\frac{r}{D}} |x| } \;.
\end{equation}
Since Eq. (\ref{eq:SS-Ansatz-ODE}) must hold for any $x$ the prefactor to the exponential in the series must vanish for any $n$ yielding the recursion relation
\begin{equation} \label{eq:NESS-recursion}
c_n \left(\frac{1}{|a|^{2n}} - 1\right) = -\frac{1}{|a|} c_{n-1} \;.
\end{equation}
Iterating the recursion in Eq. (\ref{eq:NESS-recursion}) we can express any $c_n$ in terms of $c_0$
{\color{blue}
\begin{equation}
  c_n = \left(\frac{1}{|a|} \frac{1}{\left(1 - |a|^{-2n}\right)}\right) \left(\frac{1}{|a|} \frac{1}{\left(1 - |a|^{-2(n-1)}\right)} \right) \cdots \left(\frac{1}{|a|} \frac{1}{\left(1 - |a|^{-2}\right)} \right) c_0
\end{equation}
which can be written succinctly as
}
\begin{equation} \label{eq:incomplete-cn}
c_n = \frac{1}{|a|^n} \frac{c_0}{\prod_{k = 1}^n (1 - |a|^{-2k})} \;.
\end{equation}
Using Eq. (\ref{eq:incomplete-cn}) in Eq. (\ref{eq:NESS-Ansatz}) we obtain
\begin{equation} \label{eq:incomplete-NESS}
p_{r, a}(x) = c_0 \left[\sum_{n = 1}^{+\infty} \frac{1}{|a|^n} \frac{1}{\prod_{k = 1}^n (1 - |a|^{-2k})} e^{-\frac{1}{|a|^n} \sqrt{\frac{r}{D}} |x| } + \, e^{- \sqrt{\frac{r}{D}} |x|} \right] \;,
\end{equation}
where we have separated out the $n=0$ term. The only unknown constant $c_0$ is then determined from the normalization condition \eqref{eq:NESS-normalization}, giving
\begin{equation} \label{eq:NESS-c0}
c_0^{-1} = 2 \sqrt{\frac{D}{r}} \left(\sum_{n = 1}^{+\infty} \frac{1}{\prod_{k = 1}^n (1 - |a|^{-2k}) } + 1 \right) \;.
\end{equation}
This gives the exact stationary position distribution 
\begin{align} \label{eq:full-NESS}
p_{r, a}(x) = \frac{1}{2} &\sqrt{\frac{r}{D}} \frac{1}{\left[1 + \sum_{n = 1}^{+\infty} \frac{1}{\prod_{k = 1}^n (1 - |a|^{-2k})}\right]} \\
\times &\left( e^{-\sqrt{\frac{r}{D}} |x| } + \sum_{n = 1}^{+\infty} \frac{1}{|a|^n} \frac{1}{\prod_{k = 1}^n (1 - |a|^{-2k})} e^{-\frac{1}{|a|^n} \sqrt{\frac{r}{D}} |x| } \right) \;.
\end{align}
The stationary distribution is evidently symmetric around $x=0$ and is plotted in Fig. \ref{fig:NESS} for three different values of $a$, showing a perfect agreement with numerical simulations. In the limit $a \to 0$, we recover the usual resetting result from Section \ref{sec:intro-resetting}, as expected. For $0<a<1$, this solution already appeared in Refs. \cite{P22,BCHPM23}, but for $-1<a<0$, we have not seen this solution in the literature. 

\vspace{0.2cm}

It is interesting to derive the asymptotic behaviors of $p_{r,a}(x)$ for small and large $x$. The large $x$ behavior is simple, since the term that contains the sum becomes subleading for large $x$. Hence, we get
\begin{equation} \label{Pstat_large}
p_{r, a}(x) \simeq \frac{1}{2} \sqrt{\frac{r}{D}} \frac{1}{\left[1 + \sum_{n = 1}^{+\infty} \frac{1}{\prod_{k = 1}^n (1 - |a|^{-2k})}\right]} \; e^{-\sqrt{\frac{r}{D}} |x| }  \quad, \quad {\rm as} \quad |x| \to \infty \;.
\end{equation}
In contrast, the small $x$ behavior turns out to be much more tricky. From the plot in Fig. \ref{fig:NESS}, it seems to behave as $p_{r,a}(x) \sim p_{r,a}(0) - b x^2$, as $x \to 0$, i.e. the linear term in the small expansion $x$ vanishes. Indeed, by expanding Eq. (\ref{eq:NESS-Ansatz}) up to linear order in $x$, we get
\begin{equation} \label{Pstat_small}
p_{r,a}(x) \simeq p_{r,a}(0)- \sqrt{\frac{r}{D}} |x| \sum_{n=0}^\infty \frac{c_n}{|a|^n} \;,
\end{equation}
where the coefficients $c_n$'s are given in Eqs. (\ref{eq:incomplete-cn}) and (\ref{eq:NESS-c0}). For the linear term to vanish, we must have the identity $\sum_{n=0}^\infty c_n/|a|^n = 0$. Using the explicit expressions for $c_n$'s, this amounts to the identity valid for all $0<|a| <1$
\begin{equation} \label{identity}
1 + \sum_{n=1}^\infty \frac{1}{|a|^{2n}} \frac{1}{\prod_{k=1}^n (1-|a|^{-2k})} = 0 \;.
\end{equation}
We have numerically checked with Mathematica that it is indeed true for several values of $0<|a|<1$. However, we could not prove this non-trivial identity. For $|a|>1$, the left-hand side of Eq. (\ref{identity}) can be expressed as the inverse of the Euler function; however, we cannot find an expression for it for $0<|a|<1$. Proving this identity remains an interesting number-theoretical challenge. Thus, in summary, the stationary distribution behaves as a Gaussian distribution for small $x$, while having an exponential tail for large $x$. Similar asymptotic behaviors for scaling functions also appeared in several models, e.g.,
in the time-dependent position distribution in diffusing diffusivity models~\cite{CSMS17,LG18,BB20}, for particles driven by resetting noise \cite{GMS23} and in certain experimental systems~\cite{WABG09,WGLFGH19}.

\subsection{Mean {\color{blue}first-passage} time} \label{subsec:MFPT}

For a diffusing particle starting at the origin in $d=1$ and resetting stochastically to the origin with rate $r$, we have seen in Section \ref{sec:intro-resetting} that the \MFPT $t_f(r, L)$ is not only finite, but can also be optimized with respect to $r$. In this section, we study whether the introduction of the additional parameter $a$,  with $|a|<1$, lowers the \MFPT compared to the $a=0$ case.

\vspace{0.2cm}

We consider a particle in one dimension, starting at $x$, and evolving via Eq. (\ref{eq:summary-model-dynamics}), with a target located at $x=L$. Our goal is to calculate \MFPT $\langle t_{r,a,L}(x) \rangle$ to find the target at $L$, starting at $x$.  
Eventually, for simplicity, we will focus on the case $x=0$, but for the moment we keep $x$ arbitrary, since we will derive a backward Fokker-Planck differential equation for $\langle t_{r,a,L}(x) \rangle$, with $x$ as a variable. To derive this equation, it is convenient to start with $Q(x,t)$, which denotes the survival probability
of the target up to time $t$, that is, the probability that the target is not found by the particle up to time $t$. Consequently $F(x,t) = - \partial_t Q(x,t)$ denotes the first-passage time distribution to the target. The \MFPT is just the first moment of $F(x,t)$, i.e., 
\begin{equation} \label{eq:MFPT-integral}
\langle t_{r,a,L}(x) \rangle = \int_0^{+\infty} \dd t \; \left(- \pdv{Q(x, t)}{t} \right) t = \int_0^{+\infty} Q(x, t) \dd t \;,
\end{equation}
where we performed an integration by parts and assumed that $Q(x, t) t \to 0$ as $t \to +\infty$ which can be checked a posteriori.
For the brevity of the notation, we will omit the subscripts of \MFPT $T(x) \equiv \langle t_{r, a, L}(x) \rangle$. We now consider the backward evolution equation
for $Q(x,t)$. We consider a trajectory of duration $t+\dd t$, starting at $x$ and evolving via Eq. (\ref{eq:summary-model-dynamics}). We divide the interval $[0, t + \dd t]$ into two intervals $[0, \dd t]$ and $[\dd t, t+\dd t]$. During the first interval, the walker diffuses to a new position $x + \sqrt{2 D \, \dd t} \, \eta(0)$ with probability $1-r\,\dd t$ and with complementary probability $r\,\dd t$, it resets to a new position $a\,x$. Here, $\eta(0)$ denotes the initial random jump. 
During the second interval, the evolution proceeds starting at the new position at the end of the first interval $[0, \dd t]$. Consequently, we can write 
\begin{equation} \label{eq:BFP}
Q(x, t + \dd t) = (1 - r \dd t)  \Big\langle Q(x + \sqrt{2 D \, \dd t}\,\eta(0))\Big\rangle_\eta + {r \, \dd t}  \; Q\left( {a} \,x, t \right) \;.
\end{equation}
Taking the limit $\dd t\to 0$, one arrives at the backward equation 
\begin{equation} \label{eq:survival-FP}
\pdv{Q(x, t)}{t} = D \pdv[2]{Q(x, t)}{x} - r Q(x, t) + r Q(a x, t) \;.
\end{equation}
This equation is valid in the range $x \in (-\infty, +\infty)$ with the absorbing boundary condition 
\begin{equation} \label{bc_S}
Q(x=L,t) = 0 \quad, \quad {\rm for \; all} \quad  t \geq 0 \;.
\end{equation}
This condition comes from the fact that if the particle starts exactly at $x=L$, it immediately finds the target and hence the survival
probability vanishes. Furthermore, as $x \to \pm \infty$, the survival probability must remain upper bounded by unity. The initial condition reads
\begin{equation} \label{ic_S}
Q(x,t=0) = 1 \quad {\rm for \; all \;} \quad x \neq  L \;.
\end{equation} 
Integrating Eq. (\ref{eq:survival-FP}) over $t$ from $0$ to $\infty$ and using the initial condition (\ref{ic_S}), we get, using Eq. (\ref{eq:MFPT-integral}), the backward differential equation for $T(x)$
\begin{equation} \label{eq:MFPT-ODE}
D T''(x) - r T(x) + r T(a x) = - 1\;,
\end{equation}
valid for $x \in (-\infty, +\infty)$ with the absorbing boundary condition 
\begin{equation} \label{bc_T}
T(L) = 0 \;.
\end{equation}
In addition, as $x \to \pm \infty$, the \MFPT $T(x)$ cannot diverge faster than $\sim x^2$, since
diffusion is the slowest mode of transport and the resetting can only reduce the \MFPT. Once again, this ordinary second-order equation (\ref{eq:MFPT-ODE}) is nonlocal in $x$, making it nontrivial to solve. We will see below that the solution is actually very different for $0\leq a<1$ and $-1<a \leq 0$. We discuss the two cases separately in the two subsections below.

\subsubsection{Positive rescaling: $0 \leq a < 1$} \label{sec:MFPT-positive-a}


In this subsection, our goal is to calculate the \MFPT $T(0)$ starting from the origin for the case $0 \leq a < 1$. To compute this, we need to solve the non-local
backward differential equation (\ref{eq:MFPT-ODE}) with $T(x)$ denoting the \MFPT starting from the initial position $x \leq L$. Upon finding the solution for $T(x)$ for arbitrary $x \leq L$, we will eventually set $x=0$. Since $x=0 \leq L$, we need to solve the differential equation only in the region $x \leq L$, and henceforth we will not consider the case $x > L$. Note that the non-local term in Eq. (\ref{eq:MFPT-ODE}) involves the location $a\,x$ that always stays to the left of $L$. Hence, the particle never jumps to the right of $L$, and we just need to solve the differential equation for $x \leq L$. Since there is no known general method to solve such nonlocal equations, we try below a power series solution in $x$ and show that it leads to an exact solution. We substitute
\begin{equation} \label{eq:MFPT-power-series}
T(x) = \sum_{n = 0}^{+\infty} b_n x^n 
\end{equation}
in Eq. (\ref{eq:MFPT-ODE}) and solve recursively for the $b_n$'s. This gives 
\begin{equation} \label{eq:MFPT-power-series-ODE}
-1 = \sum_{n = 0}^{+\infty} \left[D b_{n+2} (n+2)(n+1) - r b_n + r b_n a^n \right] x^n \;.
\end{equation}
This equation holds for any $x \leq L$. Therefore setting $x = 0$ sets the first even constant
\begin{equation} \label{eq:MFPT-c2}
-1 = 2 D b_2 \;.
\end{equation}
Then the remaining terms of the series must all vanish which leads to the following recursion relation
\begin{equation} \label{eq:MFPT-c-recursion}
b_{n+2} D (n+2) (n+1) = r (1 - a^n)b_n \;.
\end{equation}
Iterating Eq. (\ref{eq:MFPT-c-recursion}) we can express any $b_n$ for $n \geq 1$ as a function of $b_1$ and $b_2$
\begin{equation} \label{eq:MFPT-c}
b_{2n} = \frac{2 b_2}{(2n)!} \left(\frac{r}{D}\right)^{n-1} \prod_{j = 1}^{n-1} (1 - a^{2j}) \mbox{~and~} b_{2n+1} = \frac{b_1}{(2n+1)!} \left( \frac{r}{D} \right)^n \prod_{j = 0}^{n-1} (1 - a^{2j+1})  \;.
\end{equation}
Hence, the only constants that are still undetermined are $b_1$ and $b_0$. Putting Eqs. (\ref{eq:MFPT-c2}), (\ref{eq:MFPT-c}) and (\ref{eq:MFPT-power-series}) together we obtain
\begin{align} \label{eq:MFPT-incomplete}
T(x) = b_0 &+ b_1 \sqrt{\frac{D}{r}} \sum_{n = 0}^{+\infty} \frac{1}{(2n + 1)!} \left(\sqrt{\frac{r}{D}} x\right)^{2n + 1} \prod_{j = 0}^{n-1} (1 - a^{2j+1}) \\
&- \frac{1}{r} \sum_{n = 1}^{+\infty} \frac{1}{(2n)!} \left( \sqrt{\frac{r}{D}} x \right)^{2n} \prod_{j = 1}^{n-1} (1 - a^{2j}) \;.
\end{align}
Note that in the term $n=0$ of the sum multiplying $b_1$, the product is interpreted as unity. Using the absorbing boundary condition in Eq. (\ref{bc_T}) gives the first relation 
\begin{align} \label{eq:MFPT-c0}
b_0 = - &b_1 \sqrt{\frac{D}{r}} \sum_{n = 0}^{+\infty} \frac{1}{(2n + 1)!} \left(\sqrt{\frac{r}{D}} L\right)^{2n + 1} \prod_{j = 0}^{n-1} (1 - a^{2j+1}) \\
+& \frac{1}{r} \sum_{n = 1}^{+\infty} \frac{1}{(2n)!} \left( \sqrt{\frac{r}{D}} L \right)^{2n} \prod_{j = 1}^{n-1} (1 - a^{2j}) \;.
\end{align}
Using Eq. (\ref{eq:MFPT-c0}) we can rewrite Eq. (\ref{eq:MFPT-incomplete}) as
\begin{align}
T(x) = b_1 \sqrt{\frac{D}{r}} \sum_{n = 0}^{+\infty} \frac{1}{(2n + 1)!}  \left(\sqrt{\frac{r}{D}} \right)^{2n+1} (x^{2n+1} - L^{2n+1}) \prod_{j = 0}^{n-1} (1 - a^{2j+1}) \nonumber\\
- \frac{1}{r} \sum_{n = 1}^{+\infty} \frac{1}{(2n)!} \left(\sqrt{\frac{r}{D}} \right)^{2n} (x^{2n} - L^{2n})  \prod_{j = 1}^{n-1} (1 - a^{2j}) \;. \label{eq:MFPT-incomplete-2}
\end{align}
To simplify these expressions, we introduce the following compact notations
\begin{equation} \label{eq:def-Cn}
C_{2n} = \prod_{j = 1}^{n-1} (1 - a^{2j}) \mbox{~~for~~} n \geq 2,  C_{2n + 1} = \prod_{j = 0}^{n-1} (1 - a^{2j+1}) \mbox{~~for~~} n \geq 1 \; ,
\end{equation}
where 
\begin{equation}
  C_1 = 1 \mbox{~~and~~} C_2 = 1, 
\end{equation}
and rescaled distances
\begin{equation} \label{def_ell}
\ell = L \sqrt{\frac{r}{D}} \quad, \quad y = x  \sqrt{\frac{r}{D}} \;.
\end{equation}
In terms of these rescaled quantities, see Eq. (\ref{eq:MFPT-incomplete-2}) simplifies to
\begin{align}
T(x) = b_1 \sqrt{\frac{D}{r}} \sum_{n = 0}^{+\infty} \frac{C_{2n+1}}{(2n + 1)!}  \left(y^{2n+1} - \ell^{2n+1}\right)  - \frac{1}{r} \sum_{n = 1}^{+\infty} \frac{C_{2n}}{(2n)!} \left(y^{2n} - \ell^{2n}\right)  \;. \label{T_compact}
\end{align}
We have already used the absorbing boundary condition at $x=L$. To fix the only unknown constant $b_1$, we need to investigate the other boundary when $x \to - \infty$. As mentioned above, the \MFPT $T(x)$ should not grow faster than $\sim x^2$ as $x \to - \infty$. We now show that this condition uniquely fixes the unknown constant $b_1$. 
\begin{figure}[t]
    \centering
    \includegraphics[width = 0.5\textwidth]{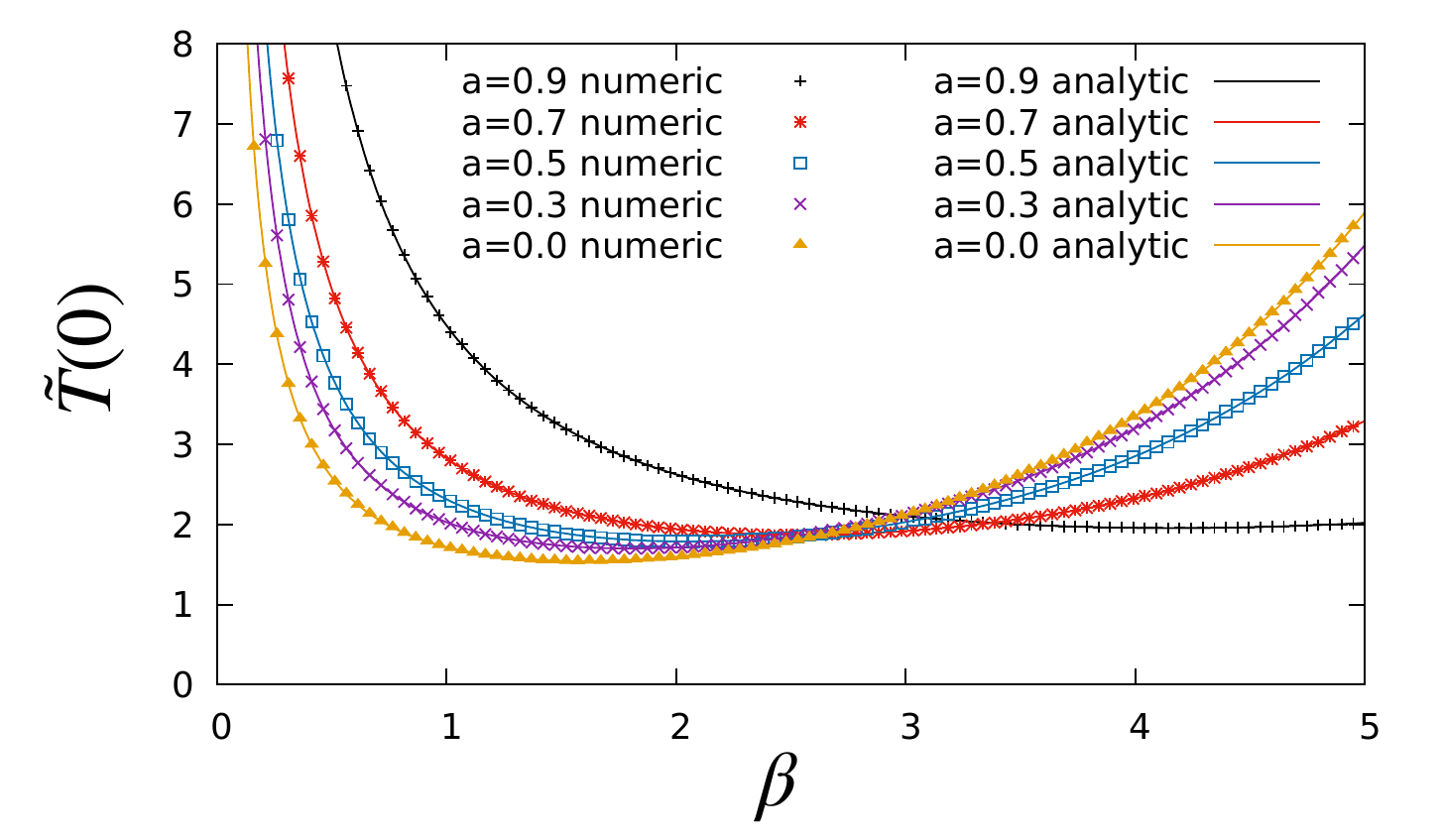}
    \caption{The dimensionless \MFPT $\tilde T(0)$ as a function of $\ell = \sqrt{r/D} L$ for different values of $0 \leq a < 1$. The analytical result in Eq. (\ref{eq:MFPT-pos-a-dimensionless}) is in excellent agreement with the results of numerical simulations for different values of $0 \leq a < 1$. For a fixed $a$, the \MFPT $\tilde T(0)$, as a function of $\ell$, has a minimum at $\ell = \ell^*(a)$.}\label{Ttilde_apos}
\end{figure}

\vspace{0.2cm}

To analyze $T(x)$ in Eq. (\ref{T_compact}) in the limit $x \to -\infty$, we first note that $\left(y^{2n+1} - \ell^{2n+1}\right)  \sim y^{2n+1}$ and similarly $\left(y^{2n} - \ell^{2n}\right) \sim y^{2n}$. Furthermore, for large $|y|$ both sums in Eq. (\ref{T_compact}) are dominated by large $n$. Hence, we can replace $C_{2n+1}$ and $C_{2n}$ by their respective $n \to \infty$ limits, i.e., 
\begin{equation} \label{lim_C}
\lim_{n \to \infty} C_{2n+1} = \prod_{j=0}^\infty (1-a^{2j+1}) \quad, \quad \lim_{n \to \infty} C_{2n} = \prod_{j=1}^\infty (1-a^{2j}) \;.
\end{equation}
Therefore, to leading order for large $|y|$, we can take out the $C$-factors outside the sums and evaluate the sums explicitly to get
\begin{align} 
T(x) \underset{x \to -\infty}{\simeq} &- b_1 \sqrt{\frac{D}{r}}  \left[\prod_{j=0}^\infty (1-a^{2j+1})\right] \sinh{(y)} \nonumber\\
&\qquad+ \frac{1}{r} \left[\prod_{j=1}^\infty (1-a^{2j})\right]\left( \cosh{(y)}-1\right) \;. \label{T_large_y}
\end{align} 
Thus, to leading order for large $|x|$, this expression diverges as $T(x) \sim \exp{|y|}$ $=$  $\exp(\sqrt{r/D}|x|)$. Since this is not allowed physically, the amplitude of this term must vanish. This fixes the constant $b_1$ uniquely as
\begin{equation}\label{eq:MFPT-c1}
-b_1 \sqrt{\frac{D}{r}} = \frac{1}{r} \frac{\prod_{j = 1}^{+\infty} (1 - a^{2j})}{\prod_{j = 0}^{+\infty} (1 - a^{2j +1})} = \frac{1}{r}\, R_a \;.
\end{equation}
where $R_a$ is given by the ratio
\begin{equation} \label{eq:def-Ra}
R(a) = \frac{\lim_{n\to+\infty} C_{2n}}{\lim_{n \to +\infty} C_{2n +1}} = \frac{\prod_{j = 1}^{+\infty} (1 - a^{2j})}{\prod_{j = 0}^{+\infty} (1 - a^{2j +1})} \;.
\end{equation}
Thus the final expression for the \MFPT $T(x)$, in terms of the rescaled coordinates (\ref{def_beta}), reads 
\begin{equation} \label{eq:MFPT-pos-a}
T\left(y \sqrt{\frac{D}{r}}\right) = \frac{1}{r} \Bigg[R(a) \sum_{n = 0}^{+\infty} \frac{C_{2n + 1}}{(2n +1)!} (\ell^{2n+1} - y^{2n+1}) + \sum_{n = 1}^{+\infty} \frac{C_{2n}}{(2n)!}  (\ell^{2n} - y^{2n})  \Bigg] \;.
\end{equation}
Setting $x = 0$ we obtain the \MFPT for a random walk starting at the origin to reach a target at $L$ 
\begin{equation} \label{eq:MFPT-pos-a-origin}
T(0) = \frac{1}{r} \left[ \sum_{n = 1}^{+\infty} \frac{C_{2n}}{(2n)!}\ell^{2n} + R(a) \sum_{n = 0}^{+\infty} \frac{C_{2n+1}}{(2n+1)!} \ell^{2n+1} \right] \;,
\end{equation}
with $R(a)$ defined in Eq. (\ref{eq:def-Ra}). Furthermore, one can also define a dimensionless \MFPT $\tilde{T}(0) = D\,T(0)/L^2$, that depends only on two dimensionless parameters: $a$ and $\ell = L \sqrt{r/D}$, which reads
\begin{equation}\label{eq:MFPT-pos-a-dimensionless}
\tilde{T}(0) = \frac{1}{\ell^2} \left[\sum_{n = 1}^{+\infty} \frac{C_{2n} \ell^{2n} }{(2n)!} + R(a) \sum_{n = 0}^{+\infty} \frac{C_{2n+1} \ell^{2n+1}}{(2n+1)!}\right] \;. 
\end{equation}
\begin{figure}[t]
\includegraphics[width = 0.9\textwidth]{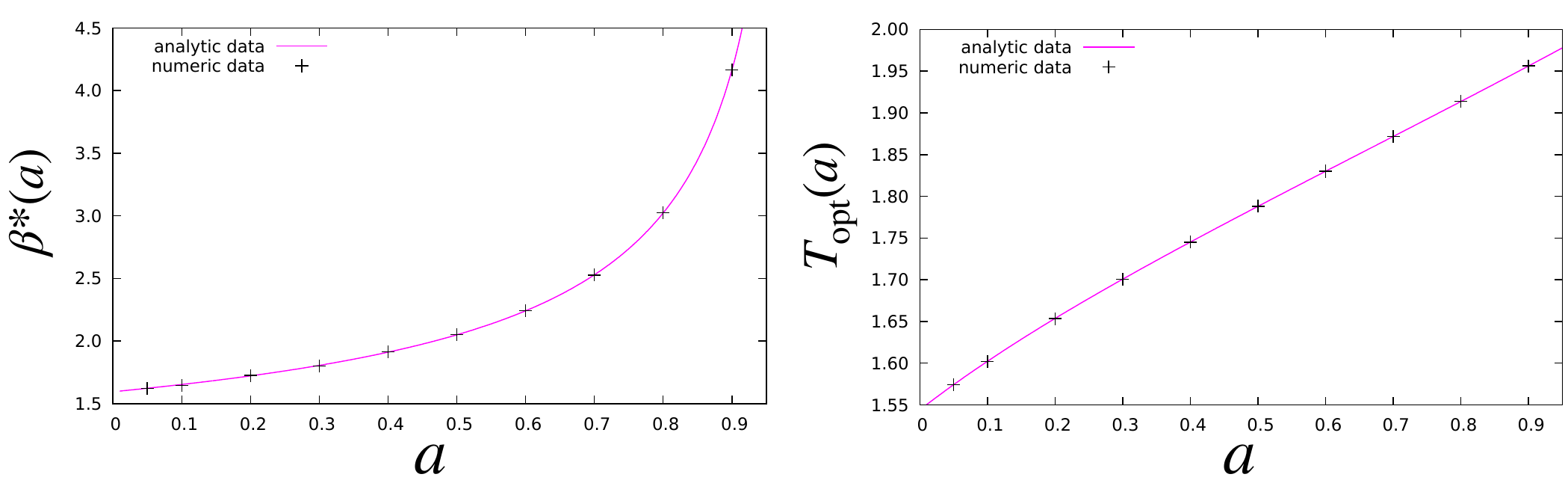}
\caption{{\bf Left:} The optimal value $\ell^*(a)$, at which $\tilde T(0)$ achieves its minimum as a function of $\ell$ for a fixed $a$, plotted as a function of $a$ for $0 \leq a < 1$. {\bf Right:} The optimal \MFPT $T_{\rm opt}(a)$, i.e., $\tilde T(0)$ evaluated at $\ell = \ell^*(a)$, plotted as a function of $0 \leq a < 1$.}\label{Figbetastarapos}
\end{figure}
For fixed $a$, the asymptotic behaviors of $\tilde{T}(0)$ for small and large $\ell$ are given by 
\begin{equation} \label{eq:MFPT-pos-a-beta-asymptotics}
\tilde{T}(0) \simeq \begin{dcases}
R(a)/\ell &\mbox{~~when~~} \ell \ll 1 \\
\frac{e^\ell}{\ell^2} \prod_{j = 1}^{+\infty} (1 - a^{2j}) &\mbox{~~when~~} \ell \gg 1 \;.
\end{dcases}
\end{equation}
The small $\ell$ {\color{blue}asymptotes} is easy to obtain from Eq. (\ref{eq:MFPT-pos-a-dimensionless}) where the $n=0$ term contributes to the dominant order for small $\ell$, leading to the first line in Eq. (\ref{eq:MFPT-pos-a-beta-asymptotics}). To derive the large $\ell$ behavior in the second line of Eq. (\ref{eq:MFPT-pos-a-beta-asymptotics}), we note that the sums in Eq. (\ref{eq:MFPT-pos-a-beta-asymptotics}) are dominated by the large values of $n$ where both $C_{2n}$ and $C_{2n+1}$ converge to the asymptotic values [see Eq. (\ref{eq:def-Cn})]
\begin{equation} \label{conv_C}
C_{2n} \underset{n \to \infty}{\longrightarrow} \prod_{j=1}^\infty (1-a^{2j}) \quad, \quad C_{2n+1} \underset{n \to \infty}{\longrightarrow} \prod_{j=0}^\infty (1-a^{2j+1}) \;.
\end{equation} 
Substituting these behaviors in Eq. (\ref{eq:MFPT-pos-a-dimensionless}) and using \eqref{eq:MFPT-pos-a-beta-asymptotics}, we get the second line of Eq. (\ref{eq:MFPT-pos-a-beta-asymptotics}). Thus we see that, as a function of $\ell$ for fixed $a$, the \MFPT $\tilde T(0)$ diverges in both limits $\ell \to 0$ and $\ell \to \infty$. Thus, it indicates that it may have a unique minimum at $\ell = \ell^*(a)$. Indeed, the analytical result in Eq. (\ref{eq:MFPT-pos-a-dimensionless}) can easily be plotted and it shows a unique minimum (see Fig. \ref{Ttilde_apos}). This minimum $\ell^*(a)$ can easily be determined by setting the derivative of Eq. (\ref{eq:MFPT-pos-a-dimensionless}) with respect to $\ell$ to zero and determining the root using Mathematica. This optimal value $\ell^*(a)$, as a function of $a$, is shown in the left panel of Fig. \ref{Figbetastarapos}. Finally, the optimal value $T_{\rm opt}(a)$ of the \MFPT, i.e., $\tilde T(0)$ evaluated at $\ell = \ell^*(a)$, is plotted as a function of $a$ in the right panel of Fig. \ref{Figbetastarapos}. Clearly, one sees that $T_{\rm opt}(a)$ is a monotonically increasing function of $a$ in the range $a \in [0,1]$.

\vspace*{0.5cm}

\noindent {\bf Small $a$ expansion.}  For small $a$, one can perform an explicit analysis. Expanding Eq. (\ref{eq:MFPT-pos-a-dimensionless}) to linear order for small $a$, we get
\begin{equation} \label{eq:MFPT-pos-a-a-asymptotics}
\tilde{T}(0) \simeq \frac{1}{\ell^2} \left[e^\ell - 1  + a \ell + O(a^2)\right] \mbox{~~when~~} a \ll 1 \;.
\end{equation}
Note that for $a=0$, we perfectly recover the known result stated in Eq. (\ref{eq:mean-first-passage-resetting}). In fact, this result in Eq. (\ref{eq:MFPT-pos-a-a-asymptotics}) for small $a$ can also be derived directly from the differential equation (\ref{eq:MFPT-ODE}) by making an expansion for small $a$ \cite{BFHMS24}. Thus, for fixed $\ell$, as $a$ increases from $0$, the \MFPT also increases. Taking a derivative of Eq. (\ref{eq:MFPT-pos-a-a-asymptotics}) with respect to $\ell$ and setting it to zero, we get the optimal $\ell^*(a)$ to linear order in $a$
\begin{equation} \label{betastar_small_a}
\ell^*(a) = \ell^*(0) + a \frac{\ell^*(0)}{\ell^*(0)-1}\,e^{-\ell^*(0)} + O(a^2) = 1.59362\ldots + a\, 0.54547\ldots  + O(a^2)\;,
\end{equation}
where $\ell^*(0) = 1.59362\ldots$ is the optimal value of $\ell$ for $a=0$. Thus as $a$ increases from $0$, the optimal $\ell^*(a)$ increases for small $a$, which is consistent with the result shown in the left panel of Fig. \ref{Figbetastarapos}. Subsequently, the optimal \MFPT $T_{\rm opt}(a)$, up to linear order in $a$, reads
\begin{align} \label{Topt_small_a}
T_{\rm opt}(a) &= T_{\rm opt}(0) + a\, \frac{[\ell^*(0)]^2 - 2 (1-e^{-\ell^*(0)})}{[\ell^*(0)]^2(\ell^*(0)-1)} + O(a^2) \\
&= 1.54413\ldots + a\, 0.627500\ldots  + O(a^2)\;.
\end{align}
Therefore, as $a$ increases from $0$, the optimal \MFPT increases linearly with $a$ for small $a$, consistent with the results reported in the right panel of Fig. \ref{Figbetastarapos}. 

\vspace{0.2cm}

In conclusion, introducing a positive value of the scaling parameter $a$ does not improve the efficiency of the search process, since $T_{\rm opt}(a)$ increases with increasing positive $a$. In the next subsection, we will see that the situation is drastically different for $-1<a\leq 0$. In this case, 
the optimal \MFPT with $-1<a\leq 0$ is {\it lower} than its value at $a=0$.


\subsubsection{Negative rescaling: $-1 < a \leq 0$} \label{sec:MFPT-negative-a}


\begin{figure}[t]
    \centering
    \includegraphics[width = 0.7 \linewidth]{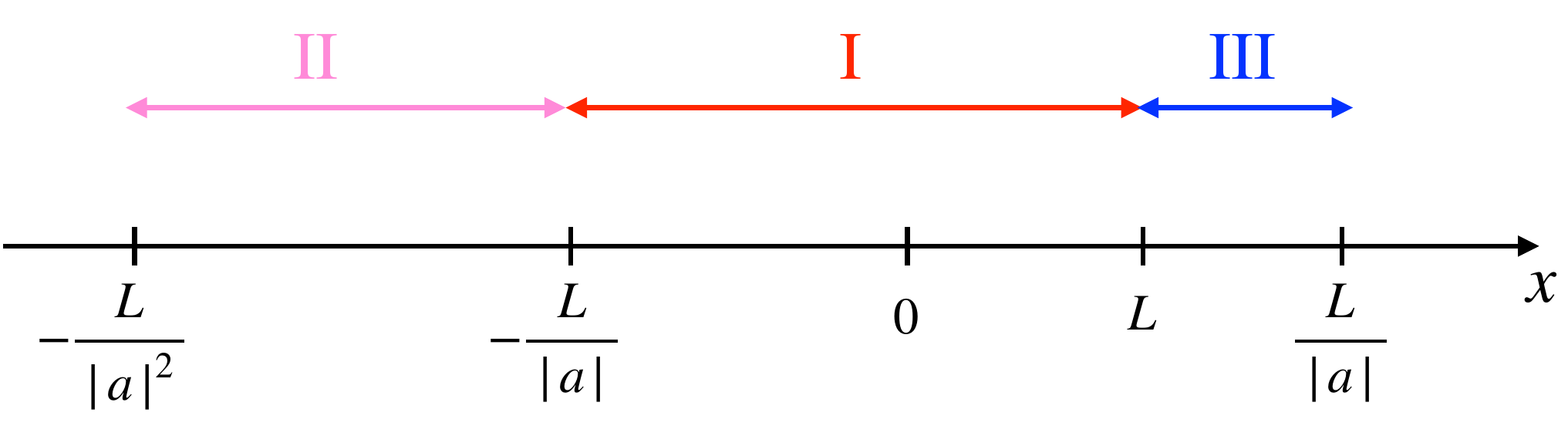}
    \caption{For $-1< a \leq 0$, the nonlocal differential equation (\ref{eq:MFPT-ODE_aneg}) needs to be solved in different segments that are interconnected. The segment ${\rm I}$ denotes the region $x \in [-L/|a|,L]$, the segment ${\rm II}$ denotes the region $[-L/|a|^2, -L/|a|]$ and the segment ${\rm III}$ denotes the region $[L,L/|a|]$.}\label{Figa<0}
\end{figure}

In this subsection, we consider the complementary case $-1<a\leq 0$. Once again, we need to solve the differential equation (\ref{eq:MFPT-ODE}) for $T(x)$ and eventually set $x=0$. However, contrary to the case $0 \leq a <1$ discussed in the previous subsection, for $-1< a \leq 0$, solving the non-local differential equation (\ref{eq:MFPT-ODE}) is much more complicated. To see why, we first rewrite Eq. (\ref{eq:MFPT-ODE}) for $-1<a\leq 0$ as
\begin{equation} \label{eq:MFPT-ODE_aneg}
-1 = D T''(x) - r T(x) + r T(-|a| x) \;.
\end{equation}
We still have the boundary conditions $T(L) = 0$ and the fact that $T(x)$ should not grow faster $\sim |x|^2$ as $x \to \pm \infty$. To solve this non-local equation \eqref{eq:MFPT-ODE_aneg} at a given point $x$, the source term $ r T(-|a| x)$ comes from the point $-|a| x$. Consequently, one needs to separate the entire line into different segments, as shown in Fig. \ref{Figa<0}. First, we note that, if $-L/|a|\leq x \leq L$ (segment I in Fig. \ref{Figa<0}), the source point $-|a| x$ also belongs to this segment. Thus, segment I closes on itself. However, we have only one known boundary condition $T(L) = 0$ and have no information on the \MFPT at the other edge at $x = - L/|a|$. For the moment, let us denote this value by $\kappa >0$. In other words, the Eq. (\ref{eq:MFPT-ODE_aneg}) in segment I satisfies the boundary conditions
\begin{equation} \label{eq:MFPT-boundary-neg}
T(L) = 0 \mbox{~~and~~} T(-L/|a|) = \kappa \;,
\end{equation}
where $\kappa$ is unknown. To determine $\kappa$, we need to solve Eq. \eqref{eq:MFPT-ODE_aneg}  in segment II, where $-L/|a|^2 \leq x \leq -L/|a|$. However, for $x$ belonging to this segment II, the source point $-|a| x$ belongs to segment III in Fig. \ref{Figa<0} where $x \in [L, L/|a|]$. Thus, the solution in segment II requires the solution in segment III, and this mechanism continues until we arrive at $x = \pm \infty$. Thus, this breaks the whole line into different segments. In Fig. \ref{Figa<0}, for simplicity, we show only three of them. Hence, we have to iteratively solve (\ref{eq:MFPT-ODE_aneg}) in each segment in order to use the boundary condition as $x \to \pm \infty$. This makes the problem much more complicated to solve. However, we see that segment I is ``closed'' onto itself (that is, it does not involve other segments), and the full function $T(x)$ for $-L/|a| \leq x \leq L$ can be fully determined, but up to only one unknown constant $\kappa$ representing the \MFPT at $-L/|a|$. Hence, our strategy would be to first analytically solve Eq. (\ref{eq:MFPT-ODE_aneg}) in segment I, that is, for $-L/|a| \leq x \leq L$ with $\kappa$ as a given parameter and then use the value of $\kappa$ obtained from numerical simulations. This will fully determine $T(x)$ in segment I, and since the origin belongs to that segment, we can set $x=0$ to find $T(0)$. In the following, we derive the solution $T(x)$ in terms of $\kappa$.    

\vspace{0.2cm}

We note that to solve $T(x)$ in segment I, that is, for $-L/|a| \leq x \leq 0$, with $-1<a\leq 0$, the procedure is identical to the previous subsection, that is, we try a series expansion as in Eq. (\ref{eq:MFPT-power-series}). Following exactly the same steps that led to the result in Eq. (\ref{eq:MFPT-incomplete-2}), we obtain
\begin{align}
T(x) = b_1 \sqrt{\frac{D}{r}} \sum_{n = 0}^{+\infty} \frac{1}{(2n + 1)!}  \left(\sqrt{\frac{r}{D}} \right)^{2n+1} (x^{2n+1} - L^{2n+1}) \prod_{j = 0}^{n-1} (1 + |a|^{2j+1}) \nonumber\\
- \frac{1}{r} \sum_{n = 1}^{+\infty} \frac{1}{(2n)!} \left(\sqrt{\frac{r}{D}} \right)^{2n} (x^{2n} - L^{2n})  \prod_{j = 1}^{n-1} (1 - |a|^{2j})  \label{eq:MFPT-aneg} \;,
\end{align}
where we used the boundary condition $T(x=L)=0$. We now use the other boundary condition $T(-L/|a|) = \kappa$. This gives 
\begin{align} \label{eq:kappa}
\kappa = T(-L/|a|) = b_1 &\sqrt{\frac{D}{r}} \sum_{n = 0}^{+\infty} \frac{C_{2n + 1}}{(2n+1)!} \left(\sqrt{\frac{r}{D}} L \right)^{2n+1} (-1/|a|^{2n+1} - 1) \\
- &\frac{1}{r} \sum_{n = 1}^{+\infty} \frac{C_{2n}}{(2n)!} \left(\sqrt{\frac{r}{D}} L \right)^{2n} (1/|a|^{2n} - 1) \;,
\end{align}
where the coefficients $C_n$'s are the same as in Eq. \eqref{eq:def-Cn}. The unknown constant $b_1$ is then determined by solving Eq.~\eqref{eq:kappa}. To express this solution in a more compact form, it is convenient to introduce two functions
\begin{equation} \label{eq:def-feven-fodd}
f_{\rm even}(x) = \sum_{n = 1}^{+\infty} \frac{C_{2n}}{(2n)!} x^{2n} \quad \mbox{~~and~~} \quad f_{\rm odd}(x) = \sum_{n = 0}^{+\infty} \frac{C_{2n+1}}{(2n+1)!} x^{2n+1} \;.
\end{equation}
Then solving Eq. (\ref{eq:kappa}) for $b_1$ we get
\begin{equation} \label{eq:c1-neg-a}
b_1 = -\sqrt{\frac{r}{D}}\, \left[  \frac{r \kappa + f_{\rm even}\left(\sqrt{\frac{r}{D}} \frac{L}{|a|}\right) - f_{\rm even}\left(\sqrt{\frac{r}{D}} L\right)}{  r f_{\rm odd}\left( \sqrt{\frac{r}{D}} \frac{L}{|a|} \right) + r f_{\rm odd}\left( \sqrt{\frac{r}{D}} L \right) } \right]  \;.
\end{equation}

\begin{figure}[t]
    \centering
    \includegraphics[width = 0.5\textwidth]{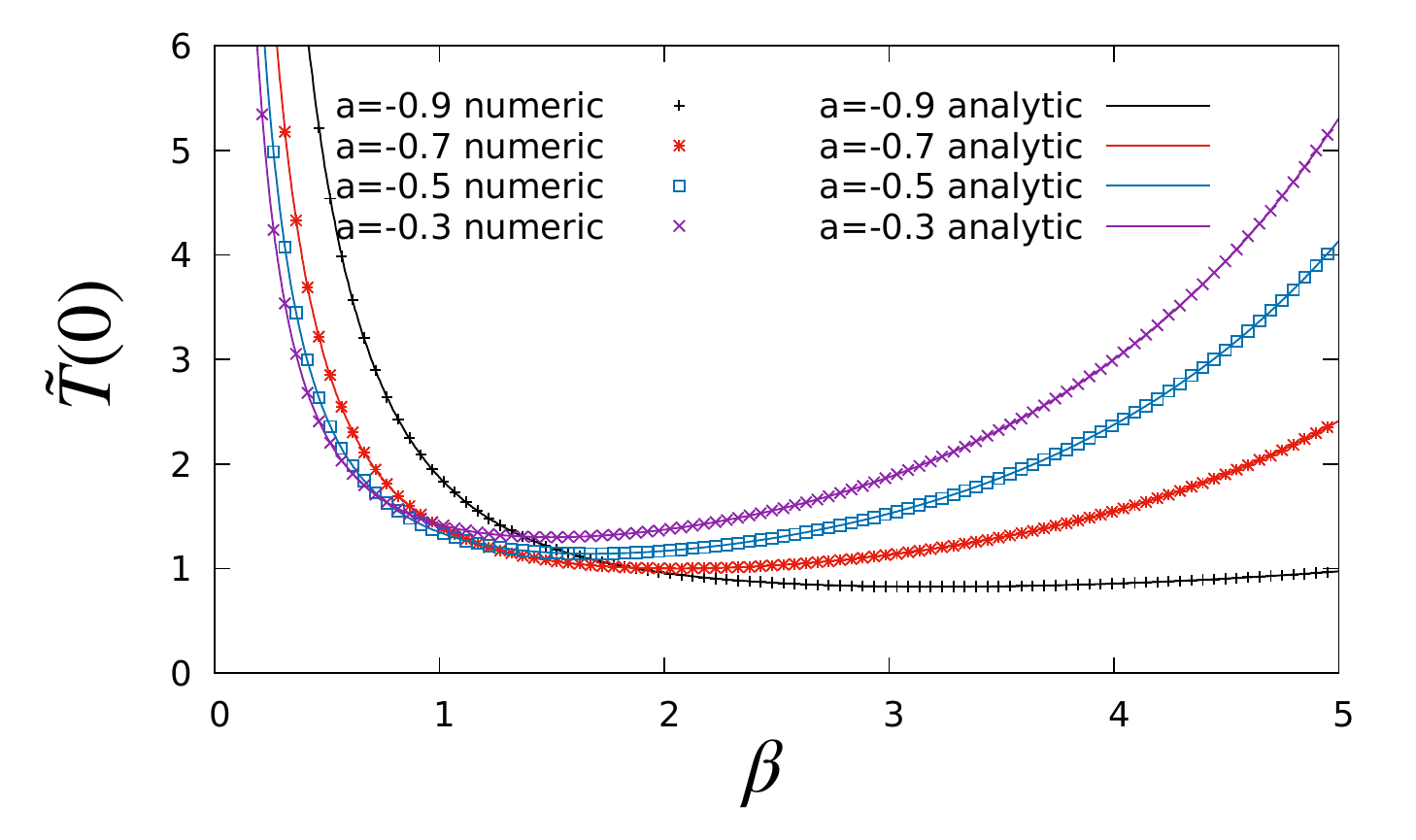}
    \caption{The dimensionless {\color{blue}MFPT} $\tilde T(0)$ as a function of $\beta = \sqrt{r/D} L$ for different values of $-1< a \leq 0$. The analytical result in Eq. (\ref{eq:MFPT-neg-a-origin-dimensionless}) (with $\tilde T(-L/|a|)$ taken as an input from numerical simulations) is in excellent agreement with numerical simulations for $\tilde T(0)$. For a fixed $a$, the \MFPT $\tilde T(0)$, as a function of $\beta$, has a minimum at $\beta = \beta^*(a)$.}\label{Ttilde_aneg}
\end{figure}
Substituting this expression for $b_1$ in Eq. (\ref{eq:MFPT-aneg}) we obtain the full \MFPT $T(x)$ in segment I (i.e., for $x \in [-L/|a|,L]$) for the case $-1 < a \leq 0$ as
\begin{align}
T(x) = \frac{f_{\rm even}\left(\sqrt{\frac{r}{D}} L\right) - f_{\rm even}\left(\sqrt{\frac{r}{D}} \frac{L}{|a|}\right) - r \kappa }{ r f_{\rm odd}\left( \sqrt{\frac{r}{D}} \frac{L}{|a|} \right) + r f_{\rm odd}\left( \sqrt{\frac{r}{D}} L \right) } \left[ f_{\rm odd}\left(\sqrt{\frac{r}{D}} x\right) - f_{\rm odd}\left(\sqrt{\frac{r}{D}} L\right) \right] \nonumber\\
- \frac{1}{r} \left[ f_{\rm even}\left( \sqrt{\frac{r}{D}} x \right) - f_{\rm even}\left( \sqrt{\frac{r}{D}} L \right) \right] \;. \label{eq:MFPT-neg-a}
\end{align}
Setting $x = 0$ we obtain the \MFPT for a walker starting from the origin to reach a target at $L$ 
\begin{align} 
T(0) =  &\frac{1}{r} \frac{1}{f_{\rm odd}\left( \sqrt{\frac{r}{D}} \frac{L}{|a|} \right) + f_{\rm odd}\left( \sqrt{\frac{r}{D}} L \right) } \nonumber\\
&\quad \times \Bigg[ f_{\rm even}\left( \sqrt{\frac{r}{D}} L \right) f_{\rm odd}\left( \sqrt{\frac{r}{D}} \frac{L}{|a|} \right) \nonumber\\
&\qquad\qquad+ f_{\rm odd}\left( \sqrt{\frac{r}{D}} L \right) \left( r \kappa + f_{\rm even}\left(\sqrt{\frac{r}{D}} \frac{L}{|a|} \right) \right) \Bigg] \;.\label{eq:MFPT-neg-a-origin}
\end{align}
As before, it is convenient to define the dimensionless \MFPT $\tilde T(x) = D\,T(x)/L^2$, and in particular, the unknown boundary value
\begin{equation} \label{TtildeL}
\tilde{T}(-L/|a|)  = \frac{D\,\kappa}{L^2} \;.
\end{equation}
The dimensionless \MFPT $\tilde{T}(0) = D T(0)/L^2$ can then be expressed as a function of two dimensionless parameters $-1 < a \leq 0$ and $\beta = L \sqrt{r/D}$ and the unknown boundary value $\tilde{T}(-L/|a|)$ in Eq. (\ref{TtildeL}), leading to
\begin{equation} \label{eq:MFPT-neg-a-origin-dimensionless}
\tilde{T}(0) = \frac{1}{\beta^2}\frac{f_{\rm even}(\beta) f_{\rm odd}(\beta/|a|) + f_{\rm odd}(\beta) \left( \beta^2 \tilde{T}(-L/|a|) + f_{\rm even}(\beta/|a|) \right)}{\left[f_{\rm odd}\left(\beta/|a|\right) + f_{\rm odd}(\beta)\right]} \;.
\end{equation}

Thus, the only unknown factor in the exact solution in Eq. (\ref{eq:MFPT-neg-a-origin-dimensionless}) is the single number $\tilde{T}(-L/|a|)  = D\,\kappa/L^2$. As explained in the beginning of this subsection, there is no simple way to determine this unknown boundary value $\kappa$ without solving for $T(x)$ on the full line, which unfortunately is rather cumbersome. Hence, our strategy is to use the numerical value of $\tilde T(-L/|a|)$ obtained from simulations and then compare the analytical solution for $\tilde T(0)$ in Eq.~(\ref{eq:MFPT-neg-a-origin-dimensionless}) with the numerical answer for $\tilde T(0)$, for different values of the parameters $\ell$ and $a$. This is reported in Fig. \ref{Ttilde_aneg}, where we see a perfect agreement between the analytical $\tilde T(0)$ (with $\kappa$ as a numerical input) and the value of $\tilde T(0)$ obtained from simulations for different $\ell$ and $-1 < a \leq 0$.  
\begin{figure}[t]
\includegraphics[width = 0.9\textwidth]{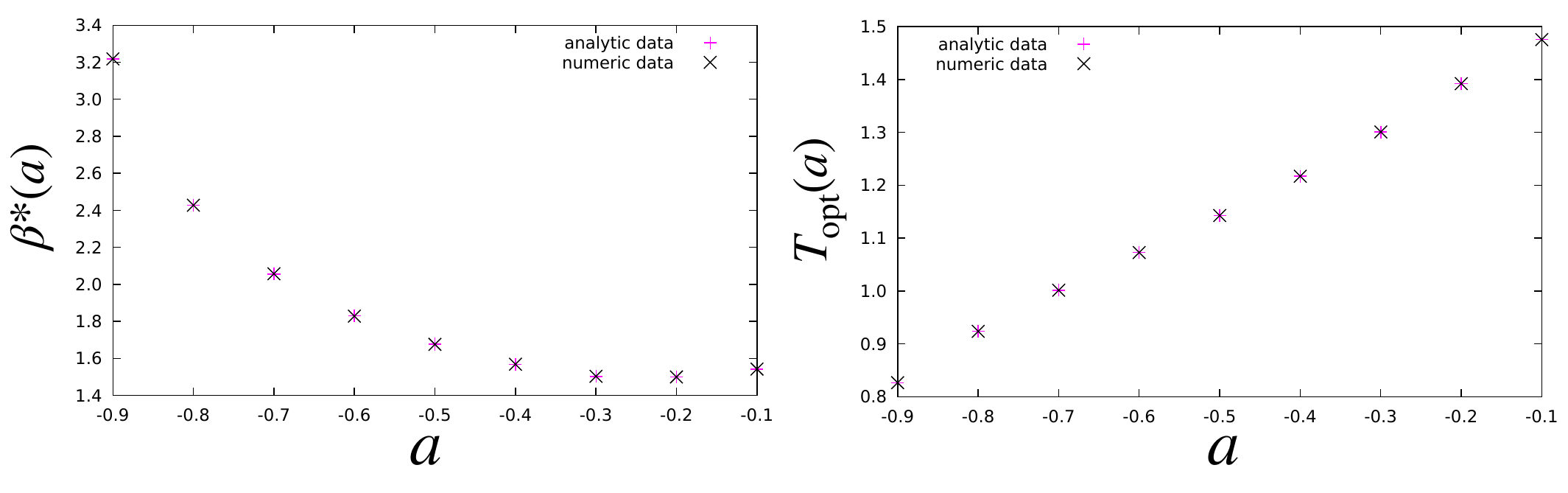}
\caption{{\bf Left:} The optimal value $\ell^*(a)$, at which $\tilde T(0)$ achieves its minimum as a function of $\ell$ for a fixed $a$, is shown for few values of $a$, for $-1 < a \leq 0$. {\bf Right:} The optimal \MFPT $T_{\rm opt}(a)$, i.e., $\tilde T(0)$ evaluated at $\ell = \ell^*(a)$, for the same values of $-1 \leq a < 1$, as in the left panel. Unlike for $0 \leq a < 1$ in Fig. \ref{Figbetastarapos}, we do not show a continuous theoretical curve of $\ell^*(a)$ vs $a$ because, for $-1<a\leq 0$, for each value of $a$, we need to take $\kappa$ as an input from numerical simulations and this is done only for few select values of $a$. The right panel shows that the optimal \MFPT $T_{\rm opt}(a)$, as a function of $a$, decreases compared to $a=0$. Thus a negative value of $a$ clearly reduces the optimal \MFPT.}\label{Figbetastaraneg}
\end{figure}

\vspace{0.2cm}

Having fixed this only unknown $\kappa$ from the simulations, we have access to the complete analytical formula for $\tilde T(0)$ as a function of $\ell$, for different values of $-1<a\leq 0$. As seen in Fig. \ref{Ttilde_aneg}, \MFPT $\tilde T(0)$ exhibits a minimum at $\ell = \ell^*(a)$. This optimal value $\ell^*(a)$ is plotted as a function of $-1 < a \leq 0$ in the left panel of Fig. \ref{Figbetastaraneg}. Finally, the optimized \MFPT $T_{\rm opt}(a)$ (that is, $\tilde T(0)$ evaluated at $\ell = \ell^*(a)$) is plotted as a function of $-1<a \leq 0$ in the right panel of Fig. \ref{Figbetastaraneg}. Contrary to the case $0\leq a < 1$, we see that for $-1< a \leq 0$, the optimal \MFPT decreases from its value in $a=0$ as $a$ decreases. This indicates that a negative $-1<a\leq 0$ actually accelerates the search process compared to the $a=0$ case, i.e. the standard resetting to the origin. 

\vspace{0.2cm}

As in the case of $0 \leq a < 1$, one can perform a small $a$ expansion of $\tilde T(0)$ in Eq. (\ref{eq:MFPT-neg-a-origin-dimensionless}) which, fortunately, does not require knowledge of the unknown $\kappa$. In fact, in the limit $a \to 0$, the term $\ell^2 \tilde T(-L/|a|) \ll f_{\rm even}(\ell/|a|)$ is on the right-hand side of the Eq. (\ref{eq:MFPT-neg-a-origin-dimensionless}). This is due to the fact that $f_{\rm even}(\ell/|a|) \sim e^{\ell/|a|}$ is as $a \to 0$ -- this follows from Eq. (\ref{eq:def-feven-fodd}) since $C_{2n} \to {1}$ as $a \to 0$. In contrast, the term $\ell^2 \tilde T(-L/|a|)$ cannot grow faster than $L^2/|a|^2$ as $a \to 0$ (this follows from the boundary condition at $x \to - \infty$ discussed earlier). Consequently, neglecting $\ell^2 \tilde T(-L/|a|)$ on the right-hand side of Eq. (\ref{eq:MFPT-neg-a-origin-dimensionless}) and taking the $a \to 0$ limit in the functions $f_{\rm even}$ and $f_{\rm odd}$, we arrive at 
\begin{equation} \label{Tsmallaneg}
\tilde{T}(0) = \frac{1}{\ell^2} \left[e^\ell - 1 - |a| \ell\right] \mbox{~~when~~} |a| \ll 1 \;.
\end{equation}
In fact, as in the $0 \leq a < 1$ case, this result for small $|a|$ can also be derived directly from the differential equation (\ref{eq:MFPT-ODE}), see the Appendix of Ref. \cite{BFHMS24} for details. Indeed, the result in Eq. (\ref{Tsmallaneg}) is identical to Eq. (\ref{eq:MFPT-pos-a-a-asymptotics}) with $a$ replaced by $-|a|$ and hence this small $a$ expansion for the whole range $-1 < a < 1$ reads the same as (\ref{eq:MFPT-pos-a-a-asymptotics}). Consequently, the analysis presented in (\ref{betastar_small_a}) and (\ref{Topt_small_a}) also holds for a small negative $a$.


\chapter{Conclusion} \label{ch:conclusion}
Extreme events —though rare— can shape the fate of physical, biological, and socio-economic systems. From natural disasters to market crashes, from reaction thresholds in chemical systems to failure points in engineered structures, the ability to predict and quantify rare fluctuations is of fundamental importance. While classical {\color{blue}extreme-value} statistics results are well-established for uncorrelated or weakly correlated systems, strongly correlated stochastic processes remain far less understood, despite their prevalence in natural systems. For strongly correlated processes, no standard tool exists to approach the description of their {\color{blue}extreme-value} {\color{blue}statistics}. In this thesis, we have {\color{blue}achieved} significant process in that direction, by identifying and universally characterizing a family of systems: conditionally independent identically distributed processes. We have shown that these {\color{blue} mathematical entities} arise naturally in physical systems and allow for their exact characterization where analytical progress would have otherwise been incredibly difficult. To study these systems we considered a wide range of physical observables, we looked at {\color{blue}extreme-value} statistics, order statistics, gap statistics and the full counting statistics. 

\vspace{0.2cm}

We focused on idealized models that can be solved exactly. However, we also showed that the universal result derived from these models could be applied to physically realistic, experimentally {\color{blue}realizable}, systems. The models we considered may be idealized however they tractability and {\color{blue}expressiveness} makes them powerful tools to describe strongly correlated out-of-equilibrium statistical systems, hence why we believe they could be useful in a variety of {\color{blue}disciplines}, not only in physics.

\vspace{0.2cm}

After recalling some known results from probability theory in Chapter \ref{ch:extreme}, in Chapter \ref{ch:ciid} we introduced the notion of conditionally independent random variables which underlies a large portion of the models studied in this thesis. For any set of random variables whose joint distribution can be written under the form
\begin{equation} \label{eq:conclusion-ciid}
    {\rm Prob.}[X_1 = x_1, \cdots, X_N = x_N] = \int \dd^M \vec{y} \; h(\vec{y}) \prod_{i = 1}^N p(x_i | \vec{y}) \;,
\end{equation}
we characterized the universal behavior of their extreme-value {\color{blue}statistics}, order statistics, gap statistics and full counting statistics in the large $N$ limit. For example, for conditional distributions $p(\cdot | \vec{y})$ belonging to the Gumbel or Weibull class the distribution of the $k$-th maximum $M_{k,N}$ is given by 
\begin{equation}
    {\rm Prob.}[M_{k, N} = w] \underset{N \to +\infty}{\longrightarrow} \int \dd^M \vec{y} \; h(\vec{y}) \delta[w - q(k/N, \vec{y})] \;,
\end{equation}
where $q(k/N, \vec{y})$ is the $k/N$-th quantile of the conditional distribution 
\begin{equation}
    \frac{k}{N} = \int_{q(k/N, \vec{y})}^{+\infty} \dd x \; p(x | \vec{y}) \;.
\end{equation}
While we have fully characterized and understood the above {\color{blue}mentioned} observables for these conditionally independent variables there remain several interesting open questions. We considered only the $N \to \infty$ limit, another possibility would be taking the number of {\color{blue}conditioning} variables $M$ to infinity, either with $N$ fixed or with the ratio $N/M$ fixed, which would probably lead to entirely new physics. Another direction, of great practical importance, would be to outline what are the necessary and sufficient conditions for this conditional structure to appear. 

\vspace{0.2cm}

In Chapter~\ref{ch:sim-reset} we presented a whole variety of stochastic systems were simultaneous resetting was used to create the above {\color{blue}mentioned} conditionally independent structure and strong long-range correlations. We looked at a simultaneously resetting gas of Brownian motions, Lévy flights, or ballistic particles. These allowed us to showcase the usefulness of the conditionally independent formalism presented in Chapter \ref{ch:ciid}. At the end of Chapter \ref{ch:sim-reset}, we looked at a gas of one-dimensional particles which simultaneously reset whenever any of them reach a certain target. {\color{blue}Although} not as apparent, the conditional independent structure is present in this model as well. 

\vspace{0.2cm}

In Chapter \ref{ch:ou-switch}, we introduce an experimental protocol that has been used to model resetting and which has preliminarily being used to recover the theoretical predictions we made in Chapter \ref{ch:ciid}. Instead of considering instantaneous resetting to a fixed position, we studied a gas of particles confined in a harmonic trap whose stiffness alternates between a wide (i.e. very free) and tight (i.e. very confined) value. Although not obvious at all a priori, conditional independence emerges in this process, allowing us to re-use the formalism introduced in Chapter \ref{ch:ciid}. Furthermore, an alternative derivation through Kesten variables, allowed us to give a physical significance to the conditionally independent structure. Namely, the fraction of time spent in one potential or the other is the physical variable containing all the information which correlates the particles. 

\vspace{0.2cm}

In Chapter \ref{ch:dyson} we moved away from this conditionally independent formalism and entered the realm of random matrix theory. We studied a gas of diffusing particles with pairwise logarithmic repulsion, on top of which, we will apply simultaneous resetting to induce strong long-range correlations. However, unlike the models studied in Chapter \ref{ch:sim-reset}, the underlying reset-free dynamics are already correlating the particles. Hence, we are not able to resort to the conditionally independent formalism from Chapter \ref{ch:ciid}. This process is a rare example of a stochastic process with competing (attractive and repulsive) strong long-range correlations, which can nevertheless be studied in great analytical detail. We studied the {\color{blue}extreme-value} statistics, the gaps and the full counting statistics, which were all completely modified from their usual known behaviors in random matrix theory. 

\vspace{0.2cm}

In Chapter \ref{ch:search} and \ref{ch:rescaling} we turned our attention to search processes. In Chapter \ref{ch:search}, we studied how the correlations induced from simultaneous resetting can affect the {\color{blue}first-passage} properties. We saw that resetting is not always optimal in a multi-searcher paradigm and that the resetting-induced correlations influence this transition from optimal to suboptimal. Finally, in Chapter \ref{ch:rescaling} we introduced a variant of stochastic resetting. Instead of resetting to the origin, the position of a diffusing particle is rescaled by a constant factor $-1 < a < 1$. We saw that for $-1 < a < 0$ we can indeed obtain a lower mean {\color{blue}first-passage} time than the usual stochastically resetting Brownian motion model.

\vspace{0.2cm}

With this work we have only begun to scratch the surface of the results that conditionally independent random variables can provide. As {\color{blue}mentioned} previously, other asymptotic limits have not been considered (such as the number of {\color{blue}conditioning} variables going to infinity). The spurious appearance of this from in the switching \OU model also begs the question of when can we guess that such a form will appear? and how many latent variables are required to decorrelate the system? All these questions and their application to physical systems are interesting avenues to explore for future research.

\bibliography{ref} 

\begin{thebibliography}{100}

\bibitem{BLMS24}
Marco Biroli, Hern{\'a}n Larralde, Satya~N Majumdar, and Gr{\'e}gory Schehr.
\newblock Exact extreme, order, and sum statistics in a class of strongly
  correlated systems.
\newblock {\em Physical Review E}, 109(1):014101, 2024.

\bibitem{BLMS23}
Marco Biroli, Hernan Larralde, Satya~N Majumdar, and Gr{\'e}gory Schehr.
\newblock Extreme statistics and spacing distribution in a brownian gas
  correlated by resetting.
\newblock {\em Physical Review Letters}, 130(20):207101, 2023.

\bibitem{BKMS24}
Marco Biroli, Manas Kulkarni, Satya~N Majumdar, and Gr{\'e}gory Schehr.
\newblock Dynamically emergent correlations between particles in a switching
  harmonic trap.
\newblock {\em Physical Review E}, 109(3):L032106, 2024.

\bibitem{BMS25}
Marco Biroli, Satya~N. Majumdar, and Gr\'egory Schehr.
\newblock Resetting dyson brownian motion.
\newblock {\em Phys. Rev. E}, 112:014101, Jul 2025.

\bibitem{BMS23}
Marco Biroli, Satya~N Majumdar, and Gr{\'e}gory Schehr.
\newblock Critical number of walkers for diffusive search processes with
  resetting.
\newblock {\em Physical Review E}, 107(6):064141, 2023.

\bibitem{BFHMS24}
Marco Biroli, Yannick Feld, Alexander~K Hartmann, Satya~N Majumdar, and
  Gr{\'e}gory Schehr.
\newblock Resetting by rescaling: Exact results for a diffusing particle in one
  dimension.
\newblock {\em Physical Review E}, 110(4):044142, 2024.

\bibitem{BCKMPS25}
Marco Biroli, Sergio Ciliberto, Manas Kulkarni, Satya~N. Majumdar, Artyom
  Petrosyan, and Gregory Schehr.
\newblock Experimental evidence for strong emergent correlations between
  particles in a switching trap, 2025.

\bibitem{K07}
Mehran Kardar.
\newblock {\em Statistical physics of particles}.
\newblock Cambridge University Press, 2007.

\bibitem{M91}
Madan~Lal Mehta.
\newblock {\em Random matrices}, volume 142.
\newblock Elsevier, 2004.

\bibitem{F10}
Peter~J Forrester.
\newblock {\em Log-gases and random matrices (LMS-34)}.
\newblock Princeton university press, 2010.

\bibitem{LNV18}
Giacomo Livan, Marcel Novaes, and Pierpaolo Vivo.
\newblock Introduction to random matrices theory and practice.
\newblock {\em Monograph Award}, 63(54):914, 2018.

\bibitem{Fi10}
Hans Fischer.
\newblock The central limit theorem from laplace to cauchy: changes in
  stochastic objectives and in analytical methods.
\newblock In {\em A history of the central limit theorem: from classical to
  modern probability theory}, pages 17--74. Springer, 2010.

\bibitem{B10}
Benoit~B Mandelbrot and Richard~L Hudson.
\newblock {\em The (mis) behaviour of markets: a fractal view of risk, ruin and
  reward}.
\newblock Profile books, 2010.

\bibitem{G58}
Emil~Julius Gumbel.
\newblock {\em Statistics of extremes}.
\newblock Columbia university press, 1958.

\bibitem{KPN02}
Richard~W Katz, Marc~B Parlange, and Philippe Naveau.
\newblock Statistics of extremes in hydrology.
\newblock {\em Advances in water resources}, 25(8-12):1287--1304, 2002.

\bibitem{KM00}
PL~Krapivsky and Satya~N Majumdar.
\newblock Traveling waves, front selection, and exact nontrivial exponents in a
  random fragmentation problem.
\newblock {\em Physical review letters}, 85(26):5492, 2000.

\bibitem{MK00}
Satya~N Majumdar and PL~Krapivsky.
\newblock Extremal paths on a random cayley tree.
\newblock {\em Physical Review E}, 62(6):7735, 2000.

\bibitem{MK02}
Satya~N Majumdar and Paul~L. Krapivsky.
\newblock Extreme value statistics and traveling fronts: Application to
  computer science.
\newblock {\em Physical Review E}, 65(3):036127, 2002.

\bibitem{MK03}
Satya~N Majumdar and PL~Krapivsky.
\newblock Extreme value statistics and traveling fronts: various applications.
\newblock {\em Physica A: Statistical Mechanics and its Applications},
  318(1-2):161--170, 2003.

\bibitem{MDK05}
Satya~N Majumdar, David~S Dean, and Paul~L Krapivsky.
\newblock Understanding search trees via statistical physics.
\newblock {\em Pramana}, 64:1175--1189, 2005.

\bibitem{MB08}
Satya~N Majumdar and Jean-Philippe Bouchaud.
\newblock Optimal time to sell a stock in the black--scholes model: comment on
  ‘thou shalt buy and hold’, by a. shiryaev, z. xu and xy zhou.
\newblock {\em Quantitative Finance}, 8(8):753--760, 2008.

\bibitem{EKM97}
Paul Embrechts, Claudia Kl{\"u}ppelberg, and Thomas Mikosch.
\newblock {\em Modelling extremal events: for insurance and finance},
  volume~33.
\newblock Springer Science \& Business Media, 2013.

\bibitem{FC15}
Jean-Yves Fortin and Maxime Clusel.
\newblock Applications of extreme value statistics in physics.
\newblock {\em Journal of Physics A: Mathematical and Theoretical},
  48(18):183001, 2015.

\bibitem{D81}
Bernard Derrida.
\newblock Random-energy model: An exactly solvable model of disordered systems.
\newblock {\em Physical Review B}, 24(5):2613, 1981.

\bibitem{BM97}
Jean-Philippe Bouchaud and Marc M{\'e}zard.
\newblock Universality classes for extreme-value statistics.
\newblock {\em Journal of Physics A: Mathematical and General}, 30(23):7997,
  1997.

\bibitem{BBP07}
Giulio Biroli, Jean-Philippe Bouchaud, and Marc Potters.
\newblock Extreme value problems in random matrix theory and other disordered
  systems.
\newblock {\em Journal of Statistical Mechanics: Theory and Experiment},
  2007(07):P07019, 2007.

\bibitem{W88}
I~Weissman.
\newblock 2. a survey of results on extremes of independent non-identically
  distributed random variables.
\newblock {\em Advances in Applied Probability}, 20(1):8--8, 1988.

\bibitem{A78}
CW~Anderson.
\newblock Super-slowly varying functions in extreme value theory.
\newblock {\em Journal of the Royal Statistical Society: Series B
  (Methodological)}, 40(2):197--202, 1978.

\bibitem{A84}
CW~Anderson.
\newblock Large deviations of extremes.
\newblock In {\em Statistical Extremes and Applications}, pages 325--340.
  Springer, 1984.

\bibitem{DR84}
Richard Davis and Sidney Resnick.
\newblock Tail estimates motivated by extreme value theory.
\newblock {\em The Annals of Statistics}, pages 1467--1487, 1984.

\bibitem{SW87}
Richard~L Smith and Ishay Weissman.
\newblock Large deviations of tail estimators based on the pareto
  approximation.
\newblock {\em Journal of applied probability}, 24(3):619--630, 1987.

\bibitem{MPS20}
Satya~N Majumdar, Arnab Pal, and Gr{\'e}gory Schehr.
\newblock Extreme value statistics of correlated random variables: a
  pedagogical review.
\newblock {\em Physics Reports}, 840:1--32, 2020.

\bibitem{RCPS01}
Subhadip Raychaudhuri, Michael Cranston, Corry Przybyla, and Yonathan Shapir.
\newblock Maximal height scaling of kinetically growing surfaces.
\newblock {\em Physical review letters}, 87(13):136101, 2001.

\bibitem{GHPR03}
G~Gy{\"o}rgyi, PCW Holdsworth, B~Portelli, and Z~R{\'a}cz.
\newblock Statistics of extremal intensities for gaussian interfaces.
\newblock {\em Physical Review E}, 68(5):056116, 2003.

\bibitem{MC04}
Satya~N Majumdar and Alain Comtet.
\newblock Exact maximal height distribution of fluctuating interfaces.
\newblock {\em Physical review letters}, 92(22):225501, 2004.

\bibitem{MC05}
Satya~N Majumdar and Alain Comtet.
\newblock Airy distribution function: from the area under a brownian excursion
  to the maximal height of fluctuating interfaces.
\newblock {\em Journal of statistical physics}, 119:777--826, 2005.

\bibitem{TW94}
Craig~A Tracy and Harold Widom.
\newblock Level-spacing distributions and the airy kernel.
\newblock {\em Communications in Mathematical Physics}, 159:151--174, 1994.

\bibitem{TW96}
Craig~A Tracy and Harold Widom.
\newblock On orthogonal and symplectic matrix ensembles.
\newblock {\em Communications in Mathematical Physics}, 177:727--754, 1996.

\bibitem{DM06}
David~S Dean and Satya~N Majumdar.
\newblock Large deviations of extreme eigenvalues of random matrices.
\newblock {\em Physical review letters}, 97(16):160201, 2006.

\bibitem{DM08}
David~S Dean and Satya~N Majumdar.
\newblock Extreme value statistics of eigenvalues of gaussian random matrices.
\newblock {\em Physical Review E—Statistical, Nonlinear, and Soft Matter
  Physics}, 77(4):041108, 2008.

\bibitem{MV09}
Satya~N Majumdar and Massimo Vergassola.
\newblock Large deviations of the maximum eigenvalue for wishart and gaussian
  random matrices.
\newblock {\em Physical review letters}, 102(6):060601, 2009.

\bibitem{MS14}
Satya~N Majumdar and Gr{\'e}gory Schehr.
\newblock Top eigenvalue of a random matrix: large deviations and third order
  phase transition.
\newblock {\em Journal of Statistical Mechanics: Theory and Experiment},
  2014(1):P01012, 2014.

\bibitem{DG86}
Bernard Derrida and E~Gardner.
\newblock Solution of the generalised random energy model.
\newblock {\em Journal of Physics C: Solid State Physics}, 19(13):2253, 1986.

\bibitem{GM64}
George~L Gerstein and Benoit Mandelbrot.
\newblock Random walk models for the spike activity of a single neuron.
\newblock {\em Biophysical journal}, 4(1):41--68, 1964.

\bibitem{SG13}
Laura Sacerdote and Maria~Teresa Giraudo.
\newblock Stochastic integrate and fire models: a review on mathematical
  methods and their applications.
\newblock {\em Stochastic biomathematical models: with applications to neuronal
  modeling}, pages 99--148, 2013.

\bibitem{T88}
Henry~Clavering Tuckwell.
\newblock {\em Introduction to theoretical neurobiology: linear cable theory
  and dendritic structure}, volume~1.
\newblock Cambridge University Press, 1988.

\bibitem{MRO14}
Ralf Metzler, Sidney Redner, and Gleb Oshanin.
\newblock {\em First-passage phenomena and their applications}, volume~35.
\newblock World Scientific, 2014.

\bibitem{VK92}
Nicolaas~Godfried Van~Kampen.
\newblock {\em Stochastic processes in physics and chemistry}, volume~1.
\newblock Elsevier, 1992.

\bibitem{B12}
William~J Bell.
\newblock {\em Searching behaviour: the behavioural ecology of finding
  resources}.
\newblock Springer Science \& Business Media, 2012.

\bibitem{AD68}
G~Adam and M~Delbr{\"u}ck.
\newblock Reduction of dimensionality in biological diffusion processes.
\newblock {\em Structural chemistry and molecular biology}, 198:198--215, 1968.

\bibitem{BC09}
Frederic Bartumeus and Jordi Catalan.
\newblock Optimal search behavior and classic foraging theory.
\newblock {\em Journal of Physics A: Mathematical and Theoretical},
  42(43):434002, 2009.

\bibitem{VDLRS11}
Gandhimohan~M Viswanathan, Marcos~GE Da~Luz, Ernesto~P Raposo, and H~Eugene
  Stanley.
\newblock {\em The physics of foraging: an introduction to random searches and
  biological encounters}.
\newblock Cambridge University Press, 2011.

\bibitem{BWH81}
Otto~G Berg, Robert~B Winter, and Peter~H Von~Hippel.
\newblock Diffusion-driven mechanisms of protein translocation on nucleic
  acids. 1. models and theory.
\newblock {\em Biochemistry}, 20(24):6929--6948, 1981.

\bibitem{CBM04}
Mathieu Coppey, O~B{\'e}nichou, R~Voituriez, and M~Moreau.
\newblock Kinetics of target site localization of a protein on dna: a
  stochastic approach.
\newblock {\em Biophysical journal}, 87(3):1640--1649, 2004.

\bibitem{GMKC18}
Soumendu Ghosh, Bhavya Mishra, Anatoly~B Kolomeisky, and Debashish Chowdhury.
\newblock First-passage processes on a filamentous track in a dense traffic:
  optimizing diffusive search for a target in crowding conditions.
\newblock {\em Journal of Statistical Mechanics: Theory and Experiment},
  2018(12):123209, 2018.

\bibitem{C19}
Debashish Chowdhury.
\newblock Laying tracks for poison delivery to “kiss of death”: Search for
  immune synapse by microtubules.
\newblock {\em Biophysical Journal}, 116(11):2057--2059, 2019.

\bibitem{B1900}
Louis Bachelier.
\newblock Th{\'e}orie de la sp{\'e}culation.
\newblock In {\em Annales scientifiques de l'{\'E}cole normale sup{\'e}rieure},
  volume~17, pages 21--86, 1900.

\bibitem{E56}
Albert Einstein.
\newblock {\em Investigations on the Theory of the Brownian Movement}.
\newblock Courier Corporation, 1956.

\bibitem{S06}
Marian By~Smoluchowski.
\newblock On the kinetic theory of brownian molecular motion and suspensions.
\newblock {\em Annals of Physics}, 326(14):756--780, 1906.

\bibitem{W51}
Eugene~P Wigner.
\newblock On the statistical distribution of the widths and spacings of nuclear
  resonance levels.
\newblock In {\em Mathematical Proceedings of the Cambridge Philosophical
  Society}, volume~47, pages 790--798. Cambridge University Press, 1951.

\bibitem{TBFDS21}
Xhek Turkeshi, Alberto Biella, Rosario Fazio, Marcello Dalmonte, and Marco
  Schir{\'o}.
\newblock Measurement-induced entanglement transitions in the quantum ising
  chain: From infinite to zero clicks.
\newblock {\em Physical Review B}, 103(22):224210, 2021.

\bibitem{EMS20}
Martin~R Evans, Satya~N Majumdar, and Gr{\'e}gory Schehr.
\newblock Stochastic resetting and applications.
\newblock {\em Journal of Physics A: Mathematical and Theoretical},
  53(19):193001, 2020.

\bibitem{EM11PRL}
Martin~R Evans and Satya~N Majumdar.
\newblock Diffusion with stochastic resetting.
\newblock {\em Physical review letters}, 106(16):160601, 2011.

\bibitem{EM11JPhysA}
Martin~R Evans and Satya~N Majumdar.
\newblock Diffusion with optimal resetting.
\newblock {\em Journal of Physics A: Mathematical and Theoretical},
  44(43):435001, 2011.

\bibitem{MC16}
Vicen{\c{c}} M{\'e}ndez and Daniel Campos.
\newblock Characterization of stationary states in random walks with stochastic
  resetting.
\newblock {\em Physical Review E}, 93(2):022106, 2016.

\bibitem{EM16}
Stephan Eule and Jakob~J Metzger.
\newblock Non-equilibrium steady states of stochastic processes with
  intermittent resetting.
\newblock {\em New Journal of Physics}, 18(3):033006, 2016.

\bibitem{HP10}
Malte Henkel and Michel Pleimling.
\newblock {\em Non-Equilibrium Phase Transitions: Volume 2: Ageing and
  Dynamical Scaling Far from Equilibrium}.
\newblock Springer Science \& Business Media, 2011.

\bibitem{HHL08}
Malte Henkel, Haye Hinrichsen, and Mette Lübeck.
\newblock {\em Non-equilibrium phase transitions: vol. 1: Absorbing Phase
  Transitions}.
\newblock Springer, 2008.

\bibitem{KMSS14}
Lukasz Kusmierz, Satya~N Majumdar, Sanjib Sabhapandit, and Gr{\'e}gory Schehr.
\newblock First order transition for the optimal search time of l{\'e}vy
  flights with resetting.
\newblock {\em Physical review letters}, 113(22):220602, 2014.

\bibitem{KG15}
{\L}ukasz Ku{\'s}mierz and Ewa Gudowska-Nowak.
\newblock Optimal first-arrival times in l{\'e}vy flights with resetting.
\newblock {\em Physical Review E}, 92(5):052127, 2015.

\bibitem{CM15}
Daniel Campos and Vicen{\c{c}} M{\'e}ndez.
\newblock Phase transitions in optimal search times: How random walkers should
  combine resetting and flight scales.
\newblock {\em Physical Review E}, 92(6):062115, 2015.

\bibitem{MSbook}
Satya~N Majumdar and Gregory Schehr.
\newblock {\em Statistics of Extremes and Records in Random Sequences}.
\newblock Oxford University Press, 2024.

\bibitem{ASD}
Alexander Kramida, Yuri Ralchenko, Joseph Reader, et~al.
\newblock Nist atomic spectra database (ver. 5.3), 2015.

\bibitem{MS24}
Satya~N Majumdar and Gregory Schehr.
\newblock {\em Statistics of Extremes and Records in Random Sequences}.
\newblock Oxford University Press, 2024.

\bibitem{ABN92}
Barry~C Arnold, Narayanaswamy Balakrishnan, and Haikady~Navada Nagaraja.
\newblock {\em A first course in order statistics}.
\newblock SIAM, 2008.

\bibitem{ND03}
Herbert~A David and Haikady~N Nagaraja.
\newblock {\em Order statistics}.
\newblock John Wiley \& Sons, 2004.

\bibitem{SM14}
Gr{\'e}gory Schehr and Satya~N Majumdar.
\newblock Exact record and order statistics of random walks via first-passage
  ideas.
\newblock In {\em First-passage phenomena and their applications}, pages
  226--251. World Scientific, 2014.

\bibitem{G43}
Boris Gnedenko.
\newblock Sur la distribution limite du terme maximum d'une serie aleatoire.
\newblock {\em Annals of mathematics}, 44(3):423--453, 1943.

\bibitem{F71}
William Feller et~al.
\newblock An introduction to probability theory and its applications.
\newblock 1971.

\bibitem{F91}
William Feller.
\newblock {\em An introduction to probability theory and its applications,
  Volume 2}, volume~2.
\newblock John Wiley \& Sons, 1991.

\bibitem{V15}
Pierpaolo Vivo.
\newblock Large deviations of the maximum of independent and identically
  distributed random variables.
\newblock {\em European Journal of Physics}, 36(5):055037, 2015.

\bibitem{T24}
Hugo Touchette.
\newblock Large deviations in statistical physics, 2024.

\bibitem{V24}
Pierpaolo Vivo.
\newblock Large deviations of spectral radius and “rightmost” particle for
  random matrices/charged fluids with logarithmic repulsion, 2024.

\bibitem{E24}
Martin~R Evans and John~C Sunil.
\newblock Stochastic resetting and large deviations.
\newblock {\em arXiv preprint arXiv:2412.16374}, 2024.

\bibitem{B64}
Simeon~M Berman.
\newblock Limit theorems for the maximum term in stationary sequences.
\newblock {\em The Annals of Mathematical Statistics}, pages 502--516, 1964.

\bibitem{R01}
Sidney Redner.
\newblock {\em A guide to first-passage processes}.
\newblock Cambridge university press, 2001.

\bibitem{M10}
Satya~N Majumdar.
\newblock Universal first-passage properties of discrete-time random walks and
  l{\'e}vy flights on a line: Statistics of the global maximum and records.
\newblock {\em Physica A: Statistical Mechanics and its Applications},
  389(20):4299--4316, 2010.

\bibitem{BMS13}
Alan~J Bray, Satya~N Majumdar, and Gr{\'e}gory Schehr.
\newblock Persistence and first-passage properties in nonequilibrium systems.
\newblock {\em Advances in Physics}, 62(3):225--361, 2013.

\bibitem{D85}
Bernard Derrida.
\newblock A generalization of the random energy model which includes
  correlations between energies.
\newblock {\em Journal de Physique Lettres}, 46(9):401--407, 1985.

\bibitem{DM01}
DS~Dean and Satya~N Majumdar.
\newblock Extreme-value statistics of hierarchically correlated variables
  deviation from gumbel statistics and anomalous persistence.
\newblock {\em Physical Review E}, 64(4):046121, 2001.

\bibitem{BC06}
Eric Bertin and Maxime Clusel.
\newblock Generalized extreme value statistics and sum of correlated variables.
\newblock {\em Journal of Physics A: Mathematical and General}, 39(24):7607,
  2006.

\bibitem{SM06}
Gr{\'e}gory Schehr and Satya~N Majumdar.
\newblock Universal asymptotic statistics of maximal relative height in
  one-dimensional solid-on-solid models.
\newblock {\em Physical Review E—Statistical, Nonlinear, and Soft Matter
  Physics}, 73(5):056103, 2006.

\bibitem{BC03}
Christian Beck and Ezechiel~GD Cohen.
\newblock Superstatistics.
\newblock {\em Physica A: Statistical mechanics and its applications},
  322:267--275, 2003.

\bibitem{BCS05}
Christian Beck, Ezechiel~GD Cohen, and Harry~L Swinney.
\newblock From time series to superstatistics.
\newblock {\em Physical Review E—Statistical, Nonlinear, and Soft Matter
  Physics}, 72(5):056133, 2005.

\bibitem{AM05}
AY~Abul-Magd.
\newblock Random matrix theory within superstatistics.
\newblock {\em Physical Review E—Statistical, Nonlinear, and Soft Matter
  Physics}, 72(6):066114, 2005.

\bibitem{BCP08}
O~Bohigas, JX~de~Carvalho, and Mauricio~Porto Pato.
\newblock Disordered ensembles of random matrices.
\newblock {\em Physical Review E—Statistical, Nonlinear, and Soft Matter
  Physics}, 77(1):011122, 2008.

\bibitem{AAV09}
AY~Abul-Magd, Gernot Akemann, and P~Vivo.
\newblock Superstatistical generalizations of wishart--laguerre ensembles of
  random matrices.
\newblock {\em Journal of Physics A: Mathematical and Theoretical},
  42(17):175207, 2009.

\bibitem{B09}
Christian Beck.
\newblock Recent developments in superstatistics.
\newblock {\em Brazilian Journal of Physics}, 39:357--363, 2009.

\bibitem{RB14}
Pau Rabassa and Christian Beck.
\newblock Extreme value laws for superstatistics.
\newblock {\em Entropy}, 16(10):5523--5536, 2014.

\bibitem{HW04}
Paulina Hetman and Karina Weron.
\newblock Extreme-value approach to the tsallis' superstatistics.
\newblock {\em ACTA PHYSICA POLONICA SERIES B}, 35(4):1375--1386, 2004.

\bibitem{DGKJS16}
Mateusz Denys, Tomasz Gubiec, Ryszard Kutner, Maciej Jagielski, and H~Eugene
  Stanley.
\newblock Universality of market superstatistics.
\newblock {\em Physical Review E}, 94(4):042305, 2016.

\bibitem{CS14}
Mykyta~V Chubynsky and Gary~W Slater.
\newblock Diffusing diffusivity: a model for anomalous, yet brownian,
  diffusion.
\newblock {\em Physical review letters}, 113(9):098302, 2014.

\bibitem{CSMS16}
Aleksei~V Chechkin, Flavio Seno, Ralf Metzler, and Igor~M Sokolov.
\newblock Brownian yet non-gaussian diffusion: from superstatistics to
  subordination of diffusing diffusivities.
\newblock {\em Physical Review X}, 7(2):021002, 2017.

\bibitem{PKR22}
Arnab Pal, Sarah Kostinski, and Shlomi Reuveni.
\newblock The inspection paradox in stochastic resetting.
\newblock {\em Journal of Physics A: Mathematical and Theoretical},
  55(2):021001, 2022.

\bibitem{GJ22}
Shamik Gupta and Arun~M Jayannavar.
\newblock Stochastic resetting: A (very) brief review.
\newblock {\em Frontiers in Physics}, 10:789097, 2022.

\bibitem{TPSRR20}
Ofir Tal-Friedman, Arnab Pal, Amandeep Sekhon, Shlomi Reuveni, and Yael
  Roichman.
\newblock Experimental realization of diffusion with stochastic resetting.
\newblock {\em The journal of physical chemistry letters}, 11(17):7350--7355,
  2020.

\bibitem{BBPMC20}
Benjamin Besga, Alfred Bovon, Artyom Petrosyan, Satya~N. Majumdar, and Sergio
  Ciliberto.
\newblock Optimal mean first-passage time for a brownian searcher subjected to
  resetting: Experimental and theoretical results.
\newblock {\em Phys. Rev. Res.}, 2:032029, Jul 2020.

\bibitem{FBPCM21}
Felix Faisant, Benjamin Besga, Artyom Petrosyan, Sergio Ciliberto, and Satya~N
  Majumdar.
\newblock Optimal mean first-passage time of a brownian searcher with resetting
  in one and two dimensions: experiments, theory and numerical tests.
\newblock {\em Journal of Statistical Mechanics: Theory and Experiment},
  2021(11):113203, 2021.

\bibitem{RUK14}
Shlomi Reuveni, Michael Urbakh, and Joseph Klafter.
\newblock The role of substrate unbinding in michaelis-menten enzymatic
  reactions.
\newblock {\em Biophysical Journal}, 106(2):677a, 2014.

\bibitem{BSS14}
Denis Boyer and Citlali Solis-Salas.
\newblock Random walks with preferential relocations to places visited in the
  past and their application to biology.
\newblock {\em Physical review letters}, 112(24):240601, 2014.

\bibitem{RRU15}
Tal Rotbart, Shlomi Reuveni, and Michael Urbakh.
\newblock Michaelis-menten reaction scheme as a unified approach towards the
  optimal restart problem.
\newblock {\em Physical Review E}, 92(6):060101, 2015.

\bibitem{MSS15}
Satya~N. Majumdar, Sanjib Sabhapandit, and Gr\'egory Schehr.
\newblock Dynamical transition in the temporal relaxation of stochastic
  processes under resetting.
\newblock {\em Phys. Rev. E}, 91:052131, May 2015.

\bibitem{PKE16}
Arnab Pal, Anupam Kundu, and Martin~R Evans.
\newblock Diffusion under time-dependent resetting.
\newblock {\em Journal of Physics A: Mathematical and Theoretical},
  49(22):225001, 2016.

\bibitem{R16}
Shlomi Reuveni.
\newblock Optimal stochastic restart renders fluctuations in first passage
  times universal.
\newblock {\em Physical review letters}, 116(17):170601, 2016.

\bibitem{MV16}
Miquel Montero and Javier Villarroel.
\newblock Directed random walk with random restarts: The sisyphus random walk.
\newblock {\em Physical Review E}, 94(3):032132, 2016.

\bibitem{NG16}
Apoorva Nagar and Shamik Gupta.
\newblock Diffusion with stochastic resetting at power-law times.
\newblock {\em Physical Review E}, 93(6):060102, 2016.

\bibitem{PR17}
Arnab Pal and Shlomi Reuveni.
\newblock First passage under restart.
\newblock {\em Physical review letters}, 118(3):030603, 2017.

\bibitem{BEM17}
Denis Boyer, Martin~R Evans, and Satya~N Majumdar.
\newblock Long time scaling behaviour for diffusion with resetting and memory.
\newblock {\em Journal of Statistical Mechanics: Theory and Experiment},
  2017(2):023208, feb 2017.

\bibitem{EM18}
Martin~R Evans and Satya~N Majumdar.
\newblock Run and tumble particle under resetting: a renewal approach.
\newblock {\em Journal of Physics A: Mathematical and Theoretical},
  51(47):475003, 2018.

\bibitem{CS18}
Aleksei Chechkin and Igor~M Sokolov.
\newblock Random search with resetting: a unified renewal approach.
\newblock {\em Physical review letters}, 121(5):050601, 2018.

\bibitem{MVB18}
Gabriel Mercado-V{\'a}squez and Denis Boyer.
\newblock Lotka--volterra systems with stochastic resetting.
\newblock {\em Journal of Physics A: Mathematical and Theoretical},
  51(40):405601, 2018.

\bibitem{PKR19}
Arnab Pal, {\L}ukasz Ku{\'s}mierz, and Shlomi Reuveni.
\newblock Time-dependent density of diffusion with stochastic resetting is
  invariant to return speed.
\newblock {\em Physical Review E}, 100(4):040101, 2019.

\bibitem{MCM19}
Axel Mas{\'o}-Puigdellosas, Daniel Campos, and Vicen{\c{c}} M{\'e}ndez.
\newblock Transport properties and first-arrival statistics of random motion
  with stochastic reset times.
\newblock {\em Physical Review E}, 99(1):012141, 2019.

\bibitem{PKR20}
Arnab Pal, {\L}ukasz Ku{\'s}mierz, and Shlomi Reuveni.
\newblock Search with home returns provides advantage under high uncertainty.
\newblock {\em Physical Review Research}, 2(4):043174, 2020.

\bibitem{BS20}
Anna~S Bodrova and Igor~M Sokolov.
\newblock Resetting processes with noninstantaneous return.
\newblock {\em Physical Review E}, 101(5):052130, 2020.

\bibitem{BS20b}
Anna~S Bodrova and Igor~M Sokolov.
\newblock Brownian motion under noninstantaneous resetting in higher
  dimensions.
\newblock {\em Physical Review E}, 102(3):032129, 2020.

\bibitem{BRR20}
B~De~Bruyne, Julien Randon-Furling, and S~Redner.
\newblock Optimization in first-passage resetting.
\newblock {\em Physical Review Letters}, 125(5):050602, 2020.

\bibitem{B20}
Paul~C Bressloff.
\newblock Diffusive search for a stochastically-gated target with resetting.
\newblock {\em Journal of Physics A: Mathematical and Theoretical},
  53(42):425001, 2020.

\bibitem{P20}
Ross~G Pinsky.
\newblock Diffusive search with spatially dependent resetting.
\newblock {\em Stochastic Processes and their Applications}, 130(5):2954--2973,
  2020.

\bibitem{BMS22}
Benjamin De~Bruyne, Satya~N Majumdar, and Gr{\'e}gory Schehr.
\newblock Optimal resetting brownian bridges via enhanced fluctuations.
\newblock {\em Physical Review Letters}, 128(20):200603, 2022.

\bibitem{MMS22}
Francesco Mori, Satya~N Majumdar, and Gr{\'e}gory Schehr.
\newblock Time to reach the maximum for a stationary stochastic process.
\newblock {\em Physical Review E}, 106(5):054110, 2022.

\bibitem{SBS22}
Ion Santra, Urna Basu, and Sanjib Sabhapandit.
\newblock Effect of stochastic resetting on brownian motion with stochastic
  diffusion coefficient.
\newblock {\em Journal of Physics A: Mathematical and Theoretical},
  55(41):414002, 2022.

\bibitem{BM23}
Benjamin De~Bruyne and Francesco Mori.
\newblock Resetting in stochastic optimal control.
\newblock {\em Physical Review Research}, 5(1):013122, 2023.

\bibitem{MOK23}
Francesco Mori, Kristian~St{\o}levik Olsen, and Supriya Krishnamurthy.
\newblock Entropy production of resetting processes.
\newblock {\em Physical Review Research}, 5(2):023103, 2023.

\bibitem{ER24}
Martin~R Evans and Somrita Ray.
\newblock Stochastic resetting prevails over sharp restart for broad target
  distributions.
\newblock {\em arXiv preprint arXiv:2410.01941}, 2024.

\bibitem{L61}
Andrew Lenard.
\newblock Exact statistical mechanics of a one-dimensional system with coulomb
  forces.
\newblock {\em Journal of Mathematical Physics}, 2(5):682--693, 1961.

\bibitem{P62}
Stephen Prager.
\newblock The one-dimensional plasma.
\newblock {\em Advances in chemical physics}, 4:201--224, 1962.

\bibitem{B63}
Rodney~James Baxter.
\newblock Statistical mechanics of a one-dimensional coulomb system with a
  uniform charge background.
\newblock In {\em Mathematical Proceedings of the Cambridge Philosophical
  Society}, volume~59, pages 779--787. Cambridge University Press, 1963.

\bibitem{DKMSS17}
Abhishek Dhar, Anupam Kundu, Satya~N Majumdar, Sanjib Sabhapandit, and
  Gr{\'e}gory Schehr.
\newblock Exact extremal statistics in the classical 1d coulomb gas.
\newblock {\em Physical review letters}, 119(6):060601, 2017.

\bibitem{VAM22}
Ohad Vilk, Michael Assaf, and Baruch Meerson.
\newblock Fluctuations and first-passage properties of systems of brownian
  particles with reset.
\newblock {\em Physical Review E}, 106(2):024117, 2022.

\bibitem{GR14}
Izrail~Solomonovich Gradshteyn and Iosif~Moiseevich Ryzhik.
\newblock {\em Table of integrals, series, and products}.
\newblock Academic press, 2014.

\bibitem{BG90}
Jean-Philippe Bouchaud and Antoine Georges.
\newblock Anomalous diffusion in disordered media: statistical mechanisms,
  models and physical applications.
\newblock {\em Physics reports}, 195(4-5):127--293, 1990.

\bibitem{HL98}
Pascal H{\'e}braud and Fran{\c{c}}ois Lequeux.
\newblock Mode-coupling theory for the pasty rheology of soft glassy materials.
\newblock {\em Physical review letters}, 81(14):2934, 1998.

\bibitem{MSM18}
B~Mukherjee, K~Sengupta, and Satya~N Majumdar.
\newblock Quantum dynamics with stochastic reset.
\newblock {\em Physical Review B}, 98(10):104309, 2018.

\bibitem{PCML21}
Gabriele Perfetto, Federico Carollo, Matteo Magoni, and Igor Lesanovsky.
\newblock Designing nonequilibrium states of quantum matter through stochastic
  resetting.
\newblock {\em Physical Review B}, 104(18):L180302, 2021.

\bibitem{SVH23}
Francisco~J Sevilla and Andrea Vald{\'e}s-Hern{\'a}ndez.
\newblock Dynamics of closed quantum systems under stochastic resetting.
\newblock {\em Journal of Physics A: Mathematical and Theoretical},
  56(3):034001, 2023.

\bibitem{KM23}
Manas Kulkarni and Satya~N Majumdar.
\newblock Generating entanglement by quantum resetting.
\newblock {\em Physical Review A}, 108(6):062210, 2023.

\bibitem{MPGOTC16}
Ignacio~A Mart{\'\i}nez, Artyom Petrosyan, David Gu{\'e}ry-Odelin, Emmanuel
  Trizac, and Sergio Ciliberto.
\newblock Engineered swift equilibration of a brownian particle.
\newblock {\em Nature physics}, 12(9):843--846, 2016.

\bibitem{GPP20}
Deepak Gupta, Carlos~A Plata, and Arnab Pal.
\newblock Work fluctuations and jarzynski equality in stochastic resetting.
\newblock {\em Physical review letters}, 124(11):110608, 2020.

\bibitem{GORKTMGM19}
David Gu{\'e}ry-Odelin, Andreas Ruschhaupt, Anthony Kiely, Erik Torrontegui,
  Sofia Mart{\'\i}nez-Garaot, and Juan~Gonzalo Muga.
\newblock Shortcuts to adiabaticity: Concepts, methods, and applications.
\newblock {\em Reviews of Modern Physics}, 91(4):045001, 2019.

\bibitem{PGOTP19}
Carlos~A Plata, David Gu{\'e}ry-Odelin, E~Trizac, and A~Prados.
\newblock Optimal work in a harmonic trap with bounded stiffness.
\newblock {\em Physical Review E}, 99(1):012140, 2019.

\bibitem{CBGOTPC18}
Marie Chupeau, Benjamin Besga, David Gu{\'e}ry-Odelin, Emmanuel Trizac, Artyom
  Petrosyan, and Sergio Ciliberto.
\newblock Thermal bath engineering for swift equilibration.
\newblock {\em Physical Review E}, 98(1):010104, 2018.

\bibitem{MBMS20}
Gabriel Mercado-V{\'a}squez, Denis Boyer, Satya~N Majumdar, and Gr{\'e}gory
  Schehr.
\newblock Intermittent resetting potentials.
\newblock {\em Journal of Statistical Mechanics: Theory and Experiment},
  2020(11):113203, 2020.

\bibitem{GPKP20}
Deepak Gupta, Carlos~A Plata, Anupam Kundu, and Arnab Pal.
\newblock Stochastic resetting with stochastic returns using external trap.
\newblock {\em Journal of Physics A: Mathematical and Theoretical},
  54(2):025003, 2020.

\bibitem{SDN21}
Ion Santra, Santanu Das, and Sujit~Kumar Nath.
\newblock Brownian motion under intermittent harmonic potentials.
\newblock {\em Journal of Physics A: Mathematical and Theoretical},
  54(33):334001, 2021.

\bibitem{XZMD22}
Pengbo Xu, Tian Zhou, Ralf Metzler, and Weihua Deng.
\newblock Stochastic harmonic trapping of a l{\'e}vy walk: transport and
  first-passage dynamics under soft resetting strategies.
\newblock {\em New Journal of Physics}, 24(3):033003, 2022.

\bibitem{GP22}
Deepak Gupta and Carlos~A Plata.
\newblock Work fluctuations for diffusion dynamics submitted to stochastic
  return.
\newblock {\em New Journal of Physics}, 24(11):113034, 2022.

\bibitem{ACB22}
Henry Alston, Luca Cocconi, and Thibault Bertrand.
\newblock Non-equilibrium thermodynamics of diffusion in fluctuating
  potentials.
\newblock {\em Journal of Physics A: Mathematical and Theoretical},
  55(27):274004, 2022.

\bibitem{MBM22}
Gabriel Mercado-V{\'a}squez, Denis Boyer, and Satya~N Majumdar.
\newblock Reducing mean first passage times with intermittent confining
  potentials: a realization of resetting processes.
\newblock {\em Journal of Statistical Mechanics: Theory and Experiment},
  2022(9):093202, 2022.

\bibitem{UO30}
George~E Uhlenbeck and Leonard~S Ornstein.
\newblock On the theory of the brownian motion.
\newblock {\em Physical review}, 36(5):823, 1930.

\bibitem{NIST}
Frank~WJ Olver.
\newblock {\em NIST handbook of mathematical functions hardback and CD-ROM}.
\newblock Cambridge university press, 2010.

\bibitem{K73}
Harry Kesten.
\newblock Random difference equations and renewal theory for products of random
  matrices.
\newblock 1973.

\bibitem{KKS75}
Harry Kesten, Mykyta~V Kozlov, and Frank Spitzer.
\newblock A limit law for random walk in a random environment.
\newblock {\em Compositio mathematica}, 30(2):145--168, 1975.

\bibitem{DH83}
Bernard Derrida and HJ715727 Hilhorst.
\newblock Singular behaviour of certain infinite products of random 2$\times$ 2
  matrices.
\newblock {\em Journal of Physics A: Mathematical and General}, 16(12):2641,
  1983.

\bibitem{KS84}
Harry Kesten and Frank Spitzer.
\newblock Convergence in distribution of products of random matrices.
\newblock {\em Zeitschrift f{\"u}r Wahrscheinlichkeitstheorie und Verwandte
  Gebiete}, 67:363--386, 1984.

\bibitem{CLNP85}
Claude de~Calan, Jean-Marc Luck, Theo~M Nieuwenhuizen, and Dimitri Petritis.
\newblock On the distribution of a random variable occurring in 1d disordered
  systems.
\newblock {\em Journal of Physics A: Mathematical and General}, 18(3):501,
  1985.

\bibitem{G91}
Charles~M Goldie.
\newblock Implicit renewal theory and tails of solutions of random equations.
\newblock {\em The Annals of Applied Probability}, pages 126--166, 1991.

\bibitem{BDMZ13}
D~Buraczewski, E~Damek, T~Mikosch, and J~Zienkiewicz.
\newblock Large deviations for solutions to stochastic recurrence equations
  under kesten’s condition.
\newblock 2013.

\bibitem{GBL21}
Tristan Gauti{\'e}, Jean-Philippe Bouchaud, and Pierre Le~Doussal.
\newblock Matrix kesten recursion, inverse-wishart ensemble and fermions in a
  morse potential.
\newblock {\em Journal of Physics A: Mathematical and Theoretical},
  54(25):255201, 2021.

\bibitem{GMS23}
Mathis Gu{\'e}neau, Satya~N Majumdar, and Gr{\'e}gory Schehr.
\newblock Active particle in a harmonic trap driven by a resetting noise: an
  approach via kesten variables.
\newblock {\em Journal of Physics A: Mathematical and Theoretical},
  56(47):475002, 2023.

\bibitem{W28}
John Wishart.
\newblock The generalised product moment distribution in samples from a normal
  multivariate population.
\newblock {\em Biometrika}, 20(1/2):32--52, 1928.

\bibitem{PB20}
Marc Potters and Jean-Philippe Bouchaud.
\newblock {\em A first course in random matrix theory: for physicists,
  engineers and data scientists}.
\newblock Cambridge University Press, 2020.

\bibitem{DN04}
Herbert~A David and Haikady~N Nagaraja.
\newblock {\em Order statistics}.
\newblock John Wiley \& Sons, 2004.

\bibitem{D62}
Freeman~J Dyson.
\newblock A brownian-motion model for the eigenvalues of a random matrix.
\newblock {\em Journal of Mathematical Physics}, 3(6):1191--1198, 1962.

\bibitem{D62b}
Freeman~J Dyson.
\newblock Statistical theory of the energy levels of complex systems. i.
\newblock {\em Journal of Mathematical Physics}, 3(1):140--156, 1962.

\bibitem{GMS21}
Jacek Grela, Satya~N Majumdar, and Gr{\'e}gory Schehr.
\newblock Non-intersecting brownian bridges in the flat-to-flat geometry.
\newblock {\em Journal of Statistical Physics}, 183(3):49, 2021.

\bibitem{RS11}
Joachim Rambeau and Gr{\'e}gory Schehr.
\newblock Distribution of the time at which n vicious walkers reach their
  maximal height.
\newblock {\em Physical Review E—Statistical, Nonlinear, and Soft Matter
  Physics}, 83(6):061146, 2011.

\bibitem{DE02}
Ioana Dumitriu and Alan Edelman.
\newblock Matrix models for beta ensembles.
\newblock {\em Journal of Mathematical Physics}, 43(11):5830--5847, 11 2002.

\bibitem{DM63}
Freeman~J Dyson and Madan~Lal Mehta.
\newblock Statistical theory of the energy levels of complex systems. iv.
\newblock {\em Journal of Mathematical Physics}, 4(5):701--712, 1963.

\bibitem{CL95}
Ovidiu Costin and Joel~L Lebowitz.
\newblock Gaussian fluctuation in random matrices.
\newblock {\em Physical Review Letters}, 75(1):69, 1995.

\bibitem{FS95}
M.~M. Fogler and B.~I. Shklovskii.
\newblock Probability of an eigenvalue number fluctuation in an interval of a
  random matrix spectrum.
\newblock {\em Phys. Rev. Lett.}, 74:3312--3315, Apr 1995.

\bibitem{MMSV14}
Ricardo Marino, Satya~N Majumdar, Gr{\'e}gory Schehr, and Pierpaolo Vivo.
\newblock Phase transitions and edge scaling of number variance in gaussian
  random matrices.
\newblock {\em Physical review letters}, 112(25):254101, 2014.

\bibitem{CLM15}
Pasquale Calabrese, Pierre Le~Doussal, and Satya~N Majumdar.
\newblock Random matrices and entanglement entropy of trapped fermi gases.
\newblock {\em Physical Review A}, 91(1):012303, 2015.

\bibitem{MMSV16}
Ricardo Marino, Satya~N Majumdar, Gr{\'e}gory Schehr, and Pierpaolo Vivo.
\newblock Number statistics for $\beta$-ensembles of random matrices:
  applications to trapped fermions at zero temperature.
\newblock {\em Physical Review E}, 94(3):032115, 2016.

\bibitem{T18}
Salvatore Torquato.
\newblock Hyperuniform states of matter.
\newblock {\em Physics Reports}, 745:1--95, 2018.

\bibitem{GVPPM23}
Gregorio Garc{\'\i}a-Valladares, Carlos~A Plata, Antonio Prados, and Alessandro
  Manacorda.
\newblock Optimal resetting strategies for search processes in heterogeneous
  environments.
\newblock {\em New Journal of Physics}, 25(11):113031, 2023.

\bibitem{S16}
Shlomi Reuveni.
\newblock Optimal stochastic restart renders fluctuations in first passage
  times universal.
\newblock {\em Physical review letters}, 116(17):170601, 2016.

\bibitem{TRR22}
Ofir Tal-Friedman, Yael Roichman, and Shlomi Reuveni.
\newblock Diffusion with partial resetting.
\newblock {\em Phys. Rev. E}, 106:054116, Nov 2022.

\bibitem{P22}
J~Kevin Pierce.
\newblock An advection-diffusion process with proportional resetting.
\newblock {\em arXiv preprint arXiv:2204.07215}, 2022.

\bibitem{BCHPM23}
Costantino Di~Bello, Aleksei~V Chechkin, Alexander~K Hartmann, Zbigniew
  Palmowski, and Ralf Metzler.
\newblock Time-dependent probability density function for partial resetting
  dynamics.
\newblock {\em New Journal of Physics}, 25(8):082002, 2023.

\bibitem{OG24}
Kristian~St{\o}levik Olsen and Deepak Gupta.
\newblock Thermodynamic work of partial resetting.
\newblock {\em Journal of Physics A: Mathematical and Theoretical},
  57(24):245001, 2024.

\bibitem{H23}
Upendra Harbola.
\newblock Stochastic walker with variable long jumps.
\newblock {\em Physical Review E}, 108(1):014135, 2023.

\bibitem{HW89}
Alistair~J Hall and GC~Wake.
\newblock A functional differential equation arising in modelling of cell
  growth.
\newblock {\em The ANZIAM Journal}, 30(4):424--435, 1989.

\bibitem{Lar04}
Hernan Larralde.
\newblock A first passage time distribution for a discrete version of the
  ornstein--uhlenbeck process.
\newblock {\em Journal of Physics A: Mathematical and General}, 37(12):3759,
  2004.

\bibitem{MK07}
Satya~N Majumdar and Michael~J Kearney.
\newblock Inelastic collapse of a ball bouncing on a randomly vibrating
  platform.
\newblock {\em Physical Review E—Statistical, Nonlinear, and Soft Matter
  Physics}, 76(3):031130, 2007.

\bibitem{CSMS17}
Aleksei~V Chechkin, Flavio Seno, Ralf Metzler, and Igor~M Sokolov.
\newblock Brownian yet non-gaussian diffusion: from superstatistics to
  subordination of diffusing diffusivities.
\newblock {\em Physical Review X}, 7(2):021002, 2017.

\bibitem{LG18}
Yann Lanoisel{\'e}e and Denis~S Grebenkov.
\newblock A model of non-gaussian diffusion in heterogeneous media.
\newblock {\em Journal of Physics A: Mathematical and Theoretical},
  51(14):145602, 2018.

\bibitem{BB20}
Eli Barkai and Stanislav Burov.
\newblock Packets of diffusing particles exhibit universal exponential tails.
\newblock {\em Physical review letters}, 124(6):060603, 2020.

\bibitem{WABG09}
Bo~Wang, Stephen~M Anthony, Sung~Chul Bae, and Steve Granick.
\newblock Anomalous yet brownian.
\newblock {\em Proceedings of the National Academy of Sciences},
  106(36):15160--15164, 2009.

\bibitem{WGLFGH19}
Patrick Witzel, Maria G{\"o}tz, Yann Lanoisel{\'e}e, Thomas Franosch, Denis~S
  Grebenkov, and Doris Heinrich.
\newblock Heterogeneities shape passive intracellular transport.
\newblock {\em Biophysical journal}, 117(2):203--213, 2019.

\end{thebibliography}


\end{document}